\pdfoutput=1
\documentclass[12pt]{thesis}
\usepackage{graphicx}

\usepackage{varioref}
\usepackage{lscape}
\usepackage{indentfirst}
\usepackage{placeins}
\usepackage{latexsym}
\usepackage{multirow}
\usepackage{tabls}
\usepackage{slashbox}
\usepackage{amsmath}
\usepackage{esint}
\usepackage{amssymb,amsmath}
\usepackage{subfigure}
\usepackage{setspace}
\usepackage{textcomp}
\usepackage{multirow}
\usepackage[pdftex,
              pdfauthor={Vlasios Vasileiou},
              pdftitle={A Blind Search for Bursts of Very High Enery Gamma Rays with Milagro},
              pdfsubject={Gamma-Ray Astrophysics},
              pdfkeywords={GRBs, Milagro, VHE},
              pdfproducer={Lyx, Kile, Latex with hyperref},
             plainpages=false,
             pdfpagelabels]{hyperref}

 \usepackage[all]{hypcap}

\providecommand{\tabularnewline}{\\}

\renewcommand{\baselinestretch}{1}
\DeclareGraphicsExtensions{.jpg,.mps,.pdf,.png}

\begin{document}
\subfigcapskip=-15pt
\subfigcapmargin=0pt
\subfigbottomskip=0pt
\subfigtopskip=0pt

\pagestyle{empty}

\clearpage 

\hbox{\ }

\renewcommand{\baselinestretch}{1}
\small \normalsize

\begin{center}
\large{{ABSTRACT}} 

\vspace{3em} 

\end{center}
\hspace{-.15in}
\begin{tabular}{ll}
Title of dissertation:    & {\large  A BLIND SEARCH FOR BURSTS OF VERY}\\
& {\large HIGH ENERGY GAMMA RAYS }\\
& {\large WITH MILAGRO} \\
\ \\
&                          {\large  Vlasios Vasileiou, Doctor of Philosophy, 2008} \\
\ \\
Dissertation directed by: & {\large  Professor Jordan A. Goodman} \\
& Department of Physics\\
\end{tabular}

\vspace{3em}

\renewcommand{\baselinestretch}{2}
\large \normalsize
Milagro is a water-Cherenkov detector that observes the extended air showers produced by cosmic gamma rays of energies $E\gtrsim100\,GeV$. The effective area of Milagro peaks at energies $E\gtrsim10\,TeV$, however it is still large even down to a few hundred $GeV$ ($\sim10\,m^2$ at $100\,GeV$). The wide field of view ($\sim2\,sr$) and high duty cycle ($>90\%$) of Milagro make it ideal for continuously monitoring the overhead sky for transient Very High Energy (VHE) emissions. This study searched the Milagro data for such emissions. Even though the search was optimized primarily for detecting the emission from Gamma-Ray Bursts (GRBs), it was still sensitive to the emission from the last stages of the evaporation of Primordial Black Holes (PBHs) or to any other kind of phenomena that produce bursts of VHE gamma rays. Measurements of the GRB spectra by satellites up to few tens of GeV showed no signs of a cutoff. Even though multiple instruments sensitive to $GeV/TeV$ gamma rays have performed observations of GRBs, there has not yet been a definitive detection of such an emission yet. One of the reasons for that is that gamma rays with energies $E\gtrsim100\,GeV$ are attenuated by interactions with the extragalactic background light or are absorbed internally at the site of the burst. There are many models that predict VHE gamma-ray emission from GRBs. A detection or a constraint of such an emission can provide useful information on the mechanism and environment of GRBs. This study performed a blind search of the Milagro data of the last five years for bursts of VHE gamma rays with durations ranging from $100\,\mu{}s$ to $316\,s$. No GRB localization was provided by an external instrument. Instead, the whole dataset was thoroughly searched in time, space, and duration. No significant events were detected. Upper limits were placed on the VHE emission from GRBs. 

\clearpage 

\thispagestyle{empty}
\hbox{\ }
\vspace{1in}
\renewcommand{\baselinestretch}{1}
\small\normalsize
\begin{center}

\large{{A BLIND SEARCH FOR BURSTS OF VERY \\
HIGH ENERGY GAMMA RAYS WITH MILAGRO}}\\
\ \\
\ \\
\large{by} \\
\ \\
\large{Vlasios Vasileiou}
\ \\
\ \\
\ \\
\ \\
\normalsize
Dissertation submitted to the Faculty of the Graduate School of the \\
University of Maryland, College Park in partial fulfillment \\
of the requirements for the degree of \\
Doctor of Philosophy \\
2008
\end{center}

\vspace{7.5em}

\noindent Advisory Committee: \\
Professor Jordan A. Goodman, Chair/Advisor \\
Professor Gregory W. Sullivan \\
Professor M. Coleman Miller \\
Dr. Andrew J. Smith\\
Dr. Julie McEnery\\
\clearpage 

\thispagestyle{empty}
\hbox{\ }

\vfill
\renewcommand{\baselinestretch}{1}
\small\normalsize

\vspace{-.65in}

\begin{center}
\large{\copyright \hbox{ }Copyright by\\
Vlasios Vasileiou  
\\
2008}
\end{center}

\vfill
%
\pagestyle{plain}
\pagenumbering{roman}
\setcounter{page}{2}

\clearpage 

\renewcommand{\baselinestretch}{2}
\small\normalsize
\hbox{\ }
 
\vspace{-.65in}

\begin{center}
\large{Preface} 
\end{center}

The primary purpose of this study is to detect the Very High Energy (VHE)\footnote{
In this work, ``very high energy'' corresponds to the $50\,GeV-100\,TeV$ energy range.}
emission from Gamma-Ray Bursts (GRBs). A simple calculation can show that an expectation of such
a detection is not unreasonable by Milagro. Let us assume that GRBs emit the same amount of isotropic
energy ($\sim10^{52}\,erg$) in the $50\,GeV-10\,TeV$ energy band as they emit in the $20\,keV-300\,keV$ band, the 
satellite detectors are sensitive to. For a GRB at a redshift $z\sim0.1$, this emission would correspond to 
an energy fluence of $\sim10^{-4}\,erg/cm^2$, approximating the absorption by the extra-galactic background light (EBL)
with a decrease of the non-absorbed fluence by $50\%$. 
If the emitted spectrum is on a power law distribution with index -2.2,
and if the absorption by the EBL can be approximated as an exponential spectral-break at an energy $\sim800\,GeV$,
then the average photon energy reaching the earth would be $\sim130\,GeV$. Therefore, the incoming energy fluence would correspond to 
a particle fluence of $\frac{10^{-4}\,erg}{130\,GeV}\,cm^{-2}\simeq5\times10^{-4}\,photons/cm^2$. The spectrally-weighted
effective area of Milagro for such a signal is $\sim10^{6}\,cm^2$. Therefore, $10^6\times5\times10^{-4}=500$ photons fro the VHE emission
of this GRB will be detected by Milagro. Let us say that the duration of the burst was $100\,s$. 
The amount of background for a simple binned search that uses a $2^o\times2^o$ square 
bin will be about $\sim400$ background events for these $100\,s$, given that the background rate of Milagro is about $1event\,deg^{-2}\,s^{1}$.
The Poisson cumulative proability of 400 background events generating a fluctuation as big as $400+500$ is $10^{-102}$ or $\sim21$ standard deviations. This probability
is more than enough to claim a detection.

The first chapter of this dissertation will
give a historical outline of GRB observations and will describe the theoretical models
explaining GRBs. 

GRB emission has been extensively observed in $MeV$ and lower energies. There are 
multiple theoretical models that predict that GRBs also emit in the $GeV/TeV$ energy range.
Observations of the GRB spectra up to tens of $GeV$ have shown no signs of a cutoff. However,
observations for their $E>50\,GeV$ emission, provide at best only hints of such an emission. 
Chapter \ref{chap:VHE} will give an overview of the theoretical models predicting such an emission
and will present the observational results that support its existence. 
One of the reasons why there has not yet been a definitive detection of the $E>100\,GeV$ emission is 
that it is strongly attenuated while it travels through the extragalactic space. Only a 
small fraction of the VHE emission reaches the earth, making it very difficult to detect. 
This important effect is described in Chapter \ref{chap:IR}. 

This search is optimized primarily for detecting the VHE emission by GRBs. However, it can also detect 
any phenomena that emit bursts of VHE gamma rays, such as the explosive last stage
of the evaporation of primordial black holes. The instrument used for this search,
the Milagro gamma-ray observatory, is not sensitive enough to detect evaporating primordial black holes at a great distance. However, the phenomenon was
exciting enough that warranted a separate chapter (chap. \ref{chap:PBH}) for its description.

The Milagro gamma-ray observatory is primarily sensitive to gamma rays of TeV energies,
but it also has some sensitivity down to $\sim40\,GeV$. Its wide field of view ($\sim2\,sr$) and high 
duty cycle ($>90\%$) make it ideal for continuously monitoring the overhead
sky for transient VHE emissions. Milagro's principle of operation and capabilities will be 
described in Chapter \ref{chap:MILAGRO}.

The Milagro data are searched in various ways for emission from GRBs. One way is to perform searches in
coincidence with GRB triggers provided by external instruments. Such searches have been
performed, and even though they resulted to no detections, they have placed upper limits on VHE emission 
on the individual externally-detected GRBs. Another way to search the Milagro data is blindly, without using
any external GRB localization. This study thoroughly searched the entire Milagro dataset in
start time, space, and duration. Chapter \ref{chap:Search} will describe the algorithm used
and compare this blind search with the other kinds of GRB searches performed on the Milagro data. 

Because of the search's blind nature and the large extent of the analyzed dataset, this search performed
a large number of trials.
When a search contains many trials, then very improbable fluctuations of the background can 
appear as real signal. In order to reduce the rate of these false positives, the requirements on 
the strength of a detected signal have to be increased. This, unfortunately, limits the 
sensitivity of the search. Chapter \ref{Chapter:Probabilities} will give an overview of the 
statistics used in this study and of the method by which false positives were avoided in this search.

This search was a simple binned search. One of the ways it was optimized was by adjusting its
bin size to the duration and the energy of the signal under search. Chapter \ref{chap:binsize} will describe
this optimization. 

After the properties of the signal under search were described, and the capabilities 
of the Milagro detector and of the search algorithm were calculated, the sensitivity of the search
could be quantified. Chapter \ref{chap:Sensitivity} will present the sensitivity of the
Milagro detector to the emission from GRBs and PBHs, and will set Milagro's prospects for 
discovering such an emission. 

The final results of the search are described in Chapter \ref{chap:RESULTS}. Unfortunately, no
bursts of VHE emission in the Milagro data were detected. A Monte Carlo simulation of the GRB
population was created, which helped to place upper limits on the VHE emission from GRBs.
Specifically, the simulation calculated the predicted number of GRBs Milagro would expect to detect
versus their VHE-emission model. By comparing the null result of the search with the prediction
of the simulation, some of the VHE-emission models were excluded. The simulation of the GRB
population and the resulting upper limits are presented in \ref{chap:GRBSim}.

\clearpage 

\renewcommand{\baselinestretch}{2}
\small\normalsize
\hbox{\ }
 
\vspace{-.65in}

\begin{center}
\vspace{4in}
\large
\textit{Dedicated to my parents.}
\end{center}
\clearpage 

\renewcommand{\baselinestretch}{2}
\small\normalsize
\hbox{\ }
 
\vspace{-.65in}

\begin{center}
\large{Acknowledgments} 
\end{center} 

\vspace{1ex}

First of all, I would like to thank my parents Dimitra and Ilias and my
grandparents Antonis and Zoi for giving me the foundations to pursue my dreams. 
Thank you for making stubborn, for making me smart, and for making me a good person. 

I would like to thank my advisor Jordan Goodman for believing in me, and for all his support
and guidance he has provided me. Jordan was always available to 
help with any kinds of problems I had. Observing him taught me how to be a better
collaborator, and a better scientist. 

I would like to give a huge thanks to Andy Smith. Andy and I spent endless hours
talking about Milagro, gamma-ray astronomy, the new macbook, and why cows don't like
corn. Andy was a patient and knowledgeable teacher, always ready to stand up, grab the
chalk, and start drawing figures and equations on the blackboard to explain
things. Working with him was one of the biggest reasons for enjoying my graduate
years working for Milagro. I owe a big part of my success to him.

I would also like to thank Brenda Dingus and Gus Sinnis for the guidance, support, and constructive
criticism they have given me. I would like to thank Brenda and Julie McEnery for
believing in me and for actively trying to help me.
Many thanks to Bob Ellsworth for his very useful
help with my PMT tests, and for all the 
knowledge he has shared with me. Thanks to Jim Linnemann for his active interest in the 
the MC simulation and for all the statistics knowledge I have learned from him.
Thank to Greg Sullivan for being my advisor during my first semesters
at Maryland. Thanks to the Milagro and HAWC collaborations for allowing me to 
present the results and the status of the experiments at national and international meetings, and at
the NSF panel review. Thanks to my undergraduate advisor in Greece, Christos Eleftheriadis for 
helping my with my first steps as a scientist. Without his help, I might had not been able to 
mention places such as CERN, UMD, Los Alamos, and NASA in my CV. 

My friends and fellow graduate students helped make my graduate life a lot easier. 
Thanks to my friend Patty and my roommate Chad, who maybe were my only social life during the first
hard years of being a graduate student in a foreign country. 
Thanks to my good friends in Greece Mpampis, Christos, Zoi, Kikh, and Foteinh who
reminded me of the good times back home, and supported me through the hard times here. 
Thanks to David Noyes for being a good friend, and for all the fun nights at Wonderland, Vegas, Los Alamos, etc. 
Thanks to my other good friend from Milagro Aous, with whom I spent a lot of fun time at random 
schools and meetings around the world (Erice, Merida, Mexico City, Los Alamos, etc). 
Greetings to Milagro grad student and office mate Buckley, with whom I enjoyed sharing the office,
and talking about that flash desktop game I forgot its name. Greetings to Milagro collaborators and friends
Curtis, Pablo, and Sabrina. Shame to Curtis for always beating me at Counterstrike.
Thanks to fellow Marylanders Helina and Dusan. Helina is a good friend, and a pleasant and funny companion.
Greetings to the other greek graduate students at UMD, and especially Christos, Kostas, and Alex.

Finally, I would like to give the biggest available thank you to Margaret. During these last $\sim$2 years, 
she was there for me spending endless amounts of time she could and could
not afford. I don't know how my emotional, mental, and physical state would have been after 300 pages of dissertation
without her. Her presence made me want to accomplish more, become a better person, and made my life beautiful.
She is the reason that I will be saying in ten years: ``Ahh, graduate school..good times...''.

%
\renewcommand{\baselinestretch}{1}
\small\normalsize
\clearpage
\tableofcontents 
\newpage
\setlength{\parskip}{1em}
\listoftables 
\newpage
\listoffigures 
\newpage
\addcontentsline{toc}{chapter}{List of Abbreviations}
 \clearpage 

\renewcommand{\baselinestretch}{1}
\small\normalsize
\hbox{\ }

\vspace{-4em}

\begin{center}
\large{List of Abbreviations}
\end{center}

\vspace{3pt}

\begin{tabular}{ll}
AS   & Air Shower \\
BATSE & Burst And Transient Source Explorer\\
CGRO & Compton Gamma-Ray Observatory\\
DAQ & Data-Acquisition System\\
EAS  & Extensive Air Shower\\
EBL  & Extragalactic Backgrould Light\\
FEB & Front-End Board\\
GCN & GRB Coordinates Network\\
GRB  & Gamma-Ray Burst\\
HA   & Hour Angle\\
HB   & Hopkins and Beacom\\
HETE-2 & High-Energy Transient Explorer\\
IACT & Imagic Atmospheric Cherenkov Telescope\\
ISM  & Interstellar Medium\\
KN04   & Kneiske \textit{et al.}\\
LST  & Local Sidereal Time\\
MU   & Muon \\
OR   & Outrigger array\\
PBH  & Primordial Black-Hole\\
PM  & Porciani and Madau\\
PMT  & Photomultiplier Tube\\
PLS  & Power-Law Segment\\
PR   & Primack \textit{et al.}\\
PSF  & Point-Spread Function\\
RA   & Right Ascension \\
SSC  & Synchrotron Self-Compton (scattering)\\
ST   & Stecker \textit{et al.}\\
TDC  & Time to Digital Converter\\
TOT  & Time Over Threshold\\
QSO  & Quasi-Stellar Object\\
SFR  & Star-Formation Rate\\
VHE  & Very-High Energy\\
VME  & Versa Module Europa
\end{tabular}

\newpage
\setlength{\parskip}{0em}
\renewcommand{\baselinestretch}{1}
\small\normalsize

\setcounter{page}{1}
 \pagenumbering{arabic}
   \clearpage \chapter{\label{chap:GRB}Gamma-Ray Bursts}

\section{Introduction}

Gamma-Ray Bursts (GRBs) are the brightest explosions in the universe. 
They are brief and bright transient emissions of $keV/MeV$ radiation, 
occurring with a rate of a few per day uniformly in the sky. Even
though they are at cosmological distances (up to Gigaparsecs or $\sim10^{28}\,cm$), 
they outshine all other sources in the gamma-ray sky, including the
sun. The energy output of a single GRB in keV/MeV gamma rays is comparable
to the emission from the sun
in all the electromagnetic spectrum, 
over $\sim10^{10}\,y$ (approximately the age of the universe), 
or to the emission of the Milky Way over few years. 

The initial (prompt) emission of GRBs is brief ($ms$ to mins), highly variable
(in time scales of $ms$ to tens of $s$), non-thermal, and observed
mostly in the keV/MeV energy range. The emission is believed to be produced
by electrons accelerated during collisionless shocks inside highly collimated relativistic
jets. For the majority of GRBs, the prompt emission is followed by
a smoothly decaying and long-lasting ``afterglow, '' observed in
longer wavelengths (from X-rays to optical), and believed to be produced by the deceleration
of the relativistic jet in the surrounding interstellar or circumburst
medium. GRBs of short durations ($\lesssim\,2s$) are believed to be produced by mergers of 
compact binaries (neutron star-neutron star or neutron star-black hole), while the
GRBs of longer durations ($\gtrsim\,2s$) are most likely produced by the collapse of the cores of
massive spinning stars. 

\begin{figure}[ht]
\begin{centering}
\includegraphics[width=0.8\columnwidth]{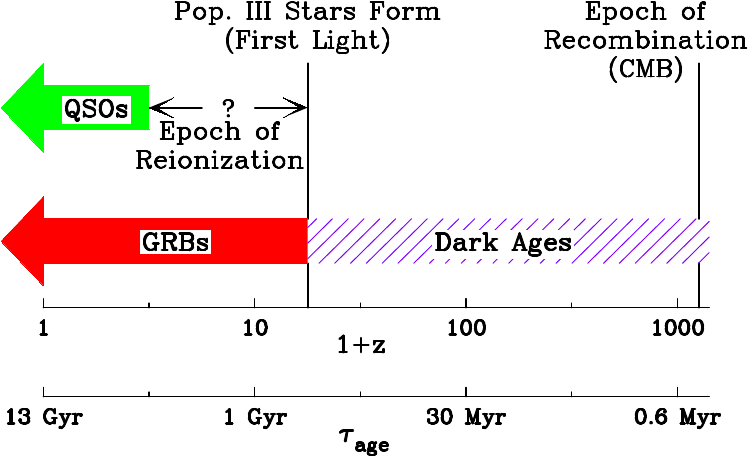}
\par\end{centering}

\caption{\label{fig:GRB_Times}Cosmological context of GRBs. Contrary to quasi-stellar objects, 
gamma-ray bursts can help us probe the properties of the universe as early in time as
the epoch of reionization. Source: \cite{GRB_Lamb_Reichart_2001}}

\end{figure}

\begin{figure}[ht]
\includegraphics[width=1\columnwidth]{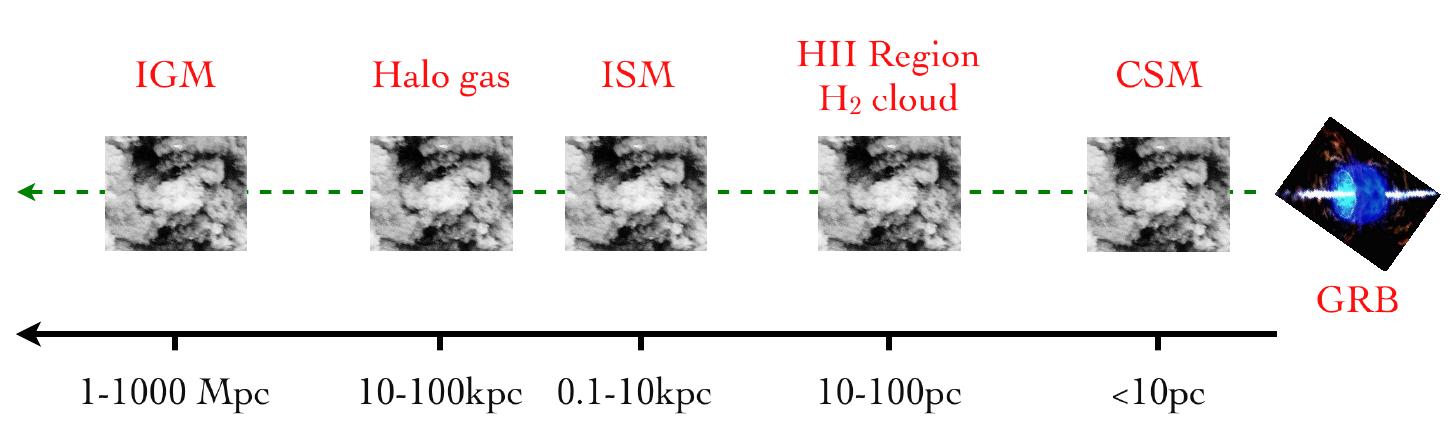}

\caption{\label{fig:GRB_Stuff}Schematic of the various environments intersected
by a GRB line of sight. The GRB emission passes through the circumstellar
medium (CSM) around the GRB progenitor; the star-forming region surrounding
the GRB (HII Region, $H_{2}$ Cloud); the ambient interstellar medium (ISM)
of the host galaxy; the baryonic halo of the galaxy (Halo gas); 
and the intergalactic medium (IGM) between the earth and the GRB.
The absorption features imprinted on the GRB signal as it passes through
these regions can reveal information about their density and composition. Source:
\cite{GRB_Prochaska_et_al2007}}
\end{figure}

High-redshift GRBs are 100 to 1000 times brighter than high-redshift
QSOs\footnote{QSO: Quasi-Stellar Object, an extremely powerful and distant active
galactic nucleus.}, and also expected to occur out to redshifts $z\simeq20$, 
while QSOs occur only out to redshift $z\sim7$ \cite{GRB_Lamb_Reichart_2001}. Therefore, GRBs are unique tools for probing
the very distant (or very young) universe as early in time as the
epoch of reionization (Fig. \ref{fig:GRB_Times}). GRBs can be used
to learn about the evolution history of the universe (star formation history, 
metallicity at different redshifts), its large-scale structure, and
the properties of the earliest generations of stars. %
The afterglow of GRBs passes through multiple regions filled with
gas or radiation fields before it reaches the earth (Fig. \ref{fig:GRB_Stuff}). 
Spectroscopic studies on the absorption features imprinted on the
GRB afterglow at each of these regions provide unique information
regarding the regions' composition and density \cite{GRB_Prochaska_et_al2007}. 
GRBs are also believed to be intense sources of neutrinos, cosmic
rays (up to ultra-high energies $10^{20}\,eV$), and gravitational waves. Observations
of these emissions can answer questions in astrophysics, 
particle physics, and general relativity. 

\begin{figure}[ht]
\begin{centering}
\includegraphics[width=0.8\columnwidth]{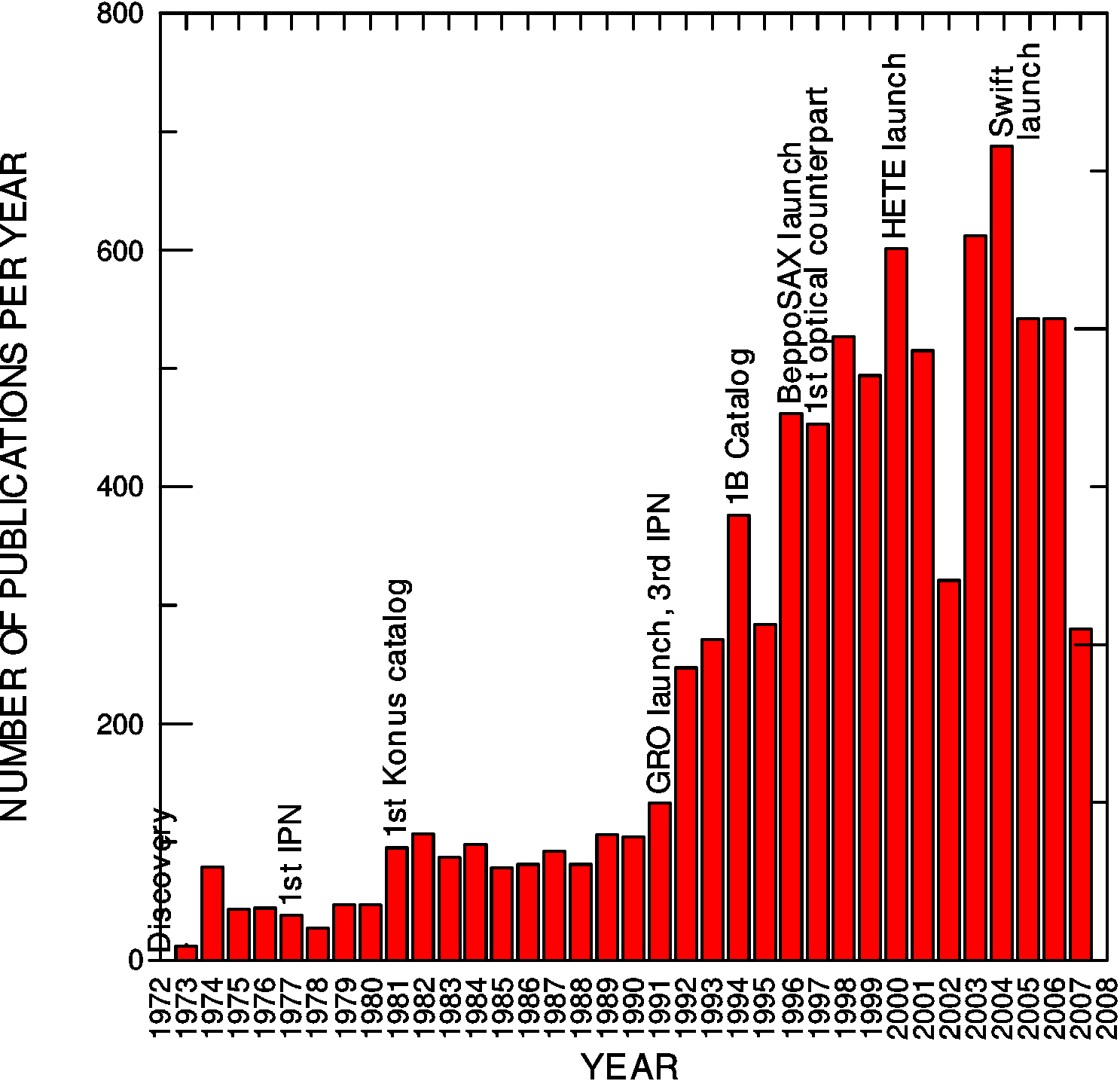}
\par\end{centering}

\caption{\label{fig:GRB_Papers}Number of GRB-related refereed publications
per year. Source: \cite{URL_hurley}}

\end{figure}

\begin{figure}[ht]
\includegraphics[angle=270, width=1\columnwidth]{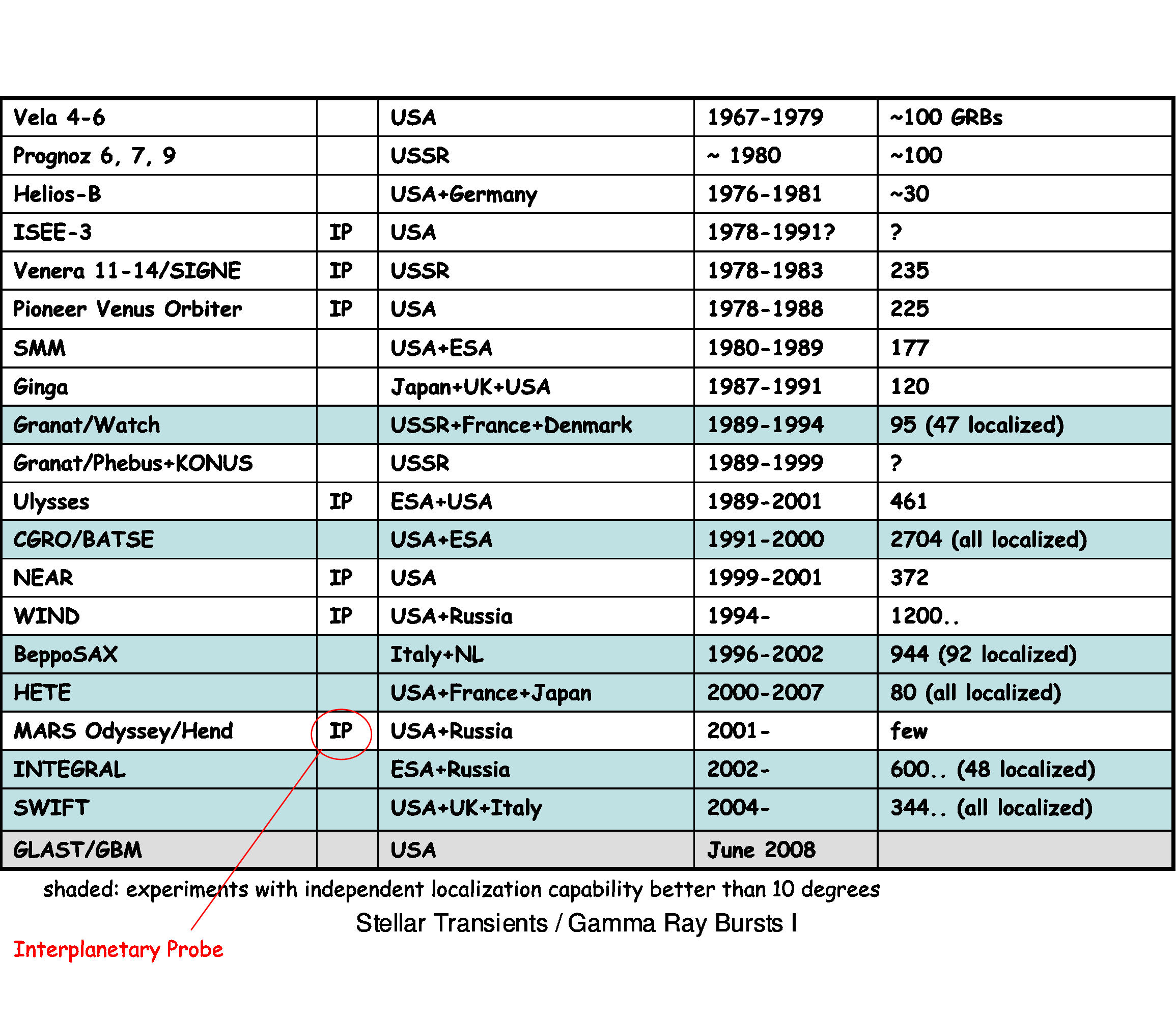}
\caption{\label{fig:GRB_Experiments}List of experiments sensitive to GRBs and approximate
numbers of detected GRBs (as of early May 2008). Source: \cite{URL_JeanZand}}
\end{figure}

The diverse and intriguing properties of GRBs make them the focus
of intense scientific research and debate (Fig. \ref{fig:GRB_Papers}), 
and the observational target of multiple instruments (Fig. \ref{fig:GRB_Experiments}). 
Despite the fact that we have known about GRBs for over thirty five years, 
and that more than eight thousand refereed papers have been written about them, 
they still continue to spark scientific interest. The commissioning
of new detectors sensitive to GRB's keV to GeV gamma-ray emission, 
such as the GBM and the LAT aboard GLAST \cite{URL_GLAST}; to their GeV/TeV
emission, such as HAWC \cite{URL_HAWC}; sensitive to their neutrino emission,
such as IceCube \cite{URL_ICECUBE}; and sensitive to their gravitational-wave emission, such as LISA \cite{URL_LISA},
will surely renew interest and raise new questions. %

This chapter will give an overview of our current knowledge regarding GRBs.
A historical overview of GRB observations will be given in section
\ref{sec:GRBs_History}, and the currently accepted
model for the progenitor and mechanism of GRBs will be presented in
section \ref{sec:GRB_Hosts}.

\section{\textmd{\label{sec:GRBs_History}}History of GRB Observations}

\subsection{The first years (1967-1991)}

Well before GRBs became publicly known, Colgate hypothesized their existence,
associating them with the ejection of
relativistic shocks from supernovae \cite{GRB_Colgate_1968}. GRBs
were accidentally discovered in 1967 by the US Vela satellites, 
operated by Los Alamos National Laboratory \cite{GRB_Dickinson_Tamarkin_1965}, 
whose purpose was to monitor from space, for violations of the nuclear-test ban
treaty. The Vela satellites carried omnidirectional detectors
sensitive to the gamma-ray pulses emitted by nuclear-weapon explosions. 
However, their detectors were also sensitive to the gamma-ray emission from GRBs. 
Soon after they were launched, they started detecting bursts of gamma rays
that were, fortunately, identified as coming from space. 
The detection of the first GRBs was immediately classified and was
not made public until seven years later \cite{GRB_Discovery}, when
sixteen GRB detections were reported in the $0.2-1.5\,MeV$ energy range. After
the Vela satellites, a number of instruments were dedicated to detecting
GRBs (Fig. \ref{fig:GRB_Experiments}). However, the number
of bursts detected was small and the angular resolution poor.

\subsection{BATSE (1991-2000)}

A breakthrough in our understanding of GRBs happened with the numerous
GRB detections from the Burst And Transient Source Explorer (BATSE) \cite{URL_BATSE}
, which flew, with other instruments, on board the Compton Gamma-Ray Observatory
 (CGRO) \cite{URL_CGRO}. BATSE operated from 1991 to 2000. 
It was sensitive to the $15\,keV-2\,MeV$ energy
range, had a wide field of view ($4\pi\,sr$ minus 30\% because of
earth obscuration), and a moderate angular resolution ($\sim4^{o}$). 
It detected 2704 GRBs \cite{GRB_Paciesas_et_al1999}, a significantly
larger number than the total number of GRBs in the pre-existing catalog
(few hundreds). In combination with the Energetic Gamma-Ray Experiment
Telescope (EGRET) \cite{URL_EGRET}, a gamma-ray detector also aboard the CGRO, GRB observations
in the extended energy range $15\,keV-30\,GeV$ were made. 

Before BATSE, the distance scale of GRBs was unknown. The
scientific community was divided among multiple theories predicting
distance scales ranging from our own galaxy to the edges of the known
universe. Even though GRBs were observed uniformly, it was
believed that they come from galactic neutron stars. It was believed that
the reason we saw a uniform-in-space distribution instead the pancake shape
of our galaxy was that the pre-BATSE instruments were not sensitive enough
to probe deep enough and see the galactic structure 

\begin{figure}[ht]
\includegraphics[width=1\columnwidth]{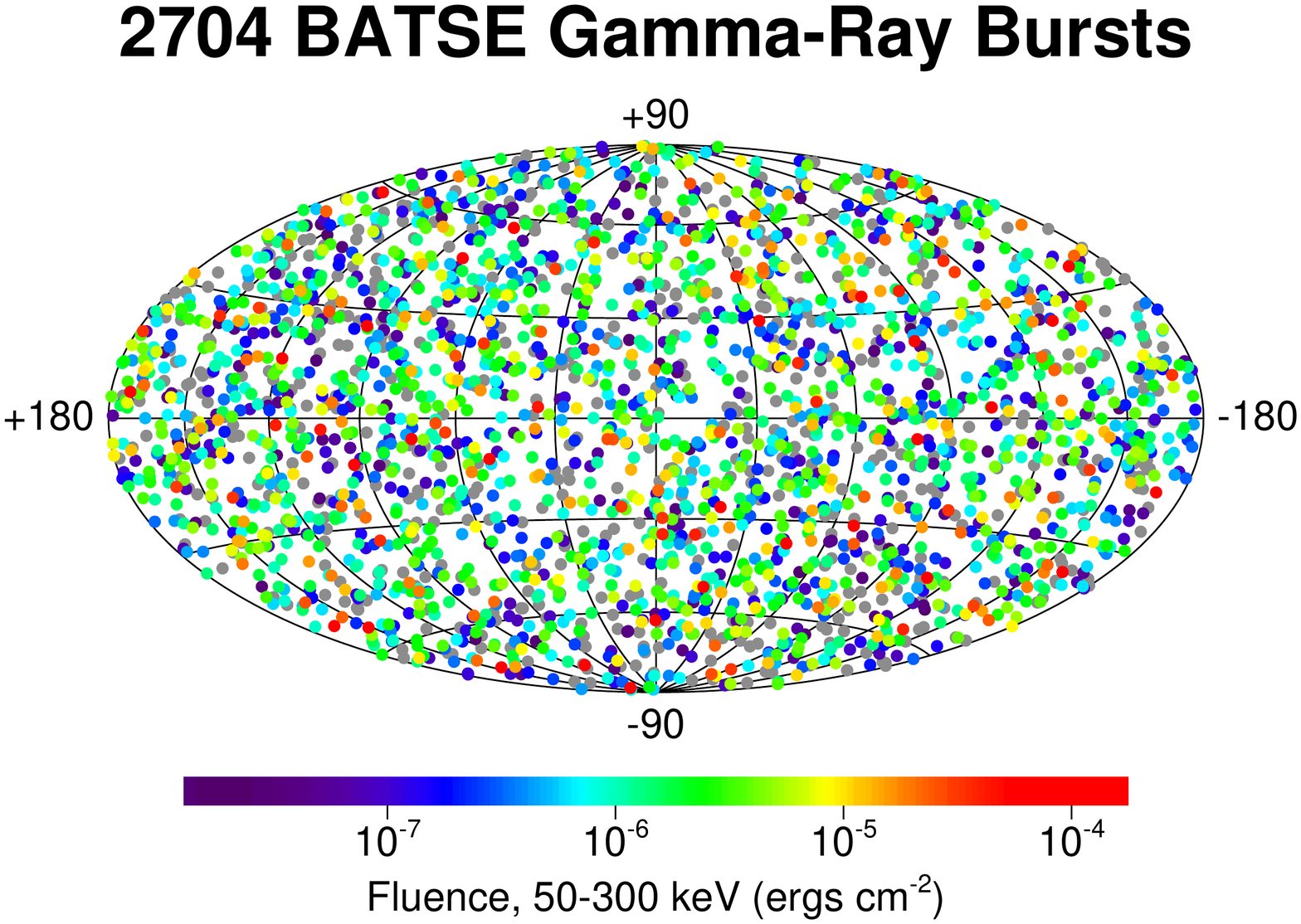}
\caption{\label{fig:GRB_Locations}Locations of all the 2704 GRBs detected
by BATSE in galactic coordinates. The plane of our galaxy is along
the horizontal line at the middle of the figure. The isotropic distribution
of GRBs in space implied that GRBs are most likely cosmological
sources. Source: \cite{URL_BATSE_Locations}}
\end{figure}

BATSE was sensitive enough to detect GRBs originating from distances 
larger than the size of our galaxy. 
Its discovery that GRBs are isotropically distributed
(Fig. \ref{fig:GRB_Locations}) narrowed down the possibilities
and suggested that GRBs are most probably located at cosmological distances
further away than our local group of galaxies. Otherwise, the GRB
spatial distribution would be correlated with the local distribution
of mass (our galaxy, the LMC, M31, globular clusters, the Virgo cluster, 
etc.) and would not be isotropic. However, the distance-scale
problem was not completely resolved, because an extended halo around our
galaxy could still generate a uniform distribution similar to the
one observed. %

\begin{figure}[ht]
\includegraphics[width=1\columnwidth]{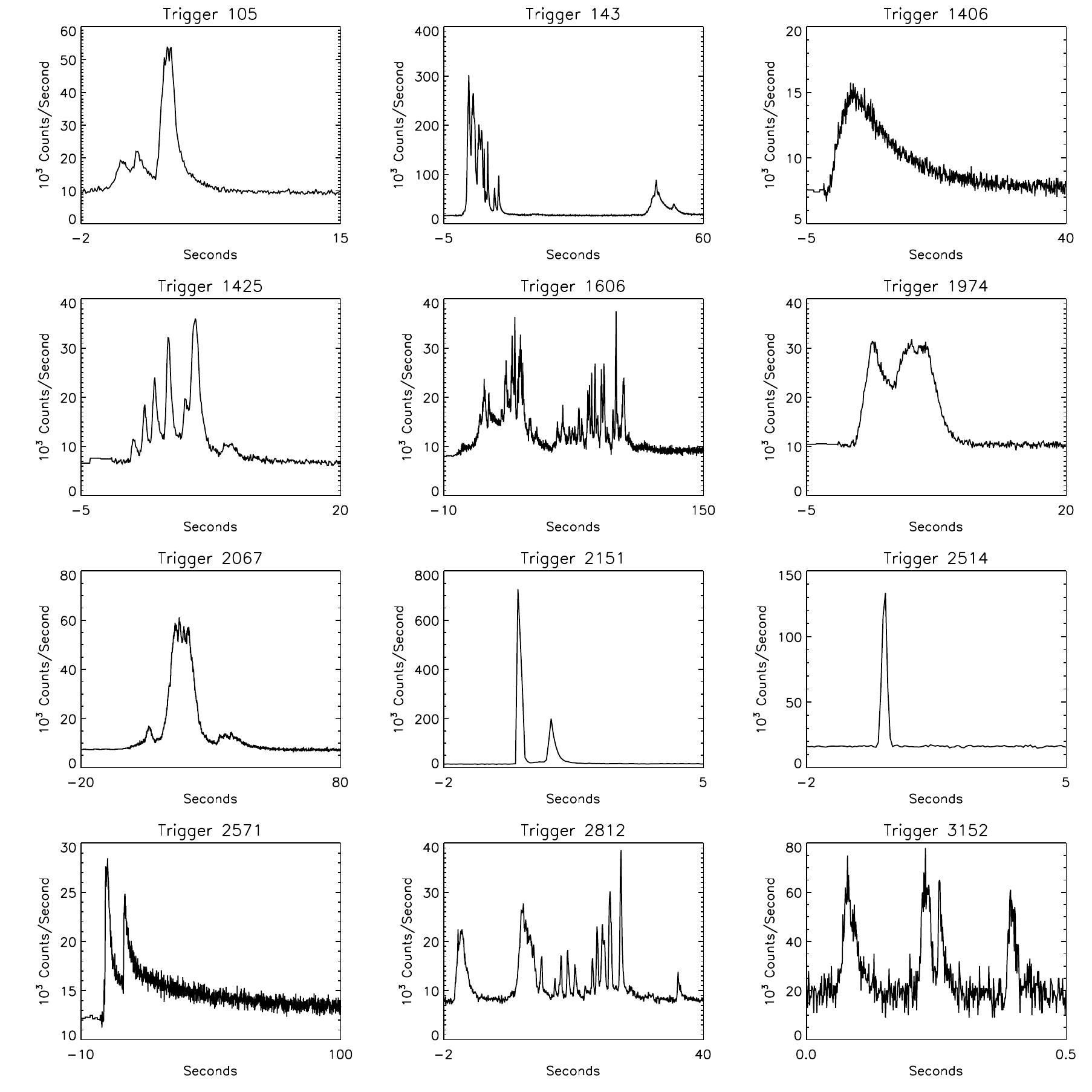}
\caption{\label{fig:GRB_TimeCurves}Light curves of some of the GRBs detected
by BATSE. The light curves of GRBs show great variability and diversity. 
Source: J. T. Bonnell (NASA/GSFC)}
\end{figure}

Some GRB properties emerged from the extensive dataset of BATSE-detected GRBs. 
The light curves of GRBs showed great morphological diversity,
ranging from smooth, fast rise, and quasi-exponential decays, 
to curves with many peaks and with a high variability, ranging
from timescales of milliseconds to many minutes (Fig. \ref{fig:GRB_TimeCurves}). 

\begin{figure}[ht]
\begin{centering}
\includegraphics[width=0.8\columnwidth]{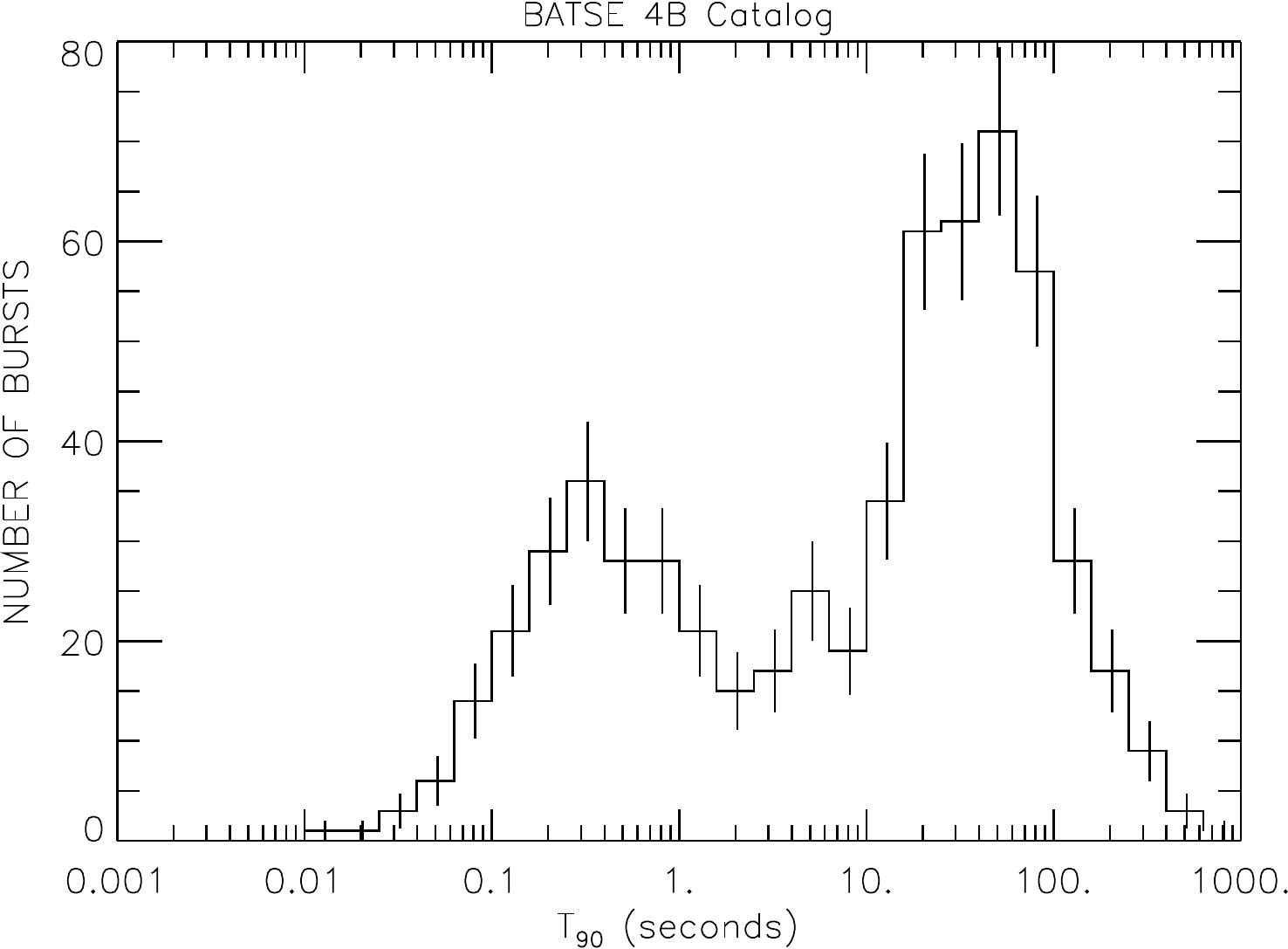}
\par\end{centering}

\caption{\label{fig:GRB_T90}Duration ($T_{90}$) distribution of 
the GRBs detected by BATSE. Two populations of bursts, separated at $T_{90}\simeq2\,s$, 
can be seen. Source: \cite{URL_BATSE_Durations}}
\end{figure}

\begin{figure}[ht]
\begin{centering}
\includegraphics[width=0.8\columnwidth]{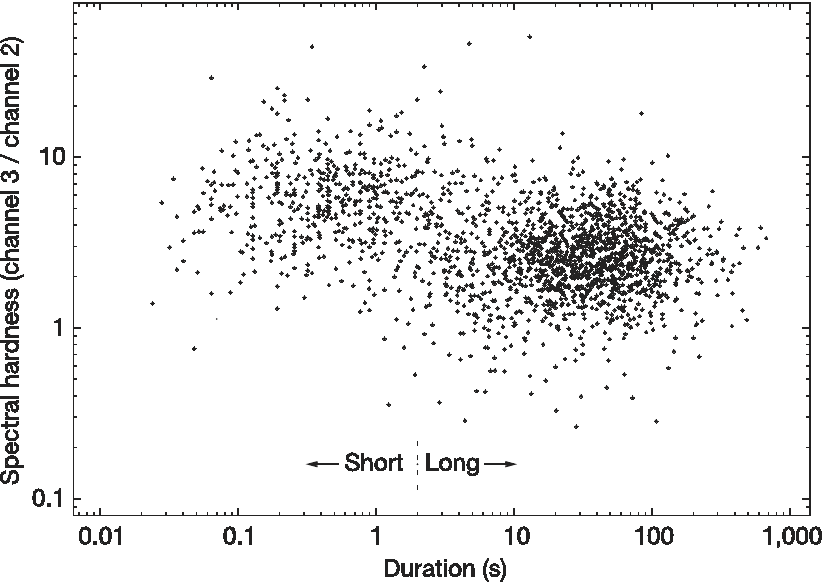}
\par\end{centering}

\caption{\label{fig:GRB_HardnessRatio}Hardness ratio versus duration $(T_{90})$
for BATSE GRBs. Short bursts ($T_{90}\lesssim2\,s$) have higher hardness 
ratios than long bursts ($T_{90}\gtrsim2\,s$), supporting the 
hypothesis that short and long bursts constitute two separate populations, 
probably originating from different progenitors. Source : \cite{GRB_Hjorth_et_al2005} }

\end{figure}

The duration of GRBs is usually described by the $T_{90}$ parameter, 
which is equal to the time over which the burst emits from 5\% to
95\% of its measured counts. The $T_{90}$ distribution of GRBs
(Fig. \ref{fig:GRB_T90}) spans a long range of durations and
has a bimodal shape. Based on that, GRBs were divided into two categories:
``short bursts'' having $T_{90}\lesssim2\,s$, and ``long bursts''
having $T_{90}\gtrsim2\,s$. Short bursts constituted $\sim30\%$
of the BATSE sample. The spectral properties of short and long GRBs were different. 
BATSE measured the fluence of a burst in different channels, each
one corresponding to a different energy range. The ``Hardness Ratio,''
defined as the ratio of the fluence in channel 3 ($100-300\,keV$)
over the fluence in channel 2 ($50-100\,keV$), was a measure of the spectral
hardness of a burst. Short bursts were found to have on average higher hardness 
ratios than long bursts, as shown on figure \ref{fig:GRB_HardnessRatio}. 
The existence of two distinct populations of bursts implied the existence of two
kinds of progenitors and inner engines.

\begin{figure}[ht]
\begin{centering}
\includegraphics[width=1.9in, height=2in]{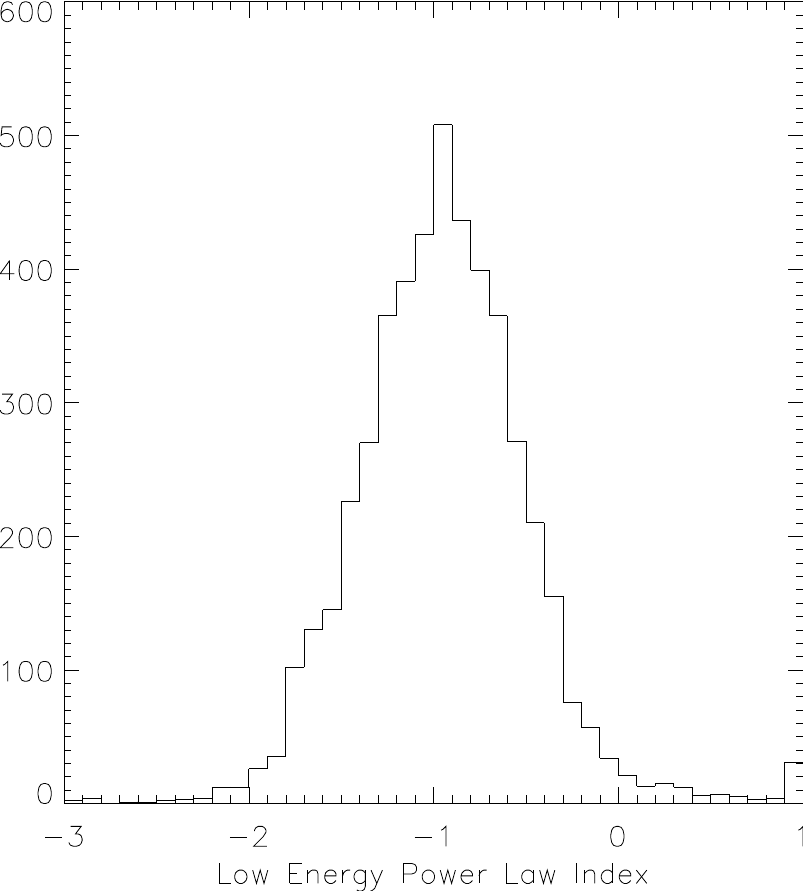} \includegraphics[width=1.9in, height=2in]{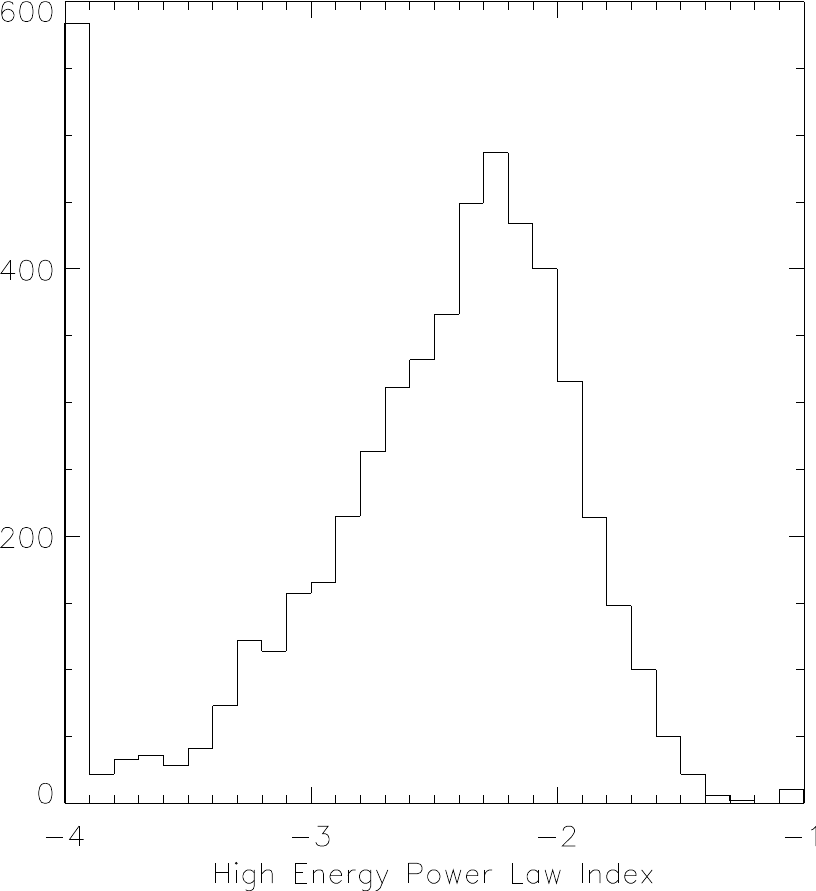}\includegraphics[width=1.9in, height=2in]{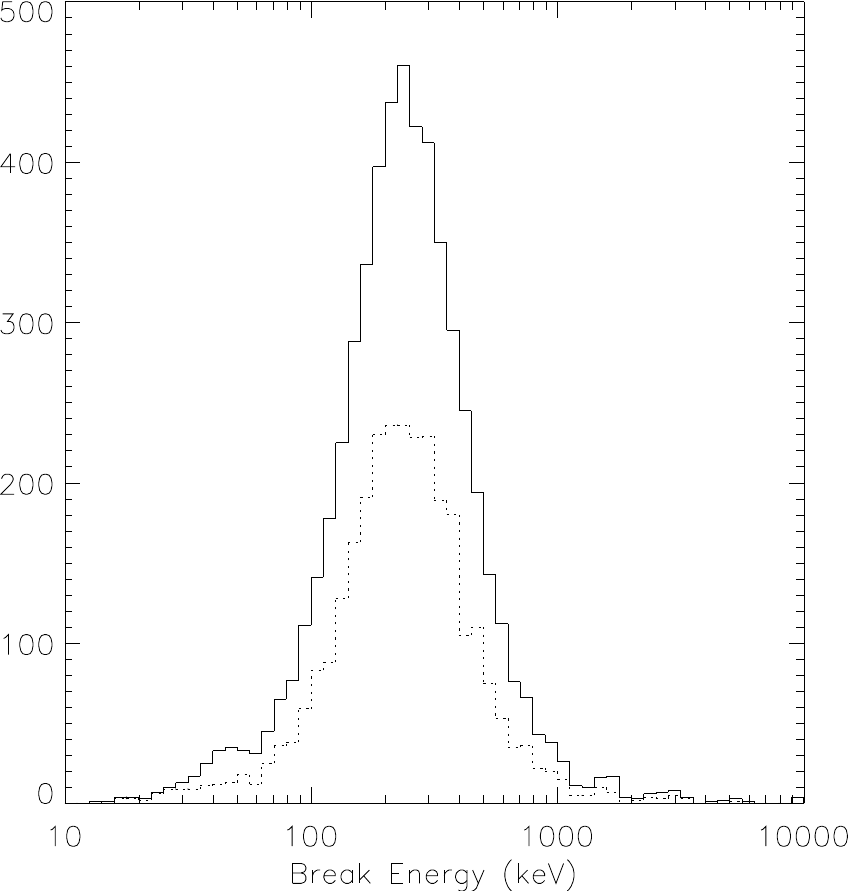}
\par\end{centering}

\caption{\label{fig:GRB_BandParameters} Distributions of the Band function parameters
from fits to the spectra of bright BATSE GRBs. 
\textit{Left}: Low-energy spectral index $a$, \textit{middle}: break energy $E_{0}$, \textit{right}: 
high-energy spectral index $\beta$. The units on the Y axes are number of bursts. Source: \cite{GRB_Preece_et_al2000} }
\end{figure}

\begin{figure}[ht]
\begin{centering}
\includegraphics[width=0.8\columnwidth]{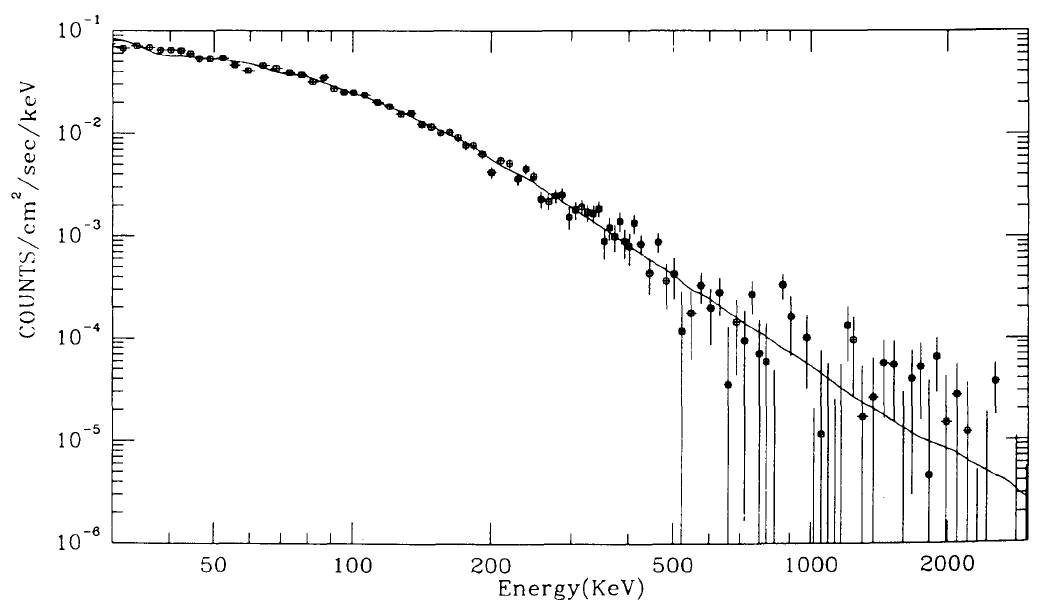}
\par\end{centering}

\caption{\label{fig:GRB_ASpectrum} Example of a fit to the spectrum of GRB911127 using the Band
function. Here $\alpha=-0.967\pm0.022$ and $\beta=-2.427\pm0.07$. Source: \cite{GRB_Band_et_al_1993} }

\end{figure}

\begin{figure}[ht!]
\begin{centering}
\includegraphics[width=0.8\columnwidth]{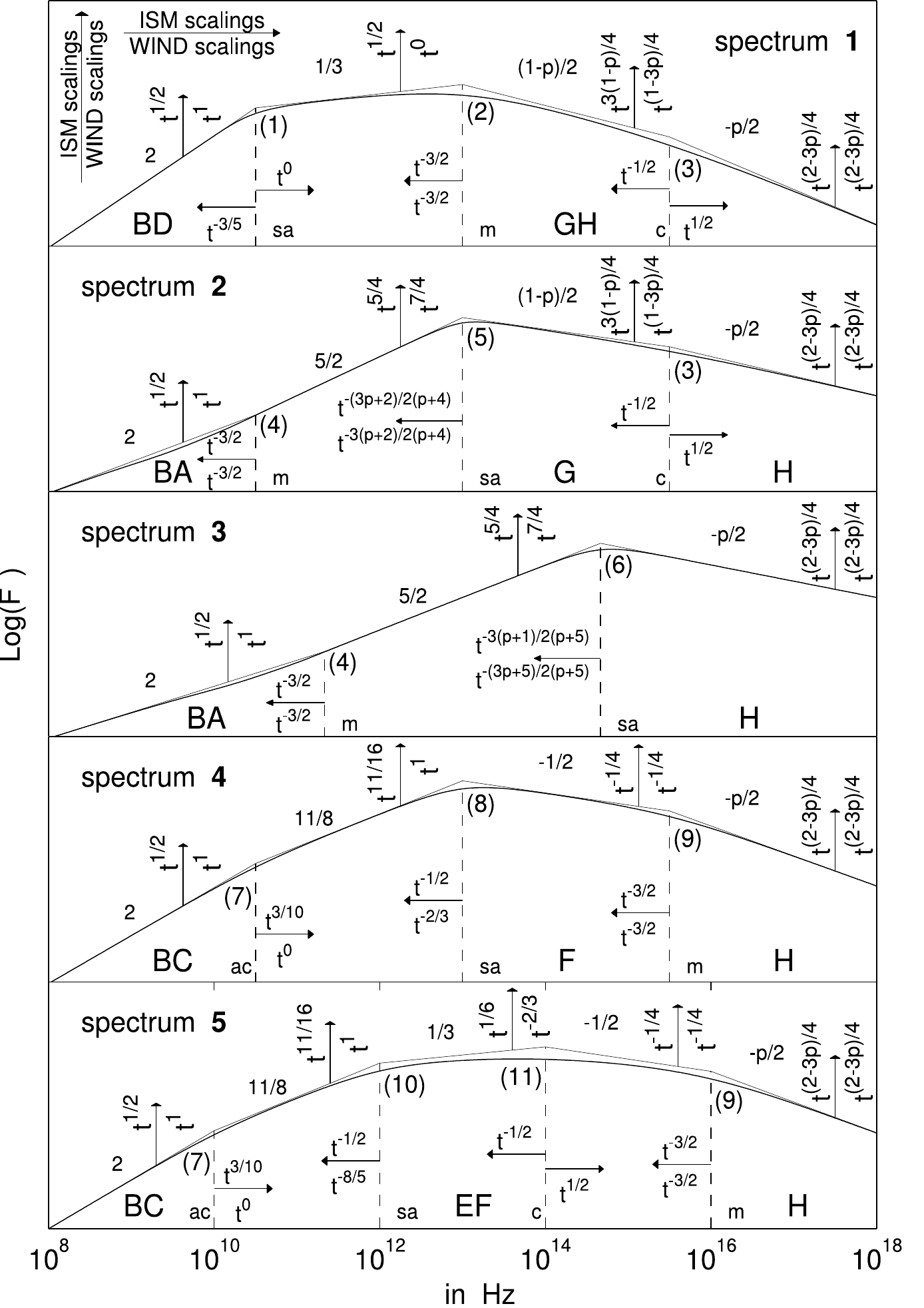}
\par\end{centering}
\caption{\label{fig:GRB_SyncSpectra}Different possible broadband synchrotron spectra from a relativistic blast wave that accelerates the electrons to a power-law distribution of energies ($N_{\gamma}\propto\gamma^{-p}$, with $\gamma$ the Lorentz factor of the electrons). The different spectra are labeled 1–5 from top to bottom. 
Different sets of physical conditions correspond to different orderings of the break frequencies: the minimal synchrotron frequency of the least energetic electron $\nu_{mu}$, the self-absorption frequency $\nu_{sa}$, and the typical synchrotron frequency of an electron whose cooling time equals the dynamical time of the system $\nu_{c}$. The similarity of these spectra to the spectrum of the prompt emission from GRBs implies that the latter is most likely synchrotron radiation. 
Source: \cite{GRB_Granot_Sari_2002}. }
\end{figure}

The measured spectra were not thermal, and evolved with time from hard to soft. 
The time-averaged spectrum followed an \textit{ad hoc} function, 
called the ``Band function'' $S(E)$ \cite{GRB_Band_et_al_1993}, 
given by:

\begin{equation}
S(E)=A\times\begin{cases}
\left(\frac{E}{100keV}\right)^{a}e^{-\frac{E}{E_{0}}} & E\le(a-\beta)E_{0}\\
\left(\frac{(a-\beta)E_{0}}{100keV}\right)^{\alpha-\beta}\left(\frac{E}{100keV}\right)^{\beta}e^{\beta-\alpha} & E\ge(a-\beta)E_{0},\end{cases}\label{eq:GRB_BandFunction}\end{equation}
where $a$ and $\beta$ are the low- and high-energy spectral indices,
respectively. The
parameters were estimated by measurements of bright BATSE bursts to
be on average $\alpha\simeq-1$, $\beta\simeq-2. 25$, and $E_{0}\simeq256keV$
(Fig. \ref{fig:GRB_BandParameters}) \cite{GRB_Preece_et_al2000}. The spectral energy distribution $\nu\,F_{\nu}$
peaked at the peak energy $E_{p}\equiv(2-a)E_{0}$. Figure \ref{fig:GRB_ASpectrum} shows an example
of a spectral fit using the Band function. 
Figure \ref{fig:GRB_SyncSpectra} shows different
possible broadband synchrotron spectra from a relativistic blast wave 
that accelerates the electrons to a power-law distribution of energies, 
as believed to happen during internal shocks. 
The similarity between figures \ref{fig:GRB_ASpectrum} and \ref{fig:GRB_SyncSpectra} implies
that the prompt emission from GRBs is mostly synchrotron radiation. 

Ghirlanda \textit{et al}. found that the spectra of short GRBs can be better fitted by a power
law combined with an exponential cutoff at high energies \cite{GRB_Ghirlanda_et_al2004}. 
Similarly to long GRBs, they also found that the $\nu F_{\nu}$ distribution for short GRBs
peaked at an energy $E_{0}=355\pm30\,keV$.

\subsection{BeppoSAX \& HETE-2 (1996-2007)}

Because gamma rays can neither be reflected
with mirrors nor refracted with lenses, 
there are not any focusing gamma-ray telescopes
used in astronomy (see \cite{MISC_Schonfelder_2004}
for a review of the detection methods
used in gamma-ray telescopes). 
As a result, the angular resolution
of gamma-ray telescopes such as BATSE was too wide to allow
optical telescopes, which needed arcmin localizations ($(1/60)^o$), 
to search for burst counterparts. This prevented the identification
of the progenitors and the environments of GRBs, as well as the
measurement of their distances. As a result, a verification of the
cosmological origins of GRBs was still lacking. A breakthrough happened
in early 1997, when the Dutch/Italian satellite BeppoSAX \cite{URL_BEPPOSAX}
(1996-2002) detected a fading X-ray emission from long GRB970228
. After a processing of a few
hours, a localization accurate enough for follow-up ground-based observations
at optical, radio, and other wavelengths was obtained. These observations
initially identified a fading optical counterpart, and, after
the burst had faded, long-duration deep imaging identified a very
distant ($z=0.498$) host galaxy at the location of the burst. This observation
was the first conclusive piece of evidence that long GRBs are cosmological
sources. It also paved the way for identifications of more host galaxies
and for redshift determinations through spectroscopy of the GRBs, 
and settled the distance-scale argument for long GRBs \cite{GRB_Metzger_et_al1997}. %

%

After BeppoSAX, the High-Energy Transient Explorer (HETE-2) (2000-2007)
\cite{URL_HETE2} performed more afterglow observations of high quality and helped identify
a new class of sources called ``X-Ray Flashes,'' similar to
softer GRBs identified earlier by BeppoSAX. HETE-2 also made the
first observations connecting long GRBs with Type Ic supernovae (see
subsection \ref{sub:GRB_GRB_SNCOnnection}). 

By 2005, although afterglows had been detected from about fifty long GRBs, there were
no such detections for short GRBs. The afterglows of short
GRBs were hard to detect because the detectors had to achieve precise
localizations using smaller numbers of photons, which required
more time than the case of long GRBS. By the time a precise
localization was achieved and an X-ray sensitive instrument was pointed toward the acquired location, 
the already weak afterglow of short GRBs had decayed to the point of becoming
undetectable. 

The first afterglow from a short GRB was detected by the Swift satellite, described next, 
due to its high sensitivity and fast slewing (re-pointing)
capabilities.

\subsection{Swift (2004 - present)}

In the instruments described above, there was an $\sim8\,hr$ or longer delay
between the initial burst detection and the follow-up
observations. This resulted in the loss of important information contained
in the burst afterglow during that delay. A new satellite, called Swift \cite{URL_SWIFT}
, that could observe the afterglow of the burst swiftly after its
detection, was launched in 2004. Swift had a GRB detector combined
with a wide field X-ray and an optical/ultraviolet telescope, 
and the ability to do automated
rapid slewing. Thus, it could localize afterglows with arcsec
accuracy a minute or so after the burst at gamma-ray, X-ray, and optical
wavelengths. 

\begin{figure}[ht]
\includegraphics[viewport=50bp 230bp 500bp 570bp, clip, width=1\columnwidth]{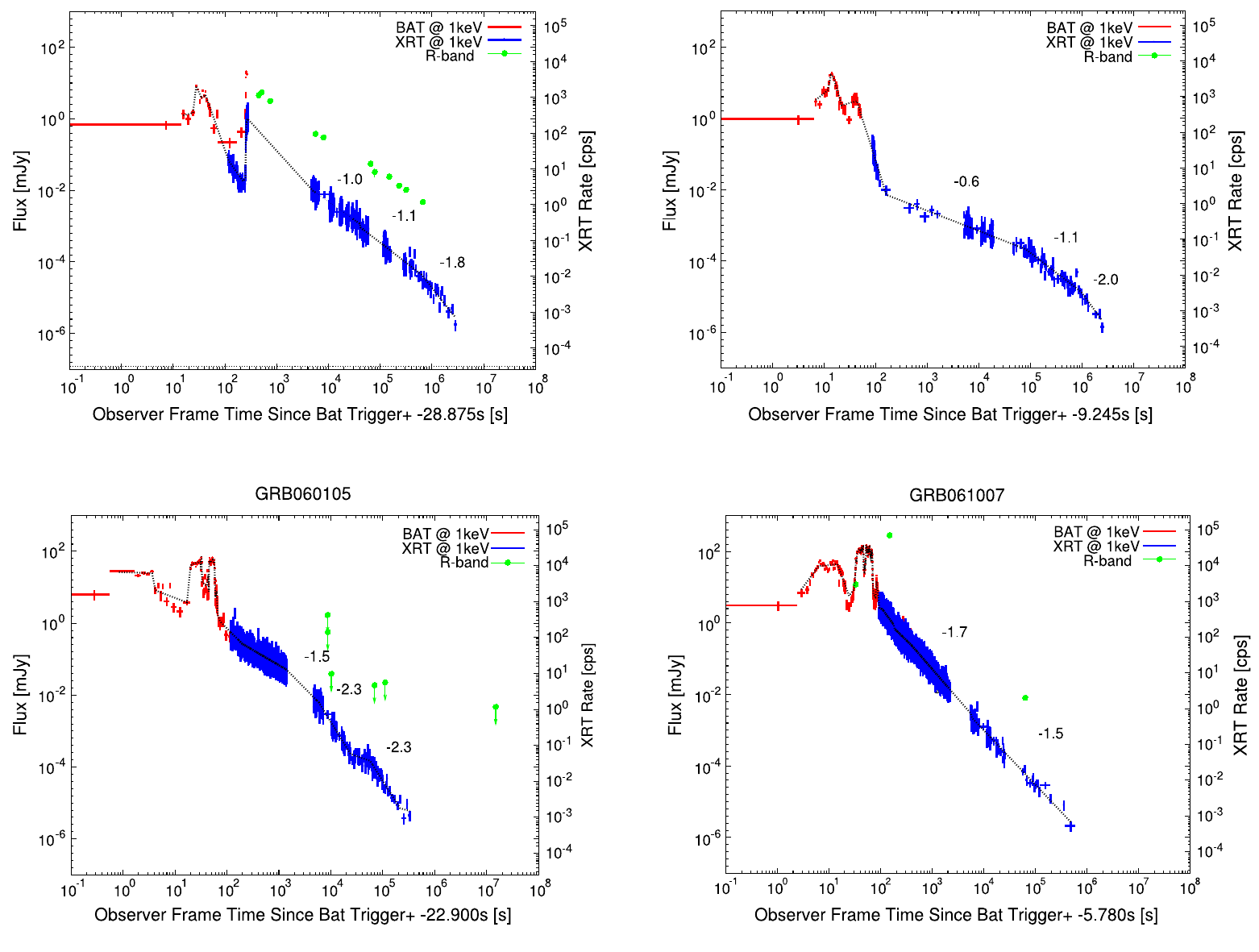}
\caption{\label{fig:GRB_Afterglow_curves}Light curves of four well-sampled
GRBs exhibiting a characteristic range of potential jet break behavior. 
The light curves are composed of measurements by two of Swift's detectors:
the Burst Alert Telescope (BAT-red) and the X-Ray Telescope (XRT
- blue). Source: \cite{GRB_Kocevski_Butler2008}}
\end{figure}

\begin{figure}[ht]
\begin{centering}
\includegraphics[width=0.8\columnwidth]{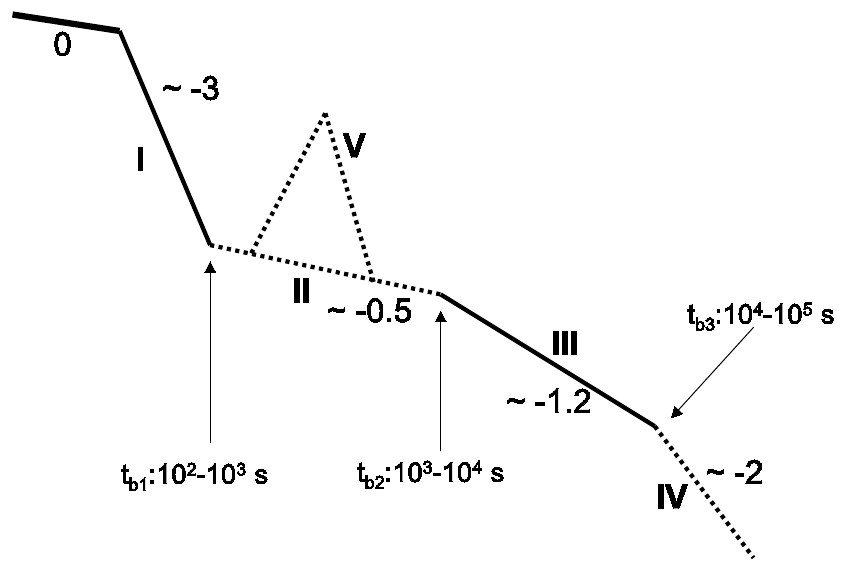}
\par\end{centering}
\caption{\label{fig:GRB_Afterglow_structure}Synthetic sketch of a light curve
based on Swift's observations. The initial phase, denoted by ``0,'' 
corresponds to the end of the prompt emission. Four power-law light-curve
segments together with a flaring component are identified in the afterglow
phase. The components marked with solid lines are the most common, 
while the ones marked with dashed lines are observed in only a fraction
of the bursts. The typical spectral indices of the power-law decay
are shown for each segment. The break between regions III and IV occurs
simultaneously for all observed frequencies (achromatic break) and is related
to the fact that the GRB emission comes from relativistic jets. Source:
\cite{GRB_Zhang_et_al_2006}}
\end{figure}

Swift's capabilities enabled us to study the transition between the
energetic and chaotic prompt emission, and the smoothly decaying softer
afterglow. These observations lead to the detection of spectral breaks
in the afterglow emission (Figs. \ref{fig:GRB_Afterglow_curves}
and \ref{fig:GRB_Afterglow_structure}), which provided support to
the collimated-emission model of GRBs and allowed us to significantly
constrain the energetics of GRBs (see subsection \ref{sub:GRB_Collimation}). 
Furthermore, it provided, for the first time, observations of the
afterglows of short $(T_{90}\lesssim2\,sec)$ bursts, which lead to redshift
measurements of short GRBs and verified the cosmological origin for
them too. 

Swift's observations of short GRBs showed that, unlike long GRBs, 
they usually originate from regions with a low star-formation rate. 
This suggested that short GRBs are related to old stellar populations, possibly
from mergers of compact-object binaries (i.e., neutron star-neutron
star or neutron star-black hole). 
Furthermore, even though supernova features such as red bumps and late-time rebrightening were detected in the afterglows
of most long GRBs close enough to allow such a detection, there were no
evidence of such features in the afterglows of long GRBs. These observations strengthened
the case for long and short GRBs having different kinds of progenitors:
compact-object binaries for short GRBs versus massive stars for long GRBs. 

The GRB afterglows, as observed by Swift, decayed on a power law and
progressively softened from X-rays, to optical, to radio. As of January
2008, Swift had detected 208 bursts in gamma rays, with almost all
of them having an X-ray afterglow. 

\begin{figure}[ht]
\begin{centering}
\includegraphics[width=0.8\columnwidth]{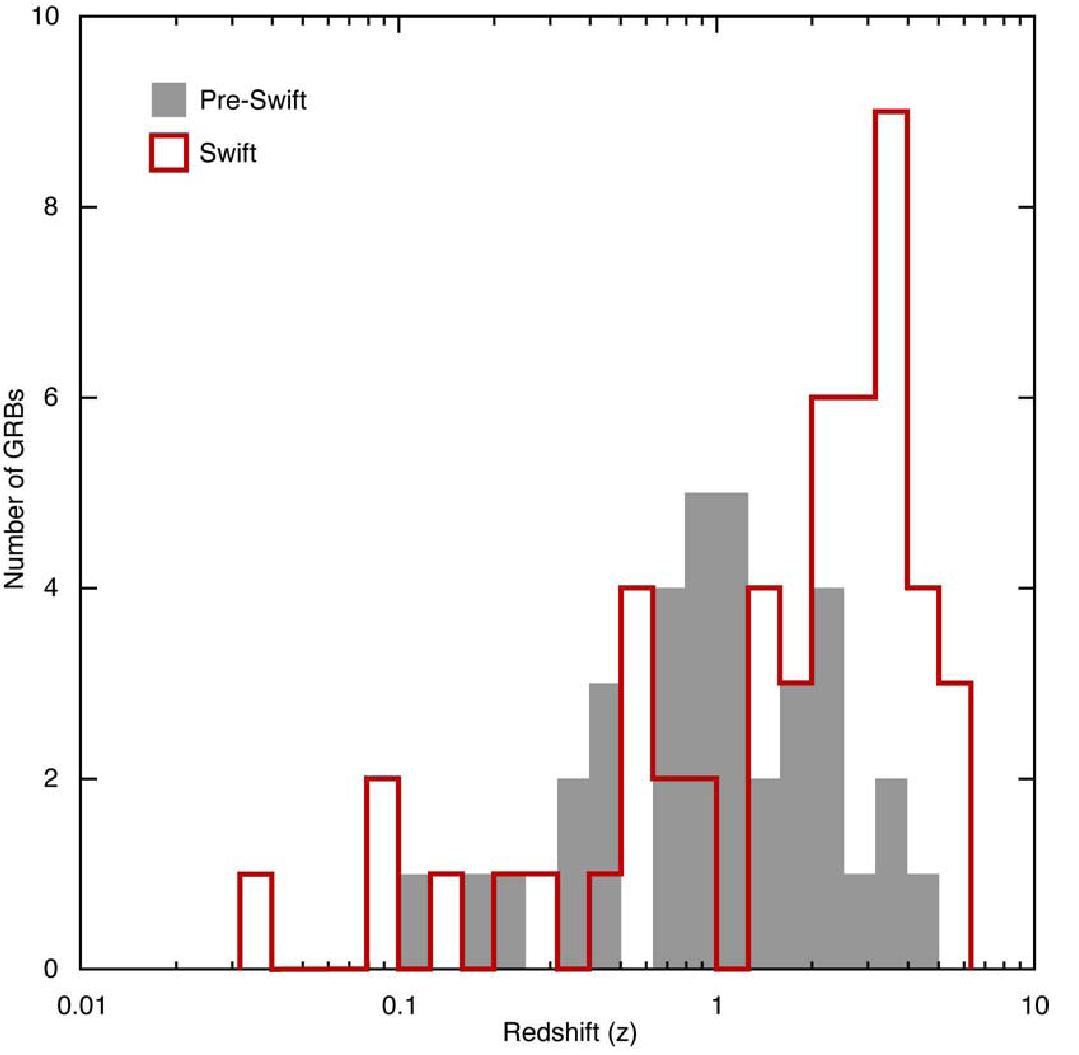}
\par\end{centering}

\caption{\label{fig:GRB_Redshifts}Redshift distributions of GRBs detected
by Swift and of GRBs detected by pre-Swift satellites. The average
redshift of the Swift sample is higher than redshift of the previous observations
(2. 3 vs 1. 2), because of the greater sensitivity of Swift to distant GRBs. Source:
\cite{GRB_Gehrels_et_al2007}}
\end{figure}

Swift is sensitive to a lower energy range ($15-150\,keV$) and to
bursts of longer durations than other detectors. 
Therefore it is more sensitive to GRBs of higher redshifts, since the signal from such GRBs is
more redshifted and time dilated. Due to its increased sensitivity to distant GRBs, 
Swift observed GRB050904, the most distant GRB ever observed. 
GRB050905 had a redshift of $z=6.295$, and when it exploded the 
age of the universe was only $\sim6\%$ of its current age. 
The redshift distribution of Swift GRBs and pre-Swift GRBs is shown in Fig. \ref{fig:GRB_Redshifts}. 
As can be seen, Swift GRBs are on average more distant than pre-Swift GRBs. 

\FloatBarrier
\section{\label{sec:GRB_Hosts}The GRB Model}
\begin{figure}[ht]
\includegraphics[width=1.0\columnwidth]{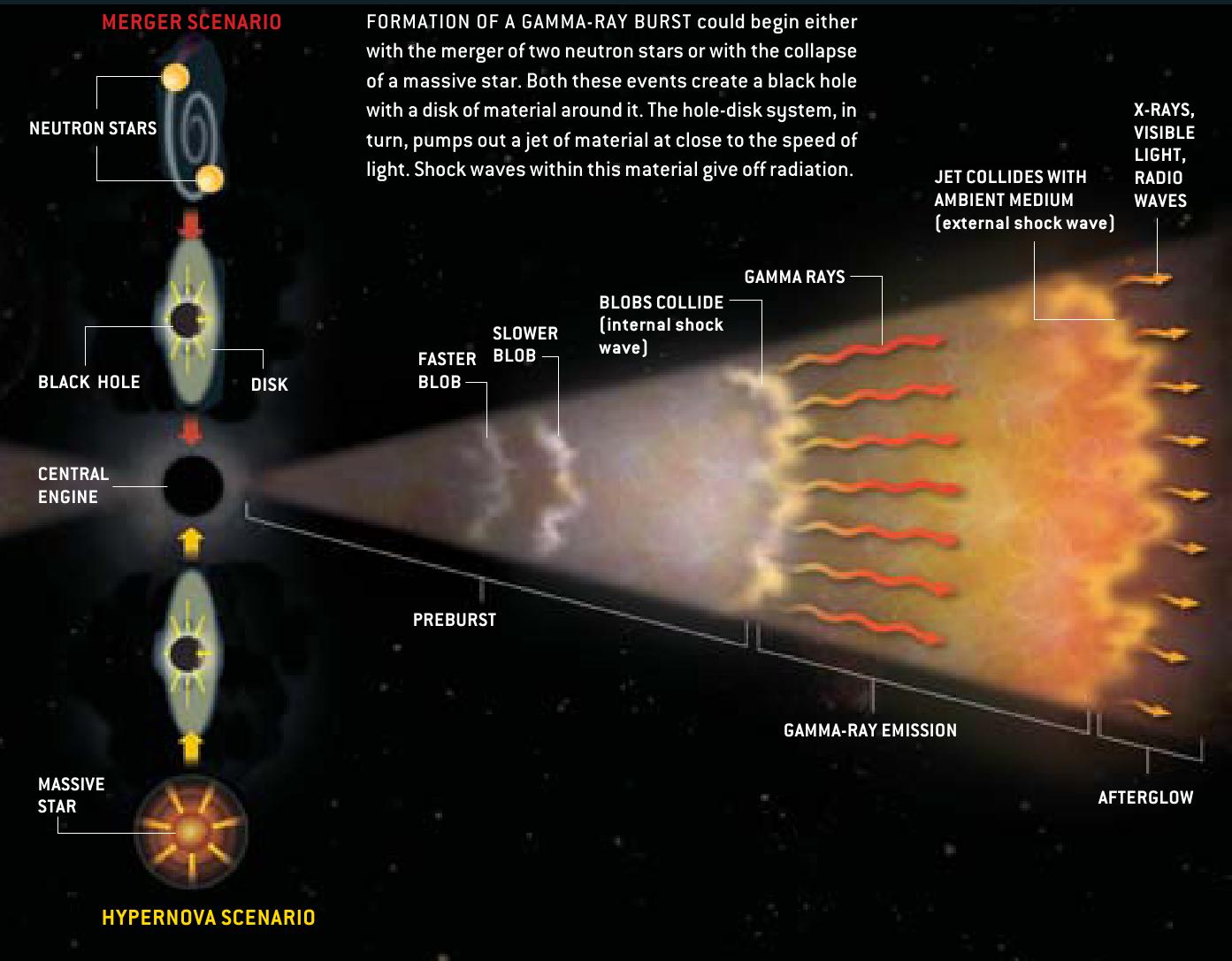}
\caption{\label{fig:GRB_Model}Sketch showing the process leading to the formation
of a GRB. Source: \cite{GRB_Gehrels_et_al2002}}

\end{figure}

According to the generally accepted model of the progenitor and the emission mechanism
of GRBs, GRBs start with a cataclysmic event, such as the merger of two
compact objects or the collapse of the core of a rotating massive
star, followed by the creation of a rapidly spinning black hole and an accreting
envelope around it (Fig. \ref{fig:GRB_Model}). This model, called the ''Collapsar model'',
was initially proposed to explain long GRBs \cite{GRB_MacFayden_Woosley1999}. However, it
was realized that the mergers of compact-object binaries that create short GRBs
also result a black hole-accretion disk system similar to the one in the collapsar model.

The collapse of the accreting material that is near the equator of the envelope is somewhat
inhibited by the strong centrifugal forces. Most of the accretion
happens through two funnels that form on the poles of the black
hole (on the axis of rotation). Large amounts of energy $(\sim10^{50}\,erg/s)$
are deposited locally on the polar regions, possibly through neutrino-driven
winds \cite{GRB_Narayan_et_al1992}, magneto-hydrodynamic processes
\cite{GRB_Blandford_Znajek1977}, magnetic instabilities in the
disk \cite{GRB_Blandford_Payne_1982}. The sources for the deposited energy are
the gravitational and rotational energy of the accreting envelope
and the spinning black hole. The relative
contribution of each source (envelope or black hole) is unknown and depends on which energy-transfer
mechanism is more efficient. It is more likely that the largest fraction
is supplied by the gravitational energy of the envelope. 

\begin{figure}[ht]
\includegraphics[width=1\columnwidth]{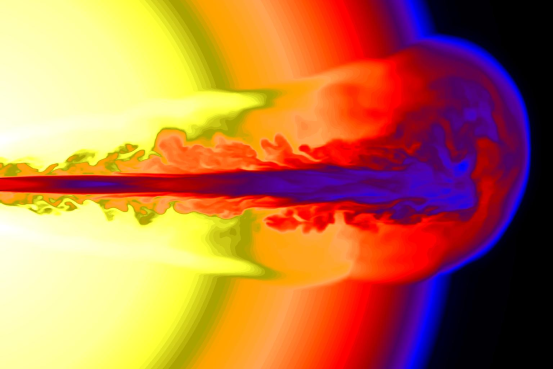}
\caption{\label{fig:GRB_JetBurst} A hyper-relativistic jet ($\Gamma\sim200$)
breaking out from the mantle of a $15M_{\odot}$ Wolf-Rayet star, 
$8\,s$ after it was launched from near the star's center. The jet's
luminosity is $3\times10^{50}\,erg/s$.  Source: \cite{GRB_Zhang_et_al_2006}}
\end{figure}

\begin{figure}[ht]
\includegraphics[width=0.50\columnwidth]{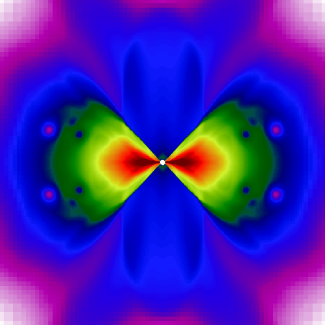}\includegraphics[width=0.50\columnwidth]{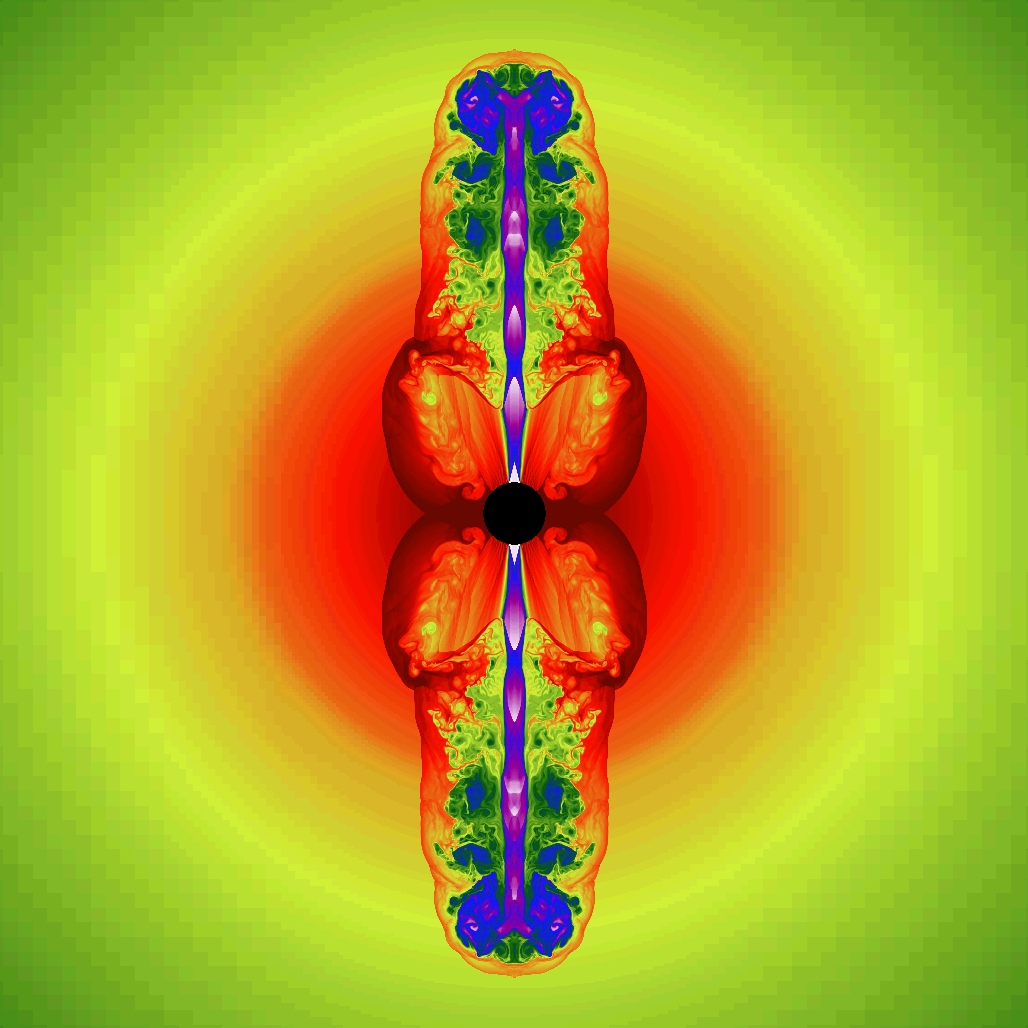}
\caption{\label{fig:GRB_JetBurst2}Collapse and explosion in a 14 solar mass
Wolf-Rayet star. \emph{Left}: The iron core of a highly evolved star
collapses to a black hole. Material along the rotational axis falls
in quickly, but matter in the equatorial plane is slowed by rotation
and piles up in a disk. Here color represents log density at a time
20 seconds after the initial collapse, and the highest density in the
equatorial plane near the black hole is $9\times10^{8}\,g/cm^{3}$. 
The figure is 1800 km across, and the inner boundary is at 13 km. 
A black hole of $4. 4M_{\odot}$ has formed and has been accreting
$0.1\,M_{\odot}/s$ for the last 15 seconds. During this time, magnetohydrodynamical
processes (not included in the simulation) would launch a jet along the rotational
axis. \emph{Right}: A GRB supernova. A two-component jet was introduced
at the origin along the top and bottom axes of a star similar to the
one above. One component had $10^{51}\,erg/s$, Lorentz factor $\Gamma=50$, 
and an internal energy to mass ratio of 3. Its opening angle was about
$10^{o}$. A second component of mildly relativistic matter extended
from $30^{o}-45^{o}$ and had power $5\times10^{50}\,erg/s$ and speed
$14\times10^{3}\,km/s$. Its composition and properties reflected
those expected for a ``disk wind''
blowing off the accretion disk by viscous processes. This figure, 
$2\times10^{5}\,km$ across, shows conditions $0.94\,s$ after the jet was launched. 
Both flows are hydrodynamically focused towards the rotational axis. 
Images and caption from \cite{GRB_Zhang_et_al_2006}.}

\end{figure}

Outward radiation and matter pressure gradually build up at the poles; however, they are initially
smaller than the pressure from the in-falling material. A point
is reached, at which the matter density over the poles and the accretion
rate are reduced to a large enough degree that they cannot counter-balance
the outward pressure. At that point an explosion occurs. 
A hot baryon-loaded $e^{-}, e^{+}, \gamma$ plasma (also called the
``fireball'') pushes outwards through the layers of the envelope. 
Matter and pressure gradients and magnetic fields collimate the
outflow, until it finally manages to erupt from the surface of the
object and break free in the form of two opposite narrow jets of half-opening
angle $\sim10^{o}$ (Figs. \ref{fig:GRB_JetBurst} and \ref{fig:GRB_JetBurst2}). 
Because the baryon load of the fireball plasma is small--$M_{b}c^{2}\ll E$
, where $M_{B}$ is the total mass of the baryons, and $E$ is the
total energy of the fireball--the fireball is quickly accelerated
to relativistic velocities. 


In the first stages following the ejection of the jet (preburst),
the density of the jet is very high, and any radiation produced in it is readily absorbed
instead of escaping. As a result, the jet accumulates energy, and
its bulk Lorentz factor increases further. However, as it expands, 
the optical depth is reduced, and radiation can escape from it. The
fact that the observed radiation is a power law and shows great
variability disfavors a model of a uniformly dense fireball 
expanding smoothly in the interstellar space and radiating on a thermal
spectrum. It was realized that the observed prompt emission and the 
afterglow could be produced during internal \cite{GRB_Kobayashi_et_al1997}
and external shocks \cite{GRB_Meszaros_Rees_1997}, respectively. The internal
shocks happen inside the jet and between shells of material moving
at different velocities. Such shells can be created if the 
energy-deposition mechanism is intermittent. During these shocks, 
the jet's electrons are accelerated to ultra-relativistic 
velocities and emit synchrotron radiation. 
Each peak of the prompt light curve is considered to be created during such an internal shock. 

The external shock occurs when the jet eventually collides with the ambient circumburst
medium, and smoothly and slowly decelerates. Similarly to internal shocks, relativistic electrons emit synchrotron radiation
observed as an afterglow that starts from gamma rays and gradually softens to longer wavelengths, down to radio as the jet is 
attenuated by the circumburst medium. 

While the general picture described by the collapsar model
is accepted by the scientific community, there is little consensus
regarding some of its details. The inner engine of GRBs is hidden
from us, so we can make only indirect inferences about its nature.
As a result, there is still uncertainty regarding many aspects of
the model, such as how exactly the jets are formed; which mechanism
transfers energy from the inner engine to the jets; the baryonic load
of the jets; the jets' bulk Lorentz factor; which physical processes
are involved in the internal shocks; what specific circumstances
lead to the creation of a GRB instead of just a supernova, etc. 

In the following, I will give a brief review of some of the observed
GRB properties. I will mention, where applicable, how these properties support
the collapsar model. For an extensive review on the physics of
GRBs, the collapsar model, and the progenitors and hosts of GRBs see,
\cite{GRB_Piran_2004, GRB_Meszaros_2006, GRB_Lee_Ruiz_2007, GRB_Woosley_Bloom_2006}.

\subsection{Inner engine of GRBs}

The light curves of the prompt emission show a variability of 
milliseconds to many minutes. These short time scales imply that a compact
object is involved in the emission, with size of the order of tens
of kilometers, typical for black holes and neutron stars. The fact
that the burst duration is usually longer than the variability suggests
a prolonged and intermittent inner-engine activity in two or three
different simultaneous time scales. This disfavors any explosive model
that releases the energy at once. The total energy emitted in gamma
rays is very high, about $10^{51}\,erg$, an amount comparable to the
energy release from supernovae. The above suggest that the inner engine
of GRBs consists of a massive object, most likely a newborn black
hole, with a massive ($mass\gtrsim0.1\,M_{\odot})$ disk accreting into
it. The accretion explains the prolonged activity and the different
time scales, and the black hole satisfies the size and energy requirements.

\subsection{Emission mechanism}

According to the collapsar model, the observed radiation is produced
at internal or external shocks. 
During these shocks, energy is transferred to the jet's electrons through a diffusive
shock acceleration mechanism \cite{GRB_Fermi_1949} in which 
magnetic field irregularities keep scattering the particles back and forth 
so they cross the same shock multiple times. During the first crossing, an electron gains an 
amount of energy of the order of $\Gamma^{2}_{sh}$, where $\Gamma_{sh}$ is the Lorentz factor
of the shock front measured in the rest frame of the jet \cite{GRB_Achterberg_et_al_2001}. 
Subsequent crossings are less efficient, and the gain is of the order of unity \cite{GRB_Achterberg_et_al_2001}. 
During these shocks, the electrons are accelerated to ultra-relativistic 
velocities ($\Gamma_{e}$ up to $\sim1000$) and emit synchrotron radiation. 
The shocks may also accelerate protons. However, the power of the synchrotron emission from protons
is considerably smaller than the power from the electrons, since an electron emits $(m_{p}/m_{e})^{2}\simeq10^{7}$ more power
through synchrotron radiation than a proton of the same Lorentz factor. Therefore the detected 
radiation is likely produced by electrons.

While the predictions of the synchrotron
model are in reasonable agreement with afterglow observations \cite{GRB_Granot_2003, GRB_Wijers_Galama1999, GRB_Panaitescu_Kumar2002}, 
there are some inconsistencies between its predictions and the observed spectral slopes \cite{GRB_Preece_et_al2002}. 
Alternative models for the emission in internal shocks include synchrotron
self-Compton \cite{GRB_Waxman_1997, GRB_Ghisellini_Celotti_1999} and
inverse Compton scattering of external light \cite{GRB_Shaviv_Dar_1995} similarly to the emission mechanism of Active Galactic Nuclei.

\subsection{Relativistic expansion}

The GRB fireball has a high radiation density, so photon pairs of center
of mass energy $\geq2\,m_{e}\,c^{2}$ should readily annihilate and create
$e^{-}e^{+}$ pairs, instead of escaping from the fireball. A calculation using
typical values yields an optical depth $\tau_{\gamma\gamma}\sim10^{15}$
\cite{GRB_Piran_1997}. In such a case, the emitted spectrum should
be thermal and should not contain an MeV or higher-energy component. 
This, in a first view creates a paradox, the ``Compactness problem,'' 
since the observed spectrum is a power law and extends up to energies of at least tens
of GeV, with no indication of a cutoff for long GRBs and up to tens of MeV for short GRBs. 

The paradox can be solved if the radiating material is moving with relativistic
velocities towards us. In such
a case, the observed GeV/MeV photons actually have a lower energy
in the fireball frame of reference. Therefore, the
optical depth of the fireball for the observed photons is actually
lower, since there is now a smaller number of photon pairs with a
center of mass energy over the annihilation threshold ($2m_{e}c^{2}$). 
If we assume that the photon energies inside the fireball are distributed
on a power law $I_{o}E^{-a}$, then this effect will decrease the
opacity by factor $\Gamma^{-2a}$, where $\Gamma$ is the bulk Lorentz factor
of the fireball \cite{GRB_Piran_2004}. Furthermore, 
because of relativistic contraction, the implied dimensions of the source
moving towards us will be smaller by a factor of $\Gamma^{2}$ than
its proper size. The power of two comes after considering the curvature of the 
emitting region (spherical-cap shape). As a result, the source's density is actually
smaller by a factor of $\Gamma^{-4}$ and the optical depth smaller
by a factor $\Gamma^{-2}$. The combined effect is
that the optical depth is actually lower by a factor of $\Gamma^{-2a-2}$
than what it would be for a non-relativistic jet, thus solving the paradox. 
Based on the above considerations and the amount of detected MeV/GeV
radiation from GRBs, lower limits on the bulk Lorentz factor of $\Gamma\gtrsim15$
were placed for short GRBs \cite{GRB_Nakar_2007} and
$\Gamma\gtrsim100$ for long GRBs \cite{GRB_Lithwick_Sari_2001}. 

Another piece of evidence supporting the case for relativistic motion
of the ejecta comes from the fact that estimates of the size of the
afterglow two weeks after the burst, independently provided by
radio scintillation \cite{GRB_Goodman_1997} and lower-frequency self
absorption \cite{GRB_Katz_Piran_1997}, can be explained only by assuming
relativistic expansion. 

\FloatBarrier
\subsection{\label{sub:GRB_Collimation}Energetics and collimated emission}

\begin{figure}[ht]
\begin{centering}
\includegraphics[height=0.8\textheight]{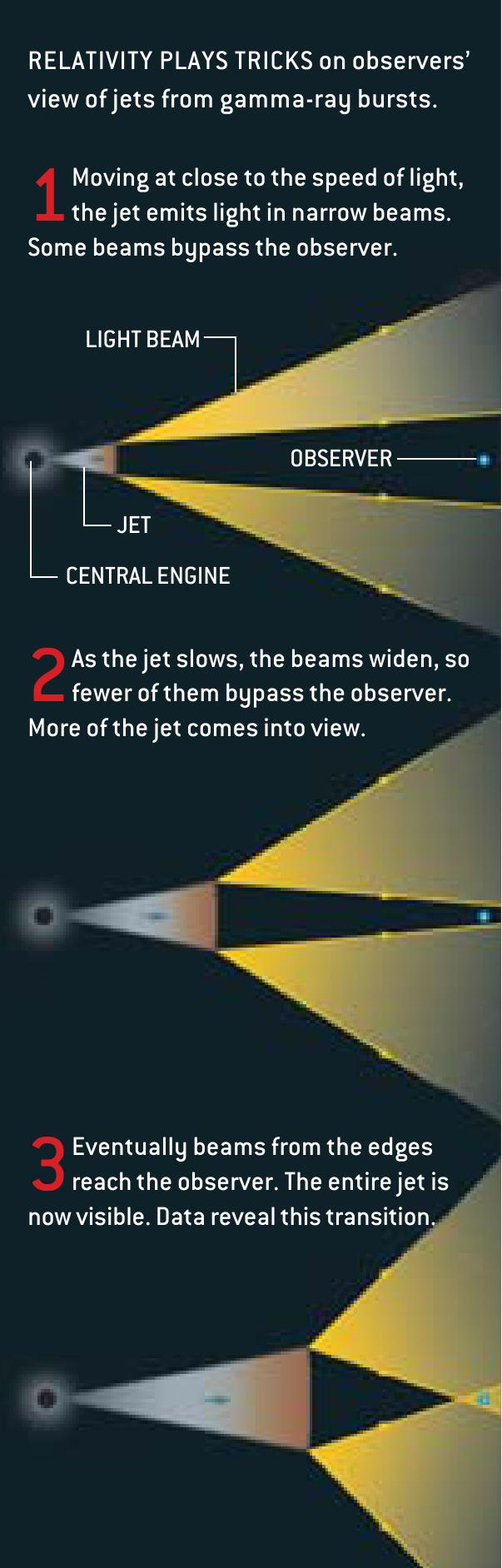}
\par\end{centering}

\caption{\label{fig:GRB_Beams}Sketch showing how the relativistic beaming combined
with  collimated matter emission produces an achromatic spectral break
in the afterglow curve of GRBs. Source: \cite{GRB_Gehrels_et_al2002}}
\end{figure}

The afterglow light curves of GRBs exhibit achromatic spectral breaks (Fig. \ref{fig:GRB_Afterglow_structure}) that
can be explained by assuming that the geometry of the ejecta
is conical (on two opposite jets) instead of spherical. Figure \ref{fig:GRB_Beams}
shows how this can happen. Because the fireball is moving with relativistic
velocities, its emission is beamed. Consider an observer that is inside
the projection of the emission cone of the fireball. Initially, when the bulk
Lorentz factor of the fireball is very high, the relativistically-beamed
radiation will be emitted in a very narrow cone. As a result, the
observer will not be able to see the emission from a part of the fireball.
Such a case is shown in the top picture of figure \ref{fig:GRB_Beams}, 
in which radiation from the sides of the fireball is clearly not visible
by the observer. As the GRB progresses, the surface of the fireball
expands (as $\propto t^{2}$), and the emitted radiation density drops
with the same rate, causing a gradual decrease in the observed brightness
of the burst. However, because of the expansion, the bulk Lorentz
factor is reduced, and the relativistic beaming becomes wider. As a
result, a larger fraction of the surface of the fireball will come
in the field of view of the observer (middle picture), reducing the
decay rate of the observed GRB brightness (now $\sim\propto t^{-1. 2}$
instead of $\sim\propto t^{-2}$). Eventually, all of the surface
of the burst becomes visible to the observer, and a gradually
increasing fraction of the fireball is no longer able to be seen.
The decay rate of the burst's brightness now depends only on the expansion
of the fireball's surface and becomes proportional to $t^{-2}$. This
transition, appearing as an achromatic break on the afterglow light curve, 
has been observed on many GRBs. For the GRB afterglows with no observed jet
breaks, it is assumed that the breaks happened at a time long after
the bursts, when no observation data exist. %

\begin{figure}[ht]
\begin{centering}
\includegraphics[width=0.8\columnwidth]{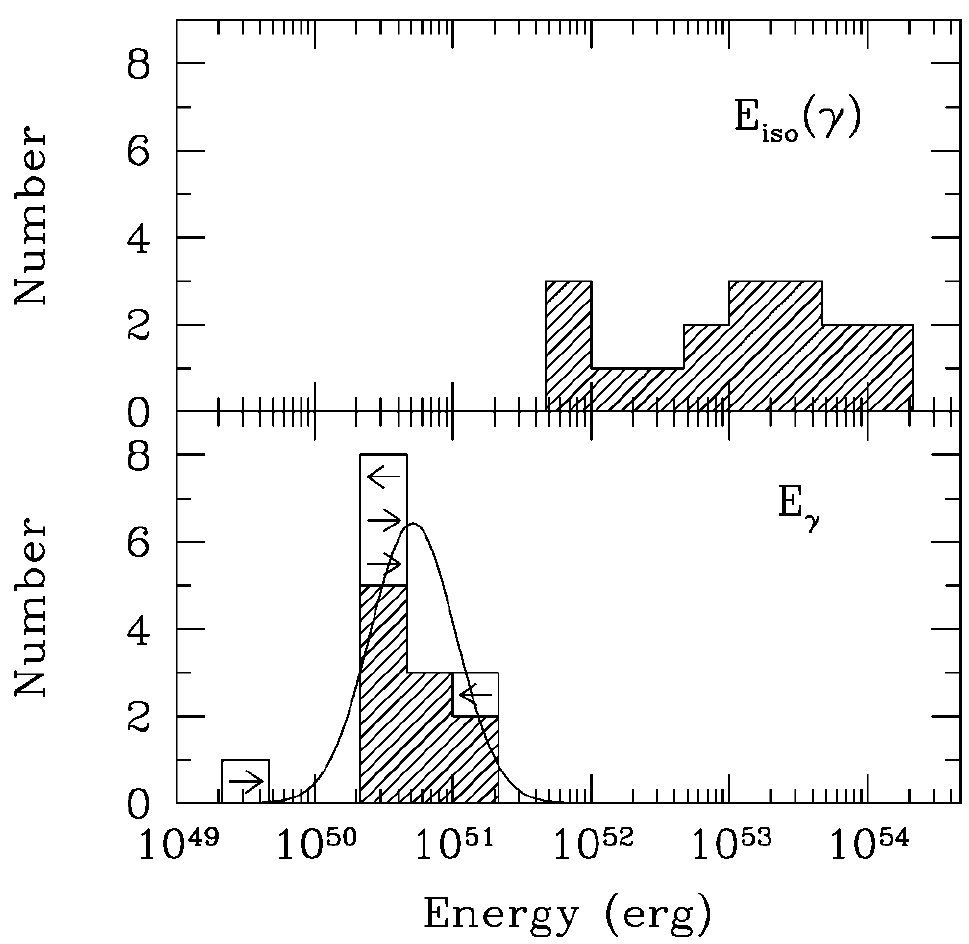}
\par\end{centering}

\caption{\label{fig:GRB_CorrectedEnergy}GRB energetics:
distribution of the isotropic-equivalent
emitted energy for a selection of GRBs with known redshifts \textit{(top)}, 
distribution of the geometry-corrected emitted energy for the same
GRBs \textit{(bottom)}. Arrows are plotted for five GRBs to indicate lower or upper
limits to the geometry-corrected energy. Source: \cite{GRB_Zhang_et_al_2006}}
\end{figure}

The typical GRB gamma-ray fluences at the earth are of the order of $10^{-5}\,erg/cm^{2}$. 
If we assumed an isotropic emission from GRBs, then this fluence, combined with 
the distance scale of GRBs (say $z=2$), would result to an isotropically-emitted energy
of $E_{iso}\simeq10^{53}\,erg$. Such an energy emission is considerably
higher than the emission from a typical supernova ($10^{51}\,erg$
in few months or $10^{49}\,erg$ in hundreds of seconds), and is difficult
to explain. However, the fact that the emission geometry is conical
ameliorates these energy requirements. 
If the emission actually happened in a solid angle $\Delta\Omega$, 
then the true amount of emitted energy is 
\begin{eqnarray*}
E_{true} & = & 2\,E_{iso}\,\Delta\Omega/4\pi\\
 & = & 2\,E_{iso}\,\frac{1-cos(\theta_{jet})}{2}\\
 & \simeq & E_{iso}\,\frac{\theta_{jet}^{2}}{2},\end{eqnarray*}
where $\theta_{jet}$ is the half opening angle of the emission cone. 
Frail \textit{et al.} \cite{GRB_Beaming_Frail2001} estimated $\theta_{jet}$
for a sample of GRBs, based on the occurrence time of the achromatic break in
their afterglow curves. Based on $\theta_{jet}$, they calculated the true amount of emitted
energy from the isotropic-equivalent amount. Their result (Fig. \ref{fig:GRB_CorrectedEnergy})
showed that even though the isotropic-equivalent emitted energy spans a wide energy range
$(4\times10^{52}-2\times10^{54}\,erg)$, the true amount of emitted energy spans a considerably
narrower energy range centered at $\sim3\times10^{50}\,erg$. This shows that the energy emission
of GRBs is comparable to that of supernovae, and suggests that GRBs have a standard energy reservoir.
The fact that the emission is conical also increases
the implied rate of GRBs by the same factor ($\simeq\theta_{jet}^{2}$), since only GRBs with their
emission cones pointing to the earth are detected. 

\subsection{Progenitors of GRBs}

There are multiple observational pieces of evidence that suggest that
not all GRBs are the same, and that there are different kinds of
progenitors and inner engines. Specifically, the duration and the hardness ratio-duration
distributions (Figs. \ref{fig:GRB_T90} and \ref{fig:GRB_HardnessRatio}) 
show that there are two kinds of bursts: short-hard
bursts and long bursts. Deep long-duration observations of the
optical afterglows of short bursts did not show any evidence of an
associated supernova \cite{GRB_Bloom_et_al2006, GRB_Hjorth_et_al2005, GRB_Fox_et_al2005}.
On the other hand, supernova-emission spectra were detected superimposed on the afterglows of most of the long
GRBs ($\sim20$ GRBs) that were close enough ($z\lesssim1$) to allow for such a detection \cite{GRB_Soderberg_et_al2006}. 

Short duration bursts are primarily observed in regions with low or
no star formation, therefore they are likely to be related to old stellar
populations. This suggests that these bursts could be the result of mergers
of compact binaries, such as neutron star-neutron star or neutron
star-black hole. The binary loses rotational energy through the
emission of gravitational radiation and eventually merges, forming
a black hole and an accretion disk surrounding it. The resulting system, 
then, produces a GRB in a way similar to
the collapsar model described above. 

Long bursts are observed in regions with high star formation and are
usually accompanied by supernovae, implying that they are related
to the death of massive stars (see subsection \ref{sub:GRB_GRB_SNCOnnection} for more details on
the GRB-supernova connection). The massive star involved is most likely
a Wolf-Rayet star,
\footnote{Massive stars ($Mass>20\,M_{\odot}$) that rapidly lose their outer envelope by
means of a very strong stellar wind.}
given that absorption features in the afterglow of long
GRBs \cite{GRB_Moeller_et_al2002} were explained by the presence
of the fast-moving wind of such a star. Furthermore, the fact that 
long-GRB counterparts are located within the blue parts of galaxies argues
against high-velocity progenitors (such as merging neutron stars). 
The above suggest that long GRBs likely come from the collapse of
the core of a Wolf-Rayet star that for some reason created a GRB instead
of just a supernova. Some of the differences between short and long GRBs come from the fact that
the engine of long GRBs operates at the center of a collapsing star, 
therefore it is covered by the mantle of the star, while the engine
of short GRBs is more or less exposed.

\subsubsection{\label{sub:GRB_GRB_SNCOnnection}Long GRB-supernova connection}

In 1998, the optical telescope ROTSE discovered a transient emission 
coincident in space and time with BeppoSAX/BATSE long GRB980425 \cite{GRB_ROTSE_980425}. 
The location, spectrum and light curve of the optical transient lead
to its identification as a very luminous Type Ic supernova%
\footnote{A Type Ic supernova has no hydrogen in its spectrum and lacks strong
lines of HE I and Si II. } (SN 1998bw) \cite{GRB_Galama_et_al1998, GRB_Kulkarni_et_al1998}. This
detection was a first of its kind, and suggested that long GRBs are
related to supernovae, and therefore to the deaths of massive stars. 
Because GRB980425 was very subenergetic comparing to other GRBs (isotropic
energy emitted was $\sim8\times10^{47}\,erg$ instead of the usual
$10^{51}-10^{54}\,erg$), the supernova-GRB connection was initially called into question. 
However, a few years later, a similar event happened. Emission from
a supernova (SN2003dh \cite{GRB_Hjorth2003}) was detected on the
afterglow of long GRB030329. This time, the associated GRB
had a normal energy. In addition to those events, there have also been red
emission ``bumps'' superimposed on the afterglows of GRBs, with
color, timing, and brightness consistent with the emission of a Type
Ic supernova similar to SN 1998bw (see \cite{GRB_Woosley_Bloom_2006}
and references therein). 

Based on the above, it is now believed that most, if not all, long
GRBs are accompanied with a Type Ic supernova. It should be noted, 
however, that not all Type Ic supernova create a GRB. The specific
conditions that lead to the creation of a GRB is one of the open questions
of the field. Observational and theoretical evidence imply that high
rotational speeds, high progenitor masses, and regions of low
metallicity \cite{GRB_Fynbo_etal2003, GRB_MacFayden_Woosley1999} favor
the creation of GRBs. 

The collapsar model of GRBs can accommodate the existence of a Type Ic supernova. 
Specifically, the GRB and the underlying supernova are powered
by different sources. The supernova and the $^{56}Ni$ that makes
it bright are produced by a sub-relativistic disk wind \cite{GRB_MacFayden_Woosley1999}. 
The wind begins as protons and neutrons, in about equal proportions, 
and after it cools, it ends up as $^{56}Ni$. The nickel comes out
in a large cone surrounding the GRB jet.

   \clearpage \chapter{\label{chap:VHE}VHE Emission from GRBs}

\section{Introduction}
The prompt and delayed emission from GRBs has been observed in many wavelengths
from $MeV$ gamma rays to optical photons. The observation of each of these components
provide unique information regarding the environment and mechanism of GRBs. However, the
$E>20\,GeV$ emission from GRBs has yet not been detected.

This chapter will present the physical processes involved in the generation and absorption
of VHE photons in GRBs, and will provide insight on which conditions favor such an emission
and what kind of information can be deduced from its detection.

Section \ref{VHE_Observations} will present 
the observational searches for VHE emission from GRBs. Then, section \ref{sub:VHE_Emission}
will give an overview of the processes that can generate VHE photons in GRBs. Lastly,
section \ref{sub:VHE_Absorption} will describe the processes that can absorb part of that radiation 
at the site of the burst. 

\section{\label{VHE_Observations}Searches for VHE Emission from GRBs}
Even though there are multiple processes that can create $E>20\,GeV$ photons
in GRBs, there has not been a definitive detection of such an emission yet.
This is likely due to the absence of an instrument capable
of detecting it, rather than an intrinsic property of GRBs. The satellite
instruments mentioned in the previous chapter\textbf{--}BATSE, EGRET,
Swift, etc.\textbf{--}were not sensitive to energies over $\sim50\,GeV$,
and therefore not capable of detecting the VHE emission from
GRBs. EGRET did have some sensitivity up to $\sim100\,GeV$, but
the flux from GRBs at that energy range is small enough that it could not
be detected by EGRET. Up until now,\footnote{
The GLAST satellite, recently launched (06/09/08), carries the the wide
field of view instrument LAT which is sensitive to gamma rays of energies
up to $300\,GeV$. %
} only ground-based detectors were sensitive to such high energies. 

The most sensitive ground-based detectors--Imaging Atmospheric Cherenkov
Telescopes (see section \ref{sec:Milagro_Milagroetal})--have a narrow field of view ($\lesssim3^{o}$)
and small duty cycles (less than $10\%$). As a result, they are not suitable
for continuously monitoring the overhead sky for GRBs (as satellite
detectors can do). Furthermore, most IACTs cannot refocus fast enough towards
the location of a GRB detected by another instrument in order to observe
its prompt emission. Based on the above, IACTs are better at searching
for VHE emission from GRBs during the afterglow phase, and this only for the
GRBs detected by external instruments. Searches for VHE emission from
GRBs by IACTs resulted in null results \cite{GRB_MAGIC_OBS,GRB_VERITAS_OBS}.

Milagro, on the other hand, has a wide
field of view ($\sim2\,sr$) and a high duty cycle ($>90\%)$, so unlike IACTs, it
can perform GRB observations both independently and in coordination
with other instruments. Milagro, similarly to most EAS arrays, has $\sim4000$ times
more exposure to the sky than IACTs. However, it is considerably less sensitive
than IACTs to gamma rays of energies $E\lesssim1\,ΤeV$. As will be
shown in Chapter \ref{chap:IR}, the higher-energy component ($E\gtrsim300\,GeV)$
of the VHE emission from GRBs is strongly attenuated before reaching
the earth. Therefore, a high sensitivity in the hundreds of $GeV$ energy
range is required in order to perform observations of the VHE emission
from GRBs. Milagro's effective area is maximal at $TeV$ energies,
so Milagro is less sensitive than IACTs to the low-energy GRB
emission reaching the earth. 

Another reason for the absence of a detection of VHE emission
from GRBs is the fact that the produced VHE radiation can be self-absorbed
before managing to escape the site of the burst (see section \ref{sub:VHE_Absorption}). 

\begin{figure}[bt]
\includegraphics[width=1\columnwidth]{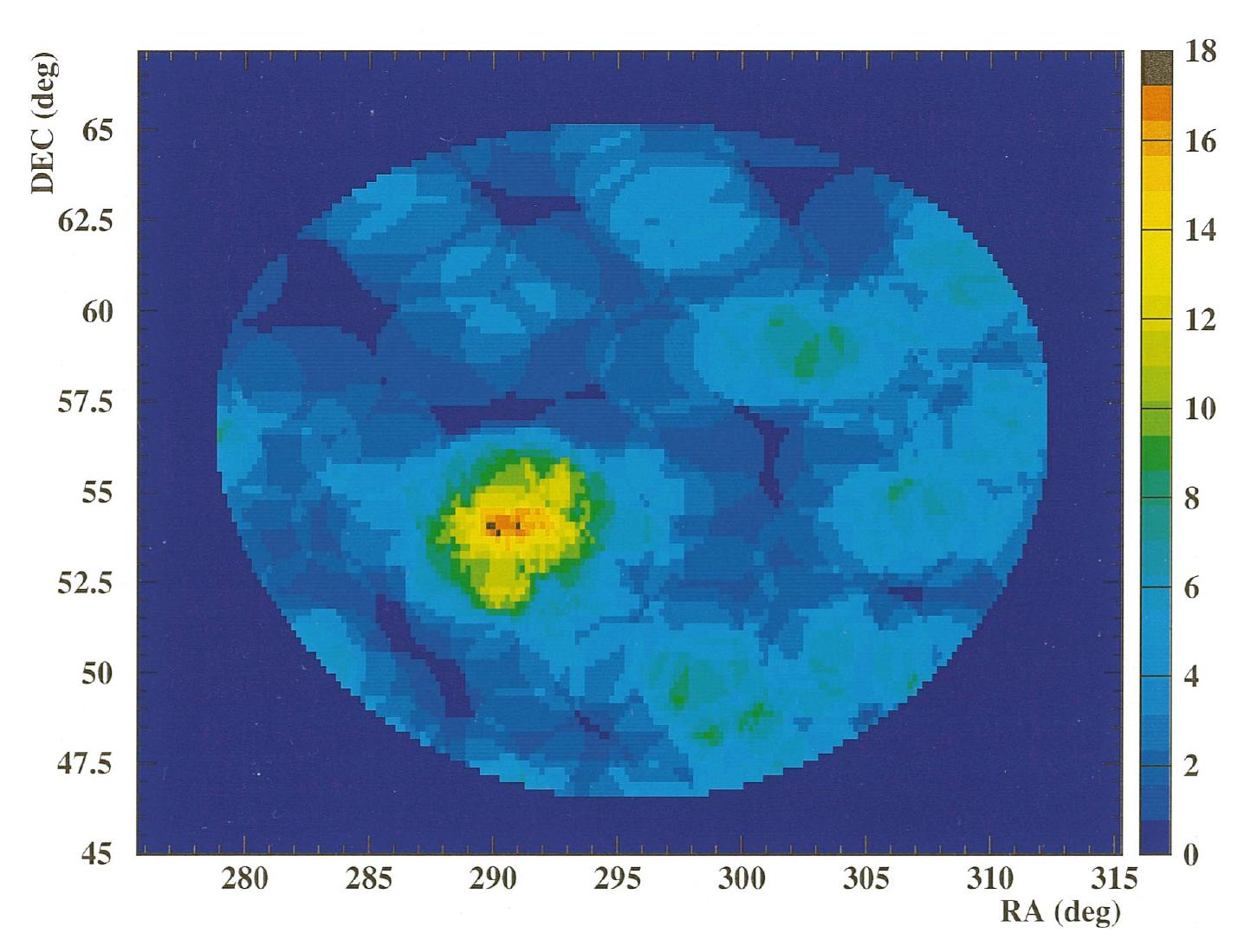}

\caption{\label{fig:GRB_Milagrito}Skymap of the events detected by Milagrito in
coincidence with GRB970417. The 18-event fluctuation in the skymap had a probability of 
only $1.5\times10^{-3}$ of being a mere fluctuation of the background, implying that 
it might had been generated by a gamma-ray emission. Source: \cite{GRB_2000_Atkins_etal} }
\end{figure}

Despite the absence of a definitive detection of VHE emission from GRBs,
there have been some hints of such emission in the observational data.
Milagrito, the prototype of Milagro, was used to perform observations
in coincidence with the 56 BATSE GRBs in its field of view. It
detected a fluctuation in coincidence in time and in the error box
of GRB970417 with a post-trials probability $1.5\times10^{-3}$ (or
$3\sigma$) (Fig. \ref{fig:GRB_Milagrito}) \cite{GRB_2000_Atkins_etal}.
However, the statistical significance was not high enough for a definitive
detection to be claimed. %

\begin{figure}[bt]
\includegraphics[width=1\columnwidth]{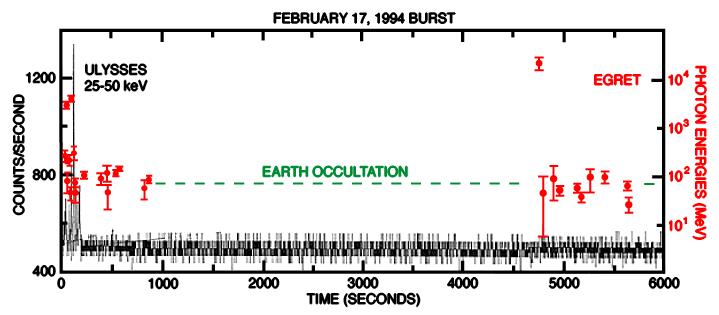}

\caption{\label{fig:GRB_Egret18Gev}Events detected by EGRET from GRB970217.
EGRET detected a $18\,GeV$ photon $\sim90\,min$ after the GRB. Source: \cite{URL_GRB970217}}
\end{figure}

BATSE, as mentioned above, was on board the CGRO along with other
instruments. One of them was EGRET, a gamma-ray detector sensitive
to energies extending past the high-energy threshold of BATSE (BATSE:
$20\,keV-2\,MeV$, EGRET: $30\,MeV-\sim30\,GeV$). EGRET performed coincident observations
on the bright GRBs detected by BATSE. In two of them, it detected
photons of $GeV$ energies. It observed two $3\,GeV$ photons from GRB970217
about the same time as the BATSE trigger, and one $18\,GeV$ photon $\sim90\,min$
later (Fig. \ref{fig:GRB_Egret18Gev}) \cite{GRB_Hurley_et_al1994}.
That $18\,GeV$ photon was the highest-energy photon ever detected from
a GRB. EGRET also detected a $10\,GeV$ photon from GRB910503 \cite{GRB_Schneid_et_al1992}.%

Gonz\'alez \textit{et al.} combined the
spectra of 26 bright GRBs measured by BATSE and EGRET \cite{GRB_Gonzalez_et_al2003}. The combined spectrum
of one of the bursts, GRB941017, had a high-energy tail extending
up to $200\,MeV$ that looked like an independent component.
That component appeared $\sim10-20\,s$ after the main burst and had
a roughly constant flux, while the lower-energy component decayed
by three orders of magnitude. The higher-energy component also had
a hard and rising spectral slope ($\sim1.0)$. Some time after the
main burst ($\sim150\,s)$, it contained more energy than the lower-energy
peak ($30\,keV-2\,MeV$). No evidence for a cutoff was seen for that
component, therefore it could continue up to $GeV/TeV$ energies. %
\begin{figure}[bt]
\begin{centering}
\includegraphics[width=0.8\columnwidth]{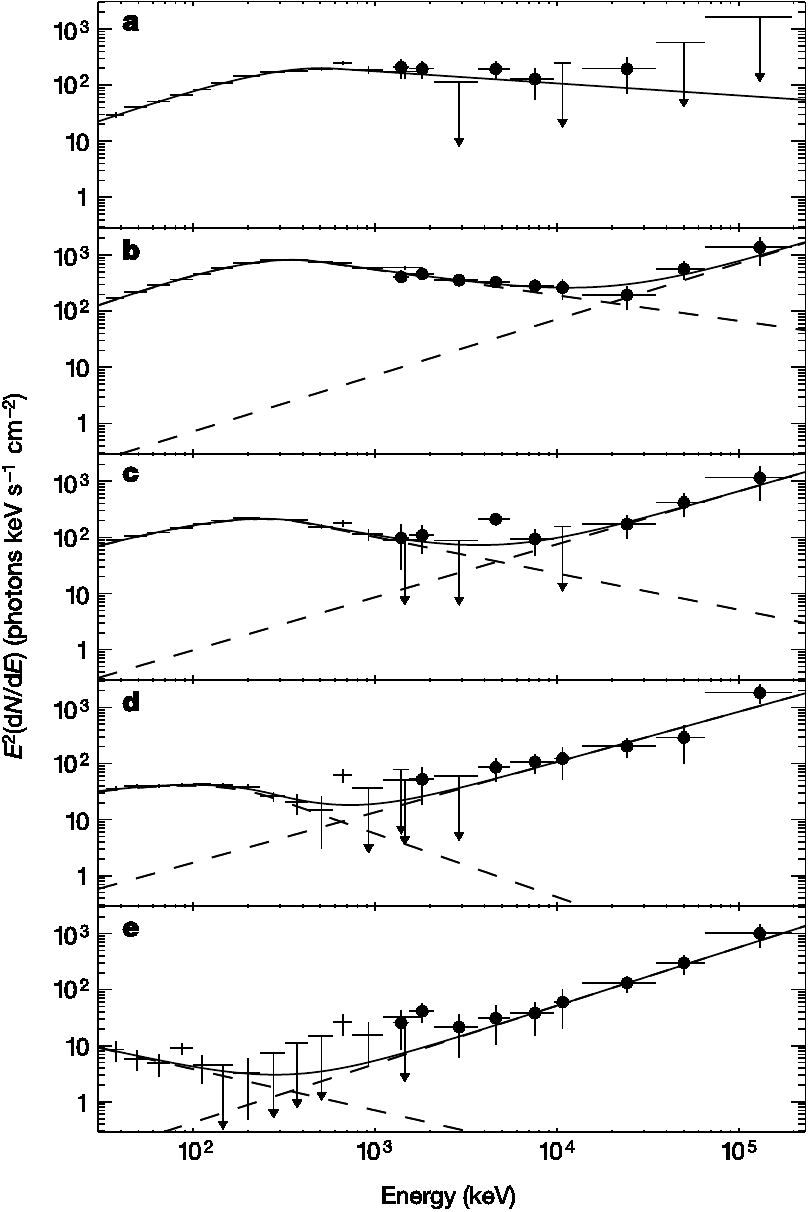}
\par\end{centering}

\caption{\label{fig:GRB_Magdas}The combined spectrum of GRB941017 measured
by BATSE and EGRET at different times relative to the BATSE trigger,
a ($-18,14\,s$), b ($14,47\,s$), c ($47,80\,s$), d ($80,113\,s$), and e ($113,211\,s$). Source:
\cite{GRB_Gonzalez_et_al2003}}

\end{figure}

\section{\label{sub:VHE_Emission}Radiation-Emission Processes}

VHE photons can be created in a GRB by both leptonic and hadronic
processes. Because the exact composition and the conditions at the
site of a burst are not known, the expected emission due to these
processes is only moderately constrained. Leptonic emission processes are
believed to produce the largest fraction of the VHE radiation, 
so they will be described in detail.

\subsection{Hadronic emission}

The protons of the GRB fireball can be accelerated to relativistic
energies (up to $10^{20}\,eV$) and emit synchrotron radiation that
can be up to $TeV$ energies \cite{GRB_Vietri_1997,GRB_Dermer_Humi_2001,GRB_Waxman_1995,GRB_Gupta_Zhang_2007}.
However, synchrotron emission from protons is weak, smaller by a factor
$(m_{e}/m_{p})^{2}$ of the synchrotron emission by electrons. 

In GRB internal and external shocks, various processes can create
neutral pions that later decay to higher energy gamma rays. These
processes are: \begin{eqnarray*}
p+p\rightarrow p+p+\pi^{0},\\
p+n\rightarrow p+n+\pi^{0},\\
p+\gamma\rightarrow\Delta\rightarrow\pi^{o}+p.
\end{eqnarray*}
The produced pions will be moving relativistically along the rest
of the fireball towards the observer. For that reason,
the energy of the pions and their decay photons will be 
relativistically boosted to higher energies by a factor of $\Gamma$, as observed by our
reference frame. Because of the high opacity of the GRB fireball,
no radiation over $\sim100\,GeV$ produced by pions at internal shocks
is expected to be observed. \cite{GRB_Razzaque_Meszaros_2006},
However, during the early afterglow and in the case of expansion into a low density
interstellar medium (ISM), these decay photons can be observed up to $TeV$ energies (Fig. 
\ref{fig:VHE_ExternalEmission_DiffDens}) \cite{GRB_Pe'er_Waxman_2005}. 

Charged pions also produce higher energy photons via synchrotron emission
of their decay electrons \cite{GRB_Gupta_Zhang_2007,GRB_Fan_Piran2008}.
These pions are produced by processes such as \begin{eqnarray*}
p+n\rightarrow n+n+\pi^{+},\\
p+p\rightarrow p+n+\pi^{+},\\
p+\gamma\rightarrow\Delta\rightarrow\pi^{+}+n,
\end{eqnarray*}
and they decay through\begin{eqnarray*}
\pi^{+} & \rightarrow & \mu^{+}+\nu_{\mu}\rightarrow e^{+}+\nu_{e}+\bar{\nu}_{\mu}+\nu_{\mu}\\
\pi^{-} & \rightarrow & \mu^{-}+\bar{\nu}_{\mu}\rightarrow e^{-}+\bar{\nu}_{e}+\nu_{\mu}+\bar{\nu}_{\mu}.\end{eqnarray*}
The energetic decay electrons and positrons emit synchrotron
radiation that could be in the $TeV$ energy range (especially the ones
from the last process) \cite{GRB_Fan_Piran2008}. However, the energy
radiated by this process is expected to be smaller by orders of magnitude
than the electron synchrotron component, unless only a very small
fraction of the thermal energy of the shocked material is carried
by the electrons $\le0.01$ \cite{GRB_Gupta_Zhang_2007,GRB_Fan_Piran2008}.

\subsection{\label{subsec:LeptonicEmission}Leptonic emission}

Because of the large amounts of energy deposited to the electrons of
the fireball during internal and external shocks, these electrons
are accelerated to ultra-relativistic velocities. There are two competing
processes through which they dissipate their energy. The first, as
already has been mentioned, is believed to be synchrotron radiation, responsible
for the observed $keV/MeV$ and lower-energy emission from GRBs. The
second process is inverse Compton scattering, in which the ultra-relativistic
electrons upscatter low-energy photons to higher energies. The energy
of a photon that underwent inverse Compton scattering is \cite{GRB_Fan_Piran2008}\begin{equation}
E_{ic}\simeq\frac{2\Gamma}{1+z}\frac{\gamma_{e}^{'2}E_{s\gamma}^{'}}{1+g},\label{eq:GRB_ICENergy}\end{equation}
where $\Gamma$ is the bulk Lorentz factor of the fireball; $z$
is the redshift of the burst; $\gamma_{e}^{'2}$ is the Lorentz factor
of the electron that caused the inverse Compton scattering; $E_{s\gamma}^{'}$
is the initial energy of the seed photon that underwent Inverse Compton
scattering, and $g\equiv\gamma_{e}^{'}E_{s\gamma}^{'}/m_{e}c^{2}$.
The parameters $\gamma_{e}^{'}$ and $E_{s\gamma}^{'}$ are for the
fireball frame of reference, and $E_{ic}$ is for the observer frame
of reference. In the Thomson regime ($g\ll1)$, equation \ref{eq:GRB_ICENergy}
becomes \begin{equation}
E_{ic,Thomson}\simeq\Gamma\gamma_{e}^{'2}E_{s\gamma}^{'}.\label{eq:GRB_ICEnergy_Thomson}\end{equation}
As can be seen, the upscattered energetic photons will have on average an
energy $\gamma_{e}^{2}$ times higher than the target photons. If
the energy of the seed photons is high ($g\gg1$) (for example, if
they have already underwent one inverse Compton scattering), then
we are in the Klein-Nishina regime, and relativistic and quantum-mechanical effects suppress the
cross section of inverse Compton scattering.

During internal shocks, the synchrotron photons generated by the electron
population can undergo inverse Compton scattering by that same electron
population. This process is called {}``Synchrotron Self-Compton''
(SSC), with {}``Self'' referring to the electrons that both produce
and upscatter the radiation. The typical Lorentz factor of the internal
shock electrons is $\gamma_{e}^{2}\sim10^{3}$ (in the fireball's
rest frame). Therefore, a typical synchrotron photon of $\sim300\,keV$
energy (from our reference frame) will be upscattered to an energy
$\sim10^{3\times2}$ times higher, equal to a few hundred $GeV$. This process
is believed to produce a second $GeV/TeV$ peak at the GRB spectra, similar to
the one observed in blazar spectra. If the X-ray flare photons ($E\sim10\,keV)$
are created by the synchrotron radiation of late internal shocks,
then they can also be upscattered to $\sim{}GeV$ energies \cite{GRB_Guetta_Granot_2003}.
In the alternative case that the X-ray flare photons are produced
by shocks between slowly moving and fast moving matter ejected simultaneously
during the onset of the prompt emission (refreshed shocks), these
photons can be upscattered to $GeV/TeV$ energies \cite{GRB_Galli_Guetta_2008}.

Synchrotron emission and inverse Compton scattering are competing
processes. The cooling time through synchrotron emission is $t_{syn}=6\pi\,m_{e}\,c/(\sigma_{T}\,B^{2}\,\gamma_{e})$,
where $\sigma_{\tau}$ is the Thomson cross section, and $B$ is the
magnetic field. The cooling time through inverse Compton scattering
can be written as $t_{IC}=t_{syn}/Y$, where $Y$ is {}``Compton
Y parameter''. $Y$ is given by \cite{GRB_Sari_et_al_1996}\begin{equation}
\label{VHEGRB_ComptonY}
Y=\begin{cases}
\frac{\epsilon_{e}}{\epsilon_{B}} & ,\,\,\frac{\epsilon_{e}}{\epsilon_{B}}\ll1\\
\sqrt{\frac{\epsilon_{e}}{\epsilon_{B}}} & ,\,\,\frac{\epsilon_{e}}{\epsilon_{B}}\gg1,\end{cases}\end{equation}
where $\epsilon_{e}$ and $\epsilon_{B}$ are the fractions of the
shocked material's energy carried by electrons and the magnetic field,
respectively. Depending on the relative magnitudes of $\epsilon_{e}$
and $\epsilon_{B}$, cooling either through synchrotron emission or through
inverse Compton scattering dominates. Cooling through inverse Compton
scattering is only important for $\epsilon_{e}>\epsilon_{B}$. 

Pe'er and Waxman \cite{GRB_Pe'er_Waxman_2004} calculated the leptonic
emission from internal shocks inside the GRB fireball. Their time-dependent
numerical calculations included all the relevant physical processes:
cyclo-synchrotron emission, synchrotron self-absorption, inverse and
direct Compton scattering, $e^{-}e^{+}$ pair production and annihilation,
and the evolution of high-energy electromagnetic cascades. 

\begin{figure}[bt]
\begin{centering}
\includegraphics[width=0.8\columnwidth]{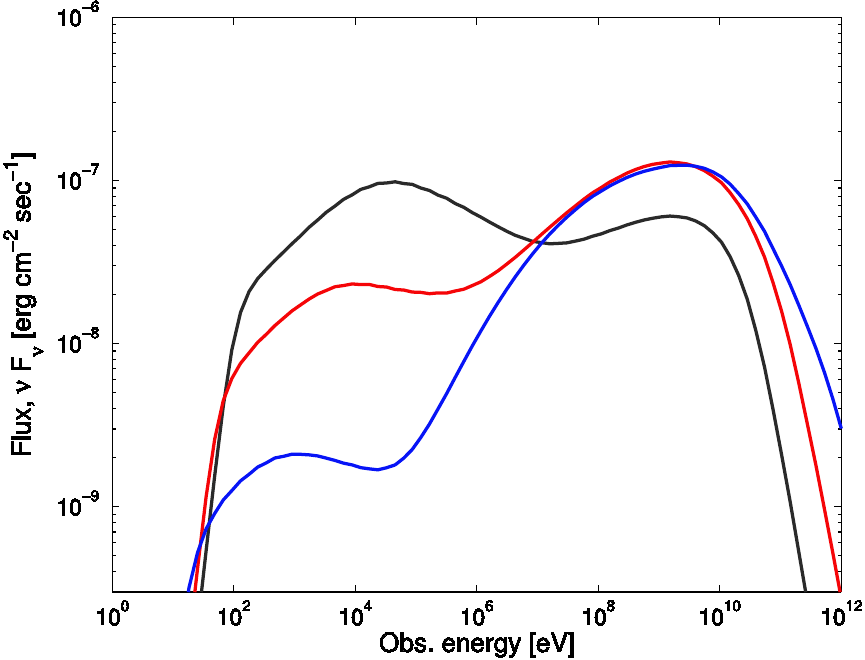}
\par\end{centering}

\caption{\label{fig:VHE_DiffEB}Synchrotron and SSC radiation produced at internal
shocks for different fractions of the jet's energy carried by the
magnetic field $\epsilon_{B}$: black $\epsilon_{B}=0.33$, red $\epsilon_{B}=10^{-2}$,
blue $\epsilon_{B}=10^{-4}$. Here $\Gamma=600$, $\epsilon_{e}=0.316$,
$\Delta{}t=10^{-3}$, $l'=0.8$, and the GRB has a redshift $z=1$.
The calculation is for a low-opacity fireball. The higher $\epsilon_{B}$ is, the smaller
the fraction of the electrons' energy dissipated through inverse Compton
scattering. Source: \cite{GRB_Pe'er_Waxman_2004}}
\end{figure}

\begin{figure}[bt]
\begin{centering}
\includegraphics[width=0.8\columnwidth]{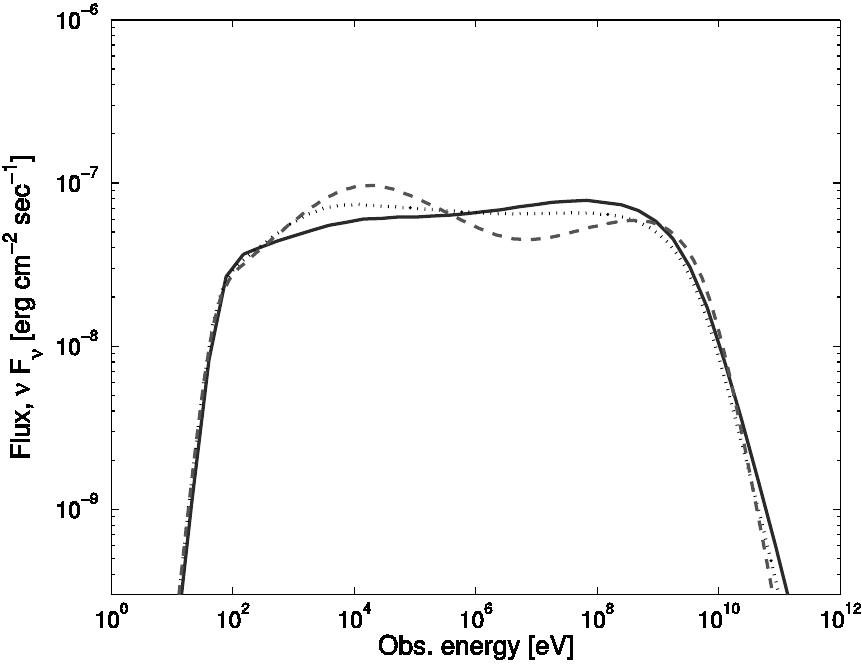}
\par\end{centering}

\caption{\label{fig:VHE_DiffP}Synchrotron and SSC radiation produced at internal
shocks for different power-law index $p$ of the accelerated electrons
($N_{e}(E)\propto E^{-p})$: \emph{solid line} ($p=2.0$), \emph{dotted
line} ($p=2.5$), \emph{dashed line} ($p=3.0$). The bulk Lorentz factor of the jet is $\Gamma=300$,
the GRB has a redshift $z=1$, and the fireball has a low opacity.
The SSC spectrum depend weakly on $p$.
Source: \cite{GRB_Pe'er_Waxman_2004}}
\end{figure}

Figure \ref{fig:VHE_DiffEB} is part of Pe'er and Waxman's results, and shows the effect
of the ratio $\epsilon_{e}/\epsilon_{B}$ on the resulting spectral
energy distribution. The first peak in the figure comes from synchrotron
emission, and the second higher-energy peak comes from inverse Compton
scattering (SSC). As can be seen, the higher $\epsilon_{B}$ is, the
larger the amount of energy dissipated by synchrotron emission. The fact that 
the emission in VHE energies can be larger than the $keV/MeV$ energies is important for 
this work. First, it implies an increased chance of Milagro detecting a GRB, and second,
it predicts a population of bursts with a very strong VHE emission that can be 
excluded in case this search produces null results. The results presented in plot \ref{fig:VHE_DiffEB}
are for a relatively transparent to VHE photons fireball. If the opacity
of the emitting region is moderate or high, most of the $E\gtrsim1GeV$
radiation is expected to be internally absorbed. 
Figure \ref{fig:VHE_DiffP} shows that the emitted spectrum of
the SSC component depends weakly on the power-law
index $p$ of the electrons $N_{e}(E)=E^{-p}$.

\begin{figure}[bt]
\begin{centering}
\includegraphics[width=0.8\columnwidth]{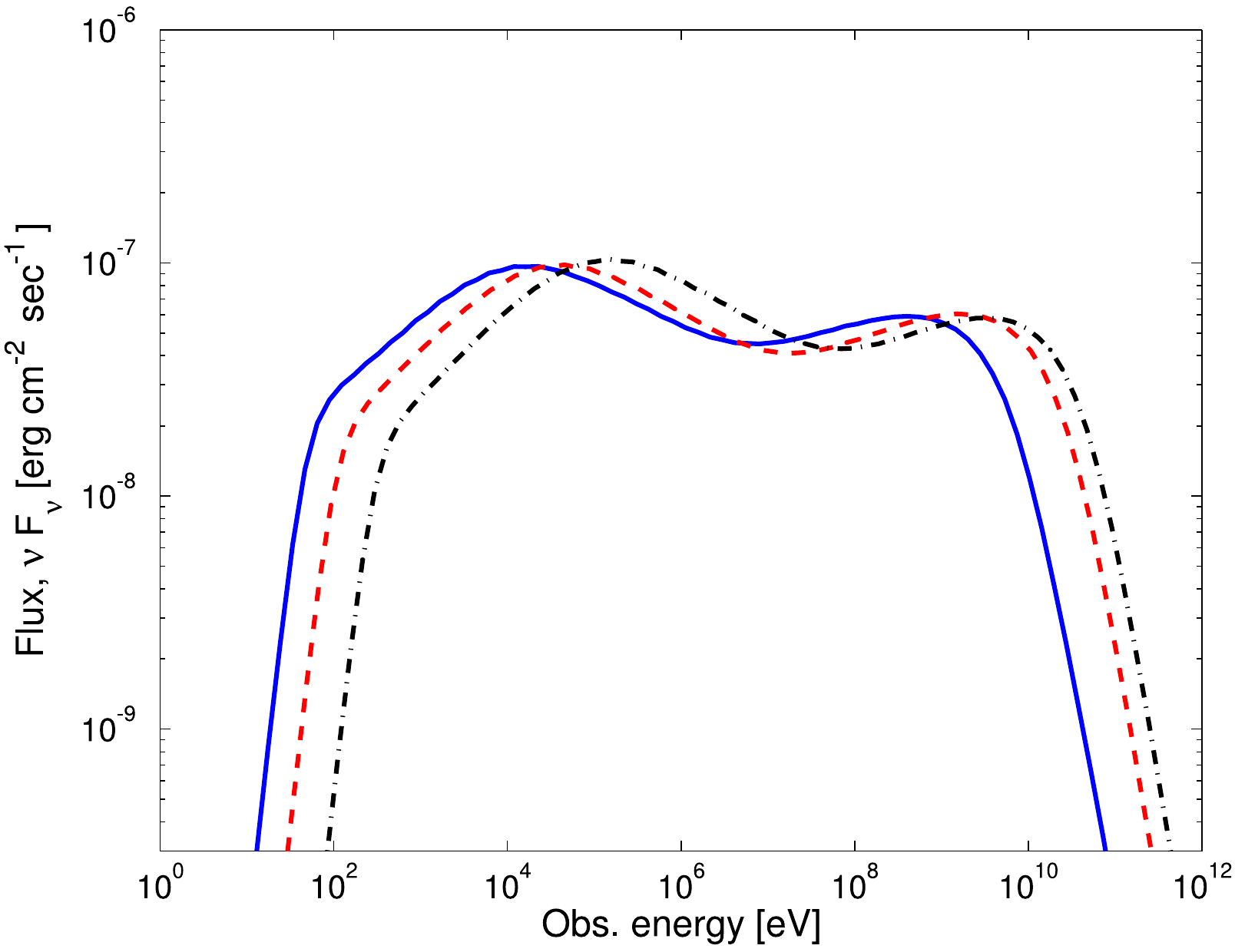}
\par\end{centering}

\caption{\label{fig:VHE_DiffGamma}Synchrotron and SSC radiation produced at
internal shocks for different bulk Lorentz factors $\Gamma$ of the
jet. All the spectra correspond to a low opacity. The time scale of
the variability $\Delta{}t$ was adjusted so that the opacity was low
for each $\Gamma$. \textit{Blue line}: ($\Gamma=300$, $\Delta t=10^{-2}s$,
$l'=2.5$), \emph{red line}: ($\Gamma=600$, $\Delta t=10^{-4}s$,
$l'=0.6$),\emph{ black line}: ($\Gamma=1000$, $\Delta t=10^{-4}s$,
$l'=0.6$). The GRB has a redshift $z=1$. The fraction of the fireball's
energy carried by the electrons $(\epsilon_{e})$ and the magnetic
field $(\epsilon_{B})$ are $\epsilon_{e}=\epsilon_{B}=0.316$. 
The higher the bulk Lorentz factor is, the higher the energy of the
GRB emission. Source: \cite{GRB_Pe'er_Waxman_2004}}

\end{figure}

\begin{figure}[bt]
\begin{centering}
\includegraphics[width=0.8\columnwidth]{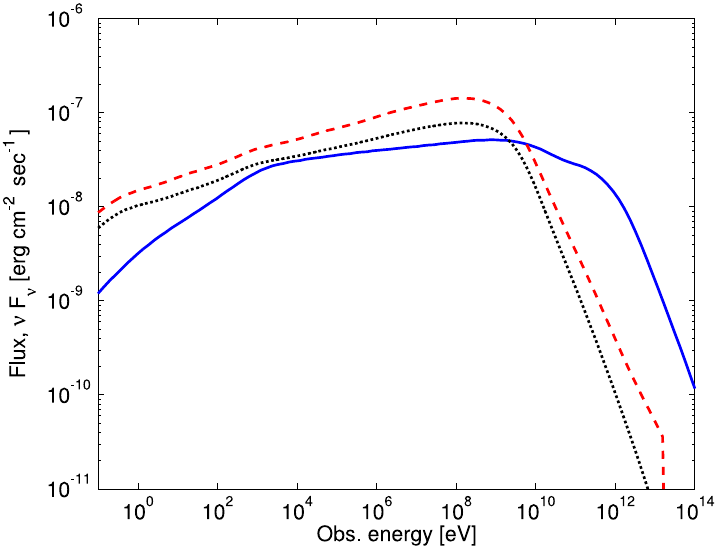}
\par\end{centering}

\caption{\label{fig:VHE_ExternalEmission_DiffDens}Radiation produced by electron
and proton synchrotron emission, inverse Compton scattering, and neutron
pion decay at external shocks for different kinds of circumburst media.
\emph{Solid}: expansion into a uniform low-density ISM ($n\simeq1/cm^{3}$), \emph{
dashed}: expansion into a wind with $A=5\times10^{11}g/cm$,
\emph{dotted}: expansion into a wind with $A=5\times10^{11}g/cm$
with the contribution from pion decays and proton synchrotron emission
omitted. The contribution of pion decays and proton synchrotron emission
is negligible for the case of expansion into an ISM. Here $\epsilon_{e}=10^{-1}$, 
$\epsilon_{B}=10^{-2}$, $p=2$, $\Gamma=10^{2.5}$, and $z=1$. No absorption effects through interactions
with the EBL were applied (see chap. \ref{chap:IR}).
Source: \cite{GRB_Pe'er_Waxman_2005}}
\end{figure}

Pe'er and Waxman also calculated the GRB emission from the early afterglow
in a time scale of tens to hundreds of seconds following the GRB.
Similarly to their other work mentioned above, they explored the dependence
of the emitted spectrum on various uncertain model parameters, in
particular the energy density of the magnetic field, the power-law
index of the accelerated particles, and the density of the circumburst
medium. As mentioned in section \ref{sub:VHE_ExternalAbs}, the density
of the surrounding medium is different for collapsars and binary mergers.
Figure \ref{fig:VHE_ExternalEmission_DiffDens} shows the emitted
spectra from synchrotron and inverse-Compton emission in the early
afterglow for the two different kinds of circumburst media. A comparison
of these spectra with the spectra of the emission from internal shocks
can be made. The red line in figure \ref{fig:VHE_DiffEB} shows a
spectrum from internal shocks for a burst with similar properties
to the ones in figure \ref{fig:VHE_ExternalEmission_DiffDens}. (The
only difference is that the plot for the external-shock case corresponds
to a lower bulk Lorentz factor: 316 vs 600). In the energy
range of interest for Milagro ($E>100\,GeV$), both kinds of shocks produce
similar spectra and energy fluxes. Perhaps external shocks produce
a larger amount of $E>1\,TeV$ radiation than internal shocks, but this
can be explained by the different bulk Lorentz factors between the
two plots.  %

\begin{figure}[bt]
\begin{centering}
\includegraphics[width=0.8\columnwidth]{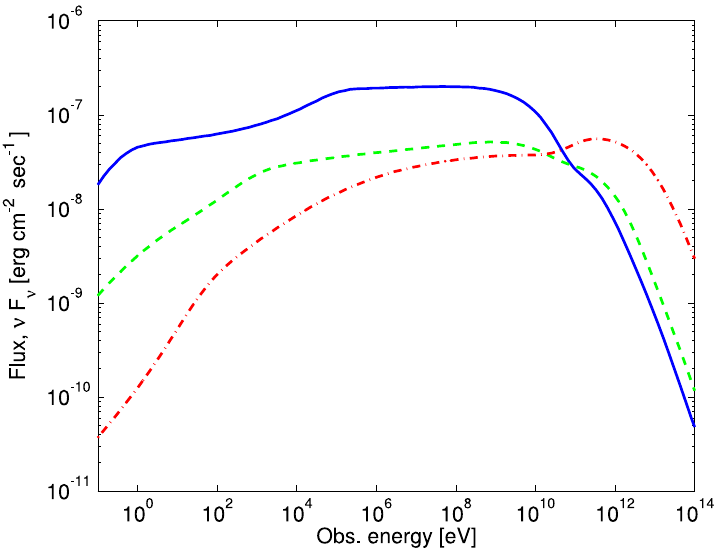}
\includegraphics[width=0.8\columnwidth]{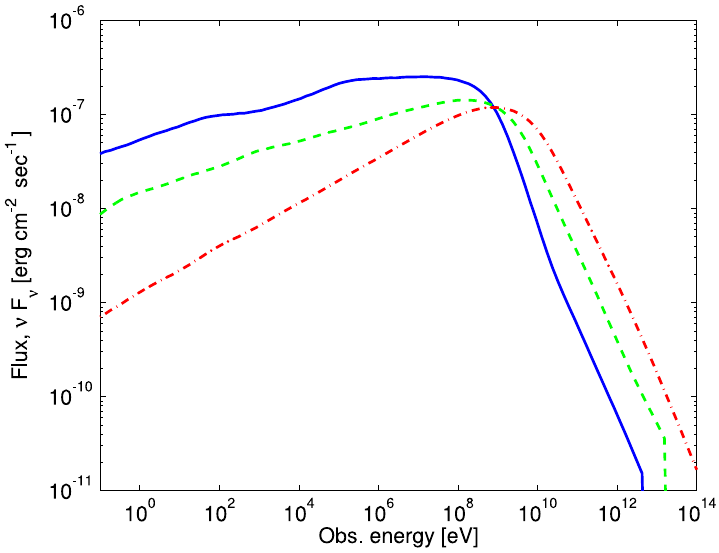}
\par\end{centering}

\caption{\label{fig:VHE_ExternalEmission_DiffEb}Synchrotron and inverse-Compton
radiation produced at external shocks for different kinds of circumburst
media and fractions of the thermal energy carried by the electrons
$\epsilon_{e}$ and the magnetic field $\epsilon_{b}$. \emph{Top}:
expansion into a low density ISM, \emph{bottom}: expansion into a stellar
wind. \emph{Solid}: ($\epsilon_{e}=10^{-5}$, $\epsilon_{b}=10^{-5})$,
\emph{dashed}: ($\epsilon_{e}=10^{-1}$, $\epsilon_{b}=10^{-2})$,
\emph{dash-dotted}: ($\epsilon_{e}=10^{-1}$, $\epsilon_{b}=10^{-4})$.
Here $p=2$, $\Gamma=10^{2.5}$, $z=1$. No absorption effects from
interactions with the EBL were applied (see chap. \ref{chap:IR}).
Source: \cite{GRB_Pe'er_Waxman_2005}}

\end{figure}

Figure \ref{fig:VHE_ExternalEmission_DiffEb} shows the emitted spectra
for different fractions of thermal energy carried by the electrons
and the magnetic field. Similar to the case of internal shocks, the
larger the relative fraction of energy carried by the electrons, the
stronger the higher-energy Inverse-Compton emission. The case of expansion
into the ISM combined with
a low fraction of thermal energy carried by the magnetic field (red
dash-dotted line in left figure), can lead to emission extending to
tens of TeV, which is easily detectable by Milagro from a nearby $(z\lesssim0.1)$
burst. 

The spectra have a small dependency on the index $p$ of the power-law distribution
of the electron's energies, and stay flat ($dN/dlog\nu\simeq constant$)
for most of the energy range.

\section{\label{sub:VHE_Absorption}Internal-Absorption Processes}

Because of the high density of the GRB fireball, the $GeV/TeV$ photons
created in it can be absorbed before managing to escape. There are
various processes that contribute to the opacity of the GRB fireball,
such as Compton scattering, $e\gamma\rightarrow e^{-}e^{+}$, $\gamma\gamma\rightarrow e^{-}e^{+}$.
The dominant process is pair creation after the scattering of the
high energy ($E>1\,MeV)$ photons with lower-energy photons of the fireball
($\gamma\gamma\rightarrow e^{-}e^{+}$). In that process, a photon
with energy $\epsilon_{\gamma}$ can annihilate with another photon
of energy $\gtrsim(m_{e}c^{2})^{2}/2\epsilon_{\gamma}$, where $m_{e}$
is the electron mass, creating an electron-positron pair. It is likely
that the opacity at the GRB site varies from burst to burst, depending
on the local conditions, making it hard to account for. For that reason,
this study searched and placed limits on the VHE signal \emph{emitted}
by a GRB, instead on the VHE signal \emph{generated} at it.

In the following two subsections, the internal absorption of the VHE emission generated at 
internal and external shocks will be described.

\subsection{Absorption of the emission from internal shocks}

The opacity of the emitting region depends on the radiation density.
A high radiation density provides an abundance of lower-energy target
photons with which the higher-energy photons can annihilate. The radiation
density is proportional to the luminosity of the emitting region,
and inversely proportional to its size. An estimate of the size of
the emitting region can be provided by the variability time scale
of the prompt emission light curve. Each spike in the prompt light
curve corresponds to one internal shock. Therefore, emission from
an internal-shock region of width $(\Delta{}R$) will create a spike
in the GRB light curve of duration $\Delta{}t$. If the fireball is
moving towards us with a bulk Lorentz factor $\Gamma$, then $\Delta{}R=\Gamma\,c\,\Delta{}t$,
with $\Delta{}R$ measured in the burst frame. Using $\Delta{}R$ and
the photon luminosity produced from the shock $L$, the ``comoving
compactness'' parameter $l'$ can be calculated as $l'=\Delta{}R\, n_{\gamma}^{'}\,\sigma_{T}$,
where $n_{\gamma}^{'}=\epsilon_{e}L/(4\pi\,m_{e}\,c^{3}\,\Gamma^{2}\,r_{i}^{2})$
is the comoving number density of photons with an energy $(E_{ph})$ that
exceeds the electron's rest mass ($E_{ph}\ge{}m_{e}\,c^{2}$); $\epsilon_{e}$
is the fraction of the post-shock thermal energy carried by the electrons;
$r_{i}\simeq2\,\Gamma^{2}\,c\,\Delta{}T$ is the radial distance of the shock
from the center of the system; and $\sigma_{T}$ is the Thomson cross
section \cite{GRB_Pe'er_Waxman_2004}. 

\begin{figure}[bt]
\begin{centering}
\includegraphics[width=0.8\columnwidth]{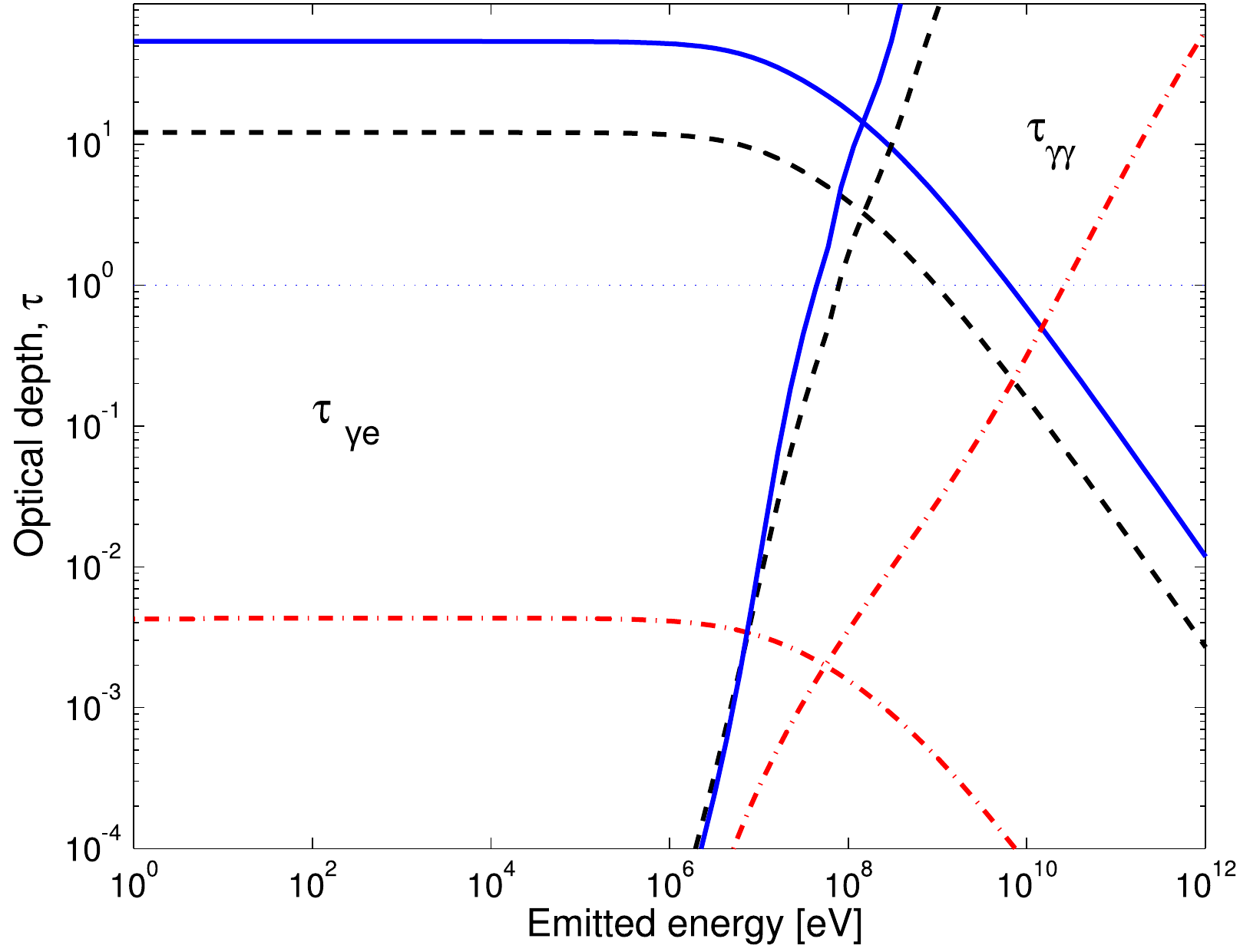}
\par\end{centering}

\caption{\label{fig:VHE_OpticalDepths}Energy-dependent optical depths for
pair production and scattering. \emph{Solid lines} ($\Delta t=10^{-5}\,s$,
$\Gamma=300$, $l'=2500$), \emph{dashed black lines }($\Delta t=10^{-4}\,s$,
$\Gamma=300$, $l'=250$), \emph{dashed red lines }($\Delta t=10^{-4}\,s$,
$\Gamma=1000$, $l'=0.6$). Source \cite{GRB_Pe'er_Waxman_2004}}

\end{figure}

The compactness parameter gives
a measure of the opacity of the burst. High-compactness conditions
($l'\gtrsim20$) will result in a suppressed VHE emission. Figure
\ref{fig:VHE_OpticalDepths} shows the optical depths for pair production
and scattering for configurations with different values of the compactness. As
can be seen, the optical depth due to pair production is expected
to be very high for photons of energies $E\gtrsim100\,MeV$ unless the
compactness is very low (red dashed curve).  %

While the opacity of the GRB fireball constitutes an important factor
that limits our prospects of detecting VHE emission from GRBs, it also
significantly increases the importance of that detection.
Based on the dependence of the opacity on $\Gamma$,
a detection can be used to place lower limits on $\Gamma$, providing
information on a key ingredient of the GRB model.

\subsection{\label{sub:VHE_ExternalAbs}Absorption of the emission from external
shocks}

\begin{figure}[bt]
\begin{centering}
\includegraphics[width=0.8\columnwidth]{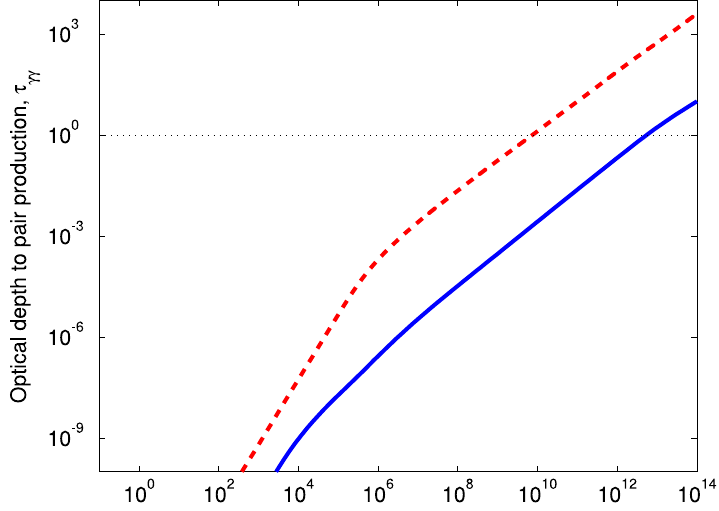}
\par\end{centering}

\caption{\label{fig:VHE_OpticalDepthsExternal}Energy-dependent optical depths
from pair production. \emph{Solid line}:\emph{ }explosion into the
inter-stellar medium, \emph{dashed line}:\emph{ }explosion into a
stellar wind. Source: \cite{GRB_Pe'er_Waxman_2005}}

\end{figure}

Pe'er and Waxman calculated the opacity to the radiation produced
during the early afterglow, tens to hundreds of seconds following
the prompt emission \cite{GRB_Pe'er_Waxman_2005}. Similarly to the case of
absorption in internal shocks, the opacity is
due primarily to pair production and depends on the density of the
medium. The emission from external shocks is produced by the jet's
interactions with the medium surrounding the burst. The density of
the circumburst medium depends on the progenitor of the GRB. In the compact-binary
merger scenario (responsible for short GRBs), a number density similar to
that of the interstellar medium ($n\simeq1/cm^{3})$ is expected.
On the other hand, in the collapsar scenario (responsible for long
GRBs), stellar winds, such as from the Wolf-Rayet star, can create a 
higher-density medium surrounding the GRB. Such a stellar wind creates a density profile $\rho(r)=A/r^{2}$,
with a typical value for $A=5\times10^{11}\,g/cm$. This corresponds
to volume densities of $n\simeq10^{3}-10^{4}/cm^{3}$, which are significantly
higher than the typical density of the ISM. Figure \ref{fig:VHE_OpticalDepthsExternal}
shows the optical depths for the two types of surrounding medium. As can be
seen, the optical depth for $GeV-TeV$ emission is considerably lower 
for expansion into a low-density interstellar medium, a case which corresponds
to short GRBs.

   \clearpage \chapter{\label{chap:IR}Absorption of the VHE Emission from GRBs by the EBL}

\section{Introduction}

In the previous chapter it was shown that GRBs can emit VHE photons 
during both their prompt and afterglow phases.
However, a part of this emission is absorbed during its passage
through the extragalactic space. Specifically, VHE photons from GRBs
annihilate with low-energy photons of the Extragalactic
Background Light (EBL) and produce $e^{-}e^{+}$ pairs: $\gamma_{VHE}\gamma_{EBL}\rightarrow e^{-}e^{+}$.
As a result, a big part of the VHE emission from GRBs is absorbed
before reaching the earth. Milagro's sensitivity to GRBs (and to any other extragalactic
source) depends on the amount of this absorption.

The purpose of this chapter is to present the different models of the
EBL, and the effects of EBL absorption on the VHE emission from GRBs. 
Section \ref{sec:IR_EBL} describes the EBL. Section \ref{sub:IR_Models} provides an
overview of the currently available models, information
on how they compare with the observational constraints, and 
justification for the model chosen to be used in this study.
Finally, section \ref{sub:IRAbs_Effects} 
presents the effects of EBL absorption on the VHE emission from GRBs. 

\section{\label{sec:IR_EBL}The EBL}
\begin{figure}[ht]
\begin{centering}
\includegraphics[width=0.8\columnwidth]{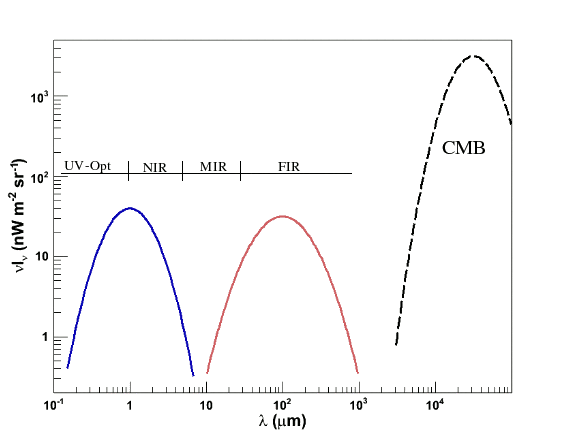}
\par\end{centering}

\caption{\label{fig:GRBSensi_SED_EBL}Schematic EBL spectrum as a function
of wavelength. The EBL consists of two kinds of light: redshifted
starlight \emph{(blue line)} and starlight absorbed and remitted by
dust \emph{(red line)}. The \emph{dashed} \emph{line} corresponds
to the cosmic microwave background, which is shown here only for comparison
purposes, since it is not considered part of the EBL. Source \cite{IR_Reyes_2007}}
\end{figure}

The EBL is a diffuse photon background which fills the space between
the galaxies. A schematic of the EBL spectrum is shown in Figure \ref{fig:GRBSensi_SED_EBL}.
The near-infrared ($\lambda\sim2-3\,\mu{}m$) and optical ($\lambda\sim0.5\,\mu{}m$)
part is redshifted starlight, and the $\lambda\sim150\,\mu{}m$
part of the spectrum is starlight absorbed by dust and re-emitted
in the far-infrared. Because of the narrowness of the pair-production
cross section, for broad-band photon spectra over half of the interactions
of a gamma ray of energy $E$ occur with a quite narrow
interval of target photons, $\Delta\lambda\sim(1\pm1/2)\lambda^{*}$,
centered on $\lambda^{*}\simeq1.5(E/1TeV)\,\mu{}m$. This means that
gamma rays of energy $100\,GeV$ ($100,TeV$) will interact with EBL
photons of wavelength about $0.15\,\mu{}m$ ($150\,\mu{}m$) which corresponds
to the ultraviolet (infrared) part of the spectrum. The magnitude
of the absorption depends on the column density of the background
photons between the source and the observer, which means that
knowledge of the EBL density for different redshifts is needed. 

\FloatBarrier
\section{\label{sub:IR_Models}EBL Models}
The measurements on the present-epoch EBL density (z=0) provides a
constraint on the integrated energy release in the universe. 
It is difficult to measure today's EBL directly, especially in the mid-infrared region,
due to strong foregrounds. Lower limits can be set by
source counts, while upper limits can be loosely set by direct measurements.
However, the density of the EBL at different redshifts cannot be
constrained by measuring the cumulative energy output only. For that
reason, there are multiple models that try to calculate the density
of the EBL at various redshifts and wavelengths. These models approach
the problem with different methods, degrees of complexity, observational
constraints and, inputs\footnote{
The following information on the various models was mostly based on
the detailed overviews in \cite{IR_Hauser_Dwek2001,IR_Reyes_2007}. %
}: 

\begin{itemize}
\item The backward-evolution models \cite{IR_Stecker_et_al2001,IR_Stecker_et_al2006} 
extrapolate present-day data or template spectra of local galaxies
to higher redshifts. They are simple and easily verifiable
since they predict quantities that can be compared with
observations. However, they do not include known processes occurring
in galaxies such as star formation, and re-emission of radiated
power by dust. 
\item The forward-evolution models predict the temporal evolution of galaxies
as a function of time, starting at the onset of star formation. 
In general, they have been proven successful
in fitting the spectra of individual galaxies, galaxy number counts
in specific bands, and the general characteristics of the EBL. However, they
do not include galaxy interactions and stochastic changes in
the star formation rate. They
fail to match the $850\,\mu{}m$ galaxy number counts without including
a new population of ultraluminous infrared galaxies. 
\item The semi-analytical models \cite{IR_Primack_et_al2005} adopt the approach of the forward-evolution
models, but they also include simulations of structure formation.
This way, they can make predictions about the observable characteristics
of galaxies and the intensity and spectrum of the EBL. These models
take into account multiple physical processes, such as the cooling
of gas that falls in the halos, the star formation, and the feedback
mechanisms that modulate the star formation efficiency. In spite of
their successes, there remain some discrepancies between their predictions
and observations. The origins of these discrepancies are difficult
to trace because of the inherent complexity of the models and the
multitude of parameters needed by them. 
\item The EBL provides an integrated over-time view of the energy release
by a wide variety of physical processes and systems that have populated
the universe. So, it is expected to be dependent mostly on the global
characteristics of cosmic history. Thus, chemical-evolution
models deal with the history of a few of the globally-averaged properties
of the universe instead of trying to model the complex processes that
determine galaxy formation, evolution, and emission. The main advantages
of these models are their global nature, their intrinsic simplicity,
and the fact that they do not require detailed knowledge of the
processes involved in the evolution of galaxies. They provide a picture
of the evolution of the mean density of stars, interstellar gas, metals,
and radiation averaged over the entire population of galaxies. They
have been successful in reproducing the generic spectral shape of
the EBL, but they fall short of some UV-optical and near-infrared measurements. 
\end{itemize}
In general, most models predict similar cosmic infrared background
spectra from $\sim5-1000\,\mu{}m$, mostly because they use similar cosmic
star formation histories. Backward evolution models assume a rising
SFR up to $z\sim1-1.5$ with a nearly constant rate at earlier times.
Forward-evolution and semi-analytical models try to reproduce the
same SFR in order to fit number counts or comoving spectral luminosity
densities at different redshifts. Larger differences occur in the
predictions regarding the UV-optical spectral range of the EBL. Backward-evolution
models do not include the physical processes that link the cosmic
infrared background and the UV-optical part of the spectrum. Some
of them try to amend this but incorporating template spectra. Other
models, naturally arrive at a doubly-peaked EBL because they explicitly
include the absorption of starlight and the following re-emission
by dust. 

Primack \textit{et al.} \cite{IR_Primack_et_al2005} (PR), use a semi-analytical
model, which in general predicts lower optical depths for nearby sources
$z\lesssim2$ than the other models. Stecker \textit{et al.} (ST) \cite{IR_Stecker_et_al2001,IR_Stecker_et_al2006}
use a backward evolution model that has been frequently updated using
new data. Their model predicts a large UV photon density and consequently
a higher gamma-ray opacity at high redshifts. Kneiske \textit{et al.} \cite{IR_Kneiske_et_al2004}
(KN04) use a chemical-evolution model for the UV-optical part of the
EBL and, based on recent
deep galaxy surveys, a backwards-evolution for the infrared part.

\begin{figure}[ht]
\begin{centering}
\includegraphics[width=1\columnwidth]{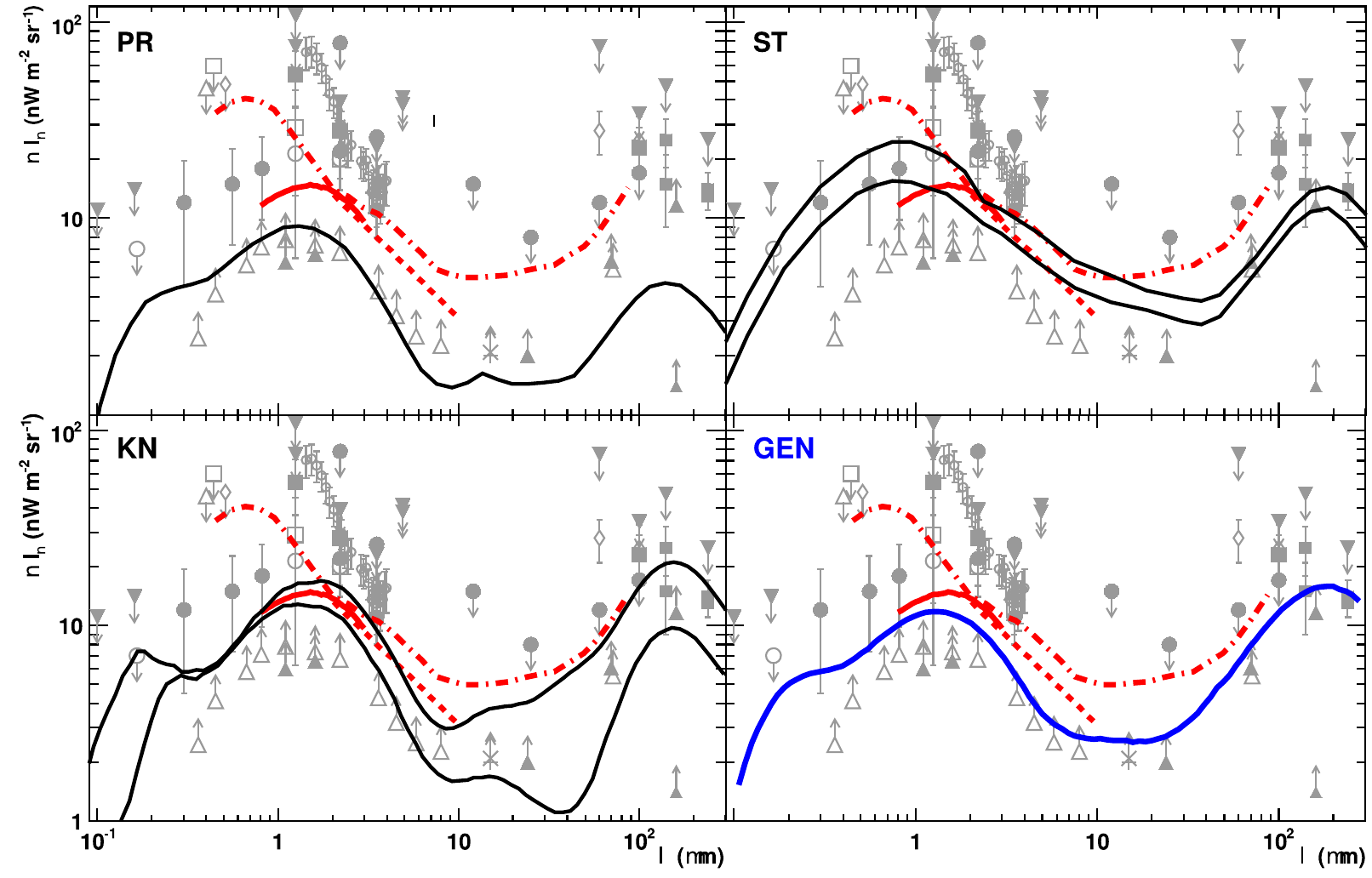}
\par\end{centering}
\caption{\label{fig:Sensi_EBLModels}Comparison between the various predictions on
the present-day EBL density and the observational constraints. Black
solid curves show the predictions from different EBL models. Grey markers are measurements
and limits from direct measurements, fluctuation analyses, and source counts.
Red curves are upper limits derived from VHE blazar
gamma-ray spectra. \emph{Upper left}: model from Primack \textit{et al.} \cite{IR_Primack_et_al2005}.
\emph{Upper right}: fast evolution and baseline EBL models from Stecker \textit{et al.} \cite{IR_Stecker_et_al2001,IR_Stecker_et_al2006}.
\emph{Lower left}: high and low 2004 models from Kneiske \textit{et al.} \cite{IR_Kneiske_et_al2004}.
\emph{Lower right}: generic EBL model from Rauen \& Mazin\cite{IR_Mazin_Raue2007_2,IR_Mazin_Raue_2007,IR_Raue_Mazin_2008}
(blue curve). All but the last one models are in disagreement with at least one observational constraint. Source \cite{IR_Raue_Mazin_2008}}

\end{figure}

The validity of the results of these models depends on how they compare
with the existing observational constraints. In 2008, Raue \& Mazin \cite{IR_Raue_Mazin_2008}
performed this comparison for most of the existing models (Fig.
\ref{fig:Sensi_EBLModels}). They found that both Stecker 2006
models \cite{IR_Stecker_et_al2006} are over the upper limits set by recent blazar measurements by HESS \cite{IR_Aharonian_et_al2006,IR_Aharonian_et_al2007,IR_Aharonian_et_al2007_2,IR_Aharonian_et_al_2007_3},
that Primack 2005 model \cite{IR_Primack_et_al2005} is under the lower limits set on the mid-infrared
component of the EBL by Spitzer \cite{IR_Aharonian_et_al2007,IR_Fazio_et_al2004}, and that Kneiske 2004 high-stellar-UV
and low-IR models \cite{IR_Kneiske_et_al2004}, respectively are either over some upper limits or under some lower limits. %

Concluding that all the EBL models they tested disagree with at least one observational
constraint, Raue \& Mazin proposed a ``generic-EBL'' shape that lies in
the yet not excluded range. Their EBL is not based on a theoretical
framework like all the above-mentioned models, but instead is just
one of the many EBL shapes that is not currently excluded by observations.
The special property of the chosen model is that it lies just above
the lower limits. Because of the lack of a theoretical framework,
they can only provide the present-epoch (z=0) EBL. They do, however,
provide a ``generic-evolution'' prescription for the EBL, in which
they calculate the EBL density at a higher redshift $z$ by scaling
the present-day EBL by $(1+z)^{3-f_{evo}}$, where $f_{evo}>0$. However, 
their generic evolution method agrees with the evolution predicted
by other complex models (PR and KN04), only up to $z\sim0.5$. 

Recently, and after private communication with the author, T. Kneiske
modified the normalization of the star-formation rate for one of her
models (warm-dust model from \cite{IR_Kneiske_et_al2004}) to make
the model agree with the latest constraints set by HESS and Spitzer.
She named the model ``Best-Fit06'' and published data
\footnote{ http://www.desy.de/\textasciitilde{}kneiske/downloads.html}
of its EBL density and optical depth versus the energy and the redshift.

Kneiske's models are accepted by the scientific community and are used by
experiments such as MAGIC. Her new model also agrees with all the
existing observational limits, including the upper limits from HESS
and the lower limits from Spitzer. Furthermore, hers is the only model 
that is valid for both the energy and redshift ranges required 
by this study. As will be seen in Chapter \ref{chap:GRBSim}, 
the calculations for setting upper limits on the VHE emission from GRBs
need the optical depths to gamma rays of energies $40\,GeV-15\,TeV$ originating
from a redshift up to 3. For the above reasons, Kneiske's Best-Fit06 model
is the one used in this study.

%
%

\FloatBarrier
\section{\label{sub:IRAbs_Effects}Effects of Absorption by the EBL}

This section will describe the effects of the EBL absorption on the VHE spectra from 
distant sources.

\begin{landscape}
\begin{figure}[ht]
\begin{centering}
\includegraphics[width=0.9\columnwidth]{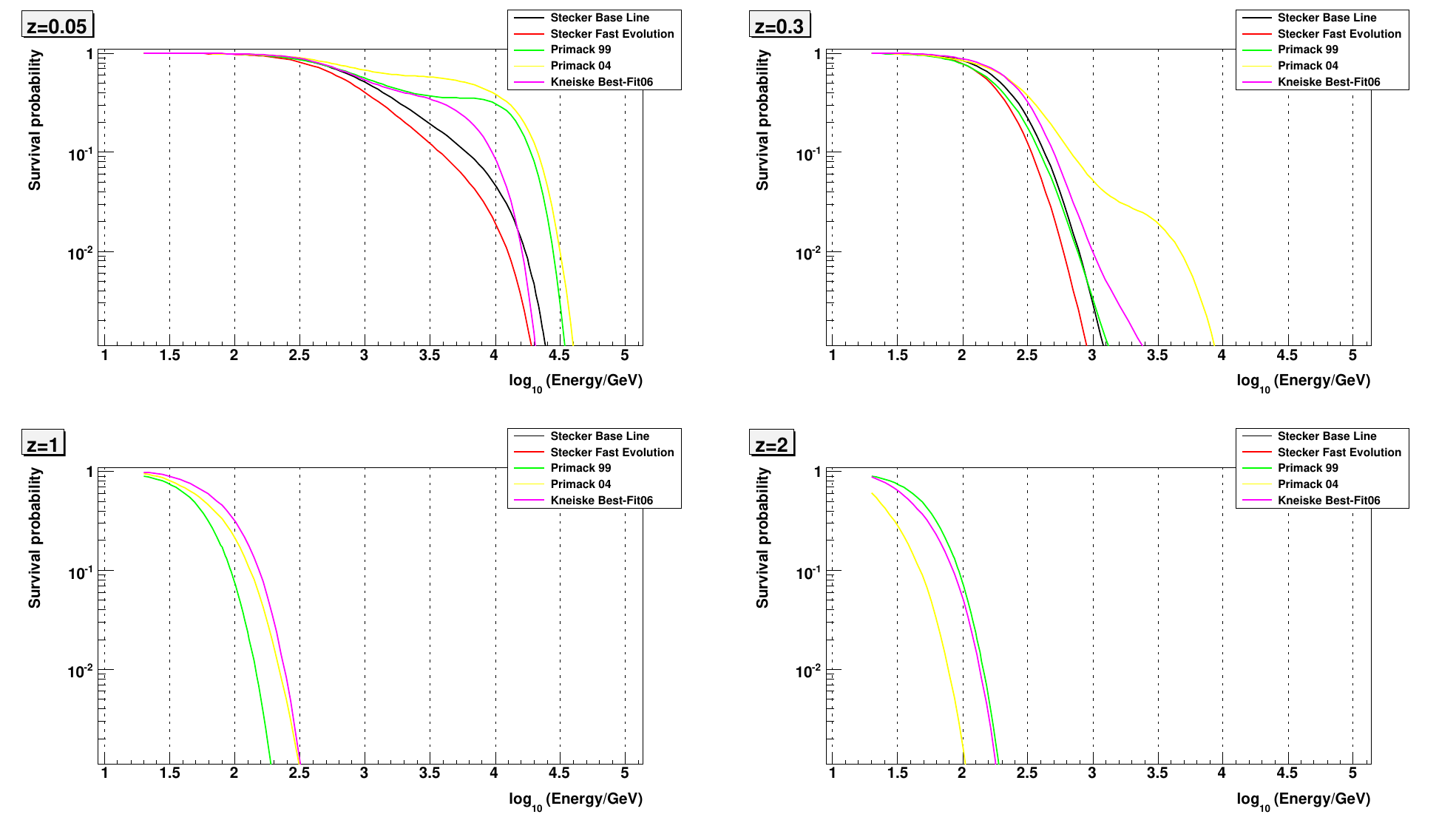}
\par\end{centering}

\caption{\label{fig:IRAbs_Att}Survival probability ($e^{-\tau(E,z)})$ due
to absorption by the EBL predicted by different EBL models versus
the photon energy and the redshift of the source. The energy is for
the observer reference frame. Stecker's models are not valid for $z>0.3$.}
\end{figure}
\end{landscape}

Figure \ref{fig:IRAbs_Att} shows the attenuation factors ($e^{-\tau(E,z)})$
versus photon energy and GRB redshift for different EBL models.
As can be seen, Primack's 2004 model predicts less absorption,
while the Stecker models predict more. 

\begin{figure}[ht]
\begin{centering}
\includegraphics[width=1\columnwidth]{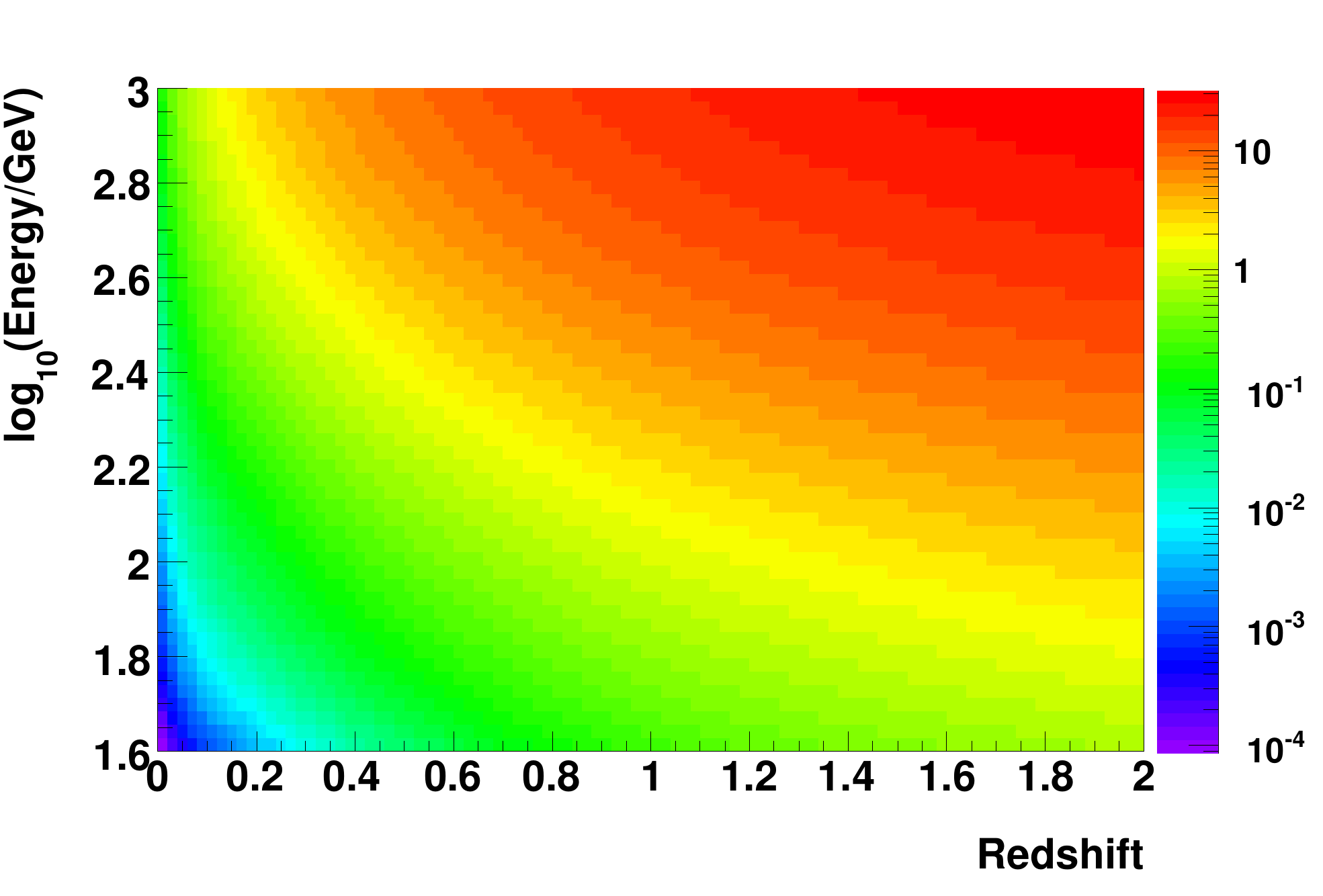}
\par\end{centering}

\caption{\label{fig:IRAbs_KneiskeTau}Optical depth $\tau(E,z)$ predicted
by Kneiske's Best-Fit06 model.}
\end{figure}

Figure \ref{fig:IRAbs_KneiskeTau} shows
the optical depth $\tau(E,z)$ predicted by Kneiske's Best-Fit06 model.
As can be seen, the attenuation is positively correlated with the redshift
of the source and the photon energy. 

\begin{figure}[ht]
\begin{centering}
\includegraphics[width=1\columnwidth]{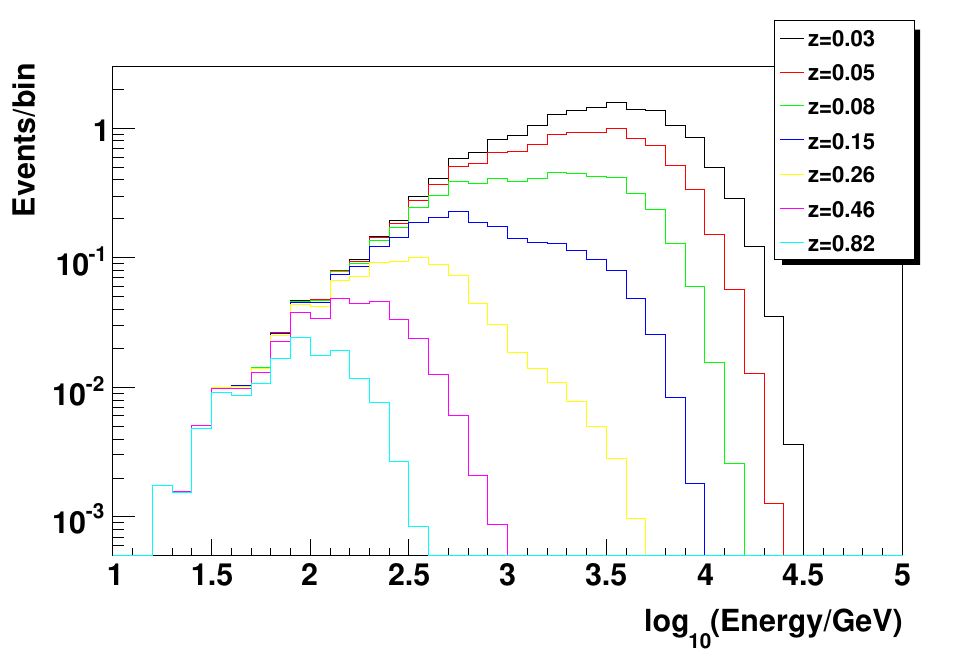}
\par\end{centering}

\caption{\label{fig:IRAbs_ETrig_Att}Energy distribution of events passing
the trigger and cut criteria versus the redshift of the source.
At least 20 PMTs were required to participate in the angular reconstruction
fits, and the maximum error in the reconstructed angle for the accepted events
was $1.5^{o}$. The energies are for the observer reference frame. The GRB is assumed
to emit VHE gamma rays in the $10\,GeV-100\,TeV$ energy range (burst
frame) on a power-law spectrum with index a=-2.2 and from a zenith
angle $10^{o}$. Kneiske's best-bit06 model was used to calculate
the absorption by the EBL. The normalization of the plot is arbitrary.}

\end{figure}

The effects of attenuation from the EBL on the energy distribution
of detected photons are shown in figure \ref{fig:IRAbs_ETrig_Att},
in which the energy distribution of triggered events caused by sources
at different redshifts is plotted.  As shown, the further the GRB
is, the higher the attenuation of the higher-energy photons. For the
larger redshifts, most of the photons that Milagro would have otherwise
detected are absorbed, limiting Milagro's sensitivity. This
effect limits the volume of the observable universe that Milagro is able to observe
for VHE emission from GRBs.

In the next figure (Fig. \ref{fig:IRAbs_Percentiles}) the median
energies of the energy distributions of detected photons are shown
for different redshifts. The edges of the error bars correspond to
the 1\% and 99\% quantiles of the same distributions. The energies
are given now for the burst frame. According to that figure, Milagro
can measure the GRB emission with energy between $\sim40\,GeV$ and
$\sim15\,TeV$. Emission of energy lower than that will not be detected
because of Milagro's limited effective area and because emission of energy
higher than that will be absorbed by the EBL before reaching the earth. 
The calculation was for a GRB that emitted on a power-law spectrum with
index -2.00 from $10\,GeV$ to $100\,TeV$. 

\begin{figure}[ht]
\begin{centering}
\includegraphics[width=1\columnwidth]{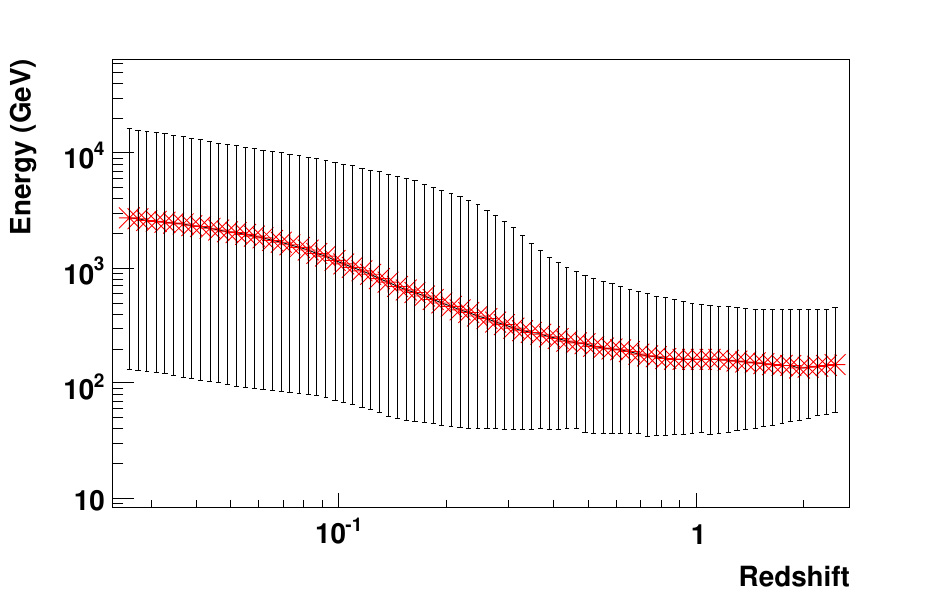}
\par\end{centering}

\caption{\label{fig:IRAbs_Percentiles}Median of the triggered energy distribution
for GRBs of different redshifts. The edges of the error bars correspond
to the 1\% and 99\% quantiles of the same distribution. The energies
are quoted in the burst reference frame. Kneiske's best-bit06 model
was used to calculate the absorption by the EBL.}
\end{figure}

    \clearpage \chapter{\label{chap:PBH}Primordial Black Holes}

\section{Introduction}

Hawking \cite{PBH_Hawking1971}
showed that fluctuations in the density of the early universe could
collapse and form mini black holes, called Primordial Black Holes (PBHs).
Primordial black holes are yet undetected relics from the first stages of the universe.
They are believed to be present in our galaxy,
and are set to evaporate all their mass through emission of Hawking radiation \cite{PBH_Hawking1974,PBH_Hawking1975}.
Their evaporation rate is progressively accelerated until it reaches explosive
degrees in the last stages of their lifetime. Some models predict that during these 
last stages, PBHs emit photons of energies that are detectable by Milagro. 
Observations of the evaporation of PBHs would allow us to probe multiple topics in physics, such as the early
universe, cosmology, gravitational collapse, particle physics, and
quantum gravity.

There have been many searches for PBHs \cite{PBH_CYGNUS,PBH_MILAGRO,PBH_RADIOSEARCH}.
However, only null results were found.
Cline \emph{et al.} \cite{PBH_CYGNUS, PBH_Cline_et_al_1992,PBH_Cline_et_al_1997}
argued that some of the very short GRBs ($<200\,ms)$ detected by
BATSE could have been created by PBH explosions. They claimed to have
found 42 such candidates, the distribution of which matches the spiral arms of our galaxy
, suggesting they are galactic. 

Because PBHs have not been detected yet and because their modeling
includes many unknown parameters, their behavior is not constrained.
For that reason it is risky to make concrete predictions on the detectability
of the emission from PBH evaporation. As will be shown in
section \ref{sec:PBH_Emission}, during
the last stages of a PBH, a short ($\sim\mu{}s-100s$) and
intense burst of emission in an energy range detectable by Milagro
could be created. Depending on the specific PBH model and on the
behavior of the poorly constrained parameter $a(M)$ (eq. \ref{eq:PBH_MssLossRate}),
that emission can be strong enough to be detected.

Section \ref{sec:PBH_HawkingRadiation} describes the mechanism with
which PBHs evaporate and gives an overview of their properties, and
section \ref{sec:PBH_Emission} describes the emission of photons
from PBHs and its detectability by Milagro.

\section{\label{sec:PBH_HawkingRadiation}Properties of PBHs}
Hawking showed that black
holes can radiate particles whose Compton wavelength is greater than
the Schwarzschild radius of the black hole. This emission arises from
the spontaneous creation of pairs of particles near the horizon of
the black hole, induced by the strong gravitational fields there.
One of the particles has positive energy and can escape to infinity,
while the other has negative energy and can tunnel into the black
hole, where particle states with negative energy with respect to infinity
exist. He showed that an uncharged, non-rotating black hole emits
particlesm that have energies in the range $E$ to $E+dE$ per state of
angular momentum and spin at a rate:
\begin{equation}
\frac{d^{2}N}{dEdt}=\frac{\Gamma_{s}}{2\pi\hbar}\left[exp\left(\frac{8\pi GME}{\hbar c^{3}}\right)-(-1)^{2s}\right]^{-1},\label{eq:PBH_SpecificRate}\end{equation}
where $M$ is the mass of the black hole, and $s$ is the spin of
the emitted species. $\Gamma_{s}$ is the absorption coefficient,
--a function of $s$, $E$, and $M$--that shows the probability that
the emitted particle would be absorbed if incident on the black hole.
In the limit $ME\gg1$, the instantaneous emission has a black-body
spectrum with temperature 
\begin{equation}
T=\frac{\hbar c^{3}}{8\pi GM}\simeq10^{-7}\left(\frac{M_{\odot}}{M}\right)K\simeq1.06\times10^{-5}\left(\frac{10^{15}g}{M}\right)TeV,\label{eq:PBH_T}\end{equation}
where $G$ is the gravitational constant, and $M_{\odot}\simeq2\times10^{33}\,g$
is the mass of the sun. Using equation \ref{eq:PBH_T}, equation \ref{eq:PBH_SpecificRate}
can also be written as:
\begin{equation}
\frac{d^{2}N}{dEdt}=\frac{\Gamma_{s}}{2\pi\hbar}\left[exp\left(\frac{E}{T}\right)-(-1)^{2s}\right]^{-1}.\end{equation}
The luminosity of a black hole is \cite{PBH_Halzen_Nature1991} \begin{equation}
L=10^{20}\left(\frac{10^{15}\,g}{M}\right)^{2}\,erg/s\end{equation}
and the photon flux is\begin{equation}
dN/dt=5.97\times10^{34}\left(\frac{1g}{M}\right)/s\end{equation}
As a black holes emits particles, its loses mass and slowly evaporates
until it disappears. Its mass-loss rate rate is inversely correlated
to its mass, therefore it's stronger at the last stages of its life:\begin{equation}
\frac{dM}{dt}=-\frac{a(M)}{M^{2}},\label{eq:PBH_MssLossRate}\end{equation}
where $a(M)$, the running constant, is a model-dependent count of
the particle degrees of freedom in the black hole evaporation \cite{PBH_Hawking1974,PBH_Halzen_Nature1991}.

As the hole radiates and loses mass, its temperature increases and
starts emitting new particle species. Every time the temperature reaches
the rest-mass of a new species, $a(M)$ smoothly rises.
For the standard model of particle physics $a(M)\geq7.8\times10^{26}\,g^{3}/s$
for $M\sim5\times10^{14}\,g$. Integrating \ref{eq:PBH_MssLossRate}
we can find the time it would take for a black hole of initial mass
$M$ to evaporate completely: \begin{equation}
\Delta{}t=\frac{M^{3}}{3\,a(M)}.\label{eq:PBH_Lifetime}\end{equation}
In order for the black hole to evaporate, rather than absorb accreting
matter, it must have a temperature greater than that of the present-day
black-body radiation of the universe ($2.7K$). This implies (from
eq. \ref{eq:PBH_T}) that its mass $M$ must be less than $10^{-7}\,M_{\odot}$.
Since in the present epoch of the history of the universe, black holes
form only through the gravitational collapse of massive bodies, and
masses less than $3\,M_{\odot}$ are stable when cold in the form of
neutron stars or white dwarfs, there is not a known process that could
create black holes of such a small mass. 

The initial mass of a PBH depends on the cosmological density at the
time of its creation. If a PBH was created a time $t_{BB}$ after the 
Big Bang, then its initial mass $M_{0}(t_{BB})$ would be
\begin{equation}
M_{0}(t_{BB})\sim\frac{c^{3}t_{BB}}{G}\simeq10^{38}\left(\frac{t_{BB}}{s}\right)\,g.\label{eq:PBH_Mass0}\end{equation}
As seen from eq. \ref{eq:PBH_Mass0}, PBHs can span an enormous
initial mass range: PBHs formed at the Planck time ($10^{-43}\,s)$ would have
an initial mass equal to the Planck Mass ($10^{-5}\,g$), while those
formed later, say at $t_{BB}=1\,s$, would have considerable larger
masses equal to $10^{5}\,M\odot$, comparable to the black holes in
the centers of Active Galactic Nuclei. Starting from equation \ref{eq:PBH_Lifetime},
Halzen \textit{et al.} \cite{PBH_Halzen_Nature1991} found that PBHs with masses
more than $M^{*}\simeq5\times10^{14}\,g$ would have evaporated entirely
by now. Using that mass limit with equation \ref{eq:PBH_Mass0}, we
find that these black holes were formed at the first $10^{-24}\,s$
of the life of the universe. PBHs with masses a bit larger than that
should be currently evaporating at a high enough rate to be observationally
significant.

\section{\label{sec:PBH_Emission}Emission from PBHs}
As the black hole radiates, its mass decreases and its temperature
increases until it becomes comparable to the Planck mass, at which
point semi-classical calculations break down and the regime of
full quantum gravity is entered. The mass-loss rate (amount of emission
integrated for all particle species) of a PBH and the types of particles
emitted by it depend on its temperature (or its mass). As the temperature
increases, becoming successively greater than the rest mass of
various particle species, the black hole starts emitting particles
of these species too. The running count $a(M)$ in equations \ref{eq:PBH_MssLossRate}
and \ref{eq:PBH_Lifetime} describes the number of particles species
emitted from a PBH of mass $M$. For lower energies, the particle species are the
ones of the standard model, while at energies over the QCD scale $\sim100\,MeV$,
multiple new species and resonances become available for emission,
causing a significant increase in $a(M)$ and therefore of the mass-loss
rate. 

Because modeling of the emission and the behavior of a PBH requires
the combination of multiple fields in physics--such as quantum gravity,
particle physics, QCD, and general relativity--the variations between
the predictions of the various models can be large. A short overview
of the various models follows:

\begin{itemize}
\item MacGibbon and Webber model (1990) \cite{PBH_MacGibbon_Webber_1990}\\
MacGibbon and Webber posited that once the temperature exceeds the
quark-hadron deconfinement temperature $\Lambda_{qh}\sim100-300\,MeV$,
individual quarks and gluons are emitted instead of hadrons and pions.
The emitted quarks and gluons then hadronize (combine) and create jets of pions
beyond the black hole horizon. The photons emitted by a black hole
are either created by the decay of the generated $\pi^{0}$ and the
fragmentation of the generated quarks, or are created directly at
energies $\sim5T$. The photon emission is far from being thermal
because the secondary photons (from the first two sources) dominate
over the direct thermal ones, since there are 72 quark and 16 gluon
degrees of freedom, while only two degrees for the direct photons.
The authors assumed that the emitted particles do not interact with
each other, therefore they move and fragment independently. 
\item Heckler model (1997) \cite{PBH_Heckler_1997b}\\
Heckler claimed that once the black hole temperature exceeds a critical
temperature $T_{crit}\sim45GeV$, QED interactions would produce an
optically thick chromosphere around it. In that case, the mean photon
energy will be reduced to $<E_{\gamma}>\simeq{}m_{e}\sqrt{T_{BH}/T_{crit}}$,
a value that is well below $T_{BH}$. The same author also proposed
that QCD effects can create a similar effect at even lower temperatures.
His arguments were disputed by MacGibbon \textit{et al.} \cite{PBH_MacGibbon_et_al_2007},
who claimed that QED and QCD interactions are never important.
\item Daghigh and Kapusta Model (2002) \cite{PBH_Daghigh_Kapusta_2002}\\
Daghigh and Kapusta is similar to the Heckler model with the difference
that they assumed that the hadronization of quarks occurs before the
onset of ($\Lambda_{qh}$), at a temperature $T_{f}\sim100-140\,MeV$,
and that all particle with masses greater than $T_{f}$ have been annihilated,
leaving only secondary photons, electrons, muons, and pions, before
free streaming occurs.\footnote{Free streaming happens when the flow is unrestricted by particle-particle interactions.}
As a result, the photon emission is either
directly produced following a boosted black-body spectrum or is the
byproduct of the $\pi^{o}$ decays. 
\end{itemize}

\begin{figure}[ht!]
\begin{centering}
\includegraphics[width=0.7\columnwidth]{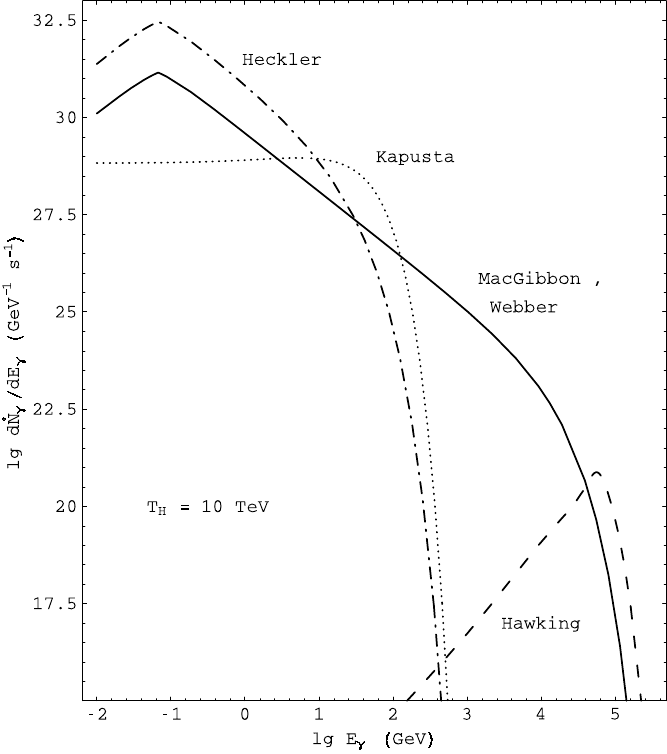}
\par\end{centering}
\caption{\label{fig:PBH_InstantaneousSpectra}Photon spectra of the instantaneous
emission from a black hole with temperature $T=10\,TeV$. The spectra
shown are based on the models by Heckler\cite{PBH_Heckler_1997},
Kapusta\cite{PBH_Daghigh_Kapusta_2002}, and MacGibbon and Webber\cite{PBH_MacGibbon_Webber_1990}.
For reference, the direct photon Hawking radiation is also shown.
Source \cite{PBH_Bugean_et_al2007}. }
\end{figure}

The instantaneous photon emission spectra from a black hole of temperature
$T=10\,TeV$ predicted by the above models are shown in figure \ref{fig:PBH_InstantaneousSpectra}.
As expected, the two last models (Heckler, and Daghigh \& Kapusta)
predict cutoffs at the higher energy part of the spectrum, effected
by absorption at the chromosphere around the black hole. On the
other hand, the first model considers the generated particles as free
and non-interacting. Thus, it does not include a chromosphere or
hadronization at lower energies, and the resulting spectrum extends to
higher energies than the other two models. For comparison, the gamma-ray
Hawking radiation (directly produced photons) is also shown in the
figure. As can be seen, the directly produced photons peak at an energy
of $5T$ and comprise only a small fraction of the total emission.
In the models including a chromosphere, these photons are absorbed
and then emitted thermally at lower energies. 

Figure \ref{fig:PBH_InstantaneousSpectra} describes the emission
spectrum at a specific instant of the black hole's lifetime. As the
black hole shrinks, the running count $a(M)$ increases and new species
of particles and resonances are emitted. The generation rate
of (direct) photons in the form of Hawking radiation is constant.
However, the generation rate of secondary photons increases strongly, as more
species that can decay or fragment to photons are starting to be emitted.
The emitted photon emission averages to higher energies
than before. 

\begin{figure}[ht!]
\begin{centering}
\includegraphics[width=0.7\columnwidth]{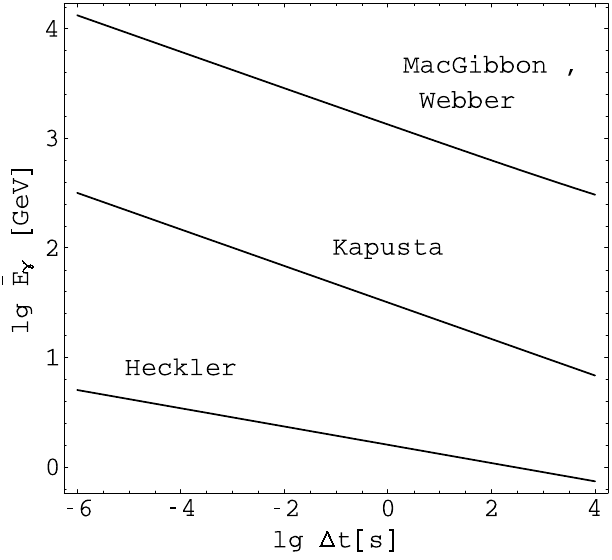}
\par\end{centering}
\caption{\label{fig:PBH_AverageE}Average energy of the photons emitted by
a black hole as a function of its remaining lifetime. The spectra
shown are based on the models by Heckler \cite{PBH_Heckler_1997},
Kapusta \cite{PBH_Daghigh_Kapusta_2002}, and MacGibbon and Webber \cite{PBH_MacGibbon_Webber_1990}.
For the MacGibbon-Webber plot, $E_{min}=100\,GeV$ was used. Source
\cite{PBH_Bugean_et_al2007} }
\end{figure}

\begin{figure}[ht!]
\begin{centering}
\includegraphics[width=0.7\columnwidth]{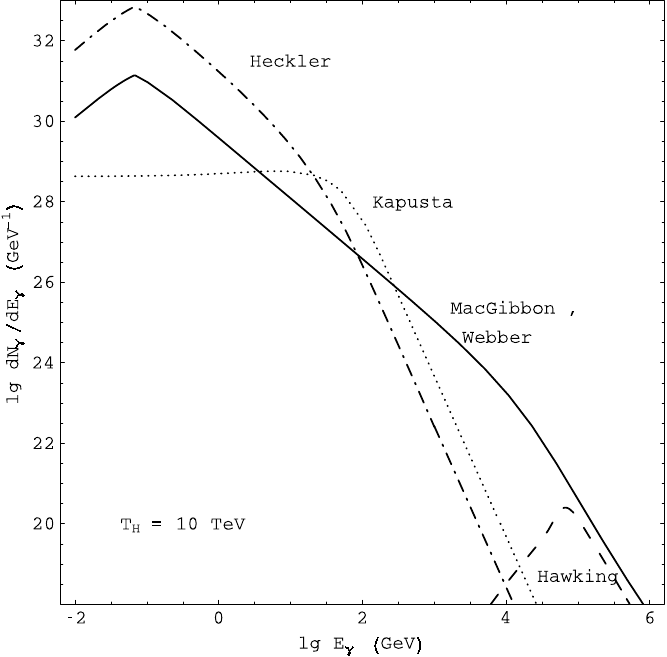}
\par\end{centering}
\caption{\label{fig:PBH_IntegratedSpectra}Photon spectra of the emission from
a black hole with initial temperature $T=10\,TeV$, integrated over its life.
The spectra shown are based on the models by Heckler \cite{PBH_Heckler_1997},
Kapusta \cite{PBH_Daghigh_Kapusta_2002}, and MacGibbon and Webber \cite{PBH_MacGibbon_Webber_1990}.
For reference, the direct photon Hawking radiation is also shown.
Source \cite{PBH_Bugean_et_al2007} }
\end{figure}

\begin{figure}[htb]
\begin{center}
\includegraphics[width=1\columnwidth]{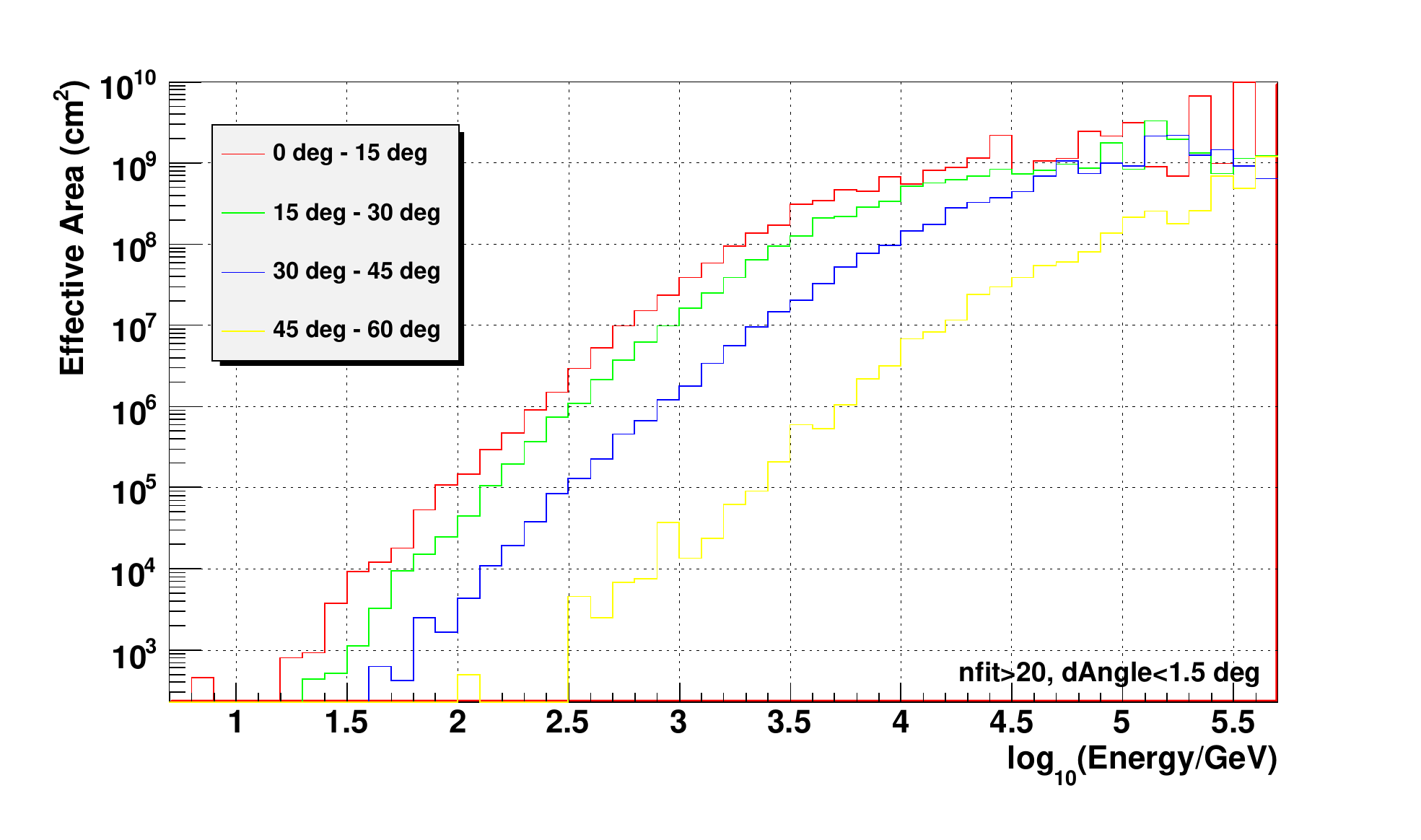}
\caption{\label{fig:Sensitivity_AEff}Effective area of Milagro for gamma rays
of different energies and from different zenith-angle ranges.}
\end{center}
\end{figure}

Figure \ref{fig:PBH_AverageE} shows the average energy
of the emitted photons as a function of the remaining lifetime of
a black hole. As the evaporation progresses, the
average energy increases at a fast rate, entering the energy range
at which Milagro is most efficient (see figure \vref{fig:Sensitivity_AEff}). 
Figure \ref{fig:PBH_IntegratedSpectra}
shows the photon spectra of the emission from a black hole with initial temperature
$T=10\,TeV$, integrated over its life. It is encouraging that even the
Daghigh-Kapusta and Heckler models, which include a chromosphere,
can create emission at an energy to which Milagro is sensitive $(\sim{}TeV)$. However,
the integrated flux might be too low to be significantly detected by
Milagro.


Unlike GRBs, PBHs can be relatively close to us. Halzen \emph{et al.} \cite{PBH_Halzen_Nature1991}
showed that, depending on their amount of clustering, PBHs of $T=1\,TeV$ can be as close as $10\,pc$ to $10\,kpc$.
This means that, contrary to GRBs, there can be emission from PBHs that is not
attenuated by interactions with the extragalactic background light.
As shown in Chapter \ref{chap:IR}, the higher-energy emission from GRBs is
strongly attenuated, resulting in only photons up to energies of few
hundreds of $GeV$ reaching the earth. Hence, most of the emission
to which Milagro is most sensitive is absorbed, which reduces Milagro's
chances of detecting a GRB. For the case of galactic PBHs, this absorption
does not occur, and Milagro can detect the full spectrum of their VHE
emission. 

It should be noted that there is great uncertainty in the predictions
of all these models. Modeling the evaporation of a PBH involves a
wide number of complicated processes and unknown parameters. One of
the big unknowns is the dependence of the running count $a(M)$ on
$M$, a function that sets the lifetime of a PBH. When the temperature
of the black hole is under the $QCD$ scale $(\sim100MeV)$, the range
of emitted particles is described by the standard model and is limited.
As the black hole temperature passes that limit, a large number of new
particles and resonances become available for emission. Then $a(M)$
starts increasing with a higher rate as $M$ becomes smaller, and
the evaporation is accelerated. At these high energies, there could
be a large number of yet undiscovered particles. If, for
example, supersymmetry is the theory that describes higher-energy
elementary particles, then $a(M)$ can increase by at least a factor
of three. Some other theories, such as from Hagedorn \cite{PBH_Hagedorn_1968},
predict an exponential increase in the number hadronic resonances.
In such a case, the final stage of a black hole evaporation would
happen in time scales of microseconds and in an explosive way. As
can be seen, both the time scale of the evaporation and the amount
of radiated energy at each stage of the process strongly depend on the
poorly constrained function $a(M)$.

   \clearpage \chapter{\label{chap:MILAGRO}	Milagro}

\section{Introduction}

The typical VHE gamma-ray emission from astrophysical sources is very
weak. For example, the integral particle flux from the Crab Nebula
above $1\,TeV$ is $1.75\times10^{-11}\,cm^{-2}\,sec^{-1}$, which corresponds
to just a little over one photon/day on an area of $100\,m^{2}$. To
be able to detect such a small signal in a reasonable amount of time,
detectors of large effective areas (at least thousands of square meter)
are needed. Placing such detectors in space is prohibited by the high
costs involved. VHE gamma rays are absorbed shortly after entering
the atmosphere, creating extended cascades of secondary electromagnetic
particles, called ``Extensive Air Showers'' (EASs). Ground-based
detectors, which can observe these EASs and have large effective areas,
can be built relatively cheaply. By measuring properties of the EAS,
such as the direction of the axis of symmetry, and the lateral and
longitudinal profiles, ground-based detectors can estimate the energy
and direction of the primary gamma ray that created the EAS. 

Milagro is a ground-based detector employing the water-Cherenkov technique to
detect gamma rays through the EASs they generate. The purpose of this chapter is to
describe the Milagro detector, its method of operation, and its capabilities. 

Initially, in order to understand how Milagro compares to the other 
ground-based detectors, a very brief review of this kind of detectors will be 
given in section \ref{sec:Milagro_Milagroetal}. Then, section \ref{sec:Milagro_Parts}
will describe the instrument Milagro, and sections 
\ref{sec:Milagro_DAQ} and \ref{sub:Milagro_Triggering} respectively
will describe its data-acquisition and triggering systems.
Section \ref{sec:Milagro_Analysis} will give an overview of the 
reconstruction, storage, and filtering of the Milagro data, and finally
section \ref{sec:Milagro_Calibration} will describe its calibration system.

\section{\label{sec:Milagro_Milagroetal}Ground-Based Gamma-Ray Detectors}

Ground-based gamma-ray detectors are divided into two broad categories: Imaging
Atmospheric Cherenkov Telescopes (IACTs) \cite{Det_IACT_Lorenz} and 
Extended Air Shower Arrays (EAS Arrays), such as Milagro. First, the principle of 
operation of IACTs will be described, followed by a description of
EAS arrays with a focus on Milagro. Where appropriate, a comparison of the
two kinds of detection techniques will be provided. It should be noted that
because EASs can also be created by cosmic rays, ground-based detectors have
to search for a signal on top of a large cosmic-ray background. 

\begin{figure}[ht]
\begin{centering}
\includegraphics[width=0.5\columnwidth]{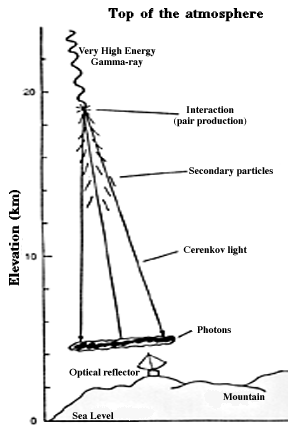}
\par\end{centering}
\caption{\label{fig:Milagro_ACT}Principle of operation of an Imaging Atmospheric
Cherenkov Telescope. A cosmic gamma ray interacts at the top of the
atmosphere creating an EAS. The air shower contains
thousands of energetic electrons and positrons that emit Cherenkov
light. Large telescopes on the ground detect this light and reconstruct
the shower and the properties of the primary gamma ray that caused
it. Source: \cite{URL_IACTPIC}}
\end{figure}

A way to measure the properties of EASs is through the Cherenkov light emitted
in the atmosphere by their energetic electrons and positrons.
IACTs accomplish this by using big ($\sim10\,m$) mirrors to focus the Cherenkov light
on sensitive photomultiplier tubes (PMTs) (Fig. \ref{fig:Milagro_ACT}).
Because IACTs observe the whole longitudinal development of the EASs,
they obtain enough information to reconstruct accurately
both the energy and the direction of the primary gamma ray. Furthermore,
by observing the shape of the shower, IACTs can efficiently distinguish
whether the shower was initiated by a gamma ray (signal) or a cosmic ray (background).
This ability, combined with the IACTs' good angular resolution, results
in a high signal to noise ratio and in a high sensitivity.
However, because of the IACTs' sensitivity to external light, they can
operate only on moonless nights, and because of their optical telescope
design, they have a small field of view (few degrees\footnote
{New IACTs with a wider field of view are currently under development: 
CTA \cite{URL_CTA} and AGIS \cite{URL_AGIS}}). As a result,
IACTs are good instruments for performing focused, high-quality observations
of selected sources, but they are not optimal for doing unbiased, whole-sky
searches, monitoring the overhead sky for transient emission, or
detecting extended sources (larger than their field of view). 

\begin{figure}[ht]
\begin{center}
\includegraphics[width=1.0\columnwidth]{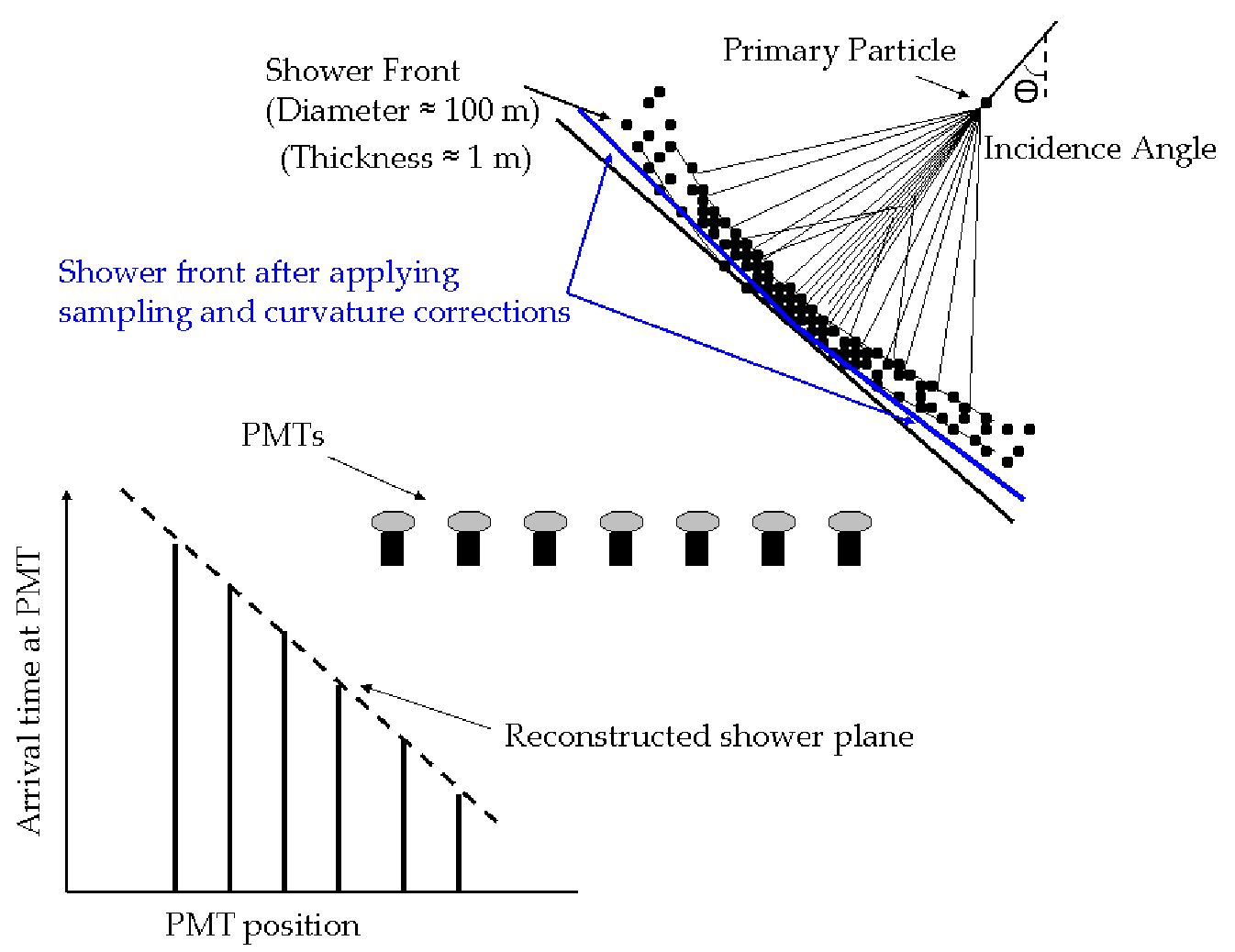}
\caption{\label{fig:Milagro_WaterCherenkovTechnique}A conceptual image depicting
the water-Cherenkov technique. Source \cite{Thesis_Aous}}
\end{center}
\end{figure}

\begin{figure}[ht]
\includegraphics[width=1\columnwidth]{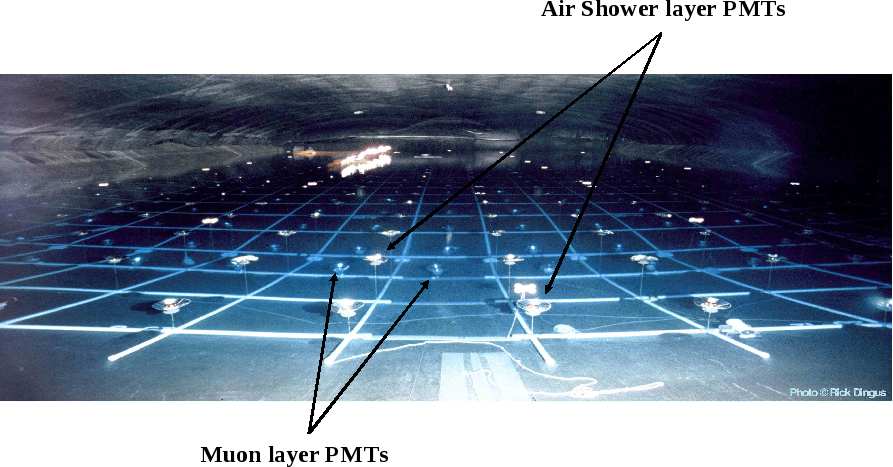}
\caption{\label{fig:Milagro_Pond}View of the inside of Milagro's pond. Photo
courtesy Rick Dingus.}
\end{figure}

Another ground-based method for observing EASs is to have an array
of detector elements set to measure the shower's lateral profile at
ground level (EAS Arrays). Milagro accomplishes this using the water-Cherenkov
technique. In this technique, shown in 
Fig. \ref{fig:Milagro_WaterCherenkovTechnique},
a primary particle (gamma ray as signal or cosmic ray as background) interacts
at the top of the atmosphere and initiates a cascade of $e^{-}, e^{+}, \gamma$
particles (the EAS). The EAS, while moving downwards with a speed
close to the speed of light, expands and develops a flat, wide, and
thin shower front. The shower front eventually reaches the ground
and enters into Milagro's water volume (Fig. \ref{fig:Milagro_Pond}).
While in the water, the EAS gamma rays convert to energetic $e^{-}e^{+}$
pairs or transfer their energy to electrons through Compton scattering.
The produced energetic electrons and positrons, and the other charged
particles of the EAS emit Cherenkov light. PMTs placed in the water, then, 
detect this light. The properties of the primary particle that caused the EAS
are reconstructed using the total amount of the detected signal,
its lateral distribution, and the relative arrival times of the detected particles.

The characteristics of an EAS's profile on the ground depend on the
height of the primary particle's first interaction. For primaries
of the same energy, the higher the first interaction is, the smaller
on average the number of EAS particles that reach the ground. As a result,
random fluctuations in the height of the first interaction cause
random fluctuations in the amount of signal that reaches the ground. EAS
arrays cannot easily account for this effect, resulting in a reduced
energy-reconstruction accuracy. IACTs, on the other hand, are not
sensitive to this effect, because they measure the EAS's integrated
light emission over the whole development of the shower in the atmosphere,
which is strongly correlated to the energy of the primary. 

EAS arrays and IACTs have similar background rejection capabilities.
However, because of the significantly better angular resolution of IACTs,
they avoid including most of the background around a source, and hence
can acquire signals with very high signal to noise ratios. As a result, for a typical
source that is similar to the Crab Nebula, IACTs are more sensitive than EAS arrays. 

Because typical gamma-ray emissions usually follow steeply-falling power-law
spectra (spectral indices usually $<-2$), the signal emitted at TeV energies has
a significantly lower intensity than the one emitted at GeV energies.
This means that longer observation times are needed in order for the
higher-energy component of some emission to be detected. As mentioned
above, IACTs have very low duty factors. Therefore they cannot afford
the extra observational time needed to detect the higher energy emission
from a source with high significance. On the other hand, EAS arrays
operate almost continuously and manage to accumulate more higher-energy signal.
As a result, despite their worse signal to noise ratio, EAS arrays 
end up being more sensitive than IACTs to signals of higher energy (usually over $\simeq10\,TeV$).

EAS arrays have a high duty factor ($>90\%$) and a wide field of
view ($\sim~2sr$). This enables them to monitor the overhead sky for
bright transient emission from GRBs or for flares from known sources,
and to perform unbiased whole-sky searches for both localized and
extended emissions. Furthermore, since space gamma-ray detectors also
share these properties, EAS arrays can successfully
collaborate with them in multi-wavelength whole-sky monitoring for
transient emissions. EAS arrays can also act complementarily to IACTs
by providing them with possible source locations, of which the IACTs
can perform high quality observations.

\FloatBarrier
\section{\label{sec:Milagro_Parts}Detector Description}

\begin{figure}[ht]
\includegraphics[width=1.0\columnwidth]{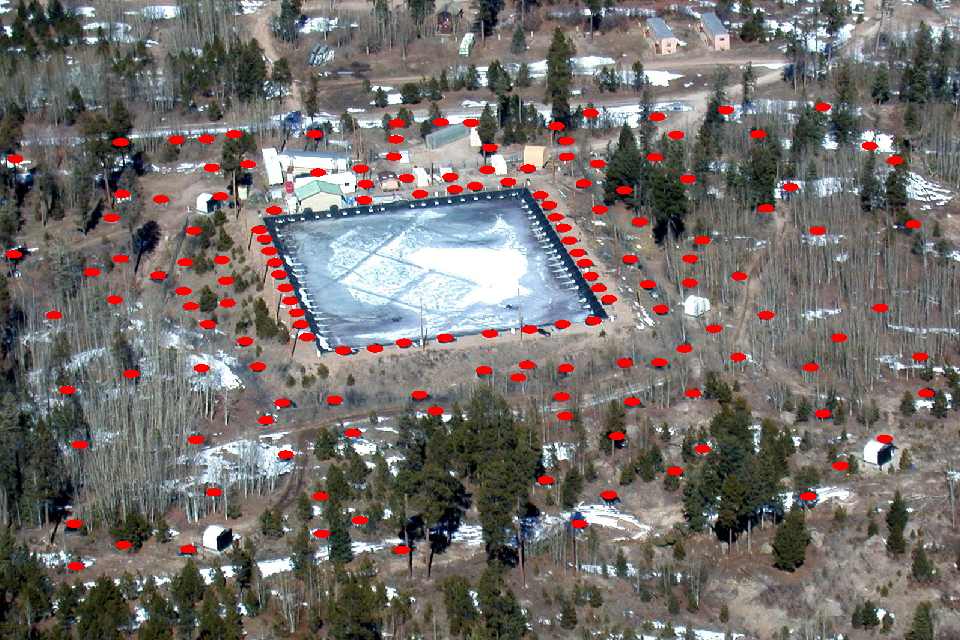}
\caption{\label{fig:Milagro_Above}Aerial view of the complete Milagro detector.
The pond, shown here with snow on the cover, is in the middle. The
buildings next to the pond house the electronics, storage spaces and
the water filtration systems. The outrigger array is also shown (red
dots) around the pond.}
\end{figure}

Milagro (Fig. \ref{fig:Milagro_Above}) is located at the Jemez Mountains
(latitude $35^{o}52'45''$ and longitude $106^{o}40'37''$ West) near
Los Alamos, New Mexico and at an altitude of $2630m$.
The main part of the Milagro detector is a rectangular artificial
reservoir (``the pond'' - Fig. \ref{fig:Milagro_Pond}) filled with water and containing two
layers of PMTs. A sparse array of water tanks, each containing a PMT,
is spread around the pond. Because the water-Cherenkov technique was
new when Milagro was proposed, the Milagro detector was built in stages,
with each stage verifying and optimizing the technique. %

\FloatBarrier
\subsection{The pond}

Milagro's pond is filled with $\sim23$ million liters of highly purified
water. The pond is $8\,m$ deep and its dimensions are $80\,m\times60\,m$
at the surface, while it slopes near the bottom to $30\,m\times50\,m$. The
Milagro pond was initially ($\sim1995)$ part of ``Hot Dry Rock,''
a geothermal experiment at Fenton Hill, New Mexico. The pond had to
be cleaned to be used as a detector. A cover and a liner made of black
polypropylene were installed to protect the pond from nature. The
cover was inflatable and was also used to block external light. The
first version of the detector, called ``Milagrissimo,'' consisted
of the pond and twenty eight PMTs arranged on a layer inside it. After encouraging
initial results, the PMT grid was expanded to contain 228 PMTs spread
over the whole surface of the bottom of the pond. The expanded version,
called ``Milagrito,'' operated from February 1997 to April 1998 \cite{Det_Milagrito_NIM}.
The water level was varied to determine the optimal depth of the PMTs
for shower reconstruction. Milagrito was later expanded by the addition
of water-filtration and calibration systems, the reconfiguration
and expansion of the existing PMT layer, and the addition of a whole
new layer over it. The new detector was named ``Milagro.'' 

Milagro's top layer of PMTs consists of 450 PMTs located $\sim1.6\,m$
under the surface of the water. This layer is called the ``Air-Shower
layer'' (AS) because it samples the light emitted by the
EAS particles shortly after they enter the pond. The hits produced
at the AS layer contain the timing information necessary for reconstructing
the shower front and, from that, the initial direction of the particle
that caused the EAS. The hits from the PMTs of the AS layer are also
used for triggering. The bottom layer consists of 273 PMTs located
$\sim6\,m$ under the surface. As will be shown later (section \ref{sub:Milagro_GammaHadron}),
the information from the hits of this layer are used to discriminate
between hadron- and gamma-induced showers. Because these hits are
primarily created by deeply penetrating muons, this layer is also
called the ``Muon layer'' (MU). 

At Milagro's altitude, EAS gamma rays outnumber the fraction of EAS
$e^{-}e^{+}$ by a factor of $\sim5$. The AS layer was placed deep
enough so that most of these gamma rays convert to $e^{-}e^{+}$
pairs or transfer their energy to electrons through Compton scattering
before they reach it. The AS layer was also placed shallow enough
so that it promptly detects the Cherenkov light with just a small 
dispersive time spreading. The horizontal
distance between the PMTs is such that, considering the $\simeq41^{o}$
half-opening angle of the Cherenkov cone and the depth of the AS
layer, the efficiency of detecting the EAS particles is high, while
the required number of PMTs is kept low. About half of the electromagnetic
EAS particles are detected by the PMTs of the AS layer, and most muons with
energy $E\gtrsim1GeV$ are detected by the PMTs of the MU layer. 

By detecting muons--particles that are mostly present in cosmic-ray initiated showers--
the muon layer of Milagro distinguishes between showers 
initiated by cosmic rays (background) versus gamma rays (signal).
These muons are usually energetic enough
to penetrate deeply and reach down to the muon layer, where they create
big hits in few adjacent PMTs. However, since the AS and MU layers are not
optically isolated, the PMTs of the MU layer are also exposed to 
the light generated by the electromagnetic component of the EAS.
If the muon layer were at a shallow depth, then the energetic EM particles could
also reach it and create big hits in its PMTs similar
to the hits from muons. If this were the case, it would be harder for the muon layer
to detect the presence of a muon, reducing the efficiency of Milagro's background
rejection. For that reason, the muon layer was placed deep enough
so that the electromagnetic EAS particles are usually absorbed well above it, and 
the light they produce creates only a few small hits on its PMTs. 

Due to the horizontal spacing and depth configuration
mentioned above, the AS layer measures the arrival times of the EAS
particles with high accuracy and statistics, allowing an accurate
reconstruction of the primary particle's direction, and the MU layer
detects and distinguishes hits from muons with high efficiency, helping
with background rejection.

\begin{figure}[ht]
\begin{centering}
\includegraphics[width=0.8\columnwidth]{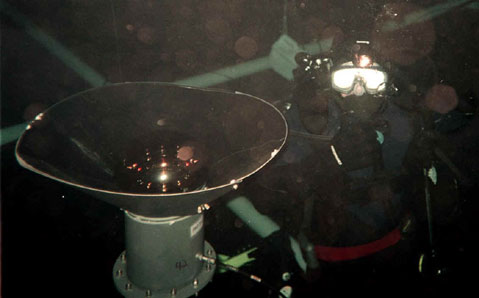}
\caption{\label{fig:Milagro_Baffle}A diver working on a PMT during a repair operation. The baffle and the PMT encapsulation can be seen.}
\end{centering}
\end{figure}

Milagro's PMTs are Hamamatsu R5912SEL PMTs. Their photocathode is
quasi semi-spherical with 8'' diameter. To protect their sensitive
electronics from water damage, each PMT is attached in a water-proof way to
a PVC cylindrical encapsulation. All the PMTs are fitted with reflective
cones, called ``baffles'' (Fig. \ref{fig:Milagro_Baffle}), in order to prevent unwanted triggers,
to improve the angular-reconstruction accuracy, and to increase the PMTs' light collection area.
The baffles block horizontally-moving light, which is primarily light
from higher zenith angle particles (usually muons), and secondarily scattered or reflected light.
The light from higher zenith-angle EAS particles can be detected by
multiple PMTs of the AS layer and can trigger the detector. Such triggers are unwanted since
they do not correspond to reconstructable gamma-ray events.
Scattered or reflected light comes long after the hits by the main EAS
particles and interferes with the angular reconstruction.
The reconstruction's accuracy depends on the reconstruction algorithms being able to fit the
PMT hit times on a plane. The fewer the late hits are, the narrower
the distribution of hit times, the better the quality of the angular-reconstruction
fit, and the more accurate the angular resolution of the detector. Another
problem caused by late hits comes from artificially inflating
the estimated number of photons detected by a PMT, a
piece of information that is used by most parts of the data
analysis and event reconstruction (see sec. \ref{sub:MILAGRO_TOT}). 
As can be seen, the baffles, by blocking a large fraction
of the late or the horizontal light, increase the quality of the Milagro
data and prevent unwanted triggers from higher zenith angle particles.
The baffles also increase the collection area of the PMTs by reflecting the
light that otherwise would have missed the PMTs towards them. However,
because most of that reflected light is directed near the sides of
the PMT's photocathode, where the PMT efficiency is reduced \cite{vasileiou-2007},
this enhanced collection area is not proportionately translated to
an increased amount of detected light. 

Initially, the baffles were made of specularly-reflecting anodized
aluminum on the inside and black polypropylene on the outside. Since
aluminum oxidizes in water, these baffles were sprayed with a protective
layer to prevent oxidation. However, if a very small spot on the surface
of the baffle was left unprotected, the oxidation would start from
that spot and slowly expand. Over the course of time, this effect
created holes in the baffles and released oxidized aluminum particles,
a white chalky powder, in the water and on the PMTs. While the holes
were not big enough to considerably affect the functionality of the
baffles, the released material had a detrimental effect on the cleanliness
of the water. As it will be shown in subsection \ref{sub:Milagro_Water}, 
there have been noticeable
reductions in the attenuation length of the pond's water coincident
in time with the decay of the baffles. The aluminum baffles were eventually
replaced in two consecutive repair operations; about half of the baffles
were replaced in 09/2003 and the rest in 09/2005. The replacement
baffles were made of diffusely-reflecting polypropylene that is white
inside and black outside. The new baffles had a similar positive 
effect on the function of the detector as the old baffles, but without
reacting with the water.

The pond is covered by a light-tight $1\,mm$ thick cover. The cover is
inflatable so that people can enter the pond during repair operations.
Most of the time, the cover floats on the surface of the water. However,
sometimes air can accumulate under it. Because the reflectivity
of the water to air interface is higher than the reflectivity of the water to cover
interface
, the accumulation of air under the cover can increase the amount of light detected by
the AS layer. Because the PMTs of the AS layer are used for triggering,
this increased reflectivity has significant effects on the response
of the detector. The effect is particularly noticeable for few days just after repair
operations, when the cover is inflated. However, the weight
of the cover slowly pushes the air under the cover away, and the detector
returns to its normal state. 
%

 \begin{figure}[ht]
 \includegraphics[width=1.0\columnwidth]{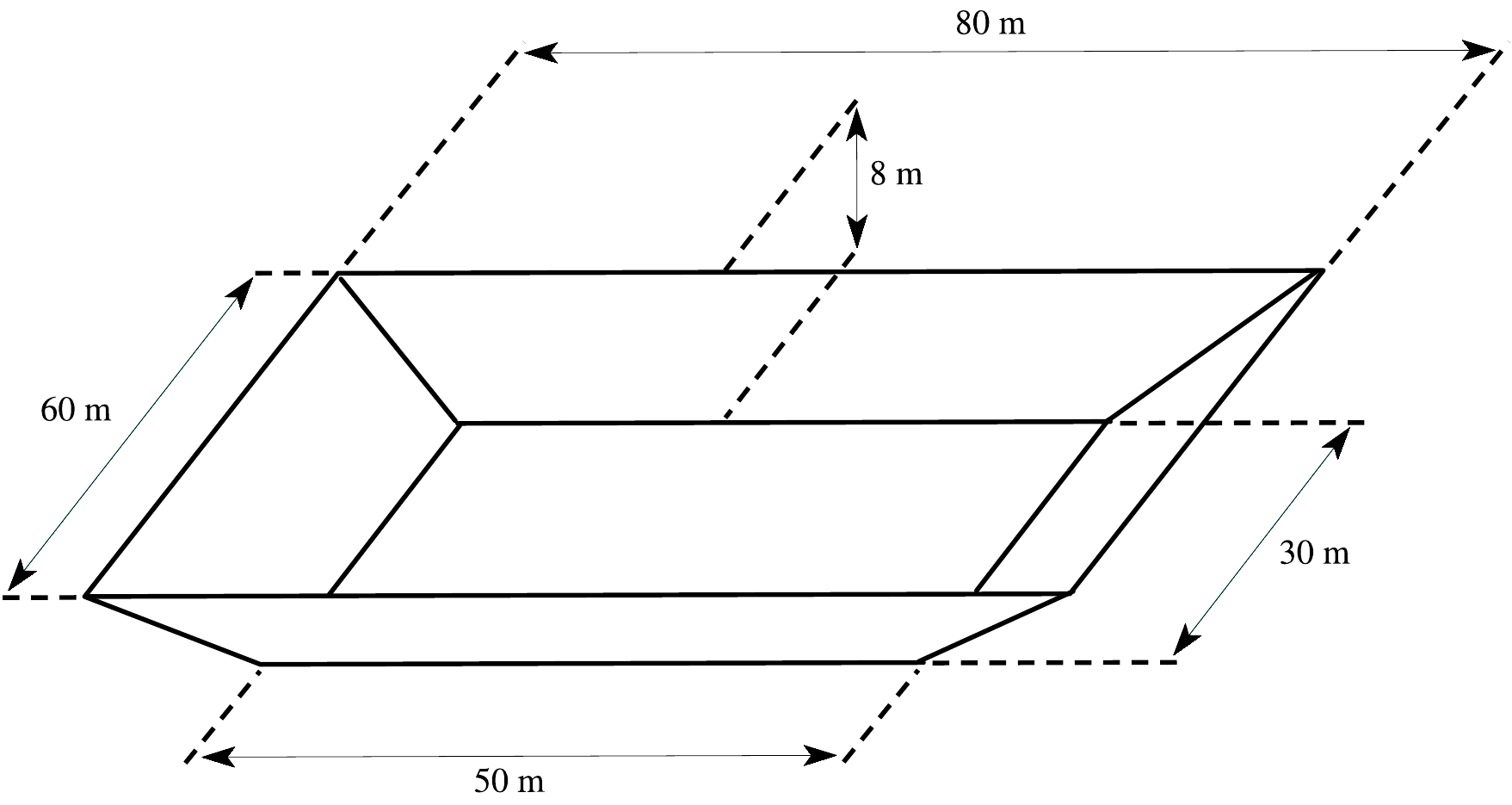}
 \caption{Schematic of the main pond. Source \cite{Thesis_Leonor}}
 \end{figure}

\FloatBarrier
\subsection{The outriggers}
\begin{figure}[ht]
\begin{centering}
\includegraphics[width=0.8\columnwidth]{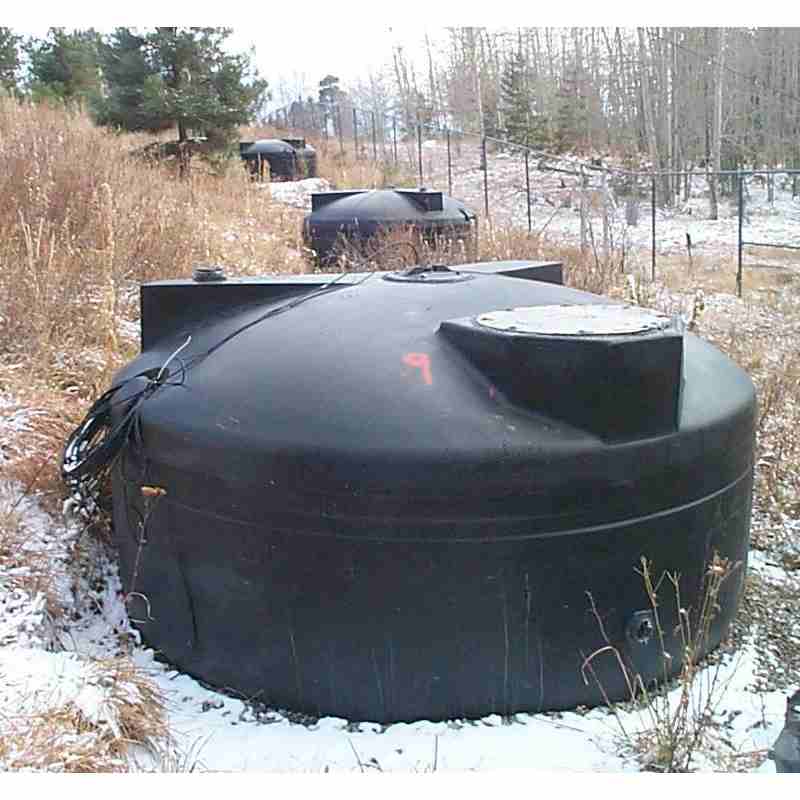}
\par\end{centering}
\caption{\label{fig:Milagro_Outrigger}Photo of an outrigger tank.}
\end{figure}

\begin{figure}[ht]
\begin{centering}
\includegraphics[width=1.0\columnwidth]{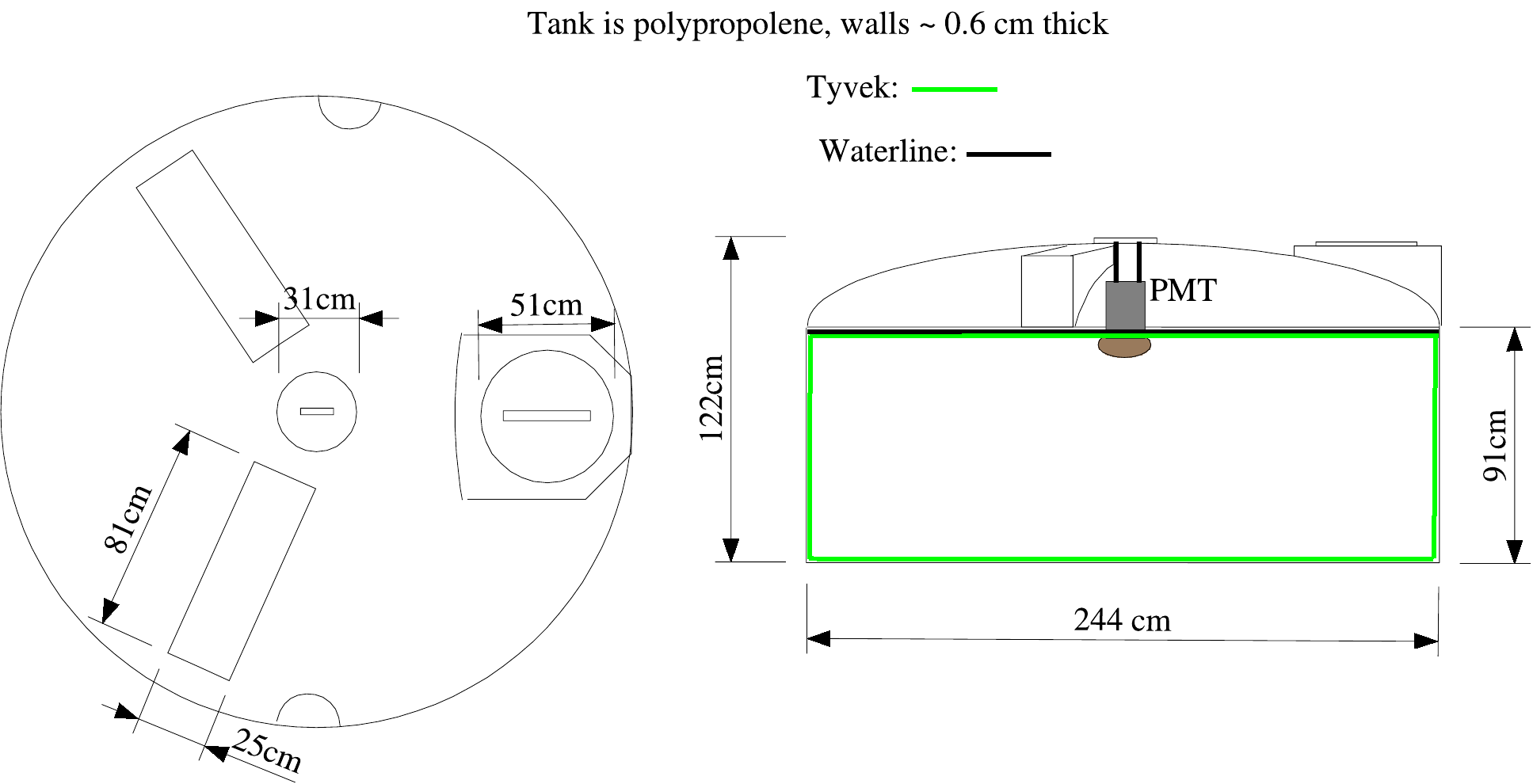}
\caption{\label{fig:Milagro_Outrigger_Schematic}Dimensions of an outrigger
tank. Source: Tony Shoup}
\end{centering}
\end{figure}

Around the pond is a sparse array of 175 water tanks called the ``Outrigger
array'' (OR). Each tank is filled with $\sim2200\,l$ of water
and contains a downwards-facing PMT (Fig. \ref{fig:Milagro_Outrigger}). The tanks are made of high-density
polyethylene and are internally lined with white, diffusely-reflecting
Tyvek to maximize the light collected by the PMT. They have a $2.4\,m$ diameter
and a $1\,m$ height (Fig. \ref{fig:Milagro_Outrigger_Schematic}).
Even though the outrigger array was part of the initial plan of the Milagro detector,
it was added later, in stages, between 1999 and 2003. 

The outrigger array significantly improved the sensitivity of the Milagro
detector by increasing its angular resolution and its background-rejection capabilities.
As I will show, the improvement on the angular reconstruction 
was primarily through the showers that landed off the pond. Milagro's AS layer is dense enough
that it samples a large fraction of the electromagnetic particles of the EASs. 
As a result, when a shower lands inside the pond, the AS layer acquires enough
information to reconstruct the showers' direction accurately.
However, when a shower lands off the pond, the AS layer by itself is not
capable of reconstructing the shower with such good accuracy. Because of
the curvature of the shower front, the angular reconstruction algorithms
need a precise localization of the shower core on the ground to 
reconstruct the direction of the shower accurately. In order for the Milagro
detector to locate the shower core with good accuracy, it has to contain the 
whole of it. Before the addition of the outriggers, the pond could not
contain the cores of showers that landed outside of it, and 
the detector had a worse angular resolution for such showers.
The addition of the outrigger array increased the physical area of
the Milagro detector from $\sim5000\,m^{2}$ to $\sim40,000\,m^{2}$,
and provided a longer lever arm by which it could reconstruct
showers. As a result, the detector now samples an area that is wide enough to 
be able to accurately reconstruct showers that land both off and on 
the pond.

Off-pond showers are especially important for Milagro's sensitivity;
they constitute the majority of Milagro triggers (since there is more effective area around the pond than
on it), and the gamma-hadron discrimination capabilities of Milagro are far better for them.
As will be shown in subsection \ref{sub:Milagro_GammaHadron},
the gamma-hadron separation of Milagro
is based on being able to detect muons or other energetic hadrons
distant from the main shower core. Milagro detects these particles
using the muon layer. If a shower lands off the pond, then
the muon layer will be properly located far from the shower core to
detect them, and thus to provide useful
information on the nature of the primary (gamma ray or cosmic ray).
Because the majority of the gamma-ray signal, correctly identified
as such by the gamma-hadron discrimination algorithms of Milagro,
is detected through off-pond showers, it is very important to
reconstruct the direction of such showers accurately. As was shown above, the addition
of the outriggers improved the angular resolution for off-pond showers 
and significantly increased Milagro's sensitivity.

The outriggers also contributed to the gamma-hadron discrimination capabilities
of Milagro by providing the background rejection algorithms with an improved
estimate of the size of the shower (see subsection \ref{sub:Milagro_GammaHadron}).


\subsection{\label{sub:Milagro_Water}Water filtration system}

\begin{figure}[ht]
\includegraphics[width=1.0\columnwidth]{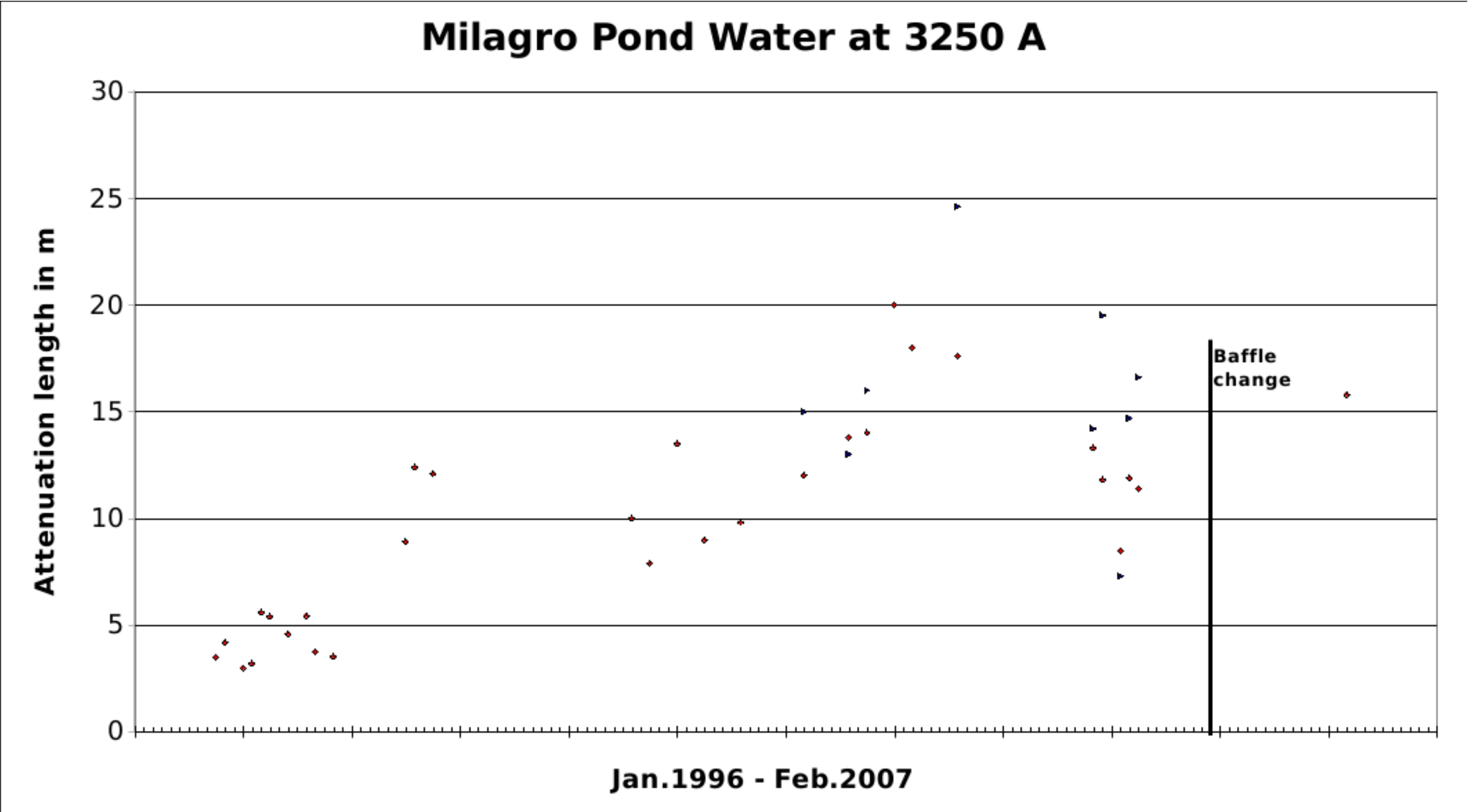}
\caption{\label{fig:Milagro_AttenuationLength}Attenuation length of Milagro's
pond water for light of wavelength $\lambda=325\,nm$. The time when
most of the aluminum baffles were replaced is also shown in the chart.
Measurements by Don Coyne and Michael Schneider.}
\end{figure}

The water of the pond is continuously circulated through a water system
that filters and disinfects it at a rate of $\sim750\,l/min$.
A set of filters ($10\,\mu{}m$, $1\,\mu{}m$ and $0.2\,\mu{}m$) remove most
of the suspended particles, and ultraviolet lamps prevent any biological growth.
During the aluminum-baffle epoch, the floating aluminum particles
clogged the finer $0.2\,\mu{}m$ filter, and the aluminum baffles had to be removed. The
water filtration system significantly improved the initial bad quality
of the water in the Milagro pond. As can be seen from figure \ref{fig:Milagro_AttenuationLength},
water filtering continuously improved the attenuation length of Milagro during the first years of Milagro's operation.
A significant deterioration, caused by the decay of the
aluminum baffles, can also be seen in the figure. The replacement of the last aluminum
baffles effected a gradual improvement in the water quality,
as the remaining aluminum-oxide particles were filtered out of the pond.

It should be noted that the outriggers are not connected to the water
system. The outriggers were filled with pond water, and their water
has not been treated since. Recent samples of outrigger water showed
attenuation lengths of the order of $\sim5\,m$, significantly shorter
than the ones in the pond. However, a short attenuation length does not
become significant until it is comparable to the dimensions of the
system that contains it. Thus, a $5\,m$ attenuation length in the $1\,m\times2.4\,m$
outrigger tanks is not worse than a $15\,m$ attenuation length
in the $60\,m\times80\,m\times8\,m$ pond. In other words, the light in the outriggers
is probably going to be absorbed from the
internal outrigger surfaces instead of from the water.%

\section{\label{sec:Milagro_DAQ}Data Acquisition System}

In order to collect and process the PMT pulses, to perform triggering,
and to generate useful information for the online reconstruction, a data
acquisition system (DAQ) was needed. The DAQ system had to satisfy some
requirements:

\begin{itemize}
\item It had to be able to extract information from the PMT output that could be used to
estimate the number of photons detected by a PMT in a fast and accurate way.
\item Because the angular resolution of the detector depended on the accuracy
of the PMT hit times, the DAQ system had to be able to measure these
times with high precision ($\lesssim1\,ns$).
\item The DAQ system had to be able to accommodate the trigger rate caused
by background cosmic rays ($\sim2\,KHz$). 
\end{itemize}

\begin{figure}[ht]
\includegraphics[viewport=0bp 0bp 500bp 600bp,clip,width=1.0\columnwidth]{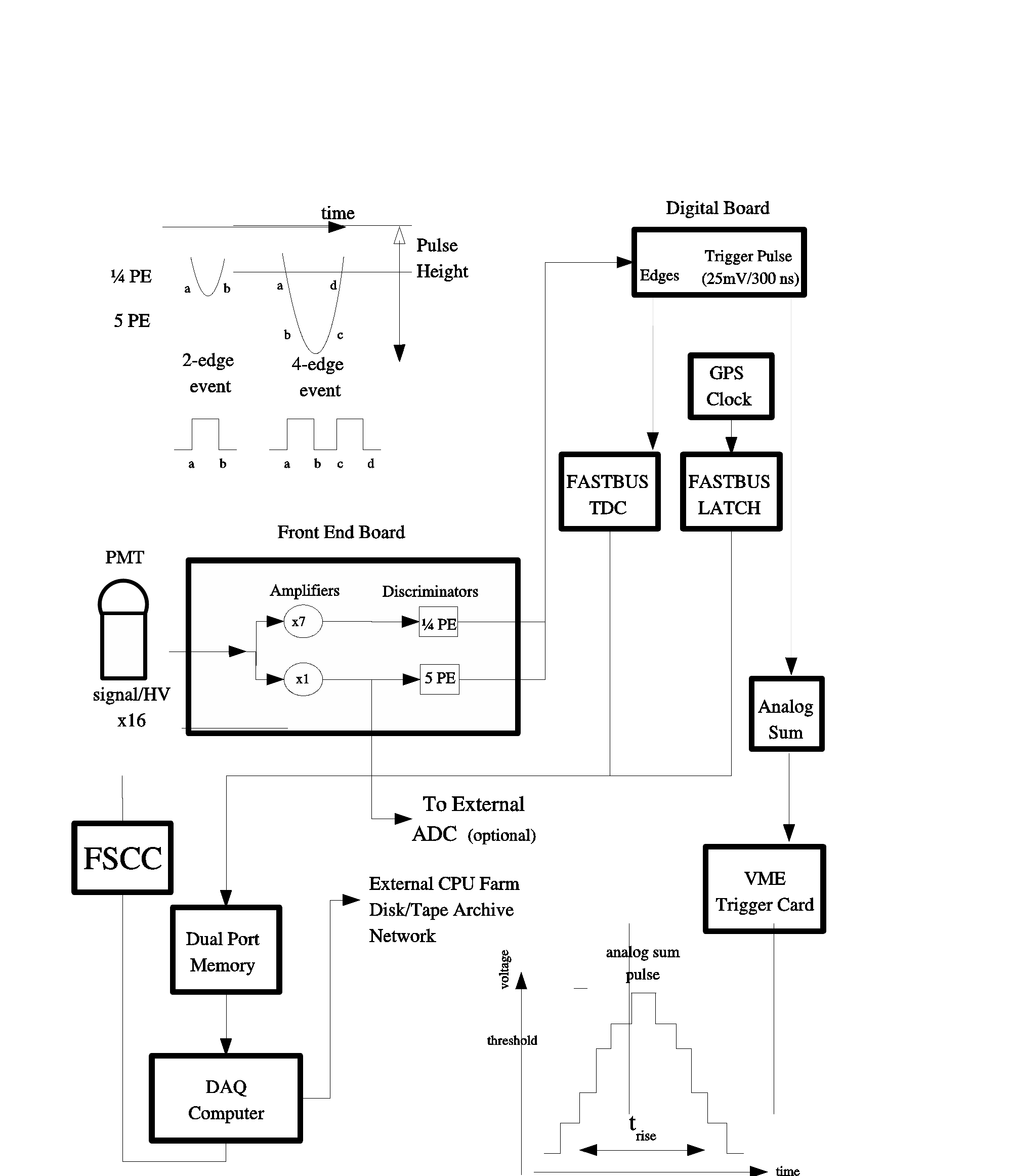}
\caption{\label{fig:Milagro_DAQ}Schematic of the Milagro's electronics system.
Source \cite{ThesisDavidNoyes}}
\end{figure}

In the following, I will describe Milagro's DAQ system and
show how it satisfied the above requirements.
A schematic of the DAQ system is shown in Fig. \ref{fig:Milagro_DAQ}.
The PMTs are divided in patches of sixteen tubes, with each patch supplied
with the same high voltage and connected to a different 16-channel
front-end board (FEB). The connection is through a single RG-59 cable
that carries both the high voltage to the PMT and the signal to the
FEB. Through capacitive coupling, the FEBs pick up the AC signal from the
PMTs to process it. 

Initially, the signal is split, and the halves are sent to two logarithmic
amplifiers (gains $\sim\times7$ and $\sim\times1$). Each amplifier
integrates all the narrow (few $ns$ width) PMT pulses and produces
a shaped pulse of $\sim70\,ns$ width and of an amplitude that depends
on the total charge of all the integrated pulses. The amplitudes of
these shaped pulses are compared to two discrete thresholds, roughly
equal to 1/4 and 5 times the average height of a pulse produced when
a PMT detects a single photon. Every time the amplitudes of the processed
PMT pulses cross any of these thresholds, an edge (square pulse) is
generated, as shown at the top left of figure \ref{fig:Milagro_DAQ}.
A shaped pulse with an amplitude between the two thresholds will trigger
the low-threshold discriminator twice and will cause the generation
of two edges, while a PMT pulse with a height greater than the high
threshold will cross both thresholds twice, generating four edges.
These edges are sent to a digital board connected to LeCroy
FASTBUS Time to Digital Converters (TDCs), which record the times of
each edge. Because of the high precision required in the measurement
of these times, the TDCs can measure the edge times with a resolution of just $0.5\,ns$. 

Every time the detector triggers (see section \ref{sub:Milagro_Triggering}
for information on how triggering is performed), a stop signal is
sent to the TDCs and the FASTBUS latch. Then, a FASTBUS Smart Crate
Controller reads out the digitized information from the TDCs and copies
it to a dual-port memory in a Versa Module Europa (VME) crate. Even though a typical event
lasts a few hundred nanoseconds, the TDC information extends a longer
time before and after the trigger ($\pm1.5\,ms$).This extended
time width allows for the inclusion of signals generated by PMTs that are far
away from the counting house (long cable lengths) and also helps the
reconstruction software with processing the edge data. Because of
hits from noise or scattered light that happened earlier or later
in time than the main shower, the pulses of some PMTs might end up
creating a number of edges different than two or four. In that case, being
able to follow the PMT signal for an extended period of time before
and after the shower helps with deciding which edges were from late
or early hits, and which from hits related to the main EAS.

The time of the trigger is given by a GPS that is connected to the
latch. The DAQ computer (PC running Linux) reads the timing information
and distributes it to a cluster of PCs (the ``workers'') for reconstruction.
After the event is reconstructed, the information
is sent back to the main DAQ computer for storage. 

\FloatBarrier
\section{\label{sub:Milagro_Triggering}Triggering}

Each time the shaped pulse of an AS-layer PMT crosses the low discriminator threshold, the generated
edge by the FEBs causes the digital board to generate a square pulse
($25\,mV$ amplitude and $180\,ns$ width) that is sent to an analog summer. That
analog sum represents a running count of the PMT hit-rate
in the pond within a $180\,ns$ time window. An EAS that lands inside
or near the pond can create hits with a rate high enough to
cause a significant increase in the amplitude of the analog sum. By
comparing the sum's amplitude to a threshold, the triggering system
decides whether to trigger the detector or not. This was the early
version of the Milagro trigger. When the analog sum crossed a threshold
corresponding to $\sim60$ AS PMTs hit in a time interval smaller
than $180\,ns$ (set by the width of the square pulses), the detector
triggered, and the data shortly before and after the trigger were
processed and recorded.

The choice of the triggering threshold effectively defined the low-energy response
of Milagro. The threshold could be reduced in order to trigger on
EASs of a lower energy. However, this would add a large number of extra
triggers caused by single large zenith-angle muons rather than by
EAS. Since Milagro was already taking data near the maximum rate permitted
by its DAQ system ($\sim1700Hz$), the inclusion of these extra events
would just increase the dead time.

\begin{figure}[ht]
\begin{centering}
\includegraphics[width=0.8\columnwidth]{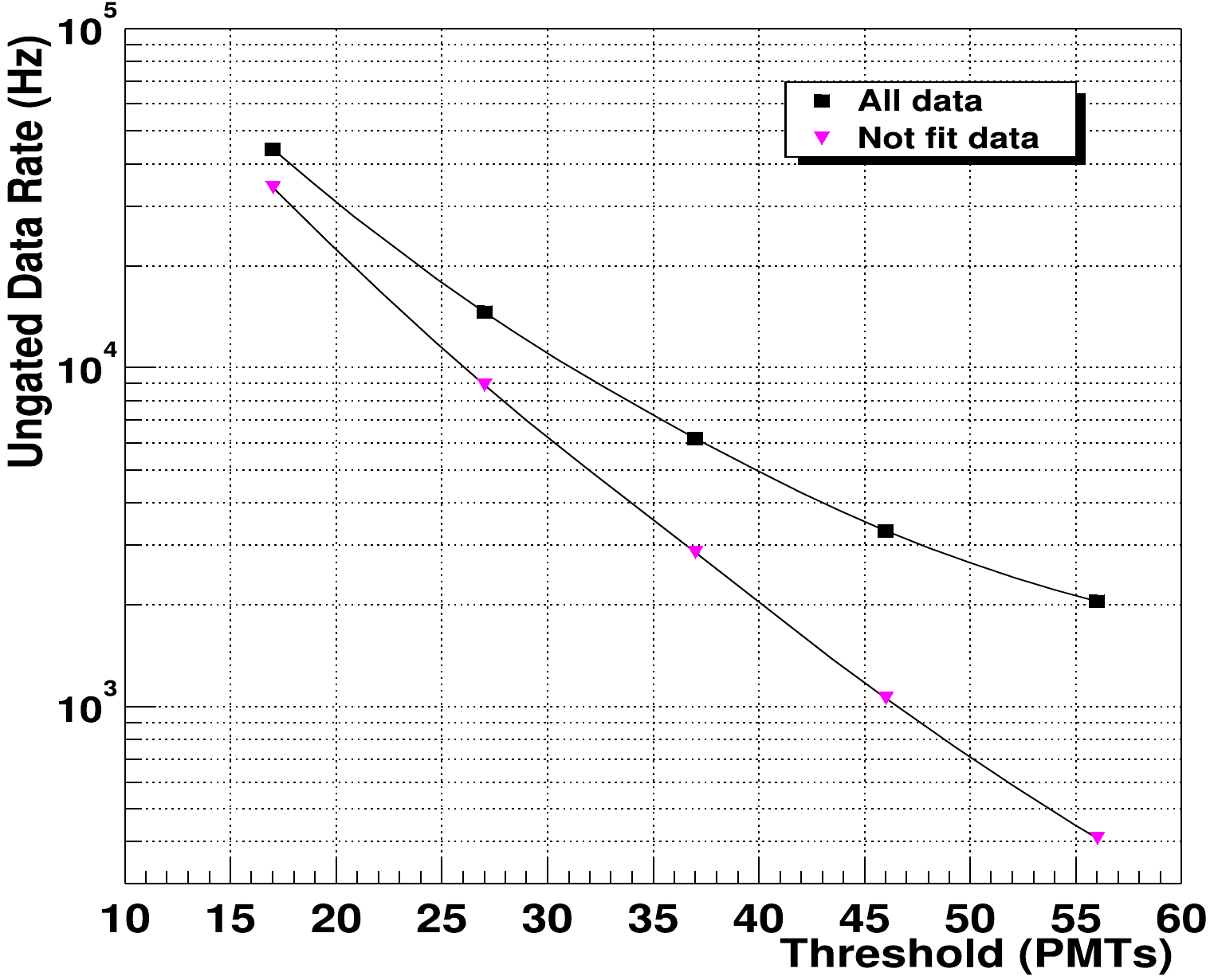}
\caption{\label{fig:Milagro_TriggerRate} Rate of events with an analog sum
over a certain threshold. The rates for all events
(black dots) and for events that could not be reconstructed (pink
triangles) are shown. Source \cite{ThesisDavidNoyes}}
\end{centering}
\end{figure}

The times of the PMT hits caused
by an EAS can be successfully fitted on a plane, while the times of the hits
by a single muon cannot. Figure \ref{fig:Milagro_TriggerRate} shows the rate that the analog
sum crosses different thresholds. For high thresholds, that rate
is dominated by reconstructable events, which, as mentioned, correspond to
EAS. As the threshold gets lower, the fraction of non reconstructable events
increases, implying that the extra events are mostly larger zenith-angle
muons. Most of these muons have zenith angles close to half of the
opening angle of the Cherenkov cone in the water ($\sim41^{o}$). Therefore,
some of their emitted light travels horizontally in the pond and creates hits
in multiple adjacent PMTs. An improved trigger had to accept
lower-energy events (smaller analog-sum amplitudes), while at the
same time, making sure not to trigger on muons.

\begin{figure}[ht]
\begin{centering}
\includegraphics[width=1.0\columnwidth]{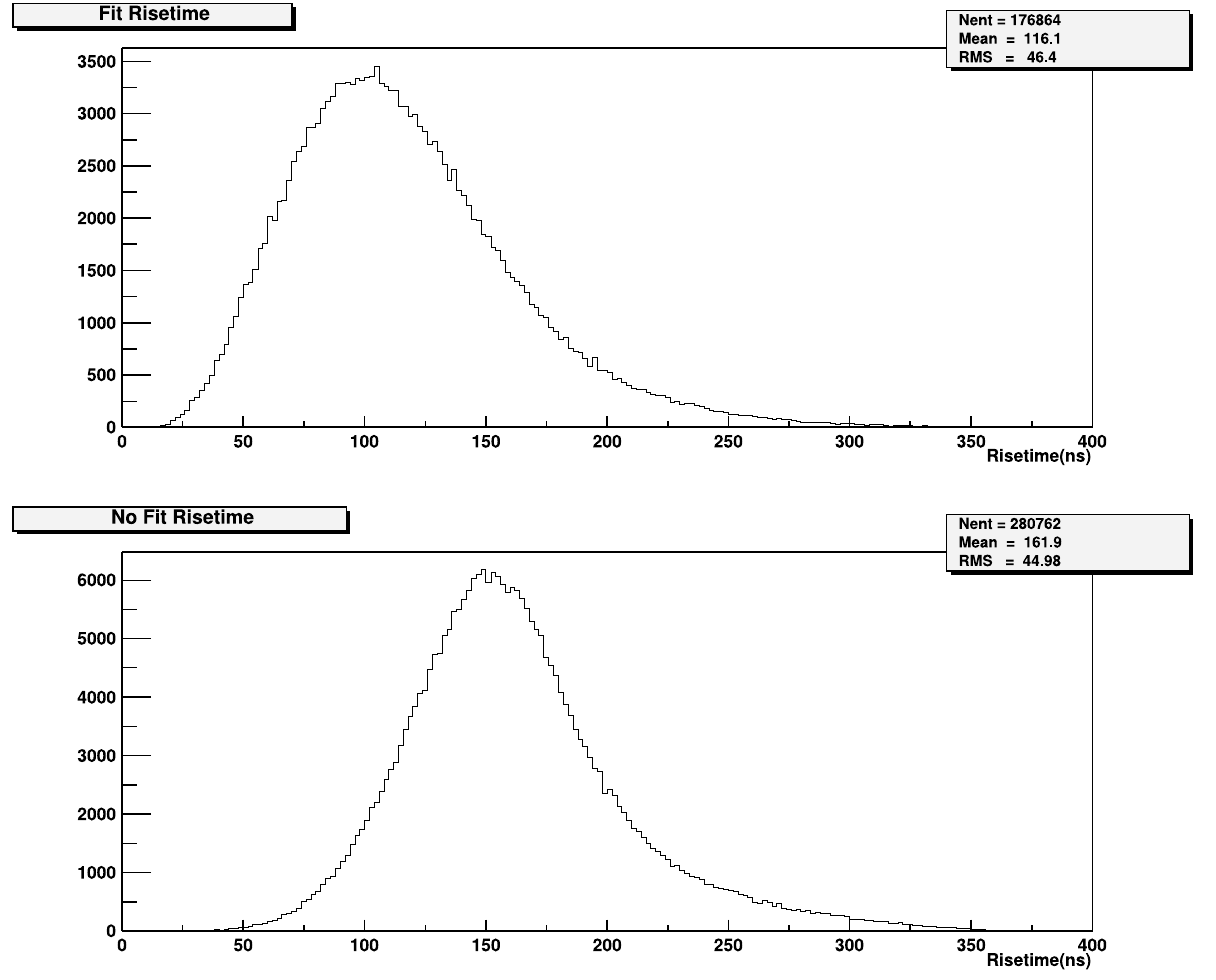}
\caption{\label{fig:Milagro_risetime} Distribution of the rise times of events
that could/could not) be reconstructed (top/bottom). Source \cite{ThesisDavidNoyes}}
\end{centering}
\end{figure}

The fact that the analog sum rises faster when the PMT
hits are caused by an EAS than when they are caused by a triggering
muon was later used for designing an improved triggering algorithm.
The horizontal light generated by a $41^{o}$ zenith-angle muon
creates hits in adjacent PMTs with time delays equal to $c/n$ times
the horizontal distance between the PMTs, where $c$ is the speed of light in
vacuum, and $n$ is the refractive index of water. 
On the other hand, an EAS moves in the water with
the speed of light and creates hits in the AS PMTs that are considerably
closer in time than the hits from a high zenith-angle muon. In the case of an EAS
that comes from zenith, most of the hits will be created almost simultaneously
(ignoring the small shower front curvature), while for the extreme
case of shower from the horizon, the hits on adjacent AS PMTs will be
created with a time delay equal to the horizontal distance of these
PMTs times the speed of light. Figure \ref{fig:Milagro_risetime}
shows the distribution of rise times from EAS events along with the
distribution from horizontal muon events. As can be seen, the EAS
distribution averages at shorter rise times. To take advantages of
this effect, a new custom VME trigger system
was built ($\sim$March 2002), which was also able to make checks on the rise
time of the analog sum. A sample trigger configuration used in the
previous years was:
\begin{itemize}
\item Trigger \#1: nAS PMTs Hit$>$72.
\item Trigger \#2: nAS PMTs Hit$>$50 \& Rise time$<87.5\,ns$.
\item Trigger \#3: nAS PMTs Hit$>$26 \& Rise time$<50\,ns$.
\end{itemize}

Here, the rise time is defined as the time it took for the analog
sum to rise from 10\% to 90\% of its full amplitude. If any of the
above three trigger conditions were true, then the detector would
trigger and record the event. The trigger system was reprogrammable
so that the trigger conditions could be adjusted to follow changes
in the detector's response caused by external factors (changes in
water quality, PMTs dying, etc). The combination of triggers was chosen
to maximize the number of selected low-energy showers, as well as
to retain an unbiased number of large showers. 

About two years ago, the VME trigger was replaced by another system
that works similarly by making checks on the rise times and the amplitude
of the analog sum, but that monitors the analog sum on a shorter time
scale ($\sim80n\,s$) than the VME trigger ($\sim180\,ns$). By doing
so, it becomes more sensitive to the direct light produced by the
EAS particles detected in a time scale of few tens of $ns$.

\begin{figure}[ht]
\begin{centering}
\includegraphics[width=0.8\columnwidth]{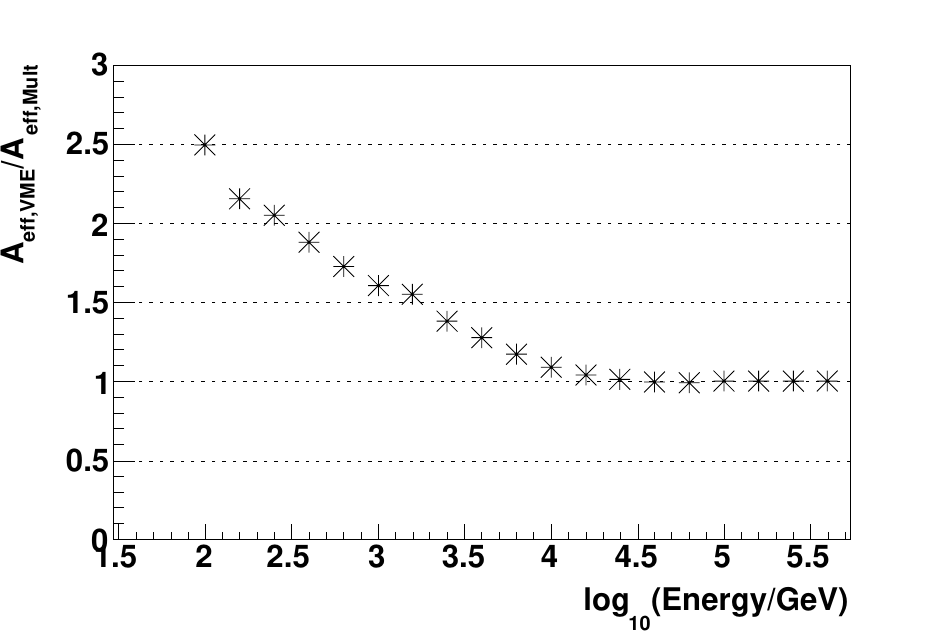}
\caption{\label{fig:Milagro_VMEImprovement} Ratio of the effective area of
Milagro with the new VME trigger system over the effective area with
the old multiplicity trigger. There is a considerable improvement
in Milagro's sensitivity to lower-energy signals when using the new
VME trigger.}
\end{centering}
\end{figure}

The implementation of the last two trigger systems brought a large increase
in the low-energy sensitivity of Milagro (Fig. \ref{fig:Milagro_VMEImprovement}). 
Because the $E\gtrsim300\,GeV$ emission from GRBs is attenuated by
interactions with the extragalactic background light (see
Chapter \ref{chap:IR}), the lower-energy response of Milagro becomes crucial
in determining Milagro's chances of detecting a GRB. The last two trigger
systems made a significant contribution towards reaching that goal.

\FloatBarrier
\section{\label{sec:Milagro_Analysis}Online Reconstruction and Filtering}

The times and charges of the PMT hits can be used to reconstruct the
direction of the primary particle that created the detected EAS. As
the EAS traverses the atmosphere, it spreads out laterally, forming
the shower front. The shower front has a spherical cap shape, whose
apex lies approximately on the extension of the trajectory of the
primary particle towards the ground. The location on the ground where
the apex of the shower front lands has the highest number and energy
density of EAS particles. The number and energy density are symmetrical
around the shower-core location and are quickly reduced with increasing distance from
it. The surface vector that passes from the apex of the shower front
is parallel to the direction of the primary particle that
caused the EAS. Therefore, by reconstructing the shower front, the direction
of the primary particle can be found (see figure \ref{fig:Milagro_WaterCherenkovTechnique}).
Based on the above, Milagro's reconstruction first
locates the shower core on the ground and then reconstructs the
shower front, and therefore the primary-particle direction, using
that core-location information and the PMT hit times.

Specifically, the online reconstruction starts by calculating the
number of photons detected by a PMT using the timing information of
the edges of its shaped pulse, provided by the TDCs (sec. \ref{sub:MILAGRO_TOT}).
Then, based on the photon multiplicity information of each PMT hit,
the reconstruction finds the location on the ground where the shower core landed (sec.
\ref{sub:MILAGRO_CoreFitting}). After that, it calculates the PMT hit times
and applies some curvature corrections on them in order to simplify
the shower-plane reconstruction process (sec. \ref{sub:MILAGRO_Corrections})
Next, it reconstructs the shower plane and, from that, the direction
of the primary particle that caused the EAS (sec. \ref{sub:MILAGRO_DirectionReconstruction}).
Finally, it calculates the values of some gamma-hadron discrimination
parameters (sec. \ref{sub:Milagro_GammaHadron}), to be used later
in the detailed offline data analyses. The following subsections
will describe in detail these steps. 

\FloatBarrier
\subsection{\label{sub:MILAGRO_TOT}Hit size estimation}

Using the time, provided by the TDCs, that a shaped PMT pulse crossed
each of the two discriminator thresholds, the duration the pulse spent
over these thresholds can be calculated. This time is called the
``Time Over Threshold'' (TOT). 
Because the TOT correlates one-to-one with the
number of photons detected by a PMT, Milagro uses it to estimate the
latter. The correlation between the two quantities is periodically
calibrated (see section \ref{sec:Milagro_Calibration}). 

It should be noted that the amplifiers that participate in the generation
of the shaped pulse are logarithmic, and therefore map a wide range of
hit sizes to a narrow range of TOTs. The dynamic range of the system
is wide enough to process, without saturating, hits with sizes ranging from 
a fraction of one photon to few thousand photons.

Even though this is a simple method, it has some disadvantages.
In the case that the detection of a number of photons from the main
shower is followed or preceded by an isolated photon detection, the
TOT of the resulting pulse can be lengthened, resulting in an artificially
inflated estimate for the number of detected photons.
Furthermore, when such late or early hits are superimposed on
the main PMT pulse, they can cause a number of edges that is different
than 2 or 4. This can make the TOT to number of photons conversion
complex, since it requires the identification of the good edges generated
by the part of the signal detected in time with the main EAS. 

\FloatBarrier
\subsection{\label{sub:MILAGRO_CoreFitting}Reconstruction of the shower core}

The knowledge of the shower-core location is needed to
reconstruct the primary particle direction accurately. There have
been many core-fitting algorithms during the lifetime of Milagro.
Initially, and before the outriggers were installed, a simple center-
of-mass fitter was used, with the number of photons detected by each
PMT being used as the weight. Because of the fitter's algorithm, it was placing
all the cores inside the boundaries of the pond, which was
usually incorrect because the majority of the triggers
are created by showers that landed off the pond. That fitter was
later improved to determine whether the shower core was likely to
have landed inside or outside the pond. If the fitter determined that 
the shower core landed outside the pond
it placed it on a fixed distance $50\,m$ away from the
center of the pond towards the direction of the center of mass. Otherwise,
it placed the core on the center of mass. The $50\,m$ distance
was determined by Monte Carlo simulations as the most probable core radius
of showers that landed outside the pond. 

After the installation of the outriggers, it was easier to determine
whether the shower core actually landed inside or outside the pond.
A reasonable parameter that could be used to determine the core location, was the
ratio of the number of hits in the OR layer over the hits in the AS
layer. If that ratio determined that the core most likely landed off
the pond, the center of mass of the outrigger hits was used to determine
the core position. Otherwise, as before, the center of mass of the
AS PMT hits was used. 

\begin{figure}[ht]
\begin{centering}
\includegraphics[width=0.8\columnwidth]{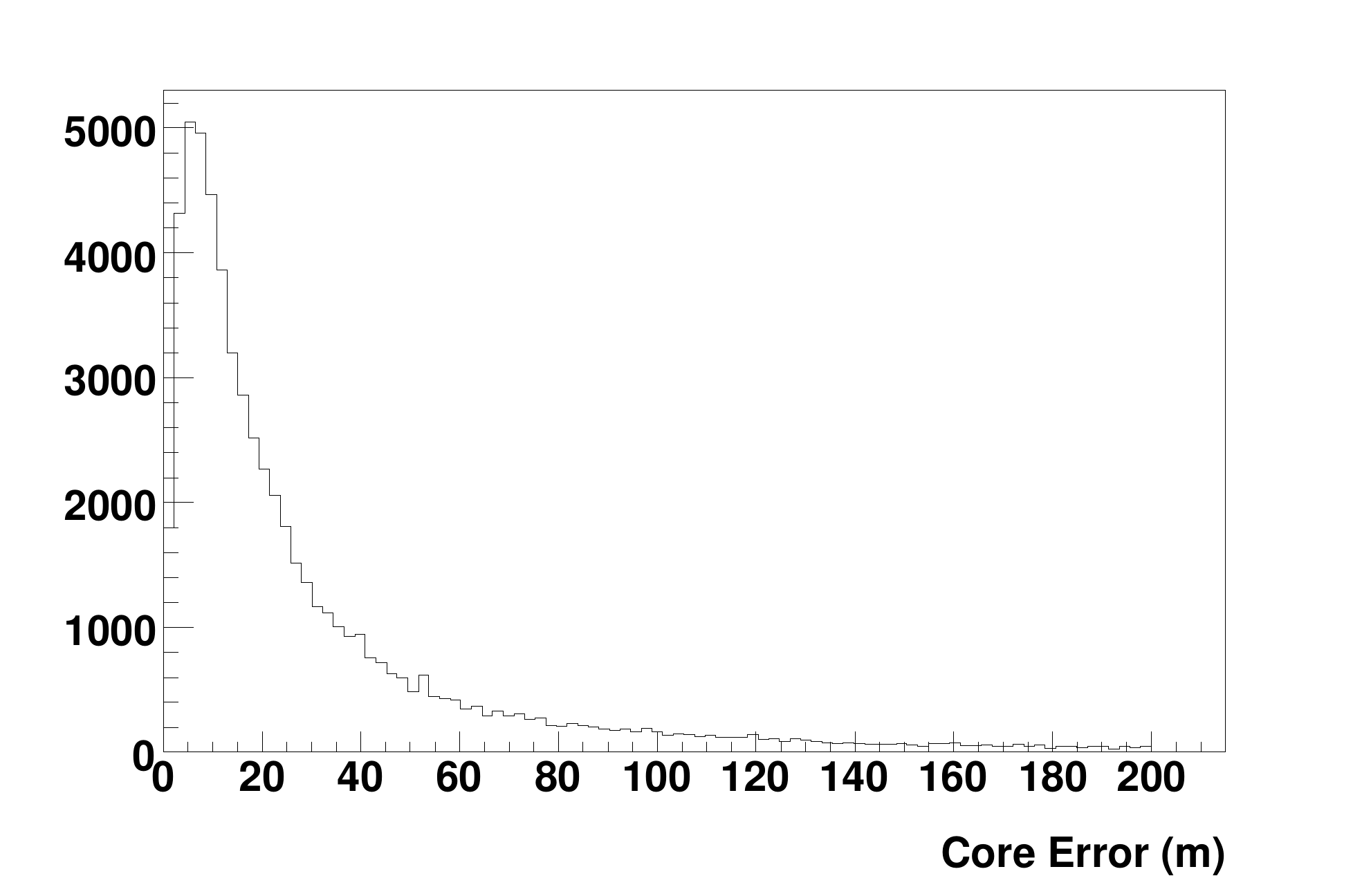}
\caption{\label{fig:Milagro_deltacore}Distribution of the errors in the core
locations for gamma rays from a Crab-like source. A soft cut has been
applied to the data requiring at least 20 PMTs to participate in the
direction-reconstruction fit.}
\end{centering}
\end{figure}

\begin{figure}[ht]
\includegraphics[width=0.8\columnwidth]{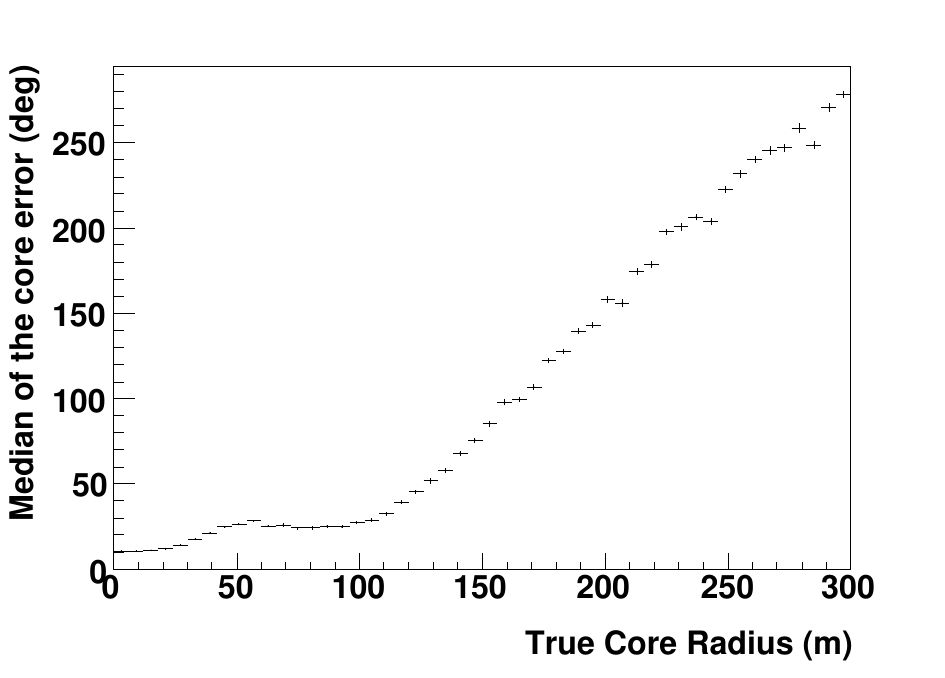}
\begin{centering}
\caption{\label{fig:Milagro_TrueCoreBadCore}Median error in the core position
versus true distance from the shower core. The error bars show statistical
errors on the median.}
\end{centering}
\end{figure}

\begin{figure}[ht]
\begin{centering}
\includegraphics[width=0.8\columnwidth]{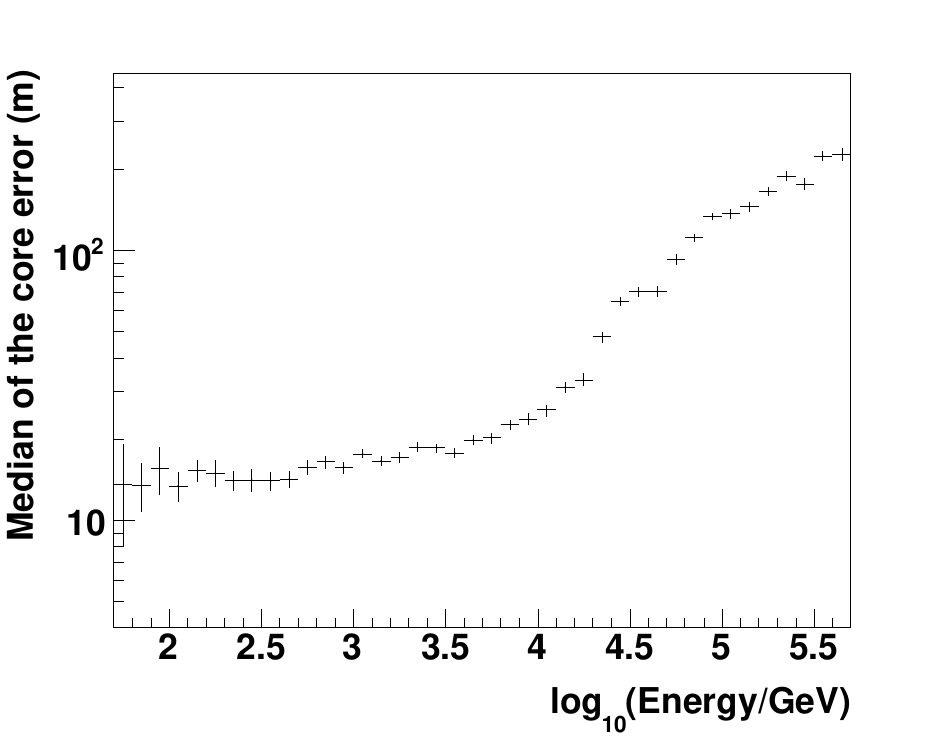}
\par\end{centering}

\begin{centering}
\includegraphics[width=0.8\columnwidth]{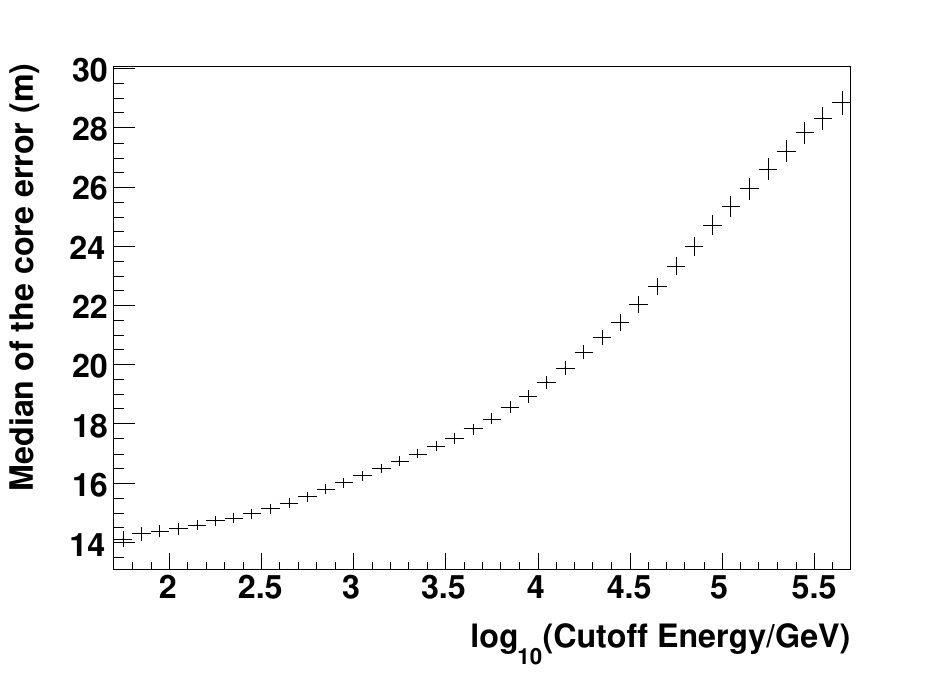}
\par\end{centering}

\caption{\label{fig:Milagro_DeltaCore_E}Effect of the primary particle's energy on the 
accuracy of the reconstructed core location. \textit{Top}: median error in the core position versus
the gamma-ray energy, \textit{bottom}: median core errors for power-law spectra with index -2.00
and different exponential cutoffs on the high energy component}
\end{figure}

The next and last version of the core fitter
did a least-squares fit to a 2D Gaussian using the hits of both the
AS and OR layers. The performance of that fitter is shown in figures
\ref{fig:Milagro_deltacore}, \ref{fig:Milagro_TrueCoreBadCore} and
\ref{fig:Milagro_DeltaCore_E}. From figure \ref{fig:Milagro_TrueCoreBadCore},
it can be seen that when the shower lands farther than about $130\,m$
from the center of the pond, a distance that roughly corresponds to
the limits of the outrigger array, the reconstruction accuracy of
the core location becomes worse. This shows that it is hard to accurately
reconstruct the shower cores of showers that landed outside the detector, a
situation that is similar to when showers landed off the
pond before the addition of the outriggers. Figure \ref{fig:Milagro_DeltaCore_E}
(top) suggests, on a first look, that the accuracy of the core fitter
is worse for higher energy showers. What actually happens is that
higher energy showers can trigger the detector from further away,
even from outside the edges of the detector. The cores of such showers,
as mentioned above, are reconstructed poorly. So, the median error
in the core location of higher energy showers (plotted in the figure)
comes from showers that landed inside the detector and were reconstructed accurately
and from showers that landed outside the detector and were reconstructed
poorly. As the shower-energy gets lower, the contrubution from showers that landed 
outside the detector becomes smaller, and the median core error is reduced. %
A sample distribution on the reconstructed core locations from data
is shown in Fig. \ref{fig:Milagro_deltacore}.%


\begin{figure}[ht]
\includegraphics[width=1.0\columnwidth]{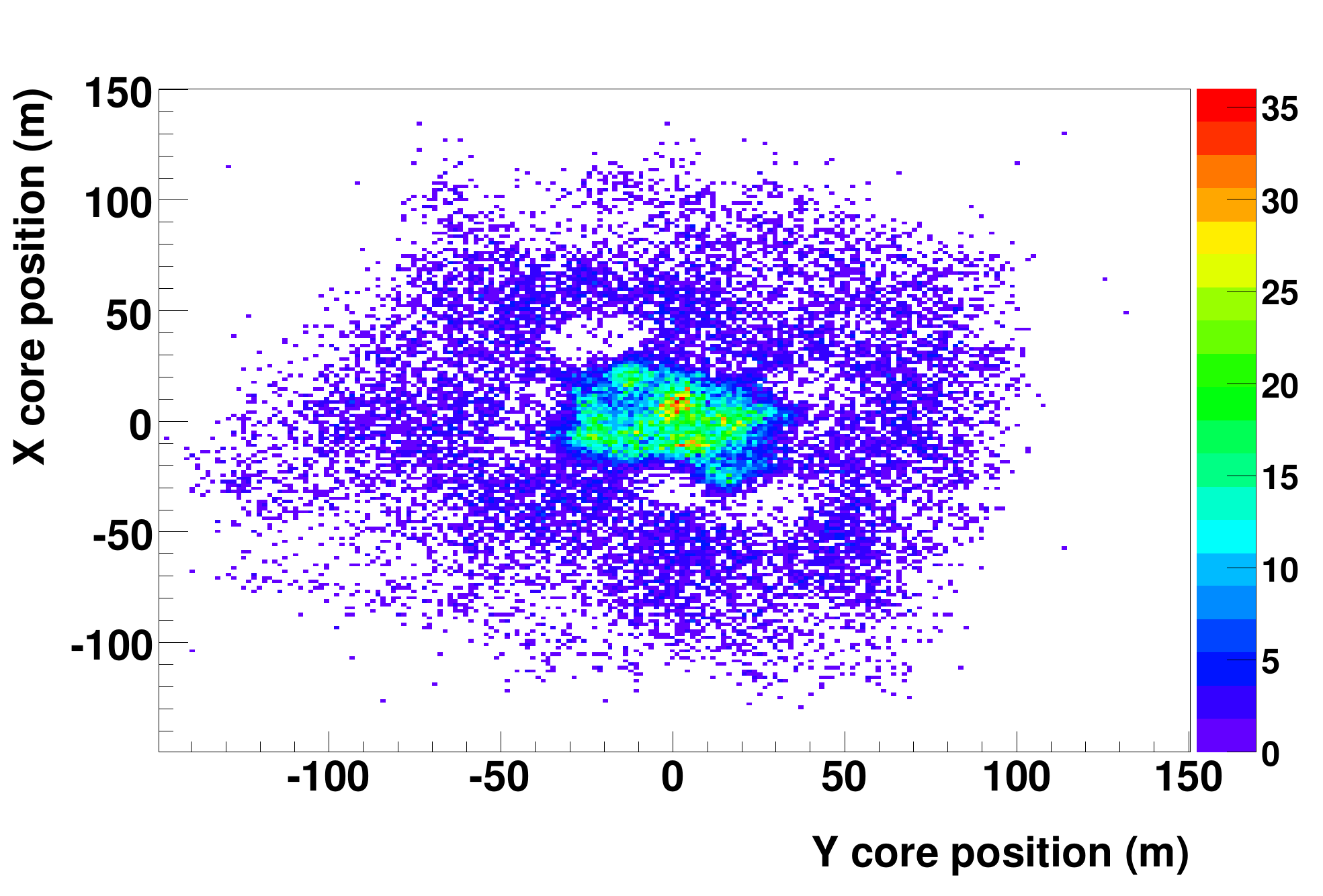}
\caption{\label{fig:Milagro_CoreDistro}Distribution of the reconstructed core
positions.}

\end{figure}

\FloatBarrier
\subsection{\label{sub:MILAGRO_Corrections}Curvature, sampling, and slewing corrections}

Since the shower front is centrally (near its apex) flat, one could
try to reconstruct it by fitting the PMT hit times on a plane. However,
this is just an approximation, and the resulting accuracy of the reconstructed
angle is not optimal. A more accurate way would be to try to fit the
curved shower front on a parabola. However, the equations involved
in such a fit are not of closed form, and the algorithms required to
solve them would be slow. A solution to this problem could be first to
remove the curvature of the shower front by adjusting the PMT hit
times and then to fit it on a plane. This can be accomplished by
adding time offsets, which depend on the distance from the shower core,
to the times of all hits. The magnitude of this ``curvature correction''
per distance from the core is of the order of $0.07\,ns/m$. Such
a correction allows the shower front to be reconstructed in a simple
yet accurate way. 

Another correction that has to be applied comes from the way the shower
front is sampled. Since the time of a PMT hit is determined by the
time of the first detected photon, instead of by the average of the times
of all detected photons, parts of the shower front with high particle
densities will create hits that, on average, preceed hits from lower
particle densities. This effect can be accounted for by applying a
``sampling correction'' to the PMT hits that depends on the amount
of detected photons. The amount of both corrections were determined
by tuning the reconstruction of simulated showers and data.

The hit time of a PMT is given by the time of the first detected photon.
Even though the latter time is not directly measured, it is correlated
to the time that the PMT shaped-pulse crossed the low threshold for the
first time. The delay between the first photon detection and the threshold
crossing depends on the number of photons detected by that PMT. Large
shaped pulses rise faster and cross the low threshold shortly after the first
photon is detected, while smaller pulses can take a longer amount of
time before crossing that threshold. This effect is called ``electronic
slewing,'' and the relation between the time delays and the TOT of
the pulses is used for correcting the calculated PMT hit times. 

\FloatBarrier
\subsection{\label{sub:MILAGRO_DirectionReconstruction}Direction reconstruction}

After the shower core has been located, and the PMT hit times are corrected
for curvature, sampling, and slewing, the shower plane is reconstructed.
The AS and OR layer PMT hit times are fit using a series of least
squares fits. The weight of each hit depends on the number of detected photons.
Because very small hits are usually from scattered light
or noise, their residuals are generally high. For that reason, small hits
(less than a two photons) are not included in the first fit. The residuals after the
first iteration are calculated, and if they are within a preset range,
they are included for the second iteration. Smaller hits than
before are also included, allowing more hits to participate in the fit.
The process of relaxing the hit-size requirements and rejecting hits
with large residuals is repeated five times. The result is that more
than $\sim95\%$ of the simulated gamma-ray showers are fit. The corresponding
percentage in data is somewhat smaller ($\sim85\%$), mostly because
some of the triggers are caused by isolated horizontal muons rather
than by EAS and because the shape of cosmic-ray showers makes them somewhat harder to fit. 
If the fit fails or if the results are unphysical (the
reconstructed direction lies below the horizon), the event is tagged as
non-reconstructable. %
For a small period, information from all three layers was used for
the fit; however, there was no improvement in the reconstruction accuracy.

\begin{figure}[ht]
\begin{centering}
\includegraphics[width=0.8\columnwidth]{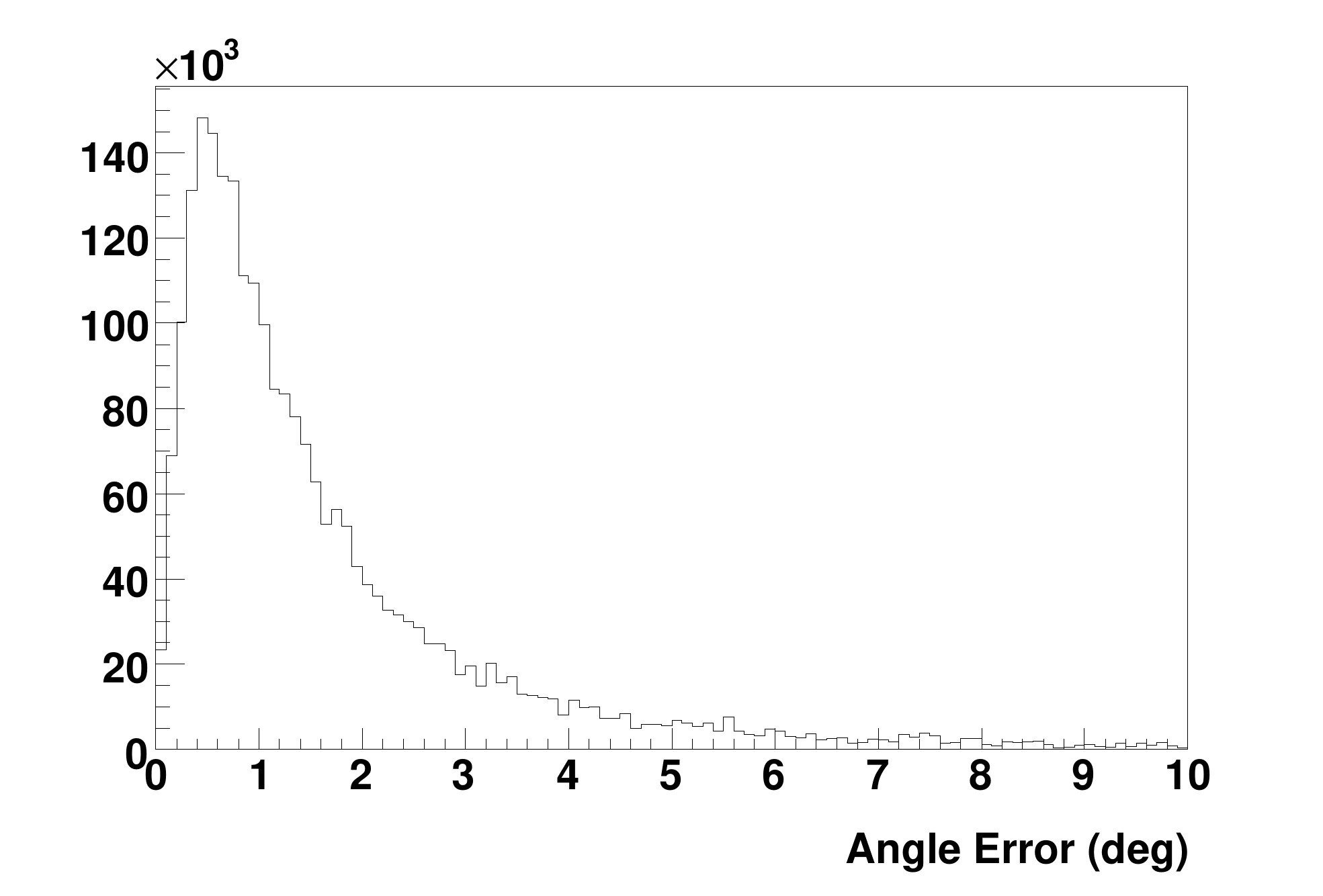}
\caption{\label{fig:Milagro_delAngle}Error in the reconstructed angle for
a simulated gamma-ray signal on a power law spectrum with index -2.00.
A soft cut has been applied to the data requiring at least 20 PMTs
to participate in the fit.}
\end{centering}
\end{figure}

\begin{figure}[ht]
\begin{centering}
\includegraphics[width=0.8\columnwidth]{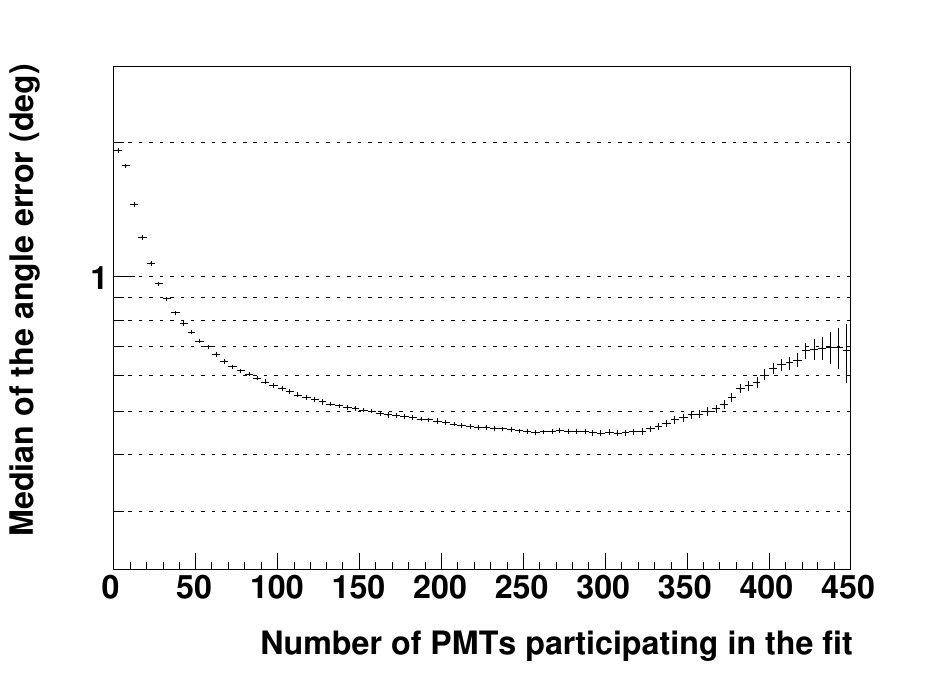}
\caption{\label{fig:Milagro_delAngle_nfit}Median error in the reconstructed
angle versus the number of PMTs participating in the direction reconstruction
fit.}
\end{centering}
\end{figure}

\begin{figure}[ht]
\begin{centering}
\includegraphics[width=0.8\columnwidth]{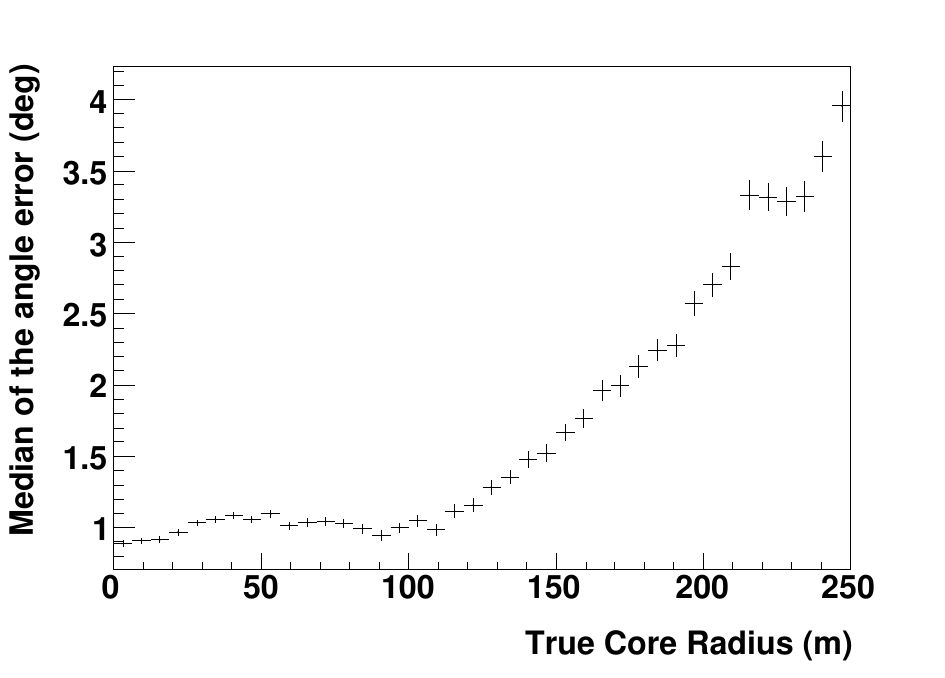}
\caption{\label{fig:Milagro_delAngle_CoreRMC}Median error in the reconstructed
angle versus the true distance of the shower core from the center
of the pond. A soft cut has been applied to the data requiring at
least 20 PMTs to participate in the fit. The error bars show the statistical
error on the median.}
\end{centering}
\end{figure}

\begin{figure}[ht]
\begin{centering}
\includegraphics[width=0.8\columnwidth]{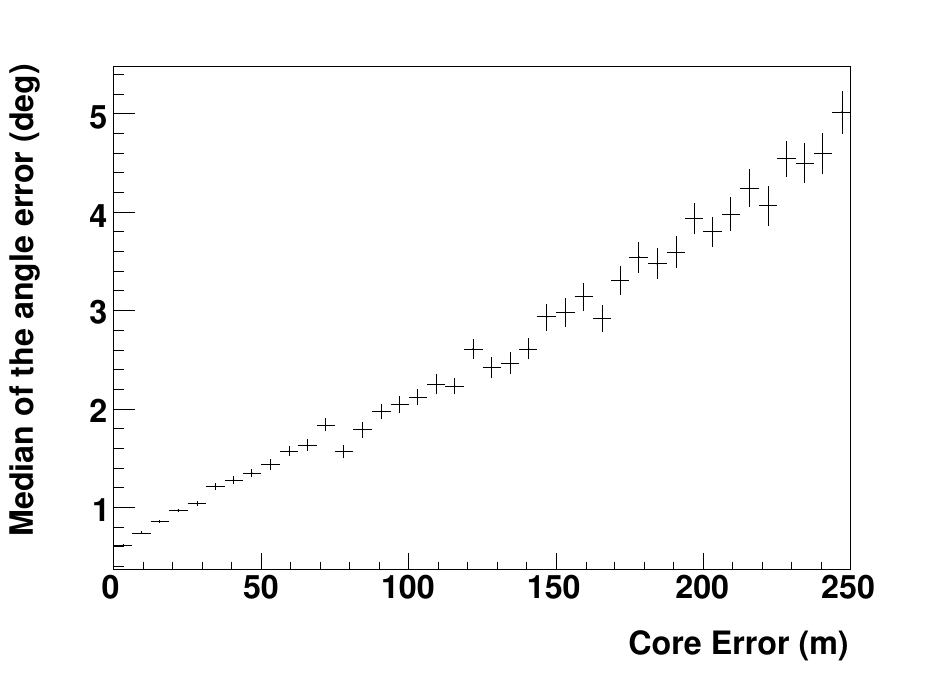}
\caption{\label{fig:Milagro_CoreE_AngleE}Median error in the reconstructed
angle vs error in the reconstructed core location. The width of the
error bars is equal to one standard deviation. A soft cut requiring
at least 20 PMTs participating in the fit has been applied. The error
bars show the statistical error on the median.}
\end{centering}
\end{figure}

The performance of the angle fitter is shown in figure \ref{fig:Milagro_delAngle},
where the distribution of errors in the reconstructed angles is plotted.
The plot in figure \ref{fig:Milagro_delAngle_nfit} was created by plotting the
median of such a distribution for different minimum
number of PMTs participating in the fit (nfit). As expected, that figure shows that
the angular resolution of the detector depends on the number of PMTs
participating in the angular reconstruction fit. Figure \ref{fig:Milagro_delAngle_CoreRMC}
shows the correlation between the median error in the reconstructed
angle and the true distance of the shower core. It can be seen that
the reconstruction is optimum for showers that have a large part of
their cores inside the boundaries of the detector. This effect is
also shown in figure \ref{fig:Milagro_CoreE_AngleE}, where the correlation
between the median error in the reconstructed angle and the error
in the reconstructed core location is plotted. Showers that landed
at the boundaries of the detector usually have poorly reconstructed
cores; therefore the angular resolution for them is worse. 

\begin{figure}[ht]
\begin{centering}
\includegraphics[width=0.8\columnwidth]{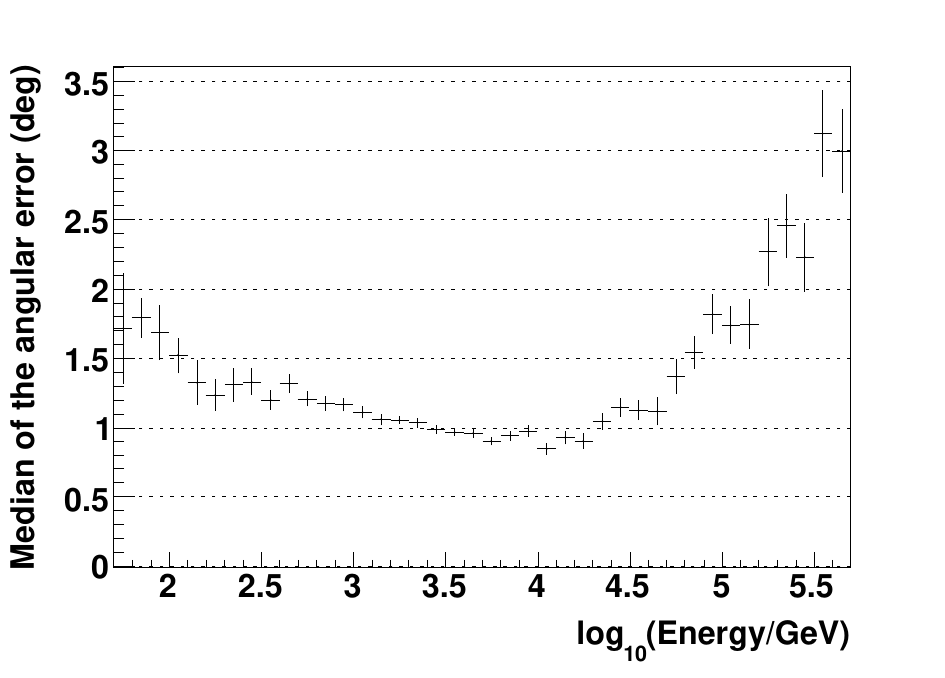}
\includegraphics[width=0.8\columnwidth]{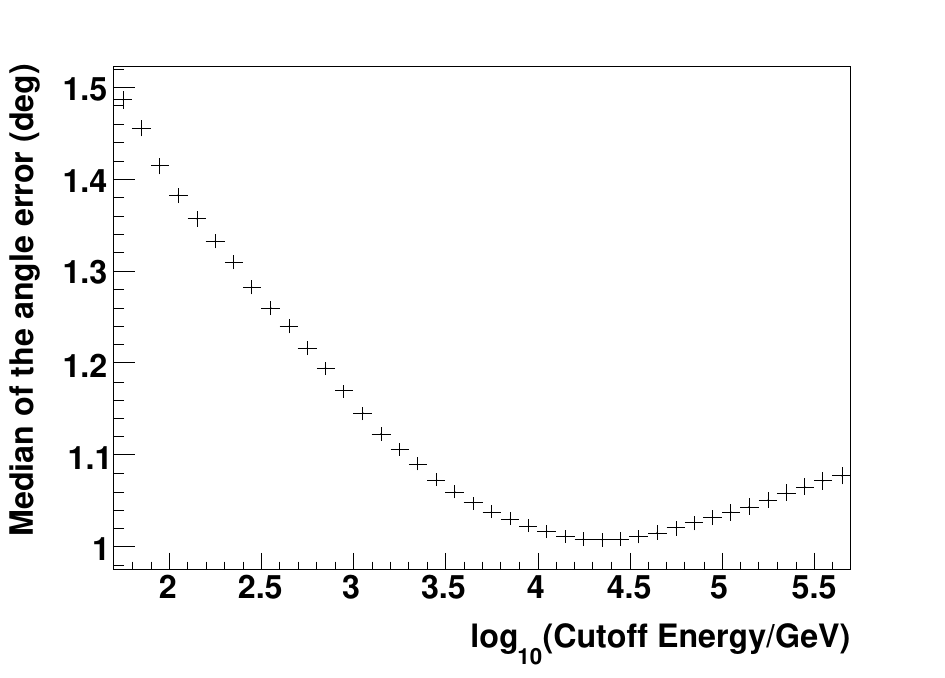}
\caption{\label{fig:Milagro_DelAngle_E}Effect of the primary particle's energy on the angular reconstruction accuracy.
\textit{Top}: median error on the reconstructed angle versus the gamma-ray
energy, \textit{bottom}: the same error for power-law spectra of index -2.00 and different exponential
cutoffs.}
\end{centering}
\end{figure}

Figure \ref{fig:Milagro_DelAngle_E} shows the median error in the reconstructed
angle versus the energy of the gamma-ray primary. The top plot shows
that the median angular resolution is worse for higher-energy showers,
since these showers have on average poorly reconstructed cores. The
bottom plot shows the median of the angular resolution for gamma-ray
signals on power law spectra with exponential cutoffs at different
energies. That plot is relevant to this study, since because of absorption
by the extragalactic background light (\ref{chap:IR}), the gamma-ray signal
emitted by a GRB arrives at the earth with its high energy component
absorbed. For example, the absorption of the signal from a GRB at redshift
$z\simeq0.3$ creates an exponential cutoff at $\sim300\,GeV$ on its
spectrum at the earth. From the bottom of figure \ref{fig:Milagro_DelAngle_E},
it can be seen that the median angle error for such a signal will
be of the order of $1.3^{o}$ instead of the optimal $\sim1^{o}$
for an unabsorbed signal. 

%

\subsection{Data storage}

There are two kinds of Milagro data: raw data and reconstructed data.
The raw data are generated by Milagro's electronics and contain information
on the times of all edges generated by the PMT pulses and the corresponding
TOTs; the time of the event; the amplitude of the analog sum; the
rise time, and other event ID bits. The reconstructed data are the
results of the processing of the raw data by the Worker computers.
They contain information such as the location of the shower core,
the direction of the primary particle that caused the EAS, values
for various gamma-hadron discrimination parameters, and brief statistics
on the number/size of hits in each layer. Even though the raw data
are saved in a compressed binary format, about 250GB of raw data are
generated each day. Because the reconstructed data contain information
about the event as a whole, rather than information on the individual
edges and hits, they take significantly less space (4GB/day). 

Milagro saves all of the reconstructed data but cannot afford to save
all of the raw data. However, because improved reconstruction algorithms
may become available in the future, a fraction of the raw data is saved.
This fraction is the one that results from the whole raw data-set
after applying a soft gamma-hadron discrimination cut ($X_{2}\gtrsim1.5$ - see next subsection)
on it. The cut value is selected so that the amount
of raw data saved per day is in the range of the capabilities of the archiving
system (roughly one third of the data). In addition, all raw data
are saved when significant events such as AGN flares and GRBs happen,
and all the time for selected sources such as the Crab Nebula, the
sun, and the moon. The data is organized into runs and subruns, with
each subrun containing about 5 minutes of data. The run number is
automatically incremented every time the data-taking process is restarted
or daily at 0:00 UT.

\subsection{\label{sub:Milagro_GammaHadron}Gamma-hadron discrimination}

The lateral distribution of particles of a gamma-ray induced EAS is
usually uniform and contains mostly $e^{-}$, $e^{+}$, and gammas. These
particles are absorbed promptly after entering Milagro's water, and
most of them do not manage to penetrate deep enough close
to the MU layer. As a result, gamma-ray induced showers usually produce
a large number of relatively small and uniformly-spread hits in the
PMTs of the MU layer. 

Because of the high transverse
momentum of hadronic interactions, the longitudinal development of
an EAS is characterized by branches of particles separated from the
main shower. These branches usually contain charged pions that later
decay to muons, which, as minimum ionizing particles, are usually
able to penetrate deeply into the pond, reach down to the bottom layer,
and create big hits in few nearby PMTs. As a result, the hit distribution
from a hadron-induced shower is described by a number of hits on the
AS and OR layers that are symmetrical around the core, accompanied
with distant from the core large hits that can show up in the MU layer. 

Milagro's first version of gamma-hadron discrimination was based on
the fact that big%
\footnote{with a high number of detected photons} 
and few hits in the MU layer imply a hadronic (cosmic ray) shower,
while many small hits imply a gamma-ray induced shower. The decision
regarding the nature of the primary was based on the value of a parameter
called $X_{2}$: \begin{equation}
X_{2}\equiv\frac{nb_{2}}{PE_{max}},\end{equation}
where $nb_{2}$ is the number of MU-layer PMTs that detected at
least two photons, and $PE_{max}$ is the largest number of photons
detected by any PMT of the same layer. 

The source of the gamma-hadron discriminating power of $X_{2}$ lies
on $PE_{max}$; hadronic showers tend to create big hits in the muon
layer (large $PE_{max}$), while gamma-ray showers tend to create small.
Thus, a large $X_{2}$ means that the primary was likely
a gamma ray. However, not all hadronic showers have large $PE_{max}$. The
smaller the primary energy of a hadronic shower is, the fewer its
muons--that create the big hits in the muon layer. For such showers,
a low $PE_{max}$ could still be consistent with a hadronic nature.
Similarly, if a hadronic shower--that creates many hits in the MU layer--is to 
be correctly identified as such (hadronic), it should also create big hits in the MU layer (large $PE_{max}$).
$X_{2}$ uses $nb_{2}$ to estimate
the size of the shower and to scale the minimum $PE_{max}$ required to
identify a shower as hadronic. 

One of the problems of $X_{2}$ is that it does not take into account
the shower size information from the outriggers. A gamma-ray shower
that landed on the outrigger array and created a large number of OR
hits and few and small MU hits would be indistinguishable by $X_{2}$
from a lower energy hadronic shower that landed in the pond and did not
create any big hits in the MU layer. By including the information
from the outriggers, we can increase our expectations for $PE_{max}$
for the first case, and decrease our expectations for the second case.
Specifically, in the first case, knowing that we had a large number
of outriggers hit, we should expect a big hit in the muon layer in
order to identify the shower as hadronic. Also, for the second case,
seeing that the detected shower had a lower size (since
it did not create many hits in the OR and AS layers), we should not
require a very large $PE_{max}$ in order to identify it as hadronic. 

\begin{figure}[ht]
\begin{centering}
\includegraphics[width=0.8\columnwidth]{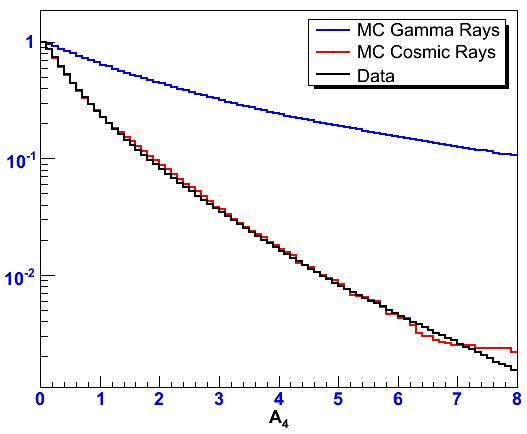}
\caption{\label{fig:Milagro_A4}Distribution of the number of events
retained from an A4 cut. The distributions predicted by the Monte Carlo simulation
of the detector for gamma rays (signal) and cosmic rays (background)
are shown with the blue and red curves respectively, and the distribution
for cosmic rays from data is shown with black. Based on the fact that the distributions for 
cosmic rays and gamma rays are very different, these two event types can be effectively 
differentiated. Source A. Abdo. }
\end{centering}
\end{figure}

The next version of Milagro's gamma-hadron discrimination parameter
successfully included the information from both the AS and OR layers.
Its name is $A_{4}$ and is given by: \begin{equation}
A_{4}\equiv\frac{f_{AS}+f_{OR}}{PE_{max}}nfit,\end{equation}
where $f_{AS}$ and $f_{OR}$ are the fractions of the AS and OR
PMTs hit, and nfit is the number of AS and OR PMTs participating in
the direction reconstruction fit. The sum $(f_{AS}+f_{OR})$ provides
a better than $nb_{2}$ measure of the size of the shower. $A_{4}$
is particularly successful at identifying higher-energy gamma-ray
showers that landed on the outrigger array. For such showers, the
large number of OR and AS hits combined with an absence of a big hit
in the muon layer would strongly support an electromagnetic nature. 
The distribution of the fraction of particles retained after applying
various A4 cuts for cosmic and gamma rays are shown in figure \ref{fig:Milagro_A4}.
As can be seen from the figure, the distributions are very different, therefore
A4 can effectively differentiate a gamma-ray from a cosmic-ray signal. 

Milagro's sensitivity to the signal from a typical gamma-ray source,
such as the Crab Nebula, is increased by factors of $\sim1.5$ and
$\sim2.5$ using $X_{2}$ and $A_{4}$ respectively. By slicing the
data based on the shower size (based on nfit) and applying different
$X_{2}$ cuts on each slice, an improvement in sensitivity comparable
or better than the one of $A_{4}$ can be obtained. Using an $A_{4}$
based gamma-hadron separation and event-weighting methods,
Milagro's observations of the Crab Nebula acquire statistical
significance with a rate of $\sim8\sigma/\sqrt{T/yr}$, where $T$ is 
the observation time; or, equivalently,
Milagro can observe the Crab Nebula with a $5\sigma$ significance in
4.6 months. 

Unfortunately, these gamma-hadron discrimination parameters are not
effective for lower-energy signals. Because the GRB emission reaching
the earth has energies usually lower than a few hundred GeV, it will
create events with only few tens of PMTs hit on average. Such
low energy gamma-ray emission cannot be easily distinguished from
lower-energy hadronic events, because such events rarely create big
hits in the PMTs of the muon layer. Furthermore, for the shorter GRB-emission
durations ($\lesssim100\,s$), the number of detected signal events is small, and the importance
of keeping all of the signal events is very high. For such durations,
only a gamma-hadron discrimination that is extremely effective in
keeping all of the signal could increase the sensitivity.
Unfortunately, such a method is not currently available in Milagro.
For the above reasons, there was no gamma-hadron separation applied on the data searched
in this study.

\section{\label{sec:Milagro_Calibration}Calibration System}

The pond PMTs are calibrated periodically to account for variations
in the PMT gain and timing caused by slow changes related to the
components of the PMT base and to the dynode structure. The quantities
calibrated are:

\begin{itemize}
\item The relation between the TOT of a PMT's shaped pulse and the number
of photons detected by the PMT,
\item the TOT's relation with the delay between the first photon detection
and the first edge generated by a PMT (slewing corrections), and 
\item the delay between the first photon detection and the time the PMT
signal arrives at the FEBs (``time pedestals''). 
\end{itemize}
The first two quantities are calibrated by illuminating a PMT with
a light intensity that creates a known average number of detected
photons at a precisely known time, and then measuring the TOT of its
shaped pulses. The third quantity is equal to the sum of several times,
such as the time it takes for the produced photoelectron by the PMTs
photocathode to be collected by the dynode structure,
the time taken for the avalanche in the dynodes to develop, and
the transit time of the PMT pulses through the cables to the FEB, etc.
The last quantity varies from PMT to PMT and can be calibrated by illuminating a
PMT and measuring the time it took for its signal
to reach the FEBs.

The calibration system consists of a fast pulsed laser, a filter wheel,
and optical fibers. The optical fibers carry the laser light into the pond or to
the outriggers and disperse it through diffusing balls glued at their
edge. There are many such illumination sources inside the pond, so
that all PMTs can be directly illuminated by at least one of them
or can be potentially cross-calibrated using illumination from different
directions. 

The calibration process starts with the laser light first passing
through the least transparent part of the filter wheel and then
through the fiber-optic system driven into the pond. Initially, the PMTs
might not be able to detect the laser pulses because the light intensity
is very low. However, after gradually passing the laser beam through
more transparent parts of the filter wheel, the PMTs will start detecting
some of them. The probability of a PMT detecting a photon from one
pulse can be approximated to be independent of detecting another photon
from the same pulse. Because the probability of detecting a photon is
very small and the number of available for detection photons (in each
laser pulse) is large, the number of detected photons per laser pulse
follows a Poisson distribution. The fraction of the pulses from which
a PMT detected at least one photon is called the ``occupancy''
($\eta$) of the PMT. If $P(n,\lambda)$ is the Poisson probability
that the PMT detected $n$ photons per pulse when the expected (average)
number of detected photons per pulse was $\lambda$, then the occupancy
is
\begin{equation}
\eta=P(n>0,\lambda)=1-P(n=0,\lambda)=1-e^{-\lambda}.\label{eq:Milagro_Occupancy}\end{equation}
Solving eq. \ref{eq:Milagro_Occupancy} for $\lambda$, we get $\lambda=-ln(1-\eta)$.
By just counting the number of laser pulses that resulted
in the detection of at least one photon ($\eta$), we calculated the
average number of photons detected per pulse ($\lambda$). The estimate
on $\lambda$ is accurate only for small light levels because a small
error in the occupancy $\Delta\eta$ will lead to a very large error
in $\lambda$:\begin{equation}
\Delta\lambda=\frac{1}{1-\eta}\Delta\eta=e^{\lambda}\Delta\eta.\end{equation}
For higher light levels, the amount of light detected by a PMT $\lambda_{i}$
can be calculated by scaling the amount of light detected at a dimmer
setting by the ratio of the filter-wheel transparencies $T_{i}$ used
in the two settings: $\lambda_{1}/\lambda_{2}=T_{1}/T_{2}$. Based on
the average number of photons a PMT detects per pulse $\lambda$,
the relations between the TOT and $\lambda$, and between the TOT and
the slewing time delays can be calibrated. 
 
    \clearpage \chapter{\label{chap:Search}The Search}
\section{Introduction}

This chapter will describe the search algorithm used in this study.
Section \ref{sec:SEARCH_Milagro's_Searches}
will describe the various methods the Milagro data are searched 
for emission from GRBs, and will explain how this
study differs from these methods. The chapter will conclude
with a detailed description of the search method used in this study.

\section{\label{sec:SEARCH_Milagro's_Searches}Milagro's Searches for GRBs}

The Milagro data can be searched in various ways for transient gamma-ray
emission. One way is to search in coincidence with GRB detections
from other instruments. In such a case, an external instrument, such
as a satellite, detects a GRB and notifies Milagro through the GRB
Coordinates Network (GCN)%
\footnote{http://gcn.gsfc.nasa.gov/%
}. Then, Milagro searches its data for signal related to the detected
event. Because the VHE component of the emission from GRBs detectable
by Milagro is produced by physical processes different than the ones producing 
the $keV/MeV$ component, to which the satellite detectors are sensitive, the emission durations of
the two components can be different or offset in time. For that reason,
Milagro's data is searched not only for emission of exactly the same
duration and starting time as that of the event detected by the external instument, but also
for emissions of multiple durations and time offsets close to the
ones of the external trigger. Milagro performs two such kinds
of searches, depending on the energy of the gamma-ray signal.

As the energy of a gamma-ray signal gets smaller, the number of air
shower particles at the Milagro altitude also becomes smaller, and
the resulting angular-reconstruction accuracy becomes worse. In the
extreme case of signals with energies of tens of $GeV$, most of the
air showers cannot even trigger the detector or, in the best case,
can trigger, but end up being reconstructed poorly. As a result, such a low-energy
gamma-ray signal usually cannot create a number of accurately reconstructed
events large enough to cause a significant detection. However, this
low-energy signal can still be detected through its effect on the
hit rates of individual PMTs. Milagro monitors these hit rates
for improbable increases in coincidence with external GRB triggers \cite{aune_icrc};
an improbably elevated hit rate close in time with a GRB localized
in Milagro's field of view would imply that a gamma-ray signal was
detected from that GRB. One disadvantage of this method is that, in
case such an increase in the PMT hit rates happens, it would not be
possible to determine where the particles that caused this increase
came from (since there will not be any reconstructed events). Thus, it
would not be possible to verify that the signal that caused the increased
hit rates actually came from the direction of the externally-detected
GRB--which could help to verify the connection between
the two events. Furthermore, because of the absence of pointing information.
the background cannot be reduced. If a specific location on the sky
were searched, then pointing would reduce the background only 
the around the probed location. As a result, the scalers are always sensitive
to the background from the whole overhead sky.
Lastly, this method is sensitive to instrumental
effects that could cause artificial spikes or drifts in the hit rates,
such as light leaks and fluctuations in the electronics' temperatures
and supply voltages. Such effects can easily interfere with
the background estimation, or, worse, manifest as a real signal. Despite
these disadvantages, this method is the only one available to Milagro
that can detect such low-energy signals, making it sensitive to more
distant GRBs than the other methods GRBs ($z\gtrsim0.5$), and to GRBs
with a highly-opaque fireball or dense surrounding material (see section \vref{sub:VHE_Absorption}). 

As the energy of the gamma-ray signal increases, the effective area
of the detector and the fraction of events with good reconstruction
fits increase too. At energies $\gtrsim100\,GeV$, a search for improbable
increases in the reconstructed event rate from a particular direction
in the sky becomes more sensitive than a search for increases in the
individual PMT hit rates (described above). In this case, the search tries
to detect bursts of events reconstructed inside the error box of an
externally-detected GRB and in coincidence or shortly later in time
after it. Milagro performs such a search, and while it has not
detected any significant events, it has published upper limits on
the VHE emission for most of the externally-detected GRBs in its field
of view \cite{pablogrb_icrc,Milagro_ShortGRB}. 

However, only a fraction of the GRBs in nature are detected by the other external
instruments, since they can be outside the instruments' field of view,
or their emission can be too short or weak to be detected. A
number of GRBs left undetected by other instruments can still be detectable
by Milagro. In this study, the Milagro data was searched in a blind
way, independently of an external localization provided by other instruments,
for emission from such GRBs. Even though such a search has the advantage
of being able to search for VHE emission from all GRBs in Milagro's
field of view, its blind nature and the large extent of Milagro's dataset
require a large number of trials (individual sub-searches - $\sim10^{15}$). As will
be shown in Chapter \ref{Chapter:Probabilities}, this large number
of trials decreases the sensitivity of such a search by only a factor of
$\sim2$. The searches in coincidence with external instruments do not
suffer from such a large decrease in sensitivity because of their considerably
smaller number of trials. An advantage of this search, in contrast
to the triggered searches described above, is that it is sensitive to any kind of
VHE transient emission, such as from primordial black hole evaporation 
or other yet undetected phenomena. 


\section{\label{sub:SEARCH_Overview}Overview of the Search}

The search essentially searched for highly improbable increases 
in the rate of reconstructed events from a particular
direction on the sky. 
\begin{landscape}
\begin{figure}[ht]
\includegraphics[height=0.9\textheight]{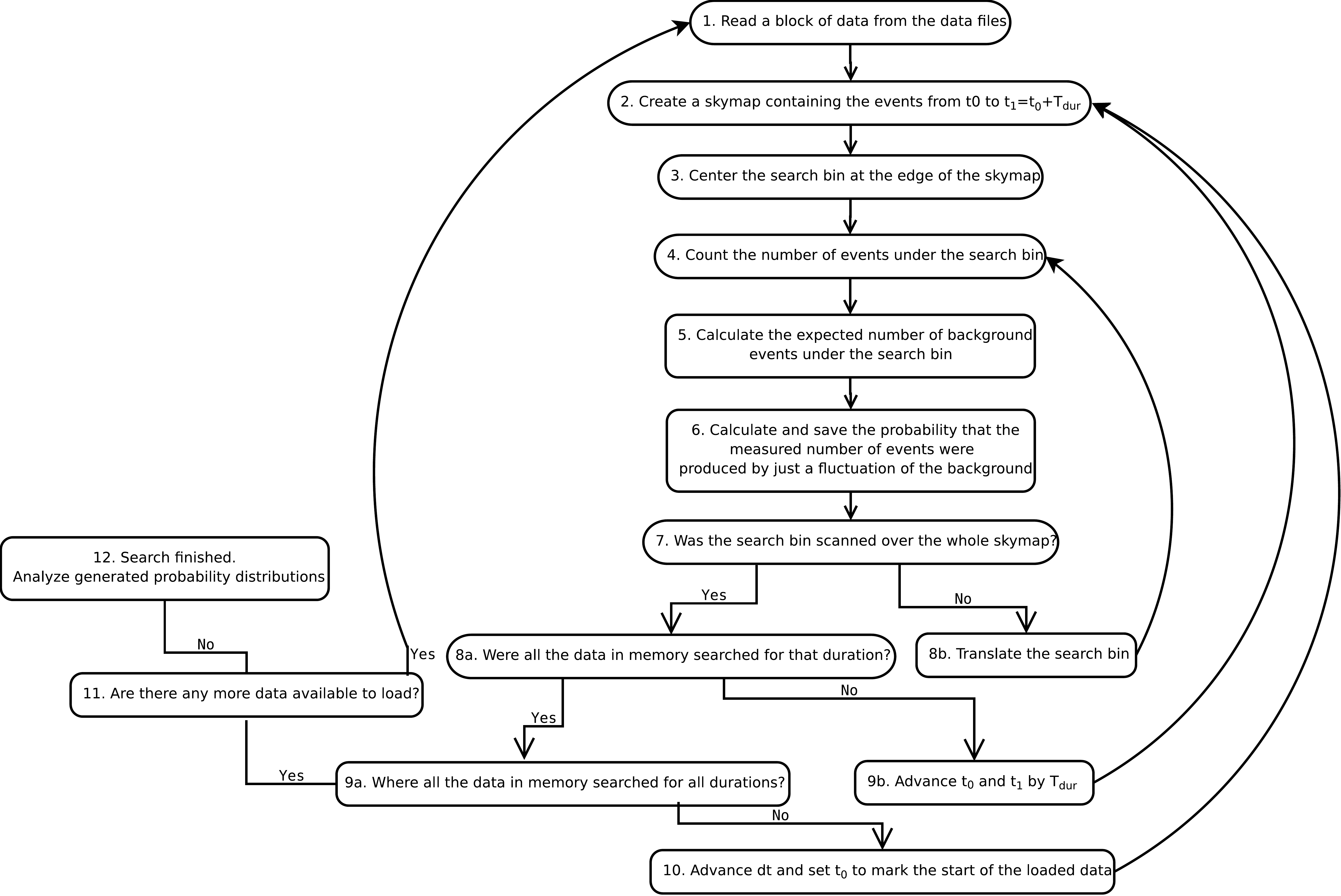}\caption{\label{fig:SEARCH_Algorithm}Overview of the algortihm used for the
search}
\end{figure}
\end{landscape}
Figure \ref{fig:SEARCH_Algorithm} shows the search algorithm
used in this study. As can be seen, the search consists of a series of
sub-searches (outer loop at the right), each one for a different emission
duration ($T_{dur}$). Each sub-search consists of the evaluation\footnote{
Evaluation of a skymap means, here, checking whether the data in the
skymap show any signs of VHE emission.
} of a series of skymaps (middle loop at the right). Skymaps are 2D
maps in Right Ascension (RA) and Declination (Dec), that contain the
reconstructed directions of the events detected in a time interval
extending from $t_{0}$ to $t_{1}=t_{0}+T_{dur}$ (step 2, see section \ref{sec:SEARCH_Skymaps}
for details). The width ($T_{dur}$) and the starting time ($t_{0}$)
of this interval are equal to the duration and the starting time of
the emission being searched. The evaluation of each skymap is performed
by scanning a rectangular ``search bin'' over the map (inner
loop at the right). 
Initially, the search bin is placed at the edge
of the skymap (step 3). The number of events under the search bin
$(N_{sig}$) are counted (step 4), and the number of the expected
background events is calculated ($N_{back})$ (step 5, see section
\ref{sec:SEARCH_Background-Calculation} for details on calculating
$N_{back}$). The probability $P(N_{sig},N_{back}$),
that the expected number of background events could create a fluctuation
as large as the measured number of events, is calculated using Poisson
statistics and these numbers (step 6):
\begin{equation}
P(N_{sig},N_{back})=\overset{\infty}{\underset{k=N_{sig}}{\sum}}\frac{N_{back}^{k}e^{-N_{back}}}{k!}.\label{eq:SEARCH_Poisson Probability}\end{equation}
The calculated probability is saved in a histogram for later processing.
After that, the search bin is translated to a nearby position on the
skymap (step 8b, see section \ref{sec:SEARCH_Skymaps} for details) and
steps 2-7 are repeated, until all of the skymap has been scanned (until
condition 7 is true). After this happens, a new skymap is created
containing events detected in a time frame of the same duration but
slightly offset from that of the previous skymap (step 9b, see
section \ref{sec:SEARCH_Organization-Time} for details on how $t_{0}$
is offset). The new skymap is evaluated like before, by repeating
steps 2-7. The process continues until all of the available data loaded
in memory have been searched for the first emission duration (until
condition 8a. is true). When this happens, the data are searched again,
this time for an emission duration $(T_{dur})$ that is slightly higher
than before (step 10, see section \ref{sec:SEARCH_Organization-Time}
on how $T_{dur}$ is advanced). The process repeats for until all the
data loaded in memory have been searched for emission of all durations
(until condition 9a is true). Then, the next chunk of data is loaded
in memory and is evaluated, like before, for all durations. This repeats
until all the data to be searched are processed (until step 11 is
false). 

The results of the search are in the form of probability distributions,
each one corresponding to a different duration emission. By examining
these distributions, the minimum probability thresholds for claiming
a discovery can be calculated, and any significant events can be detected
(step 12, see Chapter \ref{Chapter:Probabilities} for details).

\begin{figure}[ht]
\includegraphics[width=1\columnwidth]{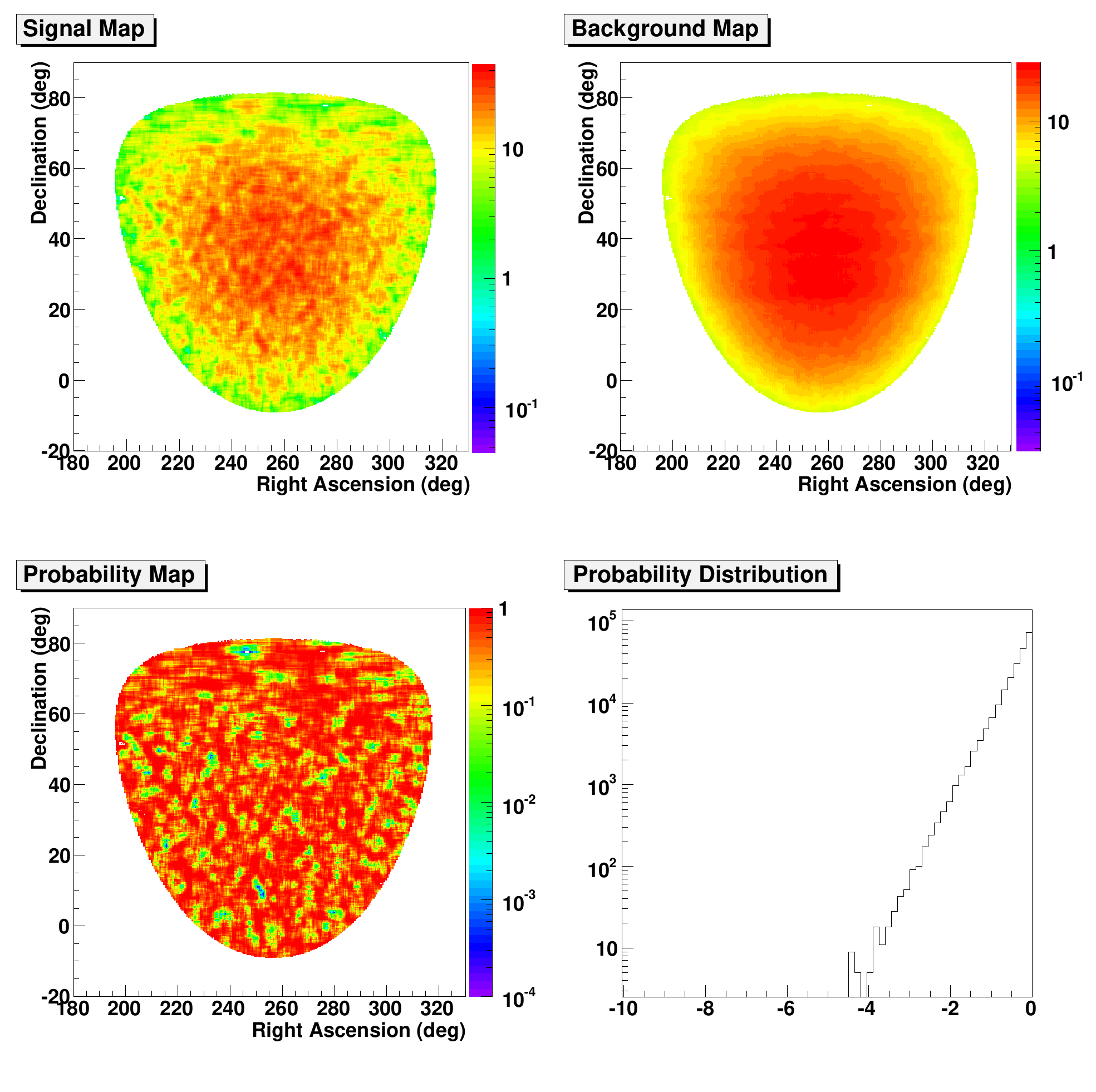}

\caption{\label{fig:SEARCH_Distros}Some of the distributions and maps involved
in the evaluation of one skymap for 10sec emission duration. \emph{Top
left}: signal map (skymap), \emph{top right}: background map, \emph{bottom
left}: probability map, \emph{bottom right}: probability distribution.}

\end{figure}

Figure \ref{fig:SEARCH_Distros} shows some of the distributions and
maps involved in the evaluation of one skymap for the $10\,s$ emission
duration. A signal map (skymap) and a background map are shown in
the first row. For each pixel of these maps, the Poisson probability
that the signal was generated by a random fluctuation of the background
is calculated using eq. \ref{eq:SEARCH_Poisson Probability}. The
calculated probabilities are shown in the probability map (bottom
left) and in the probability distribution (bottom right). By examining
such probability distributions, we can check for the existence of
any significant events expected to appear as outliers far from the
main distribution. 

\FloatBarrier
\section{\label{sec:SEARCH_Skymaps}Skymaps}

As mentioned above, a skymap is a 2D map in RA and Dec that contains
the reconstructed directions of the events detected in a predefined
time interval. The width and starting time of that time interval
are equal to the duration and starting time of the emission being
searched. The skymaps are binned more finely ($0.2^{o}\times0.2^{o}$
bins) than the angular resolution of the detector ($\sim1^{o}$) so
that in case there is some transient VHE emission in the data, there
would be at least one search bin centered on or very close to the
direction of that emission. 

By using RA/Dec instead of Hour Angle (HA)/Dec for the skymap coordinate
system, we can account for the motion of the sky relative to the earth.
The earth rotates $0.2^{o}$ in $\sim48\,s$. When searching for emission
of duration less than $48\,s$, the sky can be approximated as being
stationary, and HA and Dec can be sufficient coordinates. However,
for longer durations we have to use RA/Dec, a system that follows
the rotation of the earth so that the coordinates of objects in the
sky remain constant. Milagro's reconstruction algorithms produce the
zenith angle and azimuth of the location of an event on the sky. Using
the latitude of Milagro and its rotation relative to the North, the
HA/Dec coordinates of an event can be calculated. Then, using the
Local Sidereal Time (LST), the RA can be calculated using $RA+HA=LST$.

Because the angular resolution of Milagro is finite, the events from
a point source in the sky will be reconstructed at locations around
the true position of that source. Thus, a search for a signal from
a particular direction in the sky has to include the events from an
area around the true position, instead of including events from just
exactly that position. For that reason, when evaluating the contents
of a skymap, a search bin is centered on a potential source position,
and all the events under that bin are added to evaluate that
position. The sensitivity of the search depends on the dimensions
of that bin. A bin that is wider than the optimal one will include
more background events, thereby decreasing the sensitivity of the
search, while a smaller than optimal bin will fail to include many
of the signal events, resulting in similar reductions in sensitivity.
The optimization of the dimensions of the bin is a complicated
process and is described separately in Chapter \ref{chap:binsize}. 

As shorter durations are searched, the number of events in the skymaps
gets smaller. A point is reached at which a considerable number of
the trials correspond to zero signal events. In such a case, it is
not efficient to perform the search by scanning the search bin over
a mostly empty map. For emission durations $T_{dur}<0.2\,s$, an alternative
method is used. While the skymaps are being constructed, all the locations
with two or more nearby events are added to a list. Next, at the evaluation
stage, instead of scanning the search bin over the whole skymap, the
search bin is placed only on the locations saved in this list. This
way, the time-consuming evaluation of a large part of the empty skymap
is avoided, and the search becomes considerably faster. It should
be noted that for very short emission durations, such as the ones
of the order of milliseconds, a search that is not using this optimization
can be about a thousand times slower than the search for the one-second
emission duration. This optimization speeds up the process to such
a degree that both the millisecond and the one-second searches need
comparable amounts of time to complete.

Another speed optimization was in the evaluation of the $T_{dur}\geq0.2\,s$
skymaps. Because the search bin has dimensions of the order of a degree,
while the step between consecutive searches in space can be as
small as $0.2^{o}$, most of the data analyzed by consecutive searches
in space are the same. Because of that, every time an improbable fluctuation
happens in the data, it shows up in the results of multiple adjacent
searches. We can reduce the number of searches by making use of this
effect. A coarse search over the whole map can be made, followed by
a finer one performed only around locations with moderately small
probabilities detected by the coarse search. The configuration used
in this search was to start searching coarsely by stepping the search
bin by $0.6^{o}$ and then making a finer search with a $0.2^{o}$ step,
around all positions with probabilities less than $10^{-4}$.
With this choice, all events that are significant enough to be GRB
candidates will be detected, by only searching about one out of nine
bins of the skymaps. 

To summarize the speed optimizations mentioned in the previous two
paragraphs: for $T_{dur}<0.2$, only locations with two or more adjacent
events are evaluated, and for $T_{dur}\geq0.2$, a coarse $0.6^{o}$
search is performed on all of the skymap, followed by a finer $0.2^{o}$
search around locations found with probabilities less than $10^{-4}$.
It should be noted that the speed optimizations also contribute to
the sensitivity of the search. By having a fast algorithm, we can search
more finely in time and duration, thereby (as it will be shown in
the next section) increasing the sensitivity of the search. As a reference,
the optimized version of the search contained at least $10^{15}$ trials and
needed the equivalent of $80\times40$ modern-CPU days to complete.

\section{\label{sec:SEARCH_Organization-Time}Organization of the Search in
Time}

The duration of the emission searched for is dictated by the nature
of the phenomena one tries to detect. The $keV/MeV$ emission from GRBs
has been observed with durations ranging from few milliseconds
to many minutes (Fig. \ref{fig:GRB_T90}). Even though the duration of the prompt $GeV/TeV$
emission has not been measured yet, its time scale
should be the same as the time scale of the development of the GRB
fireball, which has already been set by the detected $keV/MeV$ emission.
The duration of the late $GeV/TeV$ emission generated by interactions
of the fireball with the circumburst medium is harder to constrain,
and probably varies significantly from burst to burst. 
The emission from PBH evaporation is currently unknown because it has not
been observed yet. However, the discussion on PBHs in Chapter \ref{chap:PBH}
(and specifically figure \ref{fig:PBH_AverageE}) shows that the relevant
timescale for PBH searches with Milagro is less than $\sim100s$. 

The duration range searched in this study was based on the duration
ranges of the $keV/MeV$ emission from GRBs (ms to mins) and also extended
to shorter timescales. The dead time of the detector ($\sim40\,\mu{}s$)
set the limit on the shortest duration searched ($100\,\mu{}s$),
and the limitations of the search algorithm (not
optimized for very long durations) set the maximum duration searched
($316\,s$)\footnote{
As will be shown later, there were 53 durations total searched, or
8 durations per decade. $316\,s$ is equal to $0.0001\times10^{52/8}\,s$.%
}. Since the chosen duration range covered many orders of magnitude,
the best way to uniformly distribute the individual durations was
to space them logarithmically (the ratio of two consecutive durations
was constant).

The sensitivity of the search is maximized when it has a sub-search
that can detect all of the signal while including as few background
events as possible. For this to happen, the search always has to have a sub-search with
a time interval of the same duration and starting time as any potential
gamma-ray emission. However, this requirement corresponds to an infinite
number of durations searched, to an infinitesimal step between consecutive
searches in time, and unfortunately, to an infinite amount of computational
time. A configuration has to be found that keeps
the sensitivity of the search as high as possible, while requiring feasible computational
resources. 

\begin{figure}[ht]
\begin{centering}
\includegraphics[width=1\columnwidth]{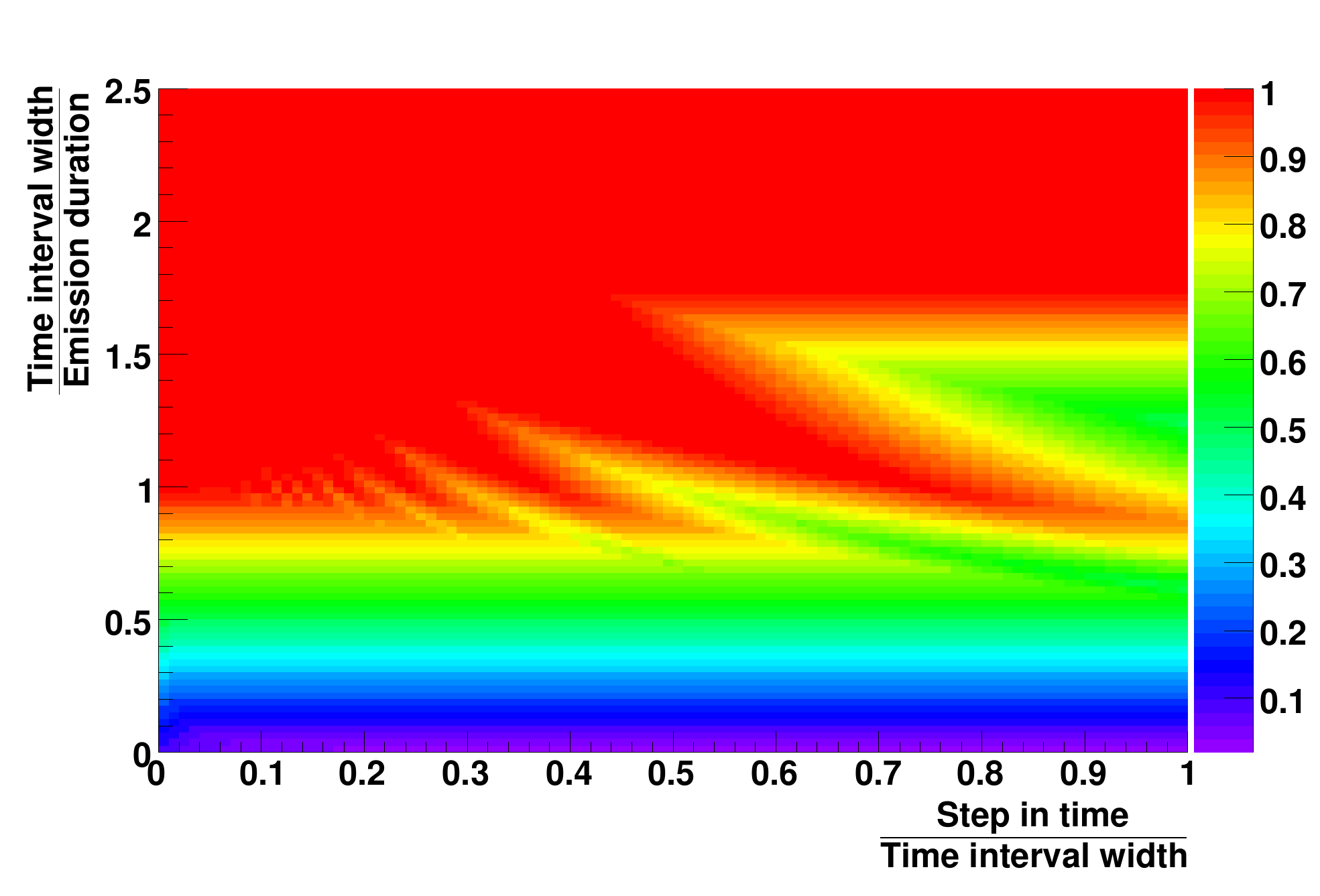}
\par\end{centering}

\caption{\label{fig:SEARCH_OptimizeDuration1}Detected fraction of signal events
(Z axis) versus the width of the time interval (Y axis), and the step
between consecutive searches in time (X axis). }

\end{figure}

Let us first examine what combination of oversampling in time and sampling
in duration is the most efficient for detecting all of the signal
events. In figure \ref{fig:SEARCH_OptimizeDuration1}, %
the detected fraction of signal events is plotted for different time
intervals and oversampling between consecutive steps in time. If the
time interval is shorter than the emission duration (Y axis value$<$1),
then some of the signal is always left out for the next or the previous
sub-search. On the other hand, if the time interval is longer than
the emission duration, then all or most of the signal, depending on
the relative offset between the emission and the time interval, is
detected by the search. As the time interval of the search gets longer
(high Y axis value), the probability of including all of the signal
increases. While this higher probability is something desirable, a very long time interval
will include much unwanted background, decreasing the sensitivity
of the search. The optimum case corresponds to the shortest time interval
that has the best chance of including all of the signal events. Based
on figure \ref{fig:SEARCH_OptimizeDuration1}, the choice for the time step between consecutive trials
(time offset between consecutive skymaps) has been selected to be
10\%. 

After the step size in time was found, the spectrum of durations searched
had to be decided. The criterion here was that the duration space
had to be sampled finely enough so that for any possible gamma-ray
emission duration, there would be a searched duration than was large
enough to include all of the signal but not so large that it included
much unwanted background. Assuming that the search, being highly oversampled
in time, always has a time interval with almost the same starting
time as that of a potential gamma-ray emission, we can examine
the effects of a mismatch between the width of the time interval and
the duration of the emission. 

For every emission duration less than the maximum duration searched
($316\,s$), there will be a number of longer searched durations, that are able
to include all of the emission's signal events. The shorter of these
durations will be the one with the highest sensitivity, since it will
include the least amount of background. The sensitivity of the search
will depend on how this best search-duration compares with the emission
duration. For example, there can be an emission duration that is just
a bit shorter than the best search-duration, resulting in a nearly
optimal sensitivity. However, if the emission duration becomes a bit
longer than that best search-duration, the latter will not be able to
include all of the signal events and will cease being the best one.
In that case, the immediately longer duration will become the best
one. However, because the new best duration can be considerably longer
than the emission duration, it may include much unwanted background.
Thus, the optimum sensitivity in this case would be lower than the
previous case, in which the emission and search durations almost matched.
It should be noted that these considerations are for a uniform-in-time signal.

\begin{figure}[ht]
\begin{centering}
\includegraphics[width=1\columnwidth]{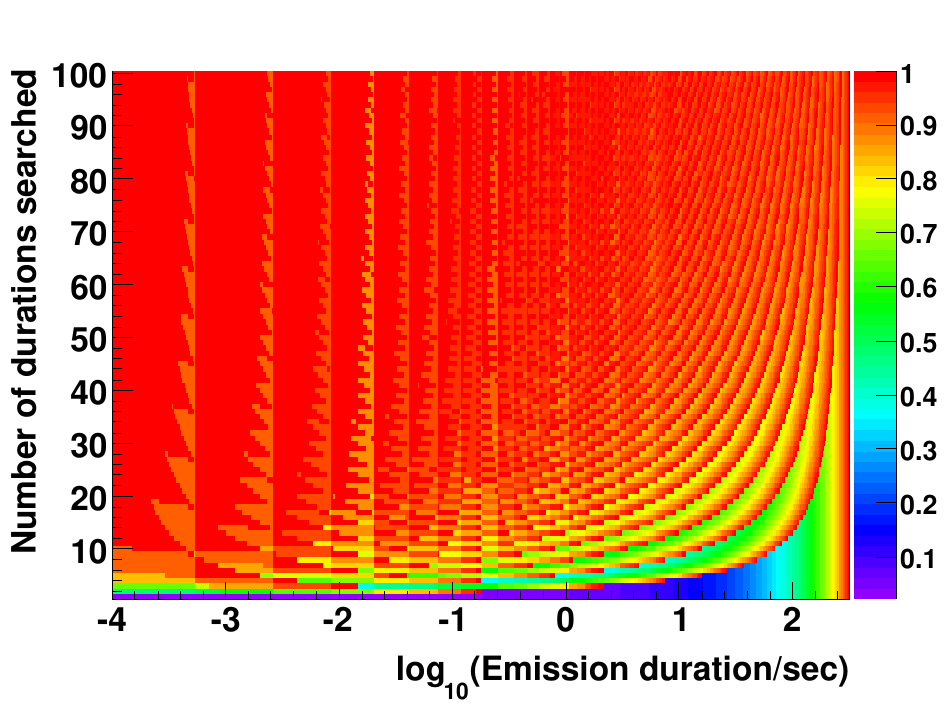}
\par\end{centering}

\caption{\label{fig:SEARCH_Duration2}Decrease in fluence sensitivity due to
the mismatch between the width of the time interval of the search
and the duration of the emission. The z axis shows the ratio of the
minimum number of events needed for a $5\sigma$ post-trials detection
with 90\% probability for matching emission and search durations,
over the minimum number of events needed when there is a mismatch.
The vertical stripes on the map are created by increases, by one,
of the small number of events needed for a detection when the emission
and search durations match (numerator of the ratio plotted on the
Z axis). }

\end{figure}

Based on effects like this, the decrease in sensitivity versus the number
of durations searched and the emission duration was calculated (Fig.
\ref{fig:SEARCH_Duration2}). %
As can be seen from the figure, when only few durations are searched,
the best search duration is usually considerably longer than the emission
duration, and the sensitivity of the search is significantly reduced.
The larger the number of durations searched, the larger the probability
that a search duration that almost matches the emission duration exists.
Taking into account the computational constraints and the need
to maximize the sensitivity, the number of durations searched was
52. This choice corresponded to 8 durations searched per decade and
a 33\% fractional increase between consecutive durations searched.
According to the figure, the sensitivity loss because of the mismatch
between the best search duration and the emission duration, for the
chosen number of durations searched, is at worst just$\sim10\%$. 

\FloatBarrier
\section{\label{sec:SEARCH_Background-Calculation}Background Estimation}

To calculate the statistical significance of a number of signal events,
the expected number of background events is needed. An accurate and
precise estimate of the background is essential for this search,
since an underestimated background will artificially increase the
significance of events, leading to false detections. On the other hand,
an overestimated background will reduce the significance of events,
possibly hiding a real signal. 

\begin{figure}[ht]
\begin{centering}
\includegraphics[width=1\columnwidth]{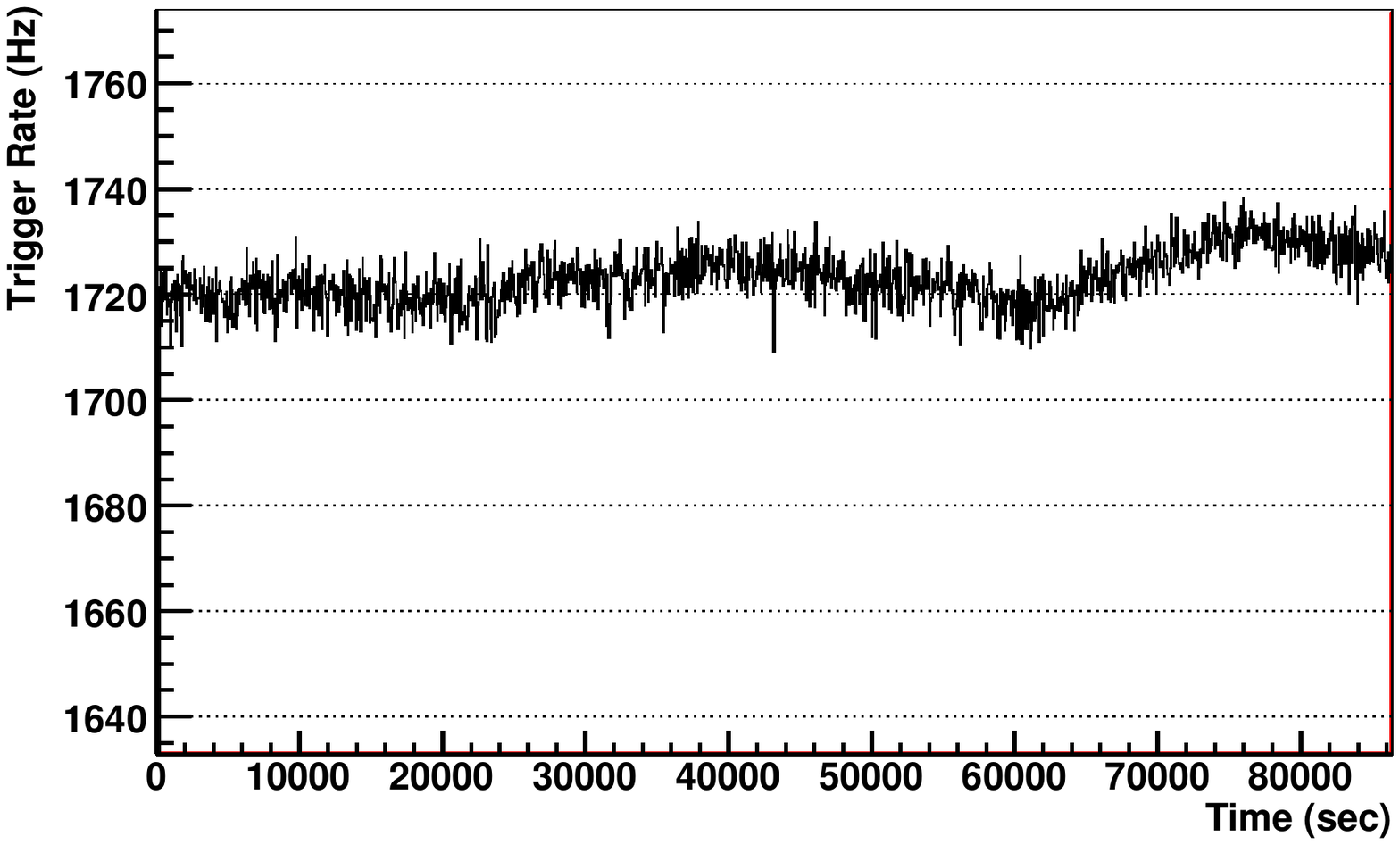}
\par\end{centering}

\caption{\label{fig:SEARCH_Trigger-rate-averaged}Trigger rate averaged over
$1s$ for modified Julian date 53699. Notice that the rate is not constant over the day.}

\end{figure}

A simple way to calculate the background rate from some particular
direction in the sky at some specific time would be to average the
event rate from that direction before and after that specific time.
The background rate of Milagro is of the order of one event per square
degree per second. To be able to calculate accurately the background
rate on a one square degree bin (say with 1\% accuracy), we would
need at least $10^4$ events (so that $\sqrt{N}/N=1\%$) or,
equivalently, we would need to average the event rate for about $10^4\,s$.
If the background rate of Milagro stayed constant in that time
period, this simple method would be enough to provide an estimate
of the background. However, as seen from figure \ref{fig:SEARCH_Trigger-rate-averaged},
the trigger rate of Milagro can fluctuate in time scales shorter than
that, causing this simple method to produce incorrect estimates for the background.
Even though the trigger rate fluctuations can be as much as few percent
over long time scales (thousands of seconds), the relative rate of
events coming from different directions in the sky is stable to a
factor of $10^{-4}$. This is because events that cause rate fluctuations,
such as changes in the atmosphere, light leaks, and temperature fluctuations
affect the detector as a whole, thereby causing fluctuations in the
all-sky event rate instead of affecting its sensitivity to signal
from a particular direction on the sky. Based on the stabilility of the 
relative event rate from different directions in the sky, the background rate from a particular direction
can be calculated in a reliable way by simply multiplying
the instantaneous all-sky rate with the probability that an event
came from that direction. Contrary to the method explained
above, in which the event rate from a particular direction in the
sky had to be averaged over a long time in order to gather good statistics,
the average all-sky rate can be calculated using only a few seconds
of data. This is because the all-sky rate averaged now is significantly
larger than the event rate from just one small patch of the sky. Specifically,
Milagro's all-sky rate of events that passed the cuts is $\sim1500Hz,$
which means that with just $10\,s$ of data, the all-sky rate average
can be calculated with high statistical accuracy ($<1\%)$. The
advantage of this background calculation method comes from the fact
that it relies on a quantity that does not fluctuate much.

Assuming an isotropic background of cosmic rays, the acceptance $E(HA,Dec)$
of the detector can be defined as the probability that a background
event comes from the differential angular element $d\Omega=dHA\,dDec$.
As mentioned in section \ref{sub:SEARCH_Overview}, during the evaluation
of a skymap, a search bin is scanned over the skymap, and the numbers
of signal $N_{sig}$ and background events $N_{back}$ under it
are calculated. If $T_{dur}$ is the time width of the
skymap (or equivalently the duration searched), $t_{c}$ is the center
of that time width, and $R(t_{c},T_{dur})$ is the all-sky rate of
events passing the cuts averaged in the time interval ($t_{c}-T_{dur}$),
then the expected number of background events under the search bin
is
\begin{equation}
N_{back}(HA,Dec,t_{c},T_{dur})=\int_{SB}E(HA,Dec)\,R(t_{c},\Delta{}T)\,T_{dur}\,d\Omega,\label{eq:TheSearch_Direct Integration}\end{equation}
where the integration $\int_{SB}d\Omega$ is performed on the area
under the search bin. 

\begin{figure}[ht]
\begin{center}
\subfigure[An acceptance map showing $E(HA,Dec)$]{\includegraphics[width=0.85\columnwidth]{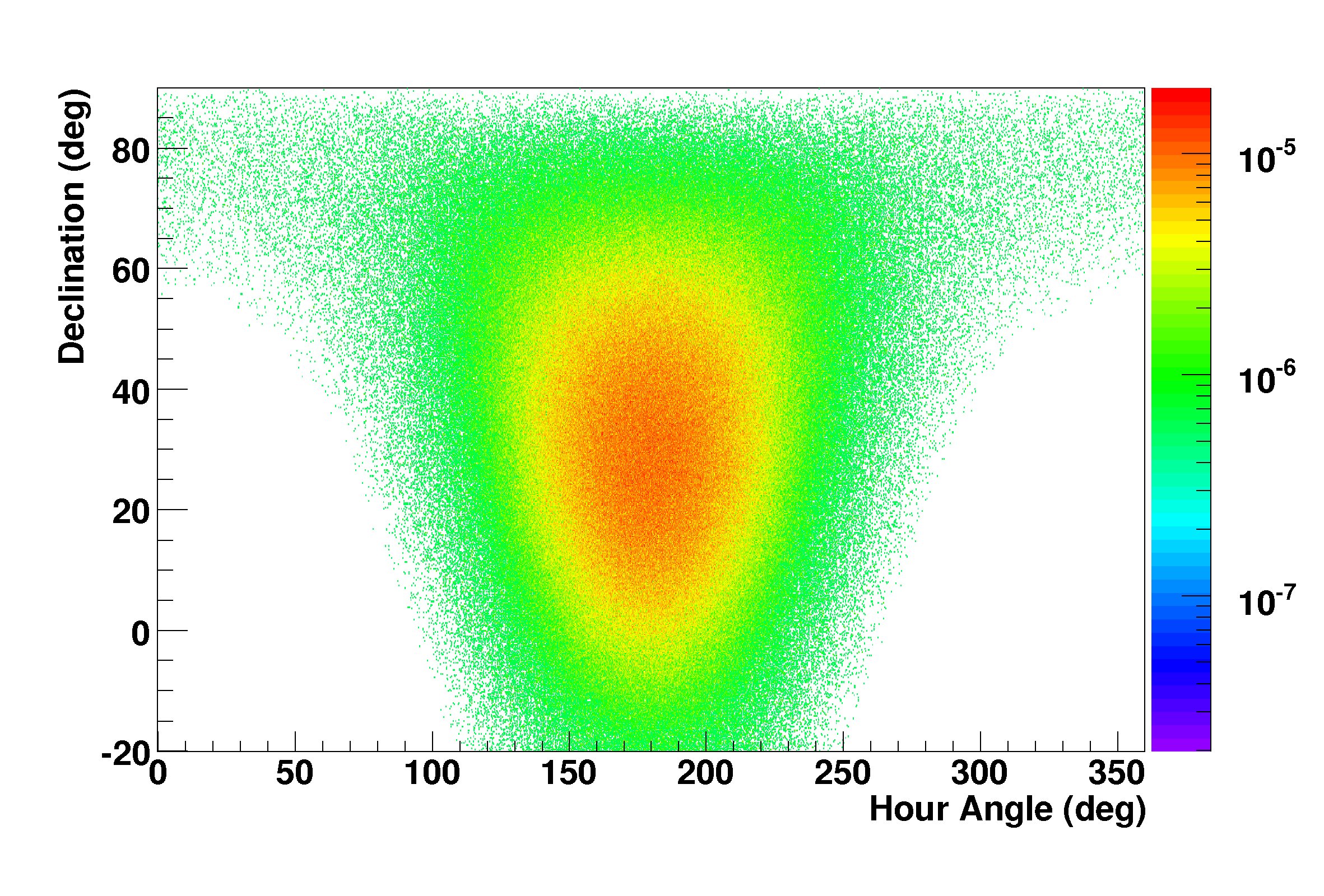}}
\subfigure[An integrated-acceptance map showing $\int{E(HA,Dec)d\Omega}$. The integration has been performed on an $1^{o}\times1^{o}/cos(Dec)$ rectangular area.]{\includegraphics[width=0.85\columnwidth]{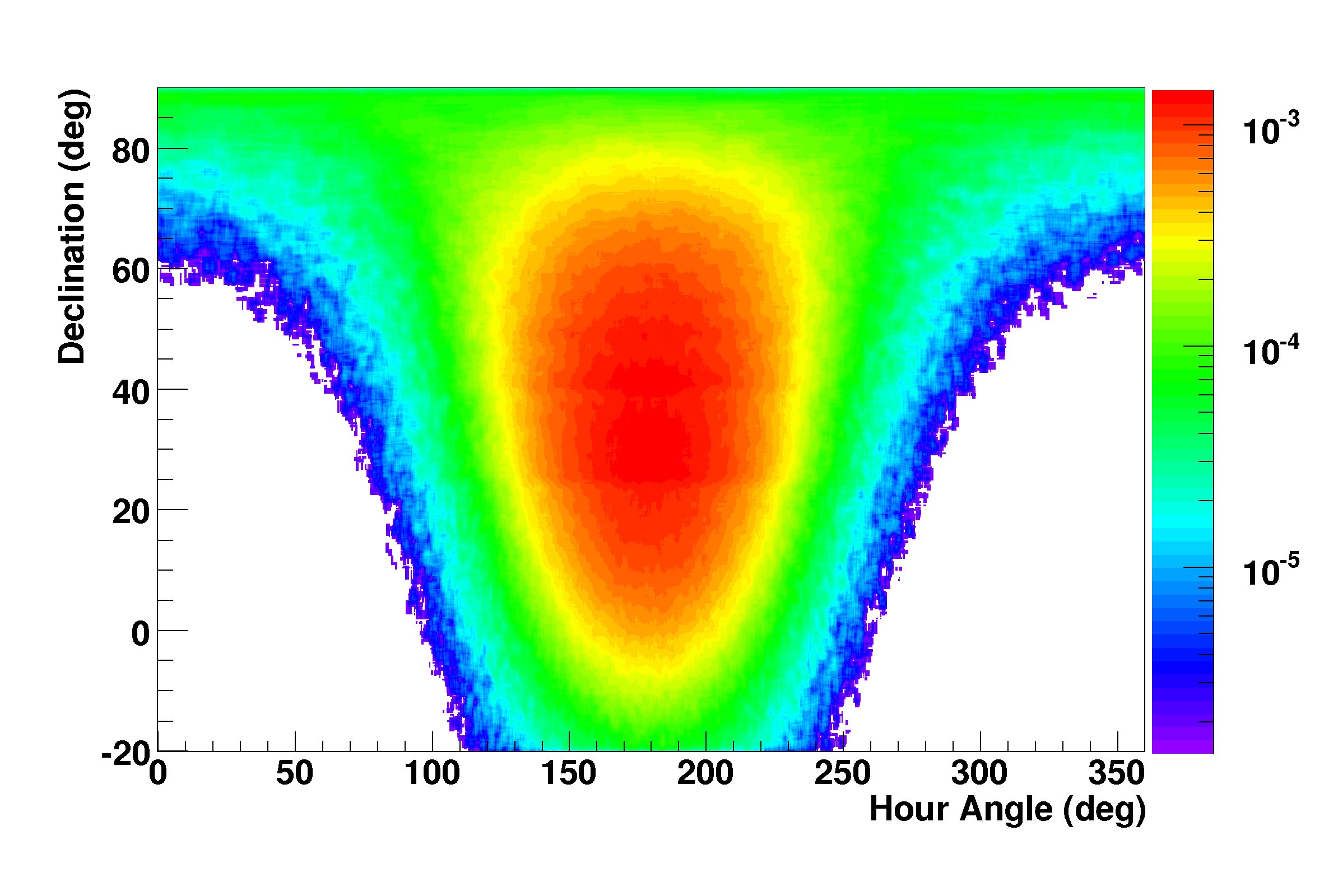}}

\caption{\label{fig:SEARCH_BackgroundMap}Maps involved in the background estimation.
The maps were created using $1200\,s$ of data. }
\end{center}
\end{figure}

To perform this calculation, an ``acceptance map'' is first created
in HA/Dec, filled with the directions of all events detected
for at least twenty minutes around $t_{c}$ and normalized to
unity. Then, the map contents around each bin of the acceptance map
are integrated and stored in a new ``integrated-acceptance map.''
This way, the acceptance map stores $E(HA,Dec)$, and the integrated-acceptance
map stores the integral $\int E(HA,dec)d\Omega$ (Fig. \ref{fig:SEARCH_BackgroundMap}).
Next, the event rate $R(t_{c},T_{dur})$ is calculated. For short
time intervals ($T_{dur}<10\,s$), $R(t_{c},T_{dur})$ is calculated
by scaling down the event rate 5s before and after the center of
the time interval: $R(t_{c},T_{dur})=R(t_{c},10\,s)\times{}T_{dur}/10$.
For longer durations, $T_{dur}>10\,s$, the rate is calculated simply
by dividing the number of events in the time interval ($t_{c},T_{dur})$
by its duration ($T_{dur}$).

\FloatBarrier

\section{Data Sample}

Milagro's low-energy sensitivity has not remained constant throughout the
$\sim7$ years it has been operating. As mentioned in Chapter
\ref{chap:MILAGRO}, the first version of Milagro did not include the outrigger
array, and hence it had a worse angular resolution than
now. Furthermore, during the first years of Milagro's operation, a
triggering system was used (multiplicity trigger) that did not accept
a large part of the lower-energy events, which are important for this
study. For these reasons, the data from those early times were not
included in this analysis. Specifically, only the last five years
of Milagro data have been analyzed: from 03/01/2003 (Modified Julian
Date 52699) to 03/01/2008 (Modified Julian Date 54526). The analysis
begins after the new VME trigger was installed (01/25/2002), and at
approximately the same time that the outriggers started being used
in the online reconstruction. It ends at approximately the same
time that the outrigger array started being dismantled (March 15th)
as part of the shutdown of Milagro.

\FloatBarrier
\section{Cuts}

Not all events of the five years analyzed were used. Cuts
were applied to the data to improve the stability, speed, and sensitivity
of the search. Specifically:
\subsubsection*{Data-quality cuts}
Multiple checks were made on the data to ensure their quality. Because
of problems in the GPS or in the DAQ system, the times of some events
were either wrong or inaccurate. Such problems could occur, for example,
when the accuracy of the GPS system was reduced because there were
not enough GPS satellites available. In the case that the worker computers
responsible for analyzing and saving the data detected that the difference
between their internal time and the GPS time was greater than a threshold,
the event was tagged as having time errors. This search rejected events
with time errors corresponding to a time difference between the GPS
and the worker computer clocks greater than $50\,\mu{}s$. Because
of problems with the electronics, the same event could be read multiple
times, the times of consecutive events could be swapped, and the events
could have the wrong dates. There were multiple checks for the presence
of such errors, and the broken events were either fixed or rejected.
Because of various problems such as power failures, patches of PMTs
going offline, or problems in the electronics and the online reconstruction
computers, there could be lost blocks of data or large changes in
the all-sky rate. Detecting such problems was very important, because
they could interfere with the background estimation, usually reducing
the estimated background. The event rate was monitored in multiple
time scales (ranging from seconds to minutes) for gaps or sudden changes.
In the case of time gaps, the search stopped at the gap, a new acceptance
map was created with the data after the gap, and the search continued
evaluating the post-gap data. In case that the event rate was low or had
a sudden change, the whole data block with the problematic rate was
rejected. 
\subsubsection*{Angular resolution and background rejection cut}
Another cut applied was on the quality of the angular reconstruction
fit. The larger the number of PMTs participating in this fit, the
better its quality, and the greater the accuracy of the results. In
figure \vref{fig:Milagro_delAngle_nfit}, the median error on the
reconstructed angle versus the number of PMTs participating in the
fit (nfit) is shown. According to that figure, events with fewer than
about 20 PMTs participating in the fit are reconstructed poorly, and therefore
should be rejected. Another reason to reject such low-nfit events
is that various instrumental effects such as light leaks, spikes in
the supply voltages, and untagged calibration runs can create bursts
of fake events that manage to be reconstructed with a very low nfit.
By applying a soft nfit cut, these kinds of events can be rejected,
improving the quality of the data and almost eliminating false detections.
For a source emitting gamma rays on an exponential spectrum with index
-2.0 and from a redshift $z=0.3$, an $nfit>20$ cut keeps about $\sim93\%$ of
the gamma-ray events that can be reconstructed accurately enough to
contribute to the significance of a detection (angle error $<2^{o}$).
The same cut also rejects $\sim74\%$ of the background signal from
cosmic-ray protons%
\footnote{Cosmic-ray protons comprise the majority $(\sim70\%$) of cosmic rays
in Milagro's energy range and can be reasonably assumed to represent
cosmic rays as a whole when making comparisons between the properties
of gamma-ray and cosmic-ray events. %
}. Thus, for the longer durations $(T_{dur}\gtrsim100s$),
where Gaussian statistics can be applied for the calculation of the probabilities,
the sensitivity of the search is improved by a factor $Q\simeq\frac{0.93}{\sqrt{0.26}}=1.8$. This
means that if a GRB were detected with a statistical significance of
$S$ standard deviations, it would be detected with a statistical significance
of $1.8\times{}S$ standard deviations after applying this cut. For
shorter durations, where Poisson statistics are applicable for calculating
the significance of a measurement, and where the importance of keeping
every single signal event is higher, the improvement is expected to
be somewhat smaller (depending on the duration). We see that an $nfit>20$
cut not only improves the quality of the data by reducing false
alerts, but also improves the sensitivity of the search by rejecting
a large part of the usually poorly reconstructed cosmic-ray events. 
\subsubsection*{Zenith-angle cut}
Because the effective path length of the air showers in the atmosphere
increases with the zenith angle, showers from large zenith angles
are attenuated more and contain smaller numbers of particles at Milagro's
altitude. For these showers, the probability of triggering the detector
is smaller, and the angular reconstruction accuracy worse. Based on
this, locations on the sky with zenith angles greater than $45^{o}$
were not searched for VHE emission, making the search faster.
\subsubsection*{Standard gamma-hadron discrimination cuts}
\begin{figure}[ht]
\includegraphics[width=0.5\columnwidth]{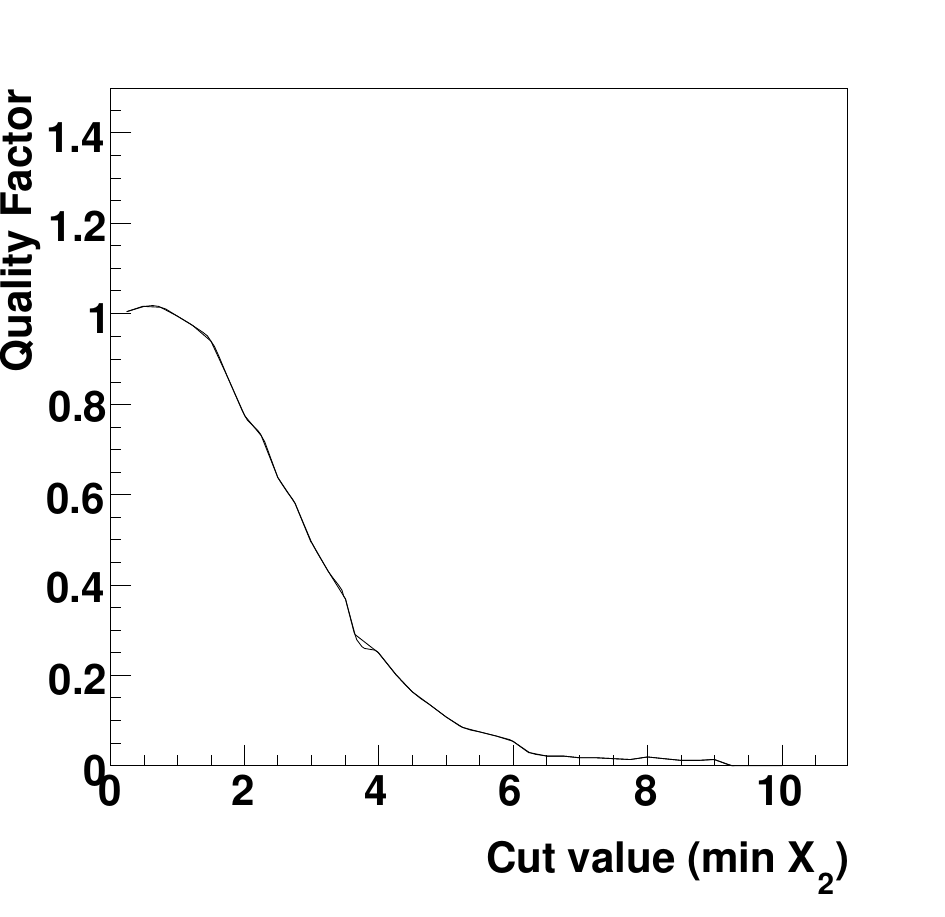}\includegraphics[width=0.5\columnwidth]{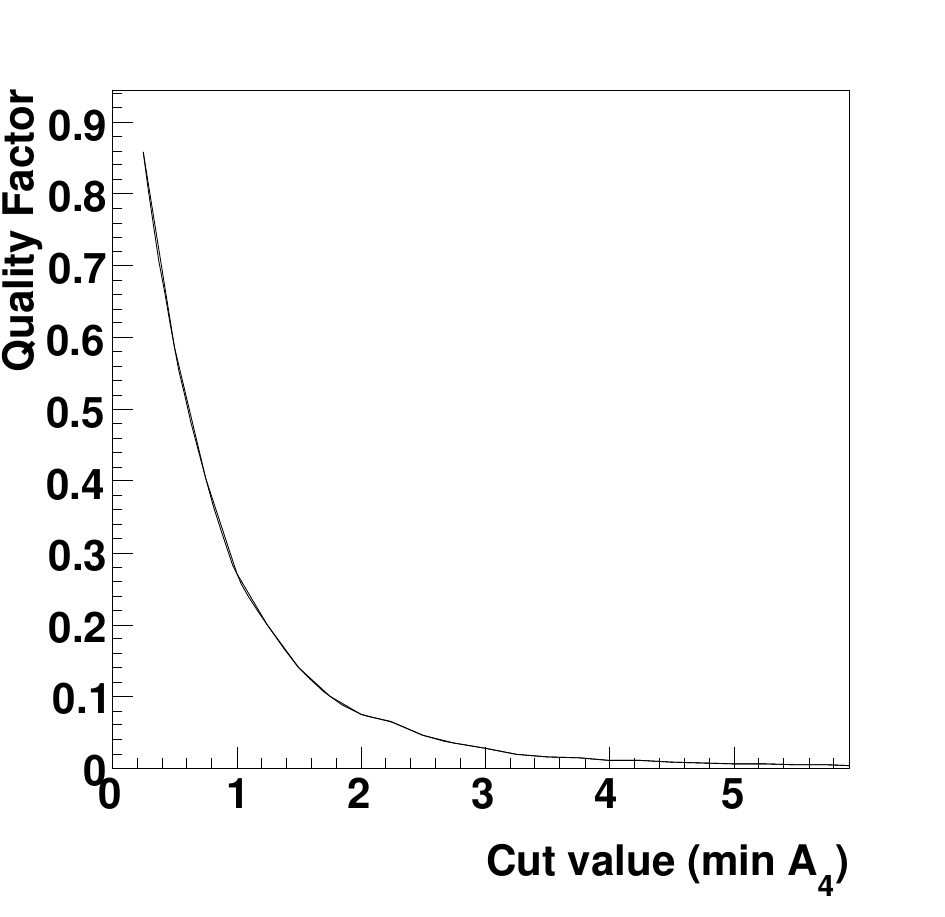}

\caption{\label{fig:TheSearch_X2A4}Effect of applying the standard gamma-hadron
discrimination cuts to the signal from a source emitting on a power-law
spectrum with index -2.00 from redshift 0.3. The quantity plotted
is the quality factor defined as $Q=\frac{f_{S}}{\sqrt{f_{B}}}$,
where $f_{S}$ and $f_{B}$ are the fractions of signal
and background that pass a specific cut. The quality factor shows
the change in the statistical significance of a signal introduced
by applying a cut. These plots are only valid for the longer durations,
where the large statistics allow Gaussian probabilities to be used. }

\end{figure}

The gamma-ray signal searched for is expected to be low in energy,
of the order of few hundreds of GeVs. As was mentioned in subsection
\vref{sub:Milagro_GammaHadron}, Milagro's standard gamma-hadron rejection parameters, namely
$X_{2}$ and $A_{4}$, are efficient only for gamma-ray signals of
higher energy. Figure \ref{fig:TheSearch_X2A4} shows the change in
the sensitivity of detecting longer-duration signals that results
from the application of Milagro's standard gamma-hadron discrimination
methods. Both plots show that these gamma-hadron discrimination parameters
cannot bring an improvement in Milagro's sensitivity to such signals. The
dataset analyzed to produce these plots had $nfit>20$, a maximum zenith
angle of $45^{o}$, and a maximum error in the reconstructed angle <$2.0^{o}$
for gammas. For the signal, simulated gamma rays on a power-law spectrum
with index -2.00 emitted from a source at redshift 0.3 were used,
while the background consisted of simulated protons on a power-law
spectrum with index -2.72, set by the proton cosmic-ray spectrum measured
by BESS \cite{BESS_CR_SPECTRUM}. For the shorter emission durations, where Poisson
statistics are applicable, the importance of keeping each one of the
signal events is very high, and the performance of these gamma-hadron
discrimination cuts is worse. For these reasons, there was no $A_{4}$
or $X_{2}$ gamma-hadron separation applied in the data searched.

   \clearpage \chapter{\label{Chapter:Probabilities}Trials and Probabilities}

\section{Introduction}

According to the previous chapters, this search essentially tries to
detect a gamma-ray signal on top of the cosmic-ray background. Because
of the background's random nature, the amount of background in each
consecutive sub-search (trial) is not the same; rather it fluctuates
about an average value in both time and space. However, most of these
fluctuations are small. The larger a fluctuation is, the less frequently
it happens. If we measured the amount of background just once (one
trial), almost surely we would find an amount almost the same as its
average. However, if we keep on searching, we would start finding
larger and less frequent fluctuations over the expected background
value. Because we do not want to mis-identify these rare, large fluctuations
as a real signal, we have to increase our expectations of the magnitude
of a result that we will claim to be a real signal. As a result, a search
that consists of many trials ends up being able to detect only large
amounts of signal, or equivalently, has a reduced sensitivity. 

This chapter will explain the above in detail, and will show the method
used to calculate the minimum threshold on the fluctuation magnitude
over which we can claim that a real signal was detected.

\section{Multi-Trial Searches}

The detection of a signal by a search is equivalent to the rejection
of the hypothesis that a signal does not exist. This hypothesis, called
the \emph{null hypothesis}, consists of the assumption that the data
comprises of only background and that the outcome of a search can
be explained by mere fluctuations of that background. A search for
a signal starts by assuming that the null hypothesis is true and then
tries to reject this assumption. To test the validity of the null
hypothesis, the search calculates the probability that the results
can be explained by it. A high probability supports the null hypothesis
and the absence of a signal. On the other hand, a very improbable
result could mean that an alternative hypothesis, one that agrees
with the data with a higher probability, is true. In that case, the
null hypothesis is rejected and replaced by the alternative hypothesis
of the existence of both signal and background. Therefore, a successful
detection of a signal requires performing a search that results in
the rejection of the null hypothesis. 

The larger the probability threshold used to determine the rejection or not
of the null hypothesis, the more likely it is a search on a background-only
dataset to end up erroneously rejecting the null hypothesis. 
The erroneous rejection of the null hypothesis is called in statistics
terminology a \emph{Type I error}. On the other hand, the smaller the probability
threshold is, the more likely it is for searches on a dataset that also includes
signal to erroneously retain the validity of the null hypothesis, thereby 
failing to detect that signal. This failure to detect the signal is called a \emph{Type
II error}. The standard in the physics community on setting this threshold
is such that the rate of Type I errors is very small and is set to
$\sim3\times10^{-7}$. This probability corresponds to the area under
a Gaussian distribution of standard deviation equal to one and mean
equal to zero, integrated from five to infinity. Based on that relation,
this probability threshold is also referred to as being equal to {}``five
standard deviations'' or $5\sigma$. 

A search can consist of either one or many sub-searches (trials)
on the data. For a search that consists of only one trial, the validity
of the null hypothesis can be verified by simply comparing the usual
$P_{thres}=5\sigma\simeq3\times10^{-7}$ probability threshold with
the probability of the result produced by its single trial (piece-wise
probability $P_{trial})$. In this case, the probability of making
a Type I error is equal to $P_{thres}$ and the probability of not
making one is $1-P_{thres}$. Consider a search that consists of two
independent trials. Now, the family-wise probability of not making
a Type I error after finishing the whole search is $(1-P_{thres})\times(1-P_{thres})=(1-P_{thres})^{2}$.
For N trials, this probability becomes $(1-P_{thres})^{N}$. This
means that the probability of erroneously rejecting the null hypothesis
(the search having a trial with $P_{trial}<P_{thres}$) has increased
from $P_{thres}$ for just one trial to $1-(1-P_{thres})^{N}$ for
N trials. What happens is that the more searches undertaken, the more
likely very improbable results will be found.

A probability that includes the effects of having taken a number of
trials is called a ``post-trials'' probability, while a probability
that does not include these effects is called a {}``pre-trials''
probability. In the example above, $P_{thres}$ is the pre-trials
probability of one trial rejecting the null hypothesis, and $1-(1-P_{thres})^{N}$
is the post-trials probability of the whole family of trials rejecting
the null hypothesis. 

The requirement on a search with any number of trials is that it should
erroneously reject the null hypothesis with probability $P_{thres}$.
In order to apply this requirement to a multi-trial search, a new
lower post-trials probability threshold for each of the trials has
to be set ($P_{thres}^{'})$, so that the family-wise Type I error
rate remains $P_{thres}$:\begin{align}
P_{thres} & =1-(1-P_{thres}^{'})^{N}\Leftrightarrow\nonumber \\
P_{thres}^{'} & =1-(1-P_{thres})^{1/N}.\label{eq:TheSearch_Sidak}\end{align}

Equation \ref{eq:TheSearch_Sidak} is called the Sidak equation and
can be used to calculate the probability threshold $P_{thres}^{'}$
that must be used in a search with $N$ trials so that the family-wise
Type I error rate stays $P_{thres}$. Because $P_{thres}\ll1$, the
exponent can be expanded, simplifying the Sidak equation:
\begin{equation}
P_{thres}^{'}\simeq1-(1-P_{thres}/N)=P_{thres}/N.\label{eq:TheSearch_Bonferonni}\end{equation}
Equation \ref{eq:TheSearch_Bonferonni} is called the Bonferonni
equation and simply says that a search with $N$ trials has a probability
$P_{thres}/N$ of having at least one event with a pre-trials probability
less than $P_{thres}$. 

The GRB search in this work contains a large number of trials. Hence
the probability thresholds for claiming a detection have to be adjusted.
It is straightforward to calculate the number of trials taken
by counting the number of searches performed on each one skymap, and
multiplying by the total number of skymaps evaluated in the search.
However, is the number of trials taken in this search the same as
the number of trials ($N$) in the Sidak or Bonferonni equations?
These equations are based on the fact that the trials are independent,
since only for independent trials the probability of their combined
outcome is equal to the products of their probabilities. In this search,
there is a significant overlap between successive trials in both space
and time. As a consequence, the same data are analyzed by multiple
successive trials. The effect of taking a number of extra \emph{independent}
trials is that the probability of making a Type I error is increased
in a predefined way as described by the Sidak equation. If these trials
were not independent, then there would still be an increase in the
Type I error rate, though somewhat smaller than in the independent case.
An \emph{effective number of trials} can be defined as the number
of independent trials that would cause the same increase in the Type
I error rate.

\section{Calculation of the Effective Number of Trials}

\subsection{\label{sub:TheSearch_old}Using the distribution of the probabilities
of all trials}

In the past, the effective number of trials was estimated 
based on the distribution of probabilities of all the trials of
the search $dN/dlog_{10}(P)$ for one duration (see figure \vref{fig:RESULTS_SomeProbs}
for an example of such a distribution). This
distribution essentially shows the chance of a trial in the
search having a specific probability. Based on this chance, the effective number
of trials per trial taken can be calculated, and from this, the effective
number of trials for the whole search. Starting from the fact that
the probability density $dN/dP$ is constant, the functional form
of that probability distribution can be calculated:\begin{eqnarray}
\frac{dN(P)}{dP} & = & N_{total}\Leftrightarrow\nonumber \\
\frac{dN(P)}{dlog_{10}(P)} & = & N_{total}\,log(10)\,P,\label{eq:TheSearchPDistro}\end{eqnarray}
where $N_{1}$ provides the normalization of the distribution and
is equal to the total number of trials. By counting how many times
a trial had a probability less than some threshold $P_{0}$ ($N_{P<P_{0}})$,
the number of effective trials corresponding to one taken trial can
be estimated:
\begin{equation}
T_{pertrial,old}\simeq\left(\frac{N_{P<P_{0}}}{N_{total}}\right)/P_{0}.\label{eq:TheSearch_Trials_Davids}\end{equation}
This equation says that if $T_{pertrial,old}$ effective trials
correspond to one trial, then the chance of one trial having a probability
less than $P_{0}$ is $N_{P<P_{0}}/N_{total}$. So, by producing a
distribution from all the trials of the search (that follows equation \ref{eq:TheSearchPDistro}),
an effective number of trials can be found using equation \ref{eq:TheSearch_Trials_Davids}.
However, this is not the case. It can be shown that this method results
in an incorrect number of effective trials. 

Starting from eq.
\ref{eq:TheSearch_Trials_Davids}, we can find: \begin{alignat*}{1}
T_{pertrial,old} & \simeq\frac{N_{total}-N_{P>P_{0}}}{N_{total}}\frac{1}{P_{0}}\overset{eq.\ref{eq:TheSearchPDistro}}{\Longleftrightarrow}\\
 & =\left(1-\frac{\int_{P_{0}}^{1}N_{total}\,dP}{N_{total}}\right)\frac{1}{P_{0}}\\
 & =\left(1-\frac{N_{total}(1-P_{0})}{N_{total}}\right)\frac{1}{P_{0}}\Longleftrightarrow\\
T_{pertrial,old} & =1.\end{alignat*}
The above equation says that if the probability distribution follows
equation \ref{eq:TheSearchPDistro}, then the effective number of trials
per trial taken is equal to one, or, equivalently, the total effective
number of trials is equal to the total number of trials. This is 
incorrect.

Equation \ref{eq:TheSearchPDistro} ignores the effects of the correlations
between trials to the probability distribution. Consider the following
example: a dice is thrown and the result is stored in a distribution.
For an unbiased experiment, the frequency distribution of the outcomes
should be flat (because each tossing result has an equal chance of occurring)
and, in the limit of many throws, the fluctuations on the number of
occurrences for each result should be Gaussian. Now let us say that
instead of throwing the dice once and filling the distribution with
one entry, we throw the dice once and fill the distribution twice
with the same entry. This would bring a correlation between subsecutive
searches. While the average number of occurrences will remain the
same, the fluctuations should now be larger since each entry now causes
a larger fluctuation. In the limit of a very large number of throws,
the fluctuations in both cases should be negligible, and the two distributions
will be almost identical (flat).

\begin{figure}[ht]
\begin{centering}
\includegraphics[width=1\columnwidth]{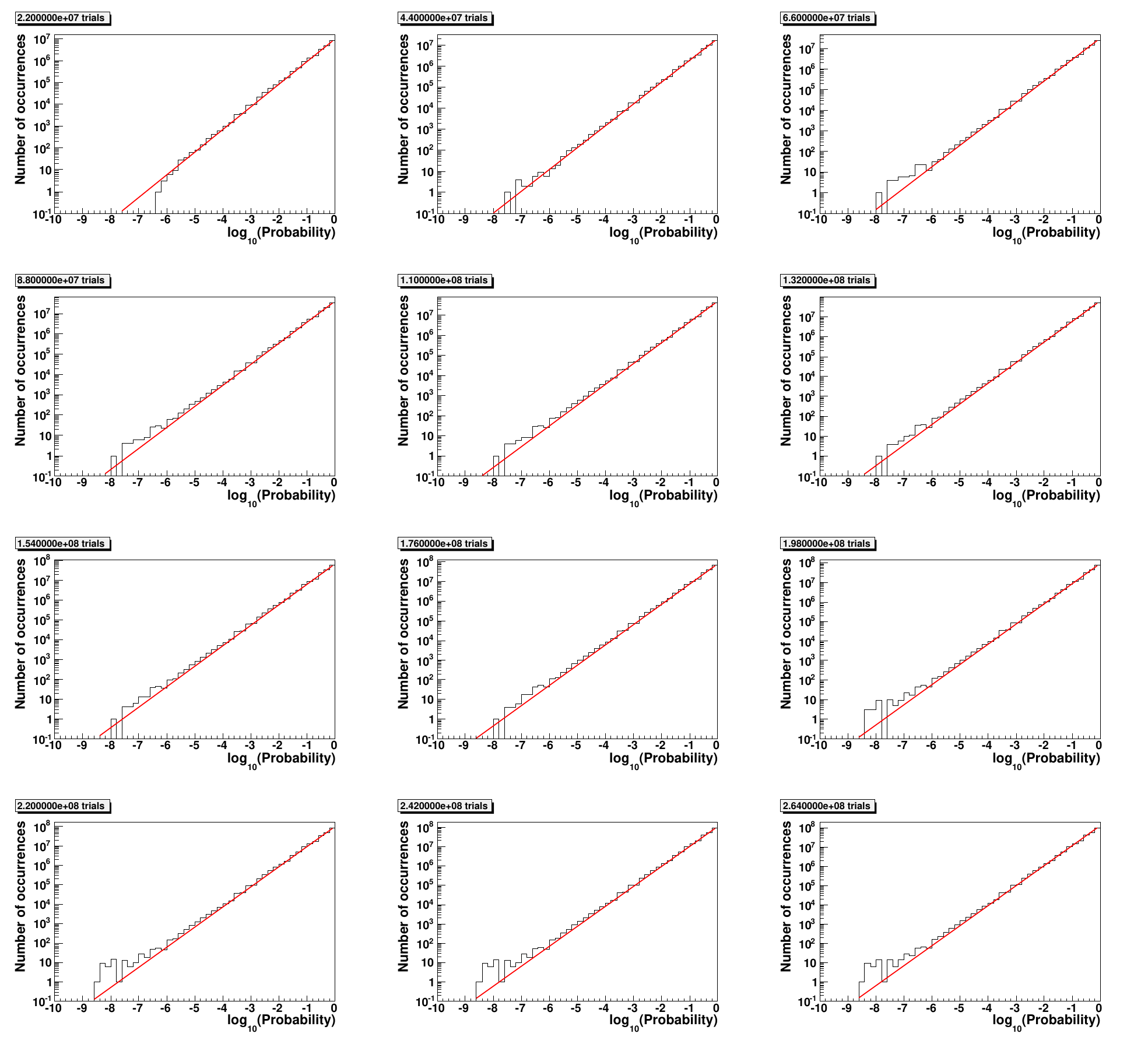}
\par\end{centering}

\caption{\label{fig:TheSearch_Distorted Probs}A probability distribution produced
by a simulation of a binned search with highly correlated trials.
The correlations between trials distort the probability distribution
away from its expected form shown by the black line.}

\end{figure}

Going back to the case of the GRB search, in short, as the search
progresses, the probability distribution shows deformities and deviations
from the form described by equation \ref{eq:TheSearchPDistro}, which
are slowly smoothed out as more statistics are gathered. These deviations
are mostly evident in the tail of the distribution, where the statistics
are always low. To illustrate this effect, a simulation of binned search,
similar to that of this study and with a high degree of correlation
between consecutive trials, was built. Figure \ref{fig:TheSearch_Distorted Probs}
shows snapshots of the probability distributions produced by this
simulation every 22 million trials. In the first snapshot (top left)
a deficit of the probability distribution (black curve) with respect
to the expected form (equation \ref{eq:TheSearchPDistro} - red line)
is evident. As the search progresses, small groups of correlated results
end up in the area of the deficit and fill it up (second plot, first
row). Sometimes a very improbable fluctuation happens (third plot,
first row), creating a number of entries in the tail of the distribution,
and, this time, an excess over the predicted shape. This excess will
be smoothed out too, as more statistics are gathered. As seen, the
tail of the distribution randomly exhibits a series of deviations (excesses
or deficits) from its expected form, which are usually smoothed out
before the next deviation happens. 

If the distribution of the probabilities of all the trials
is used for the calculation of the effective
number of trials, then the results will depend on the probability
threshold $P_{0}$ used in the calculation (eq. \ref{eq:TheSearch_Trials_Davids}).
If $P_{0}$ is large enough to be far from the tail of the distribution,
the resulting number of effective trials will be approximately equal to the total
number of trials. If, on the other hand, a $P_{0}$ close
to the tail of the distribution is used, then the resulting number
of effective trials can be a number larger or smaller than the true number
of trials, depending on whether the probability distribution had an
excess or a deficit at its tail. Usually, for very large numbers of
trials ($>10^{10}$), such as the ones of the GRB search, the probability
distribution shows a deficit in its tail. Therefore, such a method
usually predicts a random number of effective trials that is usually
lower than the total number of trials, but larger than
the true number of effective trials. 

From the above, it is shown that the probability distribution of all trials cannot
be used in a straightforward way to estimate the effective number
of trials, because the information on the degree of correlation between
the trials cannot be extracted by this distribution in a simple way. %

\subsection{\label{sub:TheSearch_new}Using the distribution of the smallest
probabilities in groups of adjacent trials}

In order to calculate the effective number of trials that correspond
to one trial taken, the probability that this trial erroneously rejects
the null hypothesis has to be calculated. To do this, the same trial
has to be repeated a number of times, and the fraction of times where
the null hypothesis was erroneously rejected has to be counted. The
fraction of the times that the trial rejected the null hypothesis
will be equal to its probability of rejecting the null hypothesis.
Using that probability and the probability threshold selected for
rejecting the null hypothesis, the effective number of trials for
this trial can be calculated using equation \ref{eq:TheSearch_Sidak}. 

In the method described in the previous subsection, the fraction of the trials resulting to a
probability less than some threshold was counted, and from the values
of the counted fraction and the selected threshold, the number of
effective trials was calculated. The problem with that method was that
every time an improbable fluctuation happened in the data, there were
multiple adjacent trials affected, having similarly reduced probabilities.
That method treated all these low probability results as independent,
since each one of them counted as a rejection of the null hypothesis.
Hence, it produced incorrectly inflated estimates of the
number of effective trials. 

It is clear that a method has to be found that counts each improbable
fluctuation that causes the erroneous rejection of the null hypothesis
only once. This can be accomplished by grouping adjacent trials together
and considering whether these trials caused the rejection of the null
hypothesis as a whole. For this search, all the trials in space and
in a time interval equal to the searched duration were grouped together,
with each group containing the trials from the evaluation of ten skymaps.
Then, a distribution containing the smallest probability found
in each group was created. To avoid double counting of any improbable fluctuations
that occurred near the borders of each group, and could possibly appear
in two adjacent groups, every other group was skipped. 

From the distribution of the smallest probabilities in every other
group, the number of effective trials per group can be calculated
by solving the Sidak equation for $T$:

\begin{equation}
eq.\ref{eq:TheSearch_Sidak}\Rightarrow T_{pergroup}=\frac{log(1-P_{0}^{'})}{log(1-P_{0})},\label{eq:TheSearch_EffectiveTrials}\end{equation}
where $P_{0}$ is a probability threshold, and $P_{0}^{'}$ is the
fraction of times a probability less than $P_{0}$ occurred. The total
number of effective trials in the search was then calculated by multiplying
the effective number of trials per group with the number of groups
in the whole search.

\section{Simulation of a 1D Binned Search}

In order to demonstrate the above, a simple simulation of an one-dimensional
binned search was made. Initially, an one-dimensional lattice was
created, composed of $61\times10=610$ bins. The bins of the lattice were
filled with a random number sampled by a Poisson distribution of average
five. Similarly to the GRB search algorithm, a search bin of length
equal to 61 lattice bins was scanned over the lattice, the contents
of the lattice elements under the search bin were added, and the probability
of this sum being produced by a mere fluctuation of the expected number
of events ($=5\times61$) was calculated. The search bin was then
translated by just one lattice bin, and the last step was repeated.
The probabilities corresponding to each evaluation were kept in a
distribution, to calculate the number of trials using the old method
described in subsection \ref{sub:TheSearch_old}. Also, to allow the
calculation of the effective number of trials using the new method
proposed in section \ref{sub:TheSearch_new}, every 61 adjacent trials
were grouped, and the smallest probability found in every other group
was recorded. The search was repeated millions of times, using different
bin contents for the lattice in each iteration. 

The consecutive trials in this search were highly correlated, because
for such trials 60 out of 61 bins under the search bin were the same.
Therefore, the number of effective trials was expected to be considerably
smaller than the total number of trials (550). The effective number
of trials was calculated using three methods:

\begin{itemize}
\item The first method calculated the true effective number of trials. The
search was repeated millions of times, and the fraction of repetitions
that a probability smaller than some threshold was found were counted.
Then, equation \ref{eq:TheSearch_EffectiveTrials} was used to calculate
the effective number of trials. 
\item The old method, that calculates an incorrect effective number of trials
from the distribution of all the probabilities.
\item The new proposed method, that calculates the effective number of trials
from the distribution of the smallest probability found in every other
group of trials. This method should give the same number of effective
trials as the correct number found by the first method. 
\end{itemize}

\begin{figure}[ht]
\includegraphics[width=1\columnwidth]{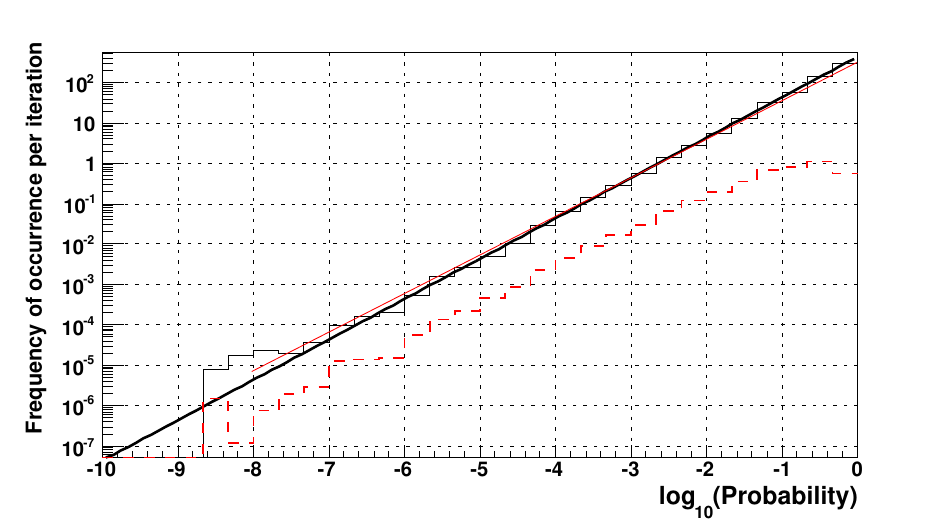}

\caption{\label{fig:TheSearch_SimProbs}Probability distributions produced
by the search simulation. \emph{Black curve}: distribution of the
probabilities of all trials, \emph{red dashed curve}: distribution
of the smallest probability found in every other group of adjacent
trials, \emph{red line}: curve described by equation \ref{eq:TheSearchPDistro}
with $N_{1}=550$, \emph{thick black line}: fit to the probability
distribution of all trials. }

\end{figure}

\begin{figure}[ht]
\includegraphics[width=1\columnwidth]{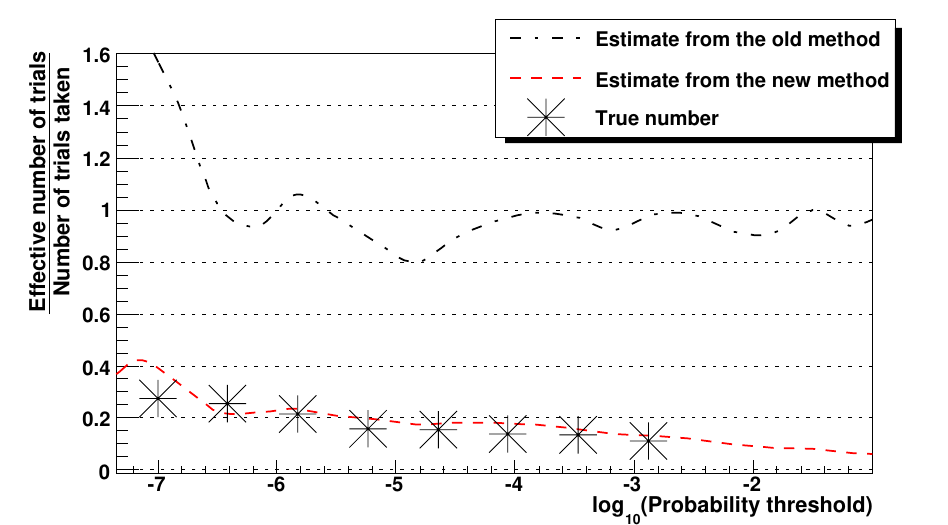}

\caption{\label{fig:TheSearch_SimTrials}Ratio of the effective number of trials
calculated by each of the three methods over the total number of trials
taken. \emph{Stars}: Ratio based on the true effective number of trials
calculated by the first method, \emph{dashed black line}: ratio based
on an incorrect estimate of the effective number of trials provided
by the second (old) method, \emph{dashed red line}: ratio based on
a correct estimate of the effective number of trials provided by the
third (new) method. }

\end{figure}

Figure \ref{fig:TheSearch_SimProbs} shows the distributions of the
probabilities of all trials and of the smallest probability found
in every other group of trials. As claimed above and shown in the
figure, the probability distribution of all trials (fit to the black
line) follows the form described by the equation \ref{eq:TheSearchPDistro}
(red line). The ratio of the effective number of trials calculated
by each of the three methods described above over the total number
of trials taken is shown in figure \ref{fig:TheSearch_SimTrials}.
As expected, the old (second) method incorrectly gives an effective
number of trials that is almost the same as the number of trials taken
(black dashed curve). The third, newly proposed, method (dashed
red line), gives an estimate that is in agreement with the true number
(stars) calculated by the first method. 

As shown above, the old method produces an incorrectly inflated
effective number of trials. As a result, the calculated
post-trials probabilities are erroneously increased, or, equivalently,
their statistical significance is decreased resulting in a reduction
in the sensitivity of the search.

The fact that the number of effective trials increases with the significance
is partially understood. The same effect has been observed
in the results of the search, although to a much smaller degree, and in
calculations of the effective number of trials in Milagro's point-source
searches \cite{ThesisChuan}. 


   \clearpage \chapter{\label{chap:binsize}Optimum Bin Size}

\section{Introduction}
One of the most important factors in maximizing the sensitivity of
the search is the proper choice of its bin size. A large bin will
include most of the signal events but will also include many background
events. On the other hand, a small bin will reduce the contamination
from background but will also fail to include a big part of the signal.
Both cases are non-optimal and correspond to a reduced sensitivity. 

As will be shown, because of statistical fluctuations involved in the process 
of converting an incoming gamma-ray flux at the earth to Milagro signal events,
the same initial flux can be detected by a specific search only a fraction 
of the times (the ``detection probability''). Every time an incoming gamma-ray 
flux creates an excess that is larger than a mere background fluctuation, a signal detection
can be claimed. The fraction of times that a detection can be claimed for the same
initial gamma-ray signal is equal to the detection probability of that signal.
The optimum bin size is the one that maximizes that probability.

The detection probability depends on how improbable are the fluctuations
created by the incoming gamma-ray flux. Fluctuation probabilities, as described in
section \ref{Chapter:Probabilities}, are calculated using Poisson statistics
(eq. \vref{eq:SEARCH_Poisson Probability}). For the search of longer-duration emissions
(duration $\gtrsim100\,s$), the mean number of background events under the search bin is
sufficiently large that the probabilities can be also calculated using Gaussian statistics.

The purpose of this chapter is to present the optimization of the search's bin size.
The optimization of the bin size was performed using Poisson statistics and is described
in section \ref{sec:BINSIZE_Poisson}. As a cross check, the optimum bin size was 
also calculated using Gaussian statistics for only the longer durations $\gtrsim100\,s$
(section \ref{BINSIZE_Gauss}). The Poisson-based optimization,
as expected, produces the same results as the Gaussian-based one in the
limit of a large number of events. 

Because the VHE emission from GRBs is absorbed by the EBL (Chapter \ref{chap:IR}), the GRB signal that reaches the earth
cuts off at energies over few hundreds of GeV.
On the other hand, because PBHs can be galactic sources, their detectable by Milagro emission extends to $>1TeV$
energies. The bin sizes calculated here are optimal
for the lower-energy signal expected from GRBs and not for the high energy
signal expected from PBHs.

\section{\label{BINSIZE_Gauss}Optimum Bin-Size for Gaussian Statistics}

In this section, the optimum bin size will be calculated for the case of large statistics 
of the background and the signal. The calculation will be performed for square and circular bins, 
and for PSFs that follow a Gaussian distribution or an arbitrary distribution.
The calculation for an arbitrary PSF and a square bin relevant 
for this search is performed in subsection \ref{subsec:BINSIZE_SQGAUSS}. The results of that 
subsection were used as verification of the Poisson-based optimization of the bin size. 
This calculation is valid for the searches of durations greater than about $100\,s$. 

In the case that:
\begin{enumerate}
\item The background distribution is uniform, 
\item the number of signal $(N_{S}$) and background $(N_{BG}$) events
follow a Gaussian distribution: $\left\{ N_{BG},N_{S}\right\} \gg1$,
\item and the statistical fluctuations of the number of background and signal
events are negligible: $\left\{ N_{BG},N_{S}\right\} \gg\left\{ \sqrt{N_{BG}},\sqrt{N_{S}}\right\},$
\end{enumerate}
the probability that the number of signal events in a bin were created
by just a background fluctuation can be approximated, similarly to equation \vref{eq:SEARCH_Poisson Probability}, by the cumulative
Gaussian probability: \begin{equation}
P_{G,C}(N_{BG},N_{S})=\int_{N_{S}}^{\infty}\frac{1}{\sqrt{2\pi N_{BG}}}e^{-\frac{\left(N_{S}-N_{BG}\right)^{2}}{2N_{BG}}}.\label{eq:GaussianCProb}\end{equation}
The measure of the consistency of a result with the null hypothesis
can be given either by the probability $P_{G,C}$ or by the ``Significance'' $\mathcal{S}$
of the measurement, which is equal to the number of signal events measured
in units of standard deviations of the background:
\begin{equation}
\mathcal{S}\equiv\frac{N_{meas.}-\hat{N}_{BG}}{\sigma(\hat{N}_{BG})}=\frac{(N_{S}+N_{BG})-\hat{N}_{BG}}{\sigma(\hat{N}_{BG})}\simeq\frac{N_{S}+\hat{N}_{BG}-\hat{N}_{BG}}{\sqrt{\hat{N}_{BG}}}=\frac{N_{S}}{\sqrt{\hat{N}_{BG}}}.\label{eq:Significance}\end{equation}
As shown in figure \vref{fig:Milagro_delAngle}, the reconstructed directions of the signal events
are distributed according to the point-spread function (PSF) of the detector,
and the reconstructed directions of the background events
are assumed to be distributed locally uniformly. Let us define the efficiency of
including the signal events in a square bin of width $w$, and area
$A=w^{2}$, as $\epsilon(w)$, the background rate per unit area as
$R$, the duration searched as $T$, and the total number of signal
events in the Milagro data as $N_{S,total}$. Then, the number of
signal events and background events included in a bin of width $w$
are $N_{S}(w)=N_{S,total}\,\epsilon(w)$ and $\hat{N}_{BG}(w)=R\,T\,A=R\,T\,w^{2}$
respectively. Equation \ref{eq:Significance} then becomes
\begin{equation}
\mathcal{S}(w,T)=\frac{N_{s}(w)}{\sqrt{\hat{N}_{BG}(w)}}=\frac{N_{S,total}}{\sqrt{R\,T}}\,\frac{\epsilon(w)}{\sqrt{w^{2}}}=\frac{N_{S,total}}{\sqrt{R\,T}}\,\frac{\epsilon(w)}{w}.\label{eq:Binsize_significance2}\end{equation}

We see that the significance is a function of the bin width $w$.
In order to maximize the significance, one has to maximize the ratio
$\frac{\epsilon(w)}{w}$. Note that the optimum bin size does not
depend on the background rate $R$ or the total amount of signal $N_{S,total}$.
Because of that, the bin size that maximizes the average significance\footnote{averaged over 
all the possible fluctuations in the conversion from an incoming gamma-ray signal
to Milagro signal events} that corresponds to an incoming gamma-ray signal is equal to
the one that maximizes the detection probability of detecting the same gamma-ray signal.
Therefore, in the case of large statistics, the bin-size optimization (maximization of the detection probability) can be 
performed in terms of finding the bin size that maximizes the average significance.

\subsection{\label{subsec:CIRC_GAUSS}Circular bin and Gaussian PSF}
Consider the case of circular bins and of a PSF that follows
a Gaussian distribution of standard deviation $\sigma$.
The numbers of expected signal and
background events within a bin of radius $R$ are proportional to:\begin{alignat}{1}
\begin{array}{ccll}
\hat{N}_{S} & \propto & \int_{0}^{R}\frac{e^{-r^{2}/2\sigma^{2}}}{\sigma\sqrt{2\pi}}r & dr\\
\hat{N}_{BG} & \propto & \int_{0}^{R}2\pi & dr.\end{array}\label{eq:BINSIZE_CircularBinsOptimization}\end{alignat}
The optical bin size $R_{0}$ maximizes the significance $\hat{N}_{S}/\sqrt{\hat{N}_{BG}}$
so that:
\begin{equation}
\frac{d}{dR}\left\{ \frac{\hat{N}_{S}}{\hat{N}_{BG}}\right\}=0.\end{equation}
Using equation \ref{eq:BINSIZE_CircularBinsOptimization} and substituting
$x=r^{2}/\sigma^{2}$ we obtain:
\begin{eqnarray}
\frac{dX}{dR}\frac{d}{dX}\left\{ \frac{\int_{0}^{X}e^{-x/2}dx}{\sqrt{\int_{0}^{X}dx}}\right\}  & = & 0.\end{eqnarray}
For non-zero bins $dX/dR\ne0$, therefore:
\begin{eqnarray}
\frac{d}{dX}\left\{ \frac{2\,(1-e^{-X/2})}{\sqrt{X}}\right\}  & = & 0\Rightarrow\\
1+X_{0} & = & e^{X_{0}/2}\Longrightarrow\\
X_{0} & = & 2.513\Longleftrightarrow\\
\mathbf{\mathrm{R_{0}}} & = & 1.585\sigma.\label{eq:BINSIZE_OptimumCircular}\end{eqnarray}

\subsection{\label{subsec:SQ_GAUSS}Square bin and Gaussian PSF}
For a square bin and a Gaussian PSF, the optimum bin
size can be calculated following a procedure similar to the one above.
In this case:
\begin{alignat}{1}
\begin{array}{ccll}
\hat{N}_{S} & \propto & \int_{-W}^{+W}\int_{-W}^{+W}\frac{e^{-x^{2}/2\sigma^{2}}}{\sigma\sqrt{2\pi}}\,\frac{e^{-y^{2}/2\sigma^{2}}}{\sigma\sqrt{2\pi}} & dxdy\\
\hat{N}_{BG} & \propto & \int_{-W}^{+W}\int_{-W}^{+W}1 & dxdy.\end{array}\label{eq:BINSIZE_SquareBinsOptimization}\end{alignat}
If the optimum half width is $W_{0}$ then:
\begin{eqnarray}
0 & = & \frac{d}{dW}\left\{ \frac{\hat{N}_{S}}{\sqrt{\hat{N}_{BG}}}\right\} \\
 & = & \frac{d}{dW}\left\{ \frac{erf\left(\frac{W}{\sqrt{2}\sigma}\right)^{2}}{2W}\right\} \\
 & = & \frac{\sqrt{2}\,erf(\frac{W_{0}}{\sqrt{2}\sigma})e^{-\frac{W_{0}^{2}}{2\sigma^{2}}}}{W_{0}\sqrt{\pi}}-\frac{1}{2}\frac{erf\left(\frac{W_{0}}{\sqrt{2}\sigma}\right)^{2}}{W_{0}^{2}}.\end{eqnarray}
This can be solved numerically giving 
\begin{equation}
W_{0}=1.40\,\sigma\simeq\sqrt{2}\,\sigma.\label{eq:BINSIZE_OptimumSquare}\end{equation}

It should be noted that equations \ref{eq:BINSIZE_OptimumCircular} and \ref{eq:BINSIZE_OptimumSquare}
say that the area of an optimal circular bin is almost the
same as the area of an optimal square bin (less than $1\%$ difference).

\subsection{\label{subsec:BINSIZE_SQGAUSS}Square bin and arbitrary PSF}

\begin{figure}[ht]
\includegraphics[width=1\columnwidth]{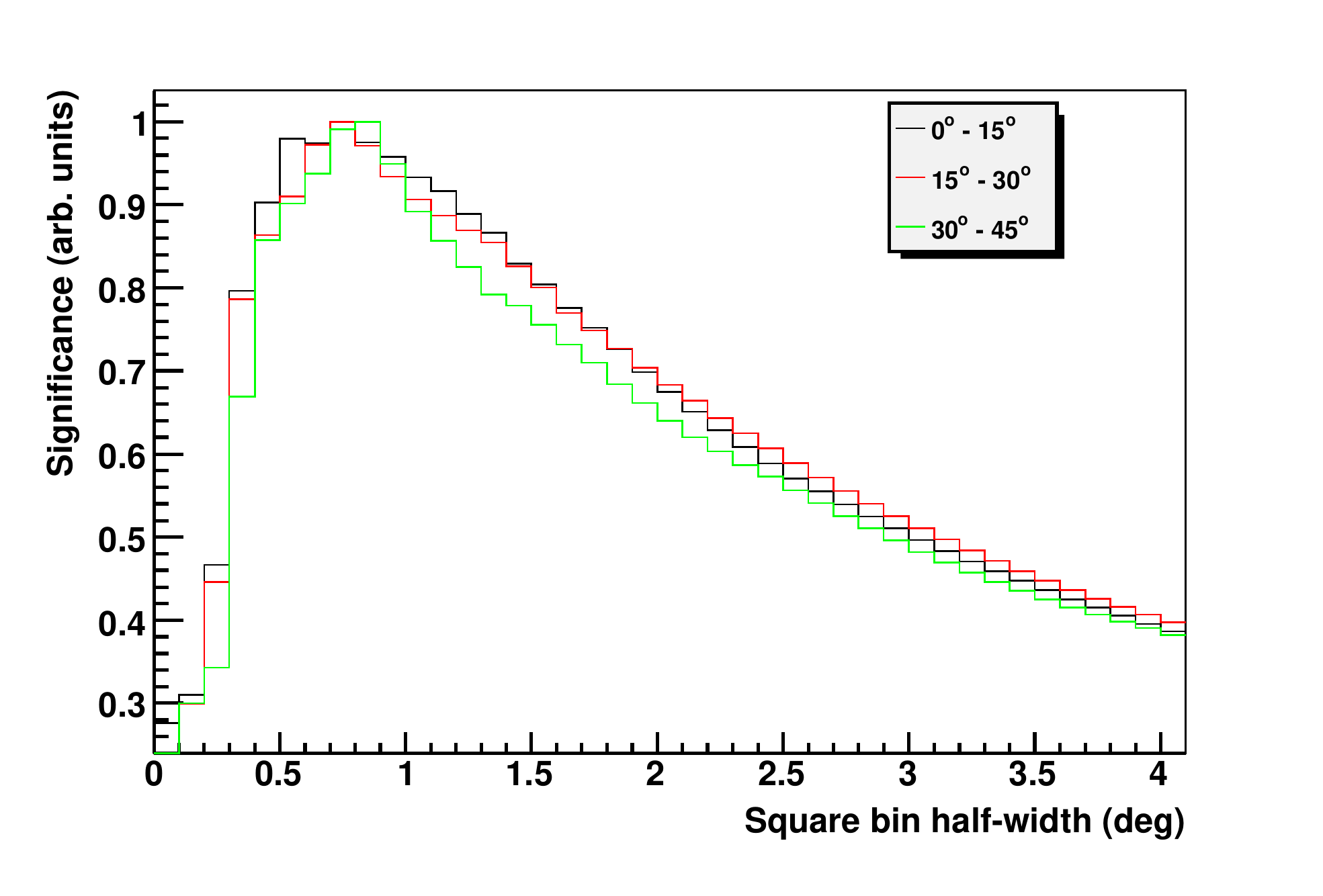}
\caption{\label{fig:binsize_gaussian}The ratio $\frac{\epsilon(w)}{w}$ found
in the significance formula (eq. \ref{eq:Binsize_significance2})
plotted versus the bin half-width $(w/2)$ for three zenith-angle
regions. The optimum bin half-width corresponds to the peaks of the curves
and is equal to $\sim0.8^{o}$ for all three curves. Data from a Monte Carlo simulation
of the detector were used to construct these plots.}
\end{figure}

For any kind of PSF, the optimum bin size can be calculated numerically
by finding the bin size that maximizes the significance ratio $\hat{N}_{S}/\sqrt{\hat{N}_{BG}}$
of equation \ref{eq:Binsize_significance2}. In figure \ref{fig:binsize_gaussian}, this ratio is plotted versus
the square bin half-width, for a detector with the background rate and PSF of Milagro. Because the angular resolution of the detector
becomes worse with increasing zenith angles, the significance acquires
a dependence on the zenith angle. However, this effect is very
small, so the optimum half-width is almost the same for all three zenith-angle
regions, and equal to $\sim0.8^{o}$. If the PSF of Milagro followed a Gaussian distribution,
this optimum bin size would correspond to a standard deviation of $\sigma=0.8/1.4=0.57^{0}$ (according to equation
\ref{eq:BINSIZE_OptimumSquare}).

The angular resolution of Milagro has a stronger dependence on the properties of the gamma-ray
signal being detected. This analysis used a gamma-ray signal on a
power-law energy spectrum with index -2.2 and with an attenuation
due to interactions with the EBL for a source at a redshift $z=0.2$.

\section{\label{sec:BINSIZE_Poisson}Using Poisson Statistics}

In the case of a small number of signal $N_{S}$ or background events $N_{B}$ ($N_{S}\not\gg1$ or $N_{B}\not\gg1$),
the fluctuations on the number of background and signal events dominate.
In this case, equations \ref{eq:GaussianCProb} and \ref{eq:Significance}
cannot be used, and a simple formula that gives the significance does not
exist, so, the chance probabilities are calculated using Poisson
statistics (eq. \ref{eq:SEARCH_Poisson Probability}), and the optimum bin size depends on the amount
of signal and background expected in the search. An optimization of the bin size
will now require the calculation of the sensitivity versus the signal
strength, the bin size, the duration, and the background rate.

Consider the random processes starting from the conversion of an incoming
gamma-ray signal at the top of the atmosphere to the final detection of it.
That signal will initially interact with the atmosphere and create EASs. 
A fraction of these EASs will be large enough
to trigger Milagro, and a fraction of the generated Milagro events 
will pass the data-quality and background-rejection cuts of a search
for emission from GRBs. Finally, only a fraction of these events will be 
reconstructed accurately enough and manage to be included under the search bin.
Because of the random fluctuations involved in these
steps, if the same series of steps is repeated on the same amount of initial signal,
the final number of events under the search bin will fluctuate.

Let us call the average number of events a source creates in the Milagro
data set (after cuts) $\hat{N}_{S,total}$ and the actual number of
events occurring in one instance of the same process $N_{S,total}$.
$N_{S,total}$ follows a Poisson distribution (or a Gaussian distribution for large $\hat{N}_{S,total}$)
with average $\hat{N}_{S,total}$.
A binned search for these events uses a bin size that includes them with some efficiency
$\epsilon(w)$. The \emph{average} number of events
ending up in the search bin is $\hat{N}_{S,bin}=\hat{N}_{S,total}\,\epsilon(w)$.
The \emph{actual} number of events ending up in the bin (number that corresponds to 
one repetition), $N_{S,bin}$, follows a Binomial distribution with average $\hat{N}_{S,bin}$
(probability of success per trial = $\epsilon(w)$ and $\hat{N}_{S,total}$ trials):
\begin{equation}
P_{Binomial}(\hat{N}_{S,bin},\epsilon(w),\hat{N}_{S,total})=\left(\begin{array}{c}
\hat{N}_{S,total}\\
\hat{N}_{S,bin}\end{array}\right)\epsilon(w)^{\hat{N}_{S,bin}}(1-\epsilon(w))^{\hat{N}_{S,total}-\hat{N}_{S,bin}}.\label{eq:Binsize_binomialPSFSPRead}\end{equation}
Because the same initial emission creates a 
fluctuating final number of signal events in the Milagro data, that emission
has a probability to be detected. The bin size optimization in this section will be 
performed in terms of maximizing the detection probability of a certain signal.

Let us define that a gamma-ray signal is called ``detected'' if it 
creates a fluctuation on the number of events under the search bin
that has a probability less than $5\sigma$ post trials of
being just a background fluctuation. All detected events are considered
equally interesting and no further effort is taken to increase
their significance.

In this bin-size optimization, the effective number of trials
is needed to calculate the post-trials probability of a fluctuation and
to decide whether a fluctuation corresponds to a signal detection or not.
However, the effective number of trials depends
on the bin size of the search. It can be seen that there is a circular dependence since 
the two quantities (bin size and effective number trials) depend on each other. 
It is not easy to analytically calculate the 
dependence of the effective number of trials on the bin size, and it is not correct
to use the easily calculable total number of trials instead. If the total number
of trials is used, then the resulting post-trials significance of the results and the 
sensitivity of the search will be erroneously lower than their true values.
To overcome the problem, the effective number of trials from an earlier search
with a fixed bin size of half width $1.5^o$ was used, and the approximation that a bin size
with a half-width close to $1.5^o$ corresponds to the same number of effective trials as that 
of a search with a bin of half width exactly $1.5^o$.

\begin{figure}[ht!]
\includegraphics[width=1\columnwidth]{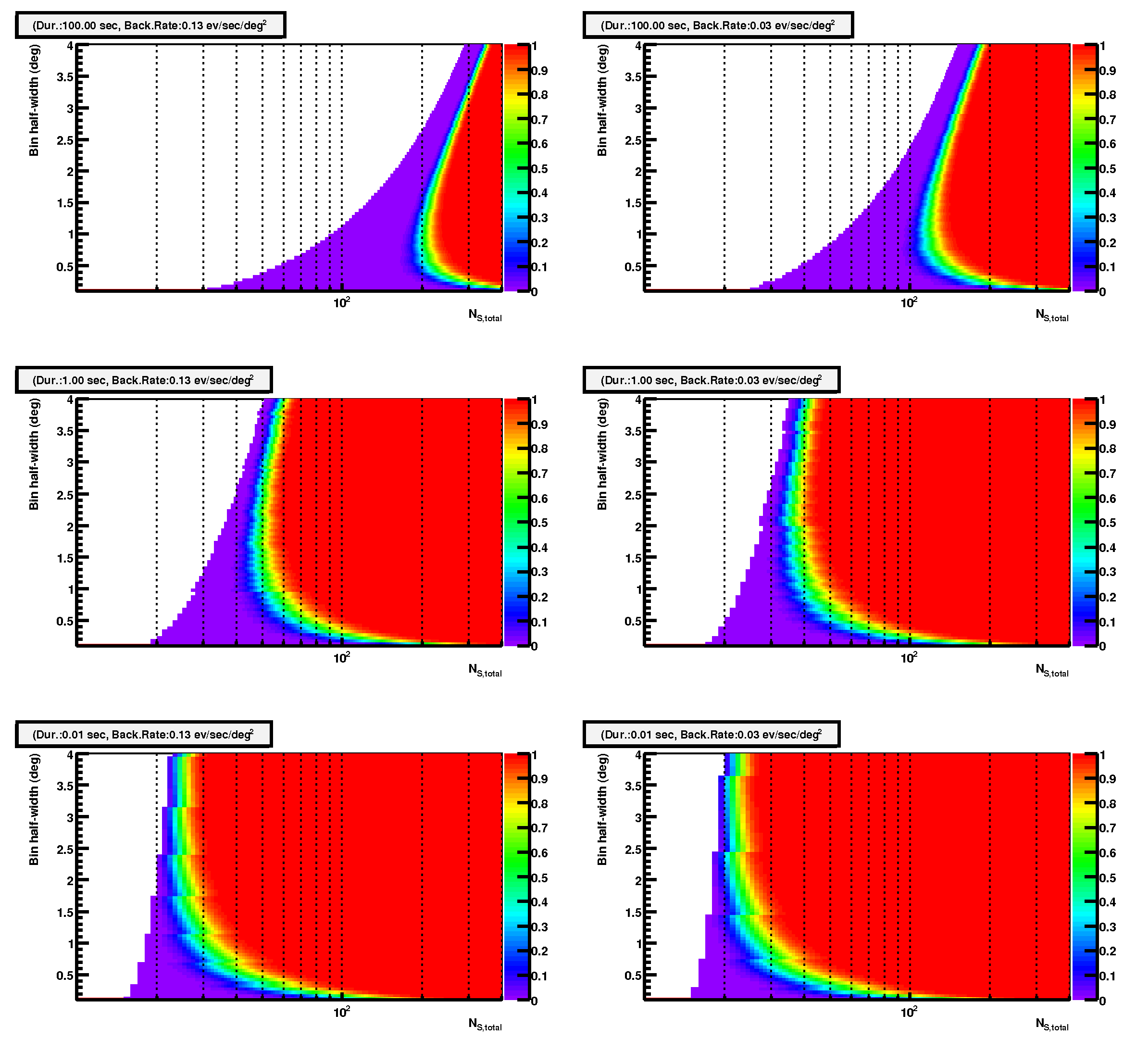}
\caption{\label{fig:SearhBin_NoSpread}Probability of detecting a signal consisting
of $N_{s,total}$ events in the Milagro data set versus the half-width
of the bin used by the search. The columns correspond to different
background rates and the rows to different durations searched. }
\end{figure}

Let us start the optimization by including only the fluctuations on
how the reconstructed directions of the events are distributed around
the true source position. In this stage, the probability of detecting
a signal $N_{S,total}$ in the Milagro dataset will be examined
versus the bin size of a search for that sgnal. The same total number
of events in the Milagro data set, $N_{S,total}$, can randomly lead
to various different numbers of events in the search bin $N_{S,bin}$,
depending on how the reconstructed event directions were distributed
around the true source position. Each of these possible $N_{S,bin}$, depending
on the number of background events for the bin size used ($N_{BG}=R\,T\,w^{2}$),
corresponds to a different chance probability $P_{P,C}$ of the measurement
being consistent with a background fluctuation. Let the Boolean outcome
of an $N_{S,bin}$ being detectable be $\Delta(N_{S,bin},N_{BG}$),
with the outcome being equal to 1 if $P_{G,C}$ is smaller than the
probability threshold set for detection (equivalent to $5\sigma$
post trials significance) and 0, if otherwise. The probability of detecting
a signal $N_{S,total}$ that can give rise to various $N_{S,bin}$,
is the weighted average of all of the final outcomes $\Delta(N_{S,bin},N_{BG})$
for each $N_{S,bin}$, with the weight being the occurrence probability
(eq. \ref{eq:Binsize_binomialPSFSPRead}) of each $N_{S,bin}$:
\begin{equation}
P_{det}^{'}(N_{S,total})=
\underset{N_{S,bin=1}}{\overset{N_{S,total}}{\sum}}P_{Binomial}(N_{S,bin,}\epsilon(w),N_{S,total})\,\Delta(N_{S,bin},N_{BG}).\end{equation}
This quantity is plotted for different $N_{S,total}$, background
rates $R$, durations $T$ and bin sizes $w$ in figures \ref{fig:SearhBin_NoSpread}.
The background rates $0.13\,ev\,s^{-1}\,deg^{-2}$ and $0.03\,ev\,s^{-1}\,deg^{-2}$
correspond to the average Milagro event rates from the zenith angle regions
$0^{o}-15^{o}$ and $30^{o}-45^{o}$, respectively. For very weak signals
($N_{S,total}$), there is not any bin size that contains enough signal
events ($N_{S,bin})$ to create a detection. As the number of total
signal events increases, there are some bin sizes that contain detectable
amounts of signal events. For these detections, the signal events
were, in an improbable way, distributed closer to the true source
position than the one predicted by the PSF. As can be seen, the small bins that do not include much background
but that contain most of the signal events are the ones generating the
detections for these signal strengths. For a higher total number of
signal events, the chance probabilities decrease even more and a larger
fraction of the events becomes detectable. In the end, almost all
of the bin sizes can generate detectable events. As the bin size increases,
the detection probabilities decrease because there is more background
included under the bin. This effect becomes stronger for longer durations
and for higher background rates. %

\begin{figure}[ht!]
\includegraphics[width=1\columnwidth]{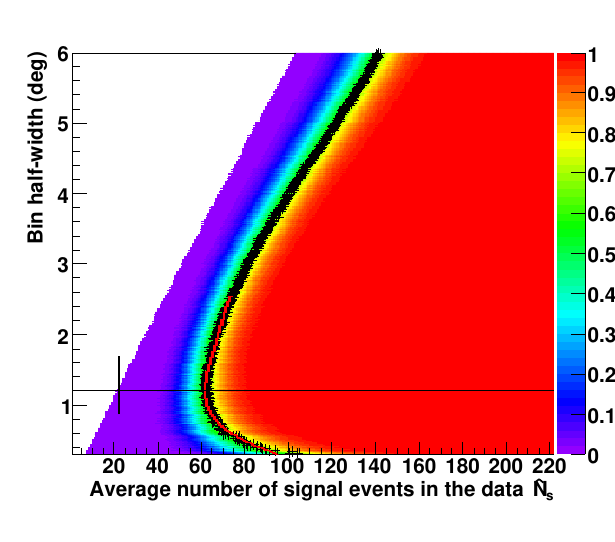}
\caption{\label{fig:SearhBin_Spread}Map showing the detection probability
of an average signal $\hat{N}_{S,total}$ versus the 
the half-width of the bin used by a search for that signal. 
\emph{Black curve}: profile of the map, \emph{red curve}: polynomial fit to the
profile, \emph{horizontal black line}: minimum of the fit, \emph{vertical black bar}: error
bars corresponding to $\pm15\%$ increase in the minimum detectable
signal over the optimum case. This map corresponds to the search for
emission with duration $100\,s$ from a zenith angle $0^{o}-15^{o}$. }
\end{figure}

Now, let us extend the calculation and include the effects
of fluctuations in the conversion between the incoming gamma-ray flux on the 
top of the atmosphere to Milagro triggered events that have passed the cuts.
As mentioned above, the actual number of events $N_{S,total}$
in the Milagro data will be distributed on a Poisson distribution
with average $\hat{N}_{S,total}$. By taking a weighted average of
the detection probability of each $N_{S,total}$, with the Poisson
probability of occurrence of each of these $N_{s,total}$ as the weight, the
detection probability of an average total signal $\hat{N}_{S,total}$
can be calculated: 
\begin{equation}
P_{det}(\hat{N}_{S,total})=\overset{\infty}{\underset{N_{s,total}=0}{\sum}}P_{Poisson}(N_{s,total},\hat{N}_{s,total})\,P_{det}^{'}(N_{S,total}).
\end{equation}
Th distribution of $P_{det}(\hat{N}_{S,total})$ is shown in the color map of figure \ref{fig:SearhBin_Spread}
for different $\hat{N}_{S,total}$ and bin sizes, for the $10\,s$ duration
search and a background rate corresponding a zenith angle $0^{o}-15^{o}$. 

The next step would be to use maps, such as the one of figure \ref{fig:SearhBin_Spread}, to find the optimum
bin size for each duration and zenith-angle range. The definition of ``optimum''
depends on the specific requirements for the search. If a known signal
is searched for, then an optimum search will be the one that will
detect it with the highest probability. In our case, this is a search
for a not-yet detected signal. Maybe the only thing that it is known
is that the signal levels present in the data are very low, otherwise
detection of VHE emission from GRBs by other experiments would have happened by now.
So the requirement for our optimum bin is to maximize the probability of
detecting the smallest signal possible. 
The distribution has a wedge shape and the optimum bin corresponds
to the tip of this wedge. In order to calculate the bin size that
corresponds to that point, a fit on the wedge has been made. Initially,
a profile of the 2D map was made by calculating the weighted average
of the contents of each row (same bin size) with weights the map-bin contents
(detection probabilities). The profile is shown with the black
curve on figure \ref{fig:SearhBin_Spread}. Map bins with a content
greater than 0.9 were ignored at the creation of the profile
in order to move it towards the edge of this wedge. The profile
essentially gives a measure of the amount of signal $\hat{N}_{s,total}$, at which
each bin size starts being efficient (starts producing detections). 
The optimum bin corresponds to the best efficiency
for the lowest signals possible, and is equal to the minimum (signal wise)
of the profile curve. That minimum was calculated by means of a polynomial bit
to the profile (shown with the red line on figure \ref{fig:SearhBin_Spread}).
The resulting minimum bin size is shown with a horizontal black line, along with error bars
that correspond to a 15\% increase over the minimum signal of the profile curve.

\begin{figure}[ht!]
\subfigure[$0^o-15^o$]{\includegraphics[width=0.5\columnwidth]{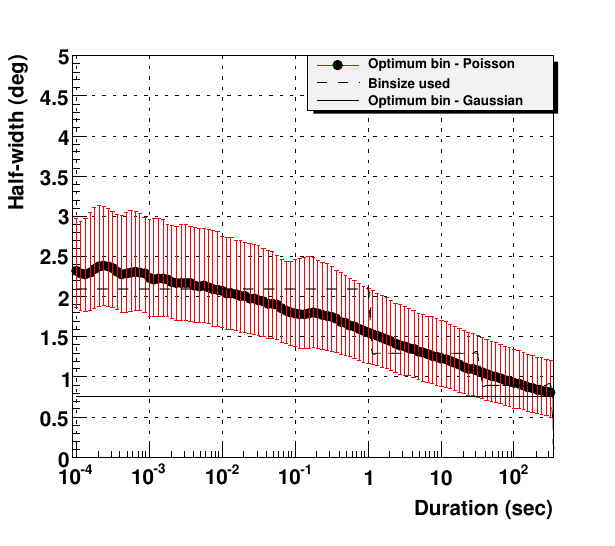}}
\subfigure[$15^o-30^o$]{\includegraphics[width=0.5\columnwidth]{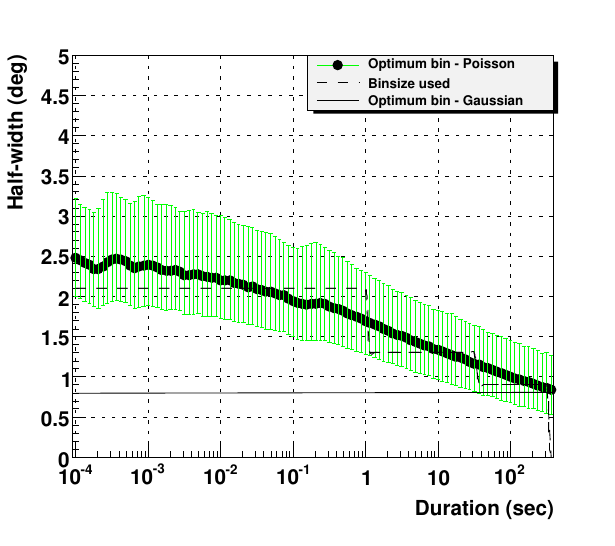}}
\subfigure[$30^o-45^o$]{\includegraphics[width=0.5\columnwidth]{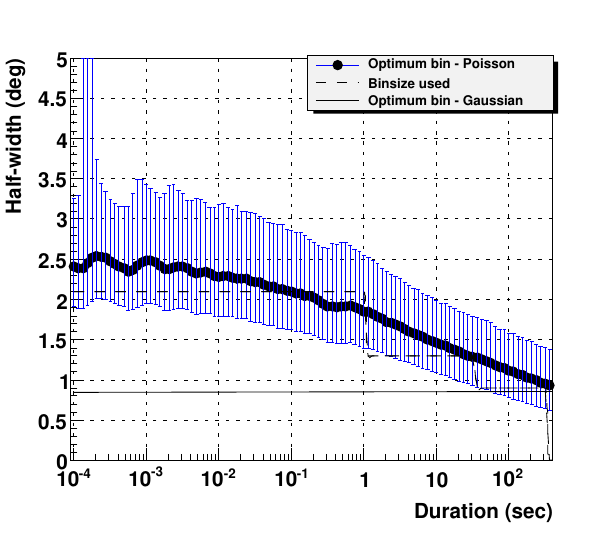}}
\caption{\label{fig:Binsize_Optimum}Optimum bin-size half-width for three
zenith-angle regions (black points with color error bars). The error
bars correspond to $\pm15\%$ increase over the minimum signal of the profile curve.
The black solid horizontal line shows the optimum bin size as calculated
using Gaussian statistics. The dashed black curve shows the bin size used in the search. }
\end{figure}

This process was repeated for all 53 durations and three zenith-angle
bands. The results are shown in figure \ref{fig:Binsize_Optimum}.
As can be seen, for the shorter durations, where the background contamination
is smaller, larger bin sizes are optimum. As the duration and the 
the background contamination increase, the optimum bin sizes become
gradually smaller until we slowly enter the large-number-of-events
Gaussian regime, where the optimum bin size does not depend on the
amount of background. For a cross check, the optimum bin size calculated using
Gaussian probabilities (subsec. \ref{subsec:BINSIZE_SQGAUSS}) is also shown with a black solid line. The
bin sizes in this section, derived using Poisson statistics, are as
expected approaching the Gaussian ones in the limit of a large number
of events (large durations).

The difference between the three zenith-angle bands results from the
different PSF and background rates. Both factors favor larger bin
sizes with increasing zenith angle. However, as seen from the figures,
the optimum bin sizes are very similar for the three zenith-angle
bands. For that reason, instead of running the search in three parts,
each one specifically optimized for a different zenith angle region,
all three regions were analyzed simultaneously using a common bin size
distribution (dashed black line). Comparing this almost optimum bin size
distribution to the error bars of the optimum bin size curves, it
can be seen that at worst, it is $\sim15\%$ less sensitive than the
optimal case. 

\begin{figure}[ht!]
\begin{centering}
\includegraphics[width=0.8\columnwidth]{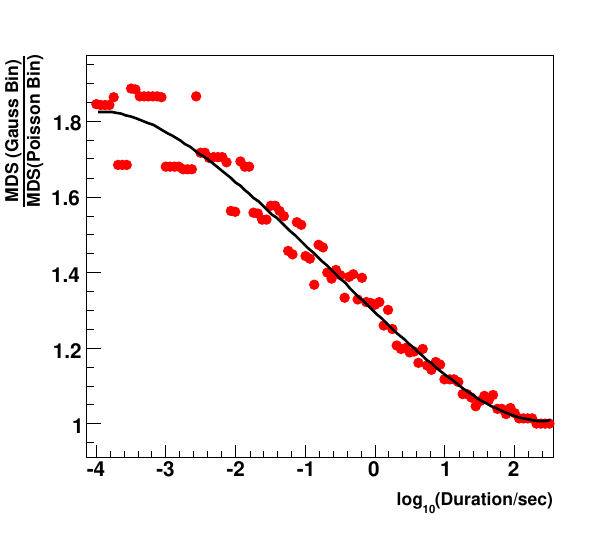}

\caption{\label{fig:Binsize_Improvement}Difference in the fluence sensitivity 
of the search when using the optimum bin size calculated using Poisson statistics
versus the optimum bin size calculated using Gaussian statistics. 
The Y axis shows the ratio of the minimum signal that can be detected with a 
90\% confidence level when using a Gaussian-optimized bin ($0.8^o$) over
the minimum signal when using a Poisson-optimized bin. The red dots show
the ratio for each duration, and the black curve is a polynomial fit to guide the eye.
The results are shown for the $0^o-15^o$ zenith angle band.
The improvement is similar for the other two zenith-angle bands.
}
\end{centering}
\end{figure}

The improvement in the sensitivity of the search using the optimized bin sizes 
is shown in figure \ref{fig:Binsize_Improvement}. 
Let the minimum detectable signal (with 90\% CL) be $N_{S,det,min}$ 
The quantity, plotted in figure \ref{fig:Binsize_Improvement} is the ratio of $N_{S,det,min}$ when
the optimum bin size from the Gaussian optimization is used over $N_{S,det,min}$
when the optimum bin size from the Poisson optimization is used.
As can be seen,
there is an improvement in the sensitivity up to a factor of $\sim1.9$
for the shorter durations. Equivalently, for the shorter durations, 
signals that are smaller by up to $1-1/1.9=\sim45\%$ can be detected
if the Poisson-optimized bin is used.
The improvement goes to zero for the larger
durations, since the optimum bin sizes for the Gaussian and Poisson
regimes become the same. The roughness of the maps comes from the fact that for the low durations,
the improvement ratios are calculated by dividing small integer numbers.

This analysis here used a gamma-ray signal on an
power-law energy spectrum with index -2.2 and with an attenuation
due to interactions with the IR background for a source at a redshift $z=0.2$. 
Signals from sources further away are expected to have a lower energy, and to be reconstructed
with a wider PSF, and vice versa. A redshift $z\sim0.2$ is representative of the 
expected redshift of the GRBs potentially detectable by this search. GRBs considerably further than that
will create a signal that is too low in energy to be easily detectable by Milagro, and GRBs
considerably closer than that are very rare.

   \clearpage \chapter{\label{chap:Sensitivity}Milagro's Sensitivity to GRBs and Evaporating PBHs}
\section{Introduction}

The purpose of this chapter is to examine Milagro's prospects of detecting
a signal from a GRB or a PBH. These prospects depend on the properties
of the signals under search and on Milagro's sensitivity to those
signals. The effective area of Milagro, a quantity essential
for any sensitivity calculations, will be calculated in section \ref{sec:GRBSensi_EFfective}.
Then, section \ref{sec:GRBSensi_GRBSensi} will present the sensitivity
of Milagro to the signal of GRBs of various redshifts and durations.
Lastly, section \ref{sub:GRBSensi_PBHEvap} will show Milagro's sensitivity
to the signal from evaporating PBHs of different temperatures (lifetimes)
and from different distances.

\section{\label{sec:GRBSensi_EFfective}Effective Area}

The effective area $A_{eff}(E,\theta)$ of Milagro describes its efficiency
of converting an incoming gamma-ray flux at the top of the atmosphere
in detected events. The effective area is calculated by simulating
the response of the detector to the EASs generated by that incoming
gamma-ray flux. The cores of the simulated EASs are distributed uniformly
on a wide area of surface $A_{throw}$ extending over and around the
detector. EAS of various energies and zenith angles are simulated.
Only a fraction of these showers manages to trigger the detector and
create signal events that pass the data-quality cuts.
If $N_{throw}(E,\theta)$ is the number of showers with energies
between $E$ and $E+dE$ thrown from a zenith angle $\theta$ to $\theta+d\theta$,
and $N_{pass}(E,\theta)$ is the number of EASs that triggered the
detector and passed the cuts, then the effective area of the detector
is \begin{equation}
A_{eff}(E,\theta)\equiv A_{throw}\frac{N_{pass}(E,\theta)}{N_{throw}(E,\theta)}.\label{eq:SensitivityEffectiveArea}\end{equation}
Figure \ref{eq:SensitivityEffectiveArea} shows the effective area
of Milagro for showers of different zenith angles. As the zenith angle
increases, the atmospheric depth increases, and the showers are attenuated
more, hence they manage to trigger Milagro with smaller efficiency.
As can be seen, Milagro's effective area is maximum at energies $E\gtrsim10\,TeV$,
and quicky decreases for lower energies. The value of the effective
area for $E\gtrsim10\,TeV$ is roughly equal to the physical area of
the detector. For that energy range, all EASs with cores landing inside
the physical area of the detector cause triggered events that have
passed the cuts. 

Even though Milagro's effective area rapidly decreases with decreasing
primary energy, Milagro has some effective area even down to $100\,GeV$
($\simeq5\,m^2)$, an energy important for GRB searches.
For comparison, the sensitive gamma-ray instrument LAT on board the
recently launched GLAST satellite has an effective area $\simeq5\times10^{3}\,cm^{2}$
for energies ($1GeV-300\,GeV)$. However, the LAT is operating
almost without any background, with the result of it being more sensitive
than Milagro. %

\section{\label{sec:GRBSensi_GRBSensi}Sensitivity to GRBs}

In this section, Milagro's sensitivity will be calculated for GRB
signals of different durations and from different redshifts. Consider
a GRB that emitted a signal with energy that follows a power-law
energy distribution with spectral index $\alpha$. The number of photons
per unit area per unit energy created by the GRB at the earth is described
by the specific flux: \begin{equation}
F_{\nu}(I_{0},\alpha,E)\equiv\frac{dN(I_{0},\alpha)}{dE}=I_{0}\,(E/E_{0})^{\alpha},\label{eq:SENSI_SpecificIntensity}\end{equation}
where $I_{0}$ provides the normalization of the distribution and
is approximately equal to the number of particles with energy $E_{0}$
per unit of area. If the signal is absorbed during its passage through
the extragalactic space, as in the case of GRBs, then the specific
flux reaching the earth from a redshift $z$ will be modified to become
\begin{equation}
F_{\nu}\left(I_{0},\alpha,E,z\right)\equiv I_{0}(E/E_{0})^{\alpha}e^{-\tau_{EBL}(E,z)},\label{eq:SENSIAbsorbedSpecificIntensity}\end{equation}
where $\tau_{EBL}(E,z)$ is the optical depth because of absorption
from interactions with the EBL. 

If the energy emission is isotropic, then for a source at redshift
$z$, the fluence (energy per unit of area) reaching the earth is
\begin{equation}
S(E_{iso},z)=\frac{1+z}{4\pi D_{l}^{2}(z)}E_{iso}(E_{min},E_{max}),\label{eq:SENSI_Fluence}\end{equation}
where $E_{iso}(E_{min},E_{max})$ is the total energy emitted isotropically
in the $[E_{min},E_{max}]$ energy range, and $D_{l}^{2}(z)$ is the
luminosity distance. Because the isotropic energy is for the distant
frame of the GRB (not redshifted), and the fluence is defined for
the frame of the observer, cosmological relativistic corrections have
to be applied in order to relate these two quantities. An amount of
energy $E_{iso}$ emitted from a redshift $z$ is observed at the
earth at a redshifted, lower by a factor of $(1+z)$, energy. However,
equation \ref{eq:SENSI_Fluence} has the factor $1+z$ in the numerator
, instead of the denominator, as would be expected based on the above
consideration. The reason for this is that the luminosity distance
is already defined in a way that takes into account all the necessary
relativistic corrections. Because the luminosity distance is in the
second power in the denominator of equation \ref{eq:SENSI_Fluence}, we
had to multiply $1/D_{l}^{2}(z)$ \ref{eq:SENSI_Fluence} by $(1+z)$
so that the relativistic redshift was applied only once.

\begin{figure}[htb]
 \begin{centering}
 \includegraphics[width=1\columnwidth]{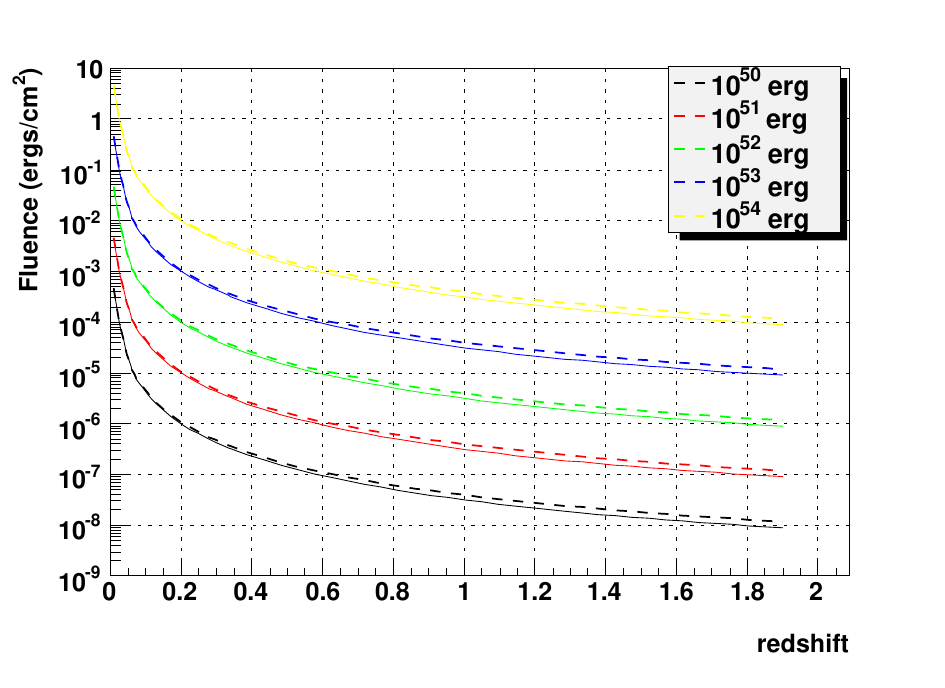}
 \par\end{centering}
 
 \caption{\label{fig:GRBSensi_Fluence_Eiso}Plot showing the effect of absorption by the EBL
 Energy fluence reaching the earth
 from GRBs emitting different isotropic energies and of different
 redshifts. The solid curves include absorption by the EBL, while the
 dashed ones do not. }
  \end{figure}

In a flat $\Lambda CDM$ model, the luminosity distance $D_{l}(z)$
is defined as: \begin{equation}
D_{l}(z)=\frac{c}{H_{0}}\int_{0}^{z}\frac{dz'}{\sqrt{\Omega_{M}(1+z^{'})^{3}+\Omega_{\Lambda}}}\end{equation}
 \begin{equation}
\frac{c}{H_{0}}=9.2516\times10^{27}h^{-1}\,cm,\end{equation}
where $h=H_{0}/100$. 
The energy fluence is also equal to:
\begin{alignat}{1}
S(E_{min},E_{max}) & \equiv\int_{E_{min}/(1+z)}^{E_{max}/(1+z)}E\,\frac{dN}{dE}dE\overset{eq.\ref{eq:SENSIAbsorbedSpecificIntensity}}{\Longleftrightarrow}\\
 & =I_{0}(z,E_{iso})\,\int_{E_{min}/(1+z)}^{E_{max}/(1+z)}E\,(E/E_{0})^{\alpha}\,e^{-\tau(E,z)}dE.\label{eq:SENSI_FluenceAtEarth}\end{alignat}
Figure \ref{fig:GRBSensi_Fluence_Eiso} shows the fluence at the earth
from GRBs at different redshifts. In this figure, two sets of curves
are shown: the curves with the solid lines show the fluence that reaches
the earth if EBL absorption is included, and the dashed lines show
the fluence that would have arrived if no absorption by the EBL existed.
For this plot and for all plots in this chapter, $\Omega_{M}=0.3$,
$\Omega_{\Lambda}=0.7$, $h=71$, and $a=-2.2$, Kneiske's ``best-fit06''
EBL model was used, $E_{min}=30\,GeV$, and $E_{max}=10\,TeV$
(energies as seen from the burst, non-redshifted). %

\begin{figure}[htb]
\begin{centering}
\includegraphics[width=1\columnwidth]{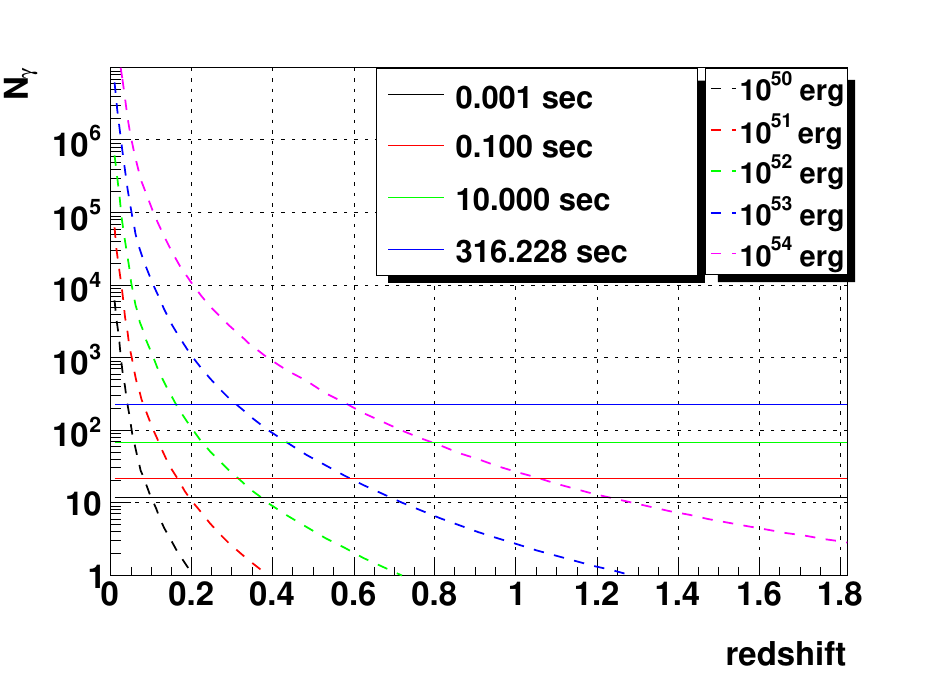}
\par\end{centering}

\caption{\label{fig:SENSI_Ng_Z.eps}Number of detected photons versus the redshift
of the GRB and the total isotropic energy emitted (curves), and minimum
number of events needed to claim a detection 99\% of the time
(horizontal lines). }

\end{figure}

To calculate the number of photons arriving at the earth from a GRB,
the spectral normalization $I_{0}(z,E_{iso})$ of the unabsorbed spectrum
(eq. \ref{eq:SENSI_SpecificIntensity}) $\tau(E,z)=0$ has to be derived
first. Combining equations \ref{eq:SENSI_Fluence} and \ref{eq:SENSI_SpecificIntensity},
we have:\begin{equation}
I_{o}(z,E_{iso})=\frac{1+z}{4\pi D_{l}^{2}}\frac{E_{iso}}{\int_{E_{min}/(1+z)}^{E_{max}/(1+z)}E\,(E/E_{0})^{\alpha}dE}.\label{eq:SENSI_SpectrumNormalizationNoABS}\end{equation}
Now that $I_{0}(z,E_{iso})$ is available, the specific flux $F_{\nu}(I_{0},\alpha,\tau_{EBL},E)$
(eq.\ref{eq:SENSIAbsorbedSpecificIntensity}) can be calculated. This
particle flux at the top of the atmosphere can be converted into a number
of Milagro signal events using the effective area. The total number
of events detected by Milagro is
\begin{alignat}{1}
N_{\gamma}(z,\theta,\alpha) & =\int_{E_{min}/(1+z)}^{E_{max}/(1+z)}A_{eff}(E,\theta)\,F_{\nu}dE\nonumber \\
 & =I_{0}(z,E_{iso})\times\int_{E_{min}/(1+z)}^{E_{max}/(1+z)}A_{eff}(E,\theta)\,(E/E_{0})^{\alpha}\,e^{-\tau_{EBL}(E,z)}dE.\label{eq:SENSI_NG}\end{alignat}
The curves of figure \ref{fig:SENSI_Ng_Z.eps} show $N_{\gamma}$
versus the redshift of a GRB at $\sim10^{o}$ zenith angle for different
values of the total isotropic energy emitted. The horizontal lines
show the minimum number of events Milagro needs to detect in order
to be able to claim a $5\sigma$ post-trials detection 99\% of the
time%
\footnote{As described in Chapter \ref{chap:binsize}, because of statistical fluctuations
involved in the detection and reconstruction of a gamma-ray signal
by Milagro, the same initial signal has a probability of being detected.%
}. Because the background level is proportional to the duration searched,
searches of longer-duration emission require more detected signal
events to make a detection. As the distance of the source increases,
the signal is attenuated by absorption from the EBL and diluted by
the geometrical $\propto1/D_{l}^{2}$ decrease, so the number of detected
events decreases. %

\begin{figure}[htb]
\begin{center}
\includegraphics[width=1\columnwidth]{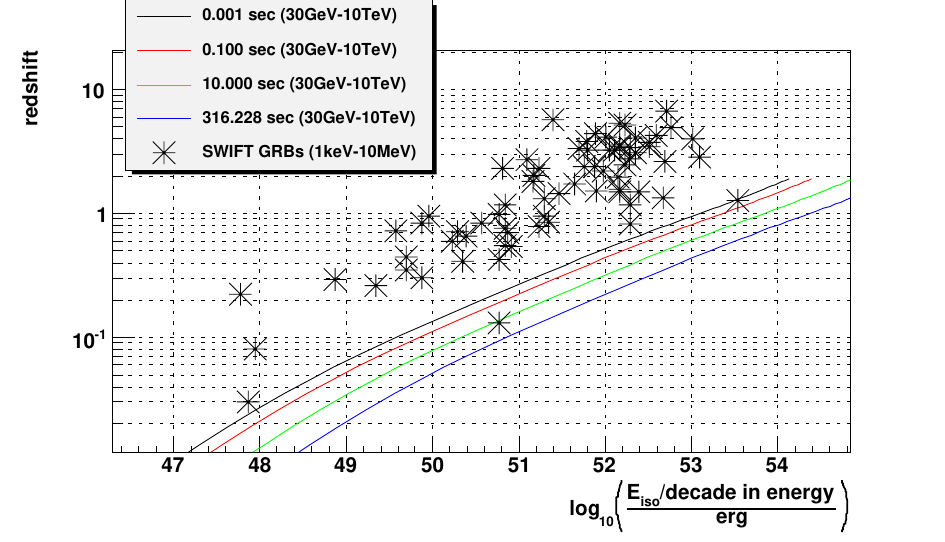}

\caption{\label{fig:SENSI_Eiso_z}Maximum detectable redshift versus the isotropic
energy released per decade in energy, and the duration of the emission
(curves). For comparison, the isotropic energies per decade in energy
and redshifts of detected Swift GRBs are also shown with crosses \cite{GRB_Floroi2007}. }
\end{center}
\end{figure}

By comparing the minimum signal needed for a detection (horizontal
lines) with the amount of signal created by a GRB (curves), we can
find the maximum redshift, at which GRBs can be detected. Figure \ref{fig:SENSI_Eiso_z}
shows the maximum detectable redshift (99\% detection probability)
for a GRB at zenith angle $\simeq10^{o}$ versus the isotropic energy
emitted per decade of energy and the duration of the emission (curves).
For comparison, the data from Swift-detected GRBs are also shown.
The Swift data correspond to an energy range ($1keV-10MeV$) that
is wider than the energy range of the VHE emission used in this analysis
($30\,GeV-10\,TeV$). Both the Swift data and the Milagro curves were
plotted versus the average energy output per decade of energy, instead
of the energy output integrated over the detector-specific energy
ranges of different widths. Based on that figure, if the GRB
energy output per decade of energy in the $30\,GeV-10\,TeV$ energy range
is comparable to that measured by Swift in the $1\,keV-10\,MeV$ energy
range ($E_{iso,Swift}/$decade of energy$\sim10^{50}-10^{53}\,erg$), then
Milagro is expected to be able to detect GRBs up to a redshift
of $z\sim0.4$. We can see from the Swift data on the figure that
GRBs with a detectable combination of energy output and redshift are
rare, implying that Milagro is expected to detect considerably fewer
GRBs than Swift. 

\begin{figure}[htb]
\begin{centering}
\includegraphics[width=1\columnwidth]{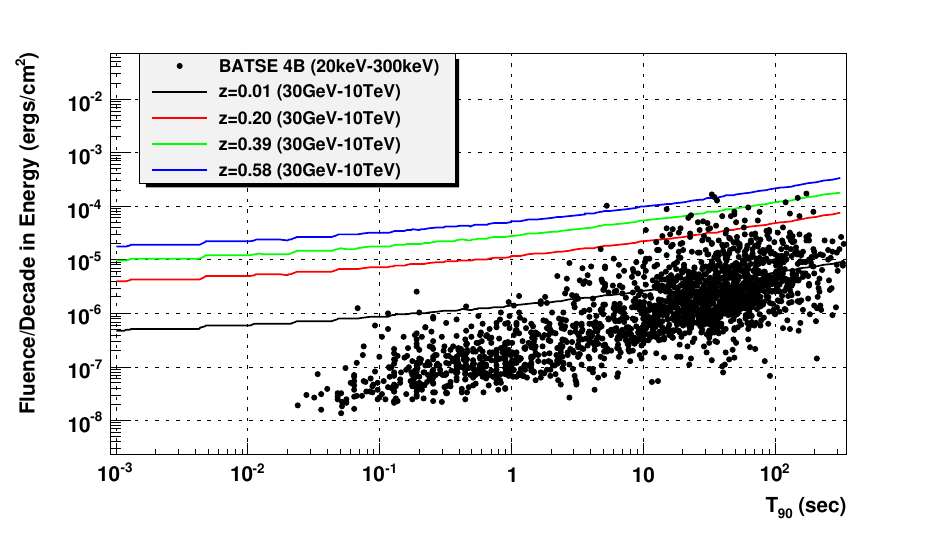}
\par\end{centering}
\caption{\label{fig:GRBSensi_MinS}Minimum detectable fluence versus the duration
and redshift (curves). Detected GRBs from BATSE's 4B catalog are also
plotted (dots) for comparison. The fluence curves correspond to no
absorption by the EBL.}

\end{figure}

Lastly, figure \ref{fig:GRBSensi_MinS} shows the minimum detectable
fluence per decade of energy versus the duration and the redshift
(curves). For comparison, the fluences of the GRBs included in
BATSE's 4B catalog\footnote{http://www.batse.msfc.nasa.gov/batse/grb/catalog/4b/} are also shown. Similarly to the previous
figure, the fluences are divided by the number of decades in the corresponding
energy ranges. In this case, the BATSE data plotted correspond to
the measurements of the first three BATSE's channels ($20\,keV-300\,keV$).
The fluence plotted is the one that would have reached the earth if
there were not any absorption by the EBL. While the calculations include
the effects of EBL absorption, the plotted fluence curves correspond
to the case of no EBL absorption. This way, a direct comparison between
the fluences of the BATSE GRBs and the Milagro fluence sensitivity
can be performed.%

\section{\label{sub:GRBSensi_PBHEvap}Sensitivity to PBH Evaporation}

In this section, Milagro's sensitivity to evaporating PBHs will be
calculated. The calculations use the favorable for Milagro model of
MacGibbon and Webber (sec. \vref{sec:PBH_Emission}), which does not include a chromosphere,
therefore predicting an unrestricted emission at $E>TeV$ energies.
According to that model, the temperature of a black hole $T$ is related
to its remaining lifetime $\tau$ as \cite{PBH_Belyanin_Kocharovsky_1996}
\begin{equation}
T=(4.7\times10^{11}/\tau)^{1/3}.\label{eq:GRBSensi_PBH_Temp}\end{equation}
The time-integrated specific flux can be parametrized as:\begin{equation}
F_{\nu}\equiv\frac{dN}{dE}\simeq3\times10^{20}\times\begin{cases}
\left(\frac{E_{0}}{T}\right)^{3}\left(\frac{T}{E}\right)^{3/2} & ,\,E<T\\
\left(\frac{E_{0}}{E}\right)^{3} & ,\,E\geq{}T\end{cases},\label{eq:GRBSensi_PBH_SpecificFlux}\end{equation}
where $E_{0}=10^{5}GeV$ and all the energies are measured in $GeV$. 

\begin{figure}[htb]
\begin{centering}
\includegraphics[width=1\columnwidth]{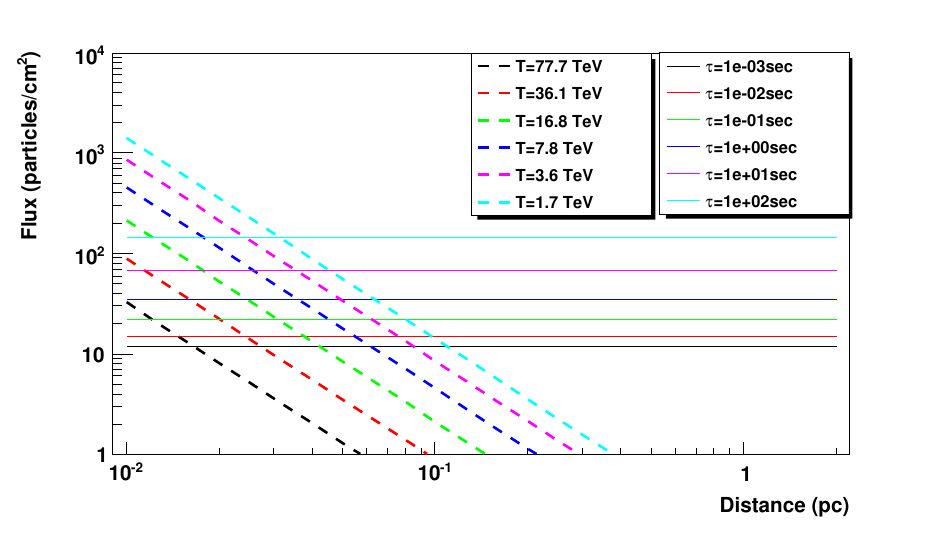}
\par\end{centering}

\caption{\label{fig:GRBSensi_PBHN}Number of detected photons versus the distance
and the temperature (duration) of an evaporating PBH (dashed lines),
and minimum number of detected photons needed to claim a detection
with 99\% of the time (horizontal solid lines). }

\end{figure}

Following a procedure similar to that of the previous section, the particle
flux generated by evaporating PBHs of different temperatures and from
different distances was calculated. For each duration, the temperature
of the PBH was calculated using equation \ref{eq:GRBSensi_PBH_Temp}, and
then the particle flux was calculated using equations \ref{eq:GRBSensi_PBH_SpecificFlux}
and \ref{eq:SENSI_NG}. Because the PBHs under consideration are galactic,
no EBL absorption effects or relativistic redshift was applied. In
these calculations, the PBH emission in the $30\,GeV-100\,TeV$ energy
range was considered. The results are shown in figure \ref{fig:GRBSensi_PBHN}.
The horizontal solid lines are the minimum number of detected photons
needed for a $5\sigma$ post-trials detection 99\% of the time,
and the dashed lines are the number of photons created by the evaporation
of PBHs of different temperatures (lifetimes) from different distances. 

\begin{figure}[htb]
\begin{centering}
\includegraphics[width=1\columnwidth]{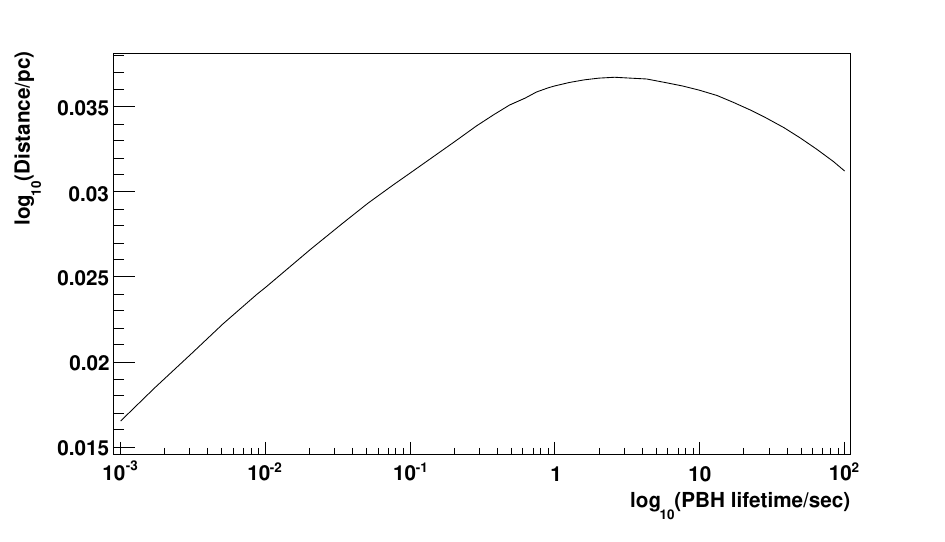}
\par\end{centering}
\caption{\label{fig:GRBSensi_PBH_MaxR}Maximum distance that an evaporating PBH
can be detected versus its remaining lifetime. The detection horizon
for PBHs by Milagro is $\sim0.037\,pc$. }
\end{figure}

The intersections of the two sets of curves correspond to the maximum
distances from which a PBH of some temperature can be detected. Figure \ref{fig:GRBSensi_PBH_MaxR}
shows these maximum distances. According to the figure, Milagro can
detect PBHs up to $\sim0.037\,pc$ \footnote{The CYGNUS EAS array, which had a 
similar field of view and duty cycle as Milagro, but a somewhat smaller effective area
, had a detection horizon of $\sim0.02\,pc$ \cite{PBH_CYGNUS}.}. For reference, $0.037\,pc$ are equal
to $\sim7200$ times the distance of the earth to the sun or $\sim3\%$
of the distance to our closest star Centauri Proxima. 
Milagro's $0.037\,pc$ horizon corresponds to an observable volume of $3.1\times10^{-5}\,pc^3$, using 
the $45^o$ field of view of the search. 

To calculate the maximum number of PBH explosions in Milagro's observable volume, an upper limit
on the rate of PBH explosions set by previous experiments has to be used. 
One way to set such a limit is to consider the diffuse particle backgrounds
created by the integrated emission from PBHs in the lifetime of the universe. 
Wright, based on the density of the gamma-ray halo of our galaxy, set an upper limit
on the explosion rate of $0.3\,pc^{-3}\,yr^{-1}$ \cite{PBH_Wright_1996}. Maki
\textit{et al.}, based on the local interstellar flux of cosmic-ray antiprotons, set
an upper limit of $0.02\,pc^{-3}\,yr^{-1}$ \cite{PBH_Maki_et_al1996}.

Another way to set upper limits on the rate of PBH explosions is by
trying to directly detect such an explosion. The sensitivity of such a technique
strongly depends on the model used for the running constant $a(M)$ (defined
in equation \vref{eq:PBH_MssLossRate}). The running constant is a measure of the number of 
particle species available for emission at a specific instant of the PBH's life.
The mass-loss rate of the PBH is proportional to the running constant.
In the case that the number of particles species
with masses over the QCD scale ($E\sim100\,MeV$) is very large, then the running
constant will considerably increase when the temperature of the PBH crosses that
scale, and the evaporation process will be accelerated to explosive degrees. 
Such an increase in the number of particle species can occur if supersymmetry
is the theory that describes elementary particles, or in a Hagedorn-type picture 
\cite{PBH_Hagedorn_1968}, where the number of hadronic resonances exponentially increases with energy. 
In such a case, the last stages of a PBH explosion would be more luminous and more easily
detectable than in the case of the Standard Model of elementary particles, where the number of
particle species does not exhibit such an increase.

The upper limits on the rate of PBH explosions set by direct searches of 
exploding PBHs can either be considerably weaker, or as strong as, the limits set by 
considerations of the diffuse backgrounds, depending on the the behaviour of $a(M)$
at energies over the QCD scale. In the case of the Standard Model,
observations with the CYGNUS experiment only placed an upper limit of $5\times10^8\,pc^{-3}\,yr^{-1}$
\cite{PBH_CYGNUS}. Porter and Weekes \cite{PBH_Porter_Weekes_1978} using IACTs,
set an upper limit of $7\times10^5\,pc^{-3}\,yr^{-1}$ for the Standard Model
and $0.04\,pc^{-3}\,yr^{-1}$ for the Hagedorn model. 
The EGRET experiment published upper limits on the explosion-rate  
density of PBHs versus the total energy emitted
by the explosion \cite{PBH_EGRET}. Page and Hawking \cite{PBH_Page_HAWKING_1976}
calculated the total emitted energy in the last stages of the PBH's lifetime 
for two extreme cases: a standard Elementary Particle
model where they found an energy $\sim10^{30}\,erg$, and a Composite Particle model
following Hagedorn, where they found $\sim10^{35}\,erg$.
EGRET's upper limits for the two cases are $5\times10^4\,pc^{-3}\,yr^{-1}$ and $\sim0.01\,pc^{-3}\,yr^{-1}$, respectively.

Multiplying the most constraining upper limit mentioned above ($0.01\,pc^{-3}\,yr^{-1}$ by EGRET for the Hagedorn picture),
times fives years of Milagro data searched,
times Milagro's observable volume calculated above, gives an upper limit of $1.6\times10^{-6}$ PBH explosions for the
duration of this search in Milagro's detectable volume. For comparison, if we use EGRET's 
upper limit for the Standard Model case ($5\times10^4\,pc^{-3}\,yr^{-1}$), we find 
an upper limit of 7.75 PBH explosions in Milagro's observable volume.

PBHs of shorter
lifetimes ($\tau<1\,s$) emit a signal of high luminosity and energy.
However the integrated number of events they can create is small so
Milagro's sensitivity to detecting them is not optimal. On the other
hand, PBHs of longer lifetimes ($\tau>10\,s$) will start by emitting
a lower luminosity and energy signal, which will eventually become
like the one of a PBH of $\tau<1\,s$. However, despite the large
integrated amount of events from such a long-lifetime PBH, the background
contamination will be large enough (because of the large duration)
that sensitivity in detecting it still will not be optimal.
We can see that there is a trade-off between the amount of accumulated
signal and background during the lifetime of the PBH. As can be seen
from figure \ref{fig:GRBSensi_PBH_MaxR}, Milagro's sensitivity in
detecting PBHs is optimal when searching for events of duration $\tau\sim2-5\,s$.

%

%

    \clearpage \chapter{\label{chap:RESULTS}Results of the Search}

\section{Introduction}

In this chapter, the results of the search will be presented.
In section \ref{sec:RESULTS_Sample}, the data set analyzed is described.
Then, in section \ref{sec:RESULTS_Trials}, the calculation of the effective
number of trials is made. Finally, in section \ref{sec:RESULTS_Probs},
the results of the search are presented, and the most significant events are evaluated.

\section{\label{sec:RESULTS_Sample}Data Sample}

Milagro's low-energy sensitivity has not been the same for all of ($\sim7$) the
years it has been operating. As mentioned in Chapter
\ref{chap:MILAGRO}, the first version of the Milagro detector did not include
the outrigger array, which resulted in a worse angular resolution.
Furthermore, during the first years of Milagro's operation, a triggering
system was used (multiplicity trigger) that did not accept a large part
of the lower-energy events, which are important for this study. For
these reasons, the data from those early times were not included in
this analysis. Specifically, only the last five years of Milagro data
have been analyzed: from 03/01/2003 (Modified Julian Date 52699) to
03/01/2008 (Modified Julian Date 54526). The starting date corresponds
approximately to the time that the new VME trigger was installed (01/25/2002)
and is after the outriggers were used in the online reconstruction. The ending
date corresponds approximately to the time that the outrigger array started
being dismantled (March 15th, 2008) as part of Milagro's shutdown.

For some of the days in the analyzed period the detector was not functioning
because of scheduled repair operations, extended power outages, or
instrumental problems. For some other days, the detector was on, but
the data were so problematic that the whole Julian day had to be rejected.
Reasons for rejecting a whole day included a large number of timing errors (events with
wrong times, swapped times) and time gaps, strongly fluctuating trigger
rates, abnormally low event rates ($<700Hz)$, and calibration runs
that were not successfully tagged as such. The data of the bad 
days were either rejected directly by the code, or were detected because they
corresponded to unphysical results. The Milagro logbook contained a satisfactory
explanation for all the dubious data. The following 57 (out of 1673) days
were rejected\footnote{the date is given in Modified Julian Date - 50000 format.}:
\begin{itemize}
\item \textbf{Repairs (16 days)}: 2890--2897, 3626--3633
\item \textbf{Power-Supply Problems (7 days)}: 2726--2729, 2861, 3239 (+
clock errors), 4495--4496
\item \textbf{Time Gaps (7 days)}: 2740--2742, 2705--2708
\item \textbf{Power Outage (7 days)}: 3060, 3588, 3651, 4099--4100, 4472,
3099 (+ other problems)
\item \textbf{Other Problems (14 days)}: 2822 (calibrations), 2827 (overheating
problems), 3101 (many clock errors), 3103 (rate fluctuations, lightnings)
3245, 3378--3380 (disk problems), 3555 (bad data), 3463--3464 (low
rates), 3961 (replacing VME trigger card), 4445-4446 (bad data).
\item \textbf{DAQ Problems (7 days)}: 3174, 3195--3197, 3215, 3229 (+ clock
errors), 3291 (too many dropped buffers, DAQ crashing).
\end{itemize}

\begin{figure}[htb]
\begin{centering}
\includegraphics[width=0.7\columnwidth]{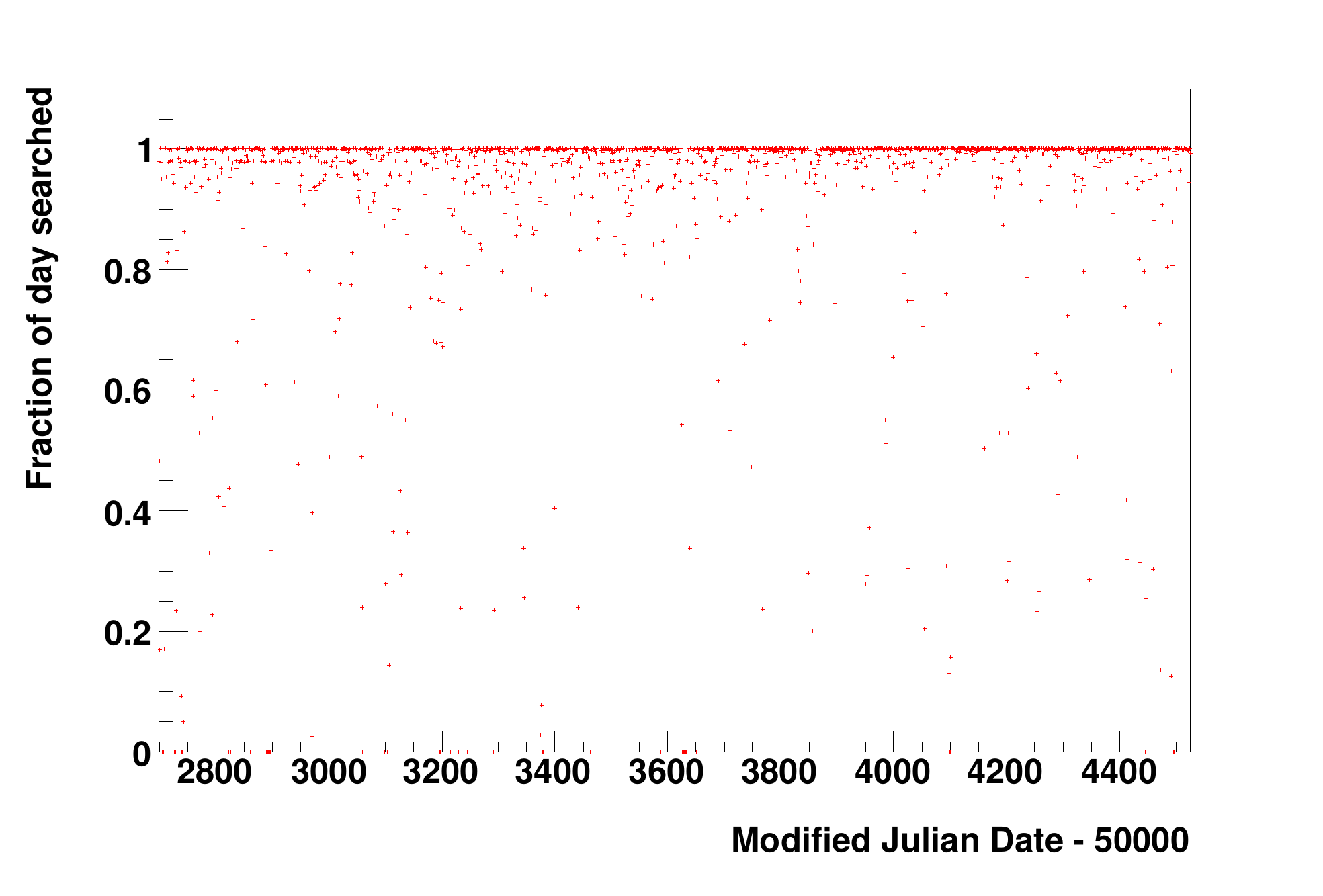}
\par\end{centering}

\caption{\label{fig:RESULTS_Fraction}Fraction of each day analyzed by this
study. Total, 92\% of the data was analyzed.}

\end{figure}

After finishing with the analysis of the data taken in each day, the
code saved the total duration of the good data segments of that day
(fig. \ref{fig:RESULTS_Fraction}). According to that information,
this analysis searched 1673 days ($4.58yrs$) worth of data, which
corresponds to $\sim93\%$ of the duration that Milagro was operating
during the five years analyzed. The Milagro detector was non-operating
because of scheduled repairs or power outages for $\sim1.3\%$ of those five years. %

\section{\label{sec:RESULTS_Trials}Effective Number of Trials}

In this section the total effective number of trials of this search
will be calculated. This number is the equal to the effective number of
trials in space, starting time, and duration. The following calculations follow the methodology described
in Chapter \ref{Chapter:Probabilities}.

Initially, the effective number of trials in space and starting time will
be calculated based on the distributions of the minimum probabilities
found in groups of successive searches. For the searches of emission
duration under $56\,s$, each group consisted of the searches in a
time width equal to twice the duration searched (20 skymaps per group, with 
each skymap offset in time by 10\% the duration),
while for the searches of emission duration over $56\,s$, each group
consisted of all the searches in starting time and space made in a time width
equal to the duration searched (10 skymaps per group).

The first step in the calculation is to choose the integration threshold
$P_{0}$ of equation \vref{eq:TheSearch_EffectiveTrials}. $P_{0}$ has to be
large enough so that the relative statistical error of the integral
from $\-\infty$ to $P_{0}$ of the distribution is small: 
\begin{equation}
\sqrt{\int_{-\infty}^{P_{0}}(dN/dP)dP}/\int_{-\infty}^{P_{0}}(dN/dP)dP\ll1.
\end{equation}
$P_{0}$ also has to be small enough so that it is far from the peak
of the distribution, otherwise the trials factors will be erroneous.
$P_{0}$ was selected so that the integral has a value of at least
$10^{4}$. This way the relative statistical error on the value of
the integral is just $1\%$, and the relative statistical error in
the number of trials per group ($T$) is \begin{eqnarray*}
\frac{\sigma_{T}}{T} & = & \sigma_{P_{0}^{'}}\times\frac{dT(P_{0}^{'})}{dP}\times\frac{1}{T}\\
 & = & \sigma_{P_{0}^{'}}\times\frac{1}{(1-P_{0}^{'})\,log(1-P)}\times\frac{1}{T}\\
 & \simeq & 10^{-2},\end{eqnarray*}
where $P_{0}^{'}=\int_{-\infty}^{P_{0}}(dN/dP)dP/\int_{-\infty}^{1}(dN/dP)dP$;
$\sigma_{P_{0}^{'}}=\sqrt{\int_{-\infty}^{P_{0}}(dN/dP)dP}/\int_{-\infty}^{1}(dN/dP)dP$
is the variance of $P_{0}^{'}$, $T=\frac{log(1-P_{0}^{'})}{log(1-P_{0})}$;
is the effective number of trials; and $\sigma_{T}$ is
the variance of $T$. 

\begin{figure}[ht!]
\noindent \begin{centering}
\includegraphics[width=0.8\columnwidth]{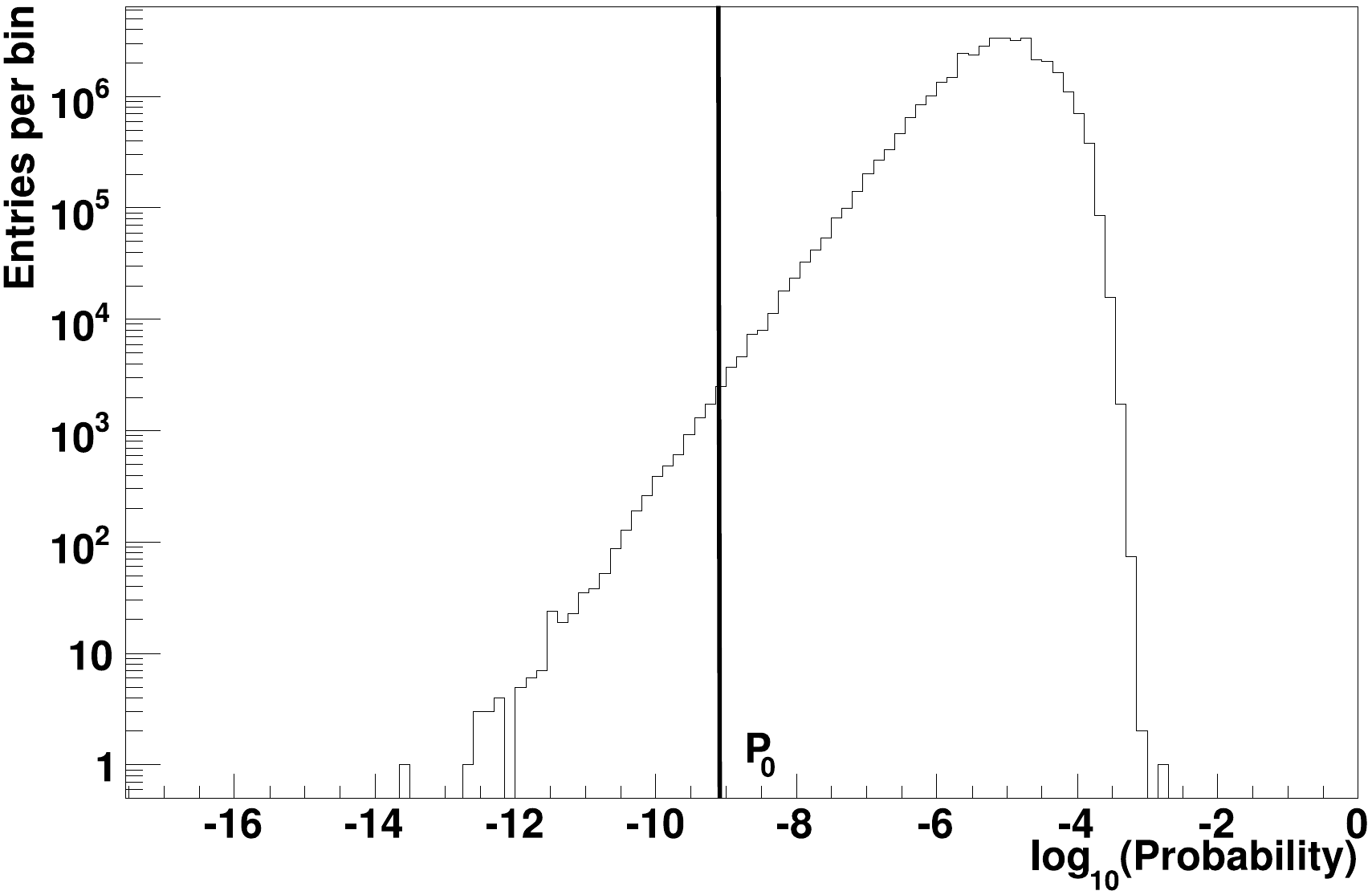}
\par\end{centering}

\caption{\label{fig:RESULTS_MinProb1sec}Distribution of the minimum probabilities
found in groups of successive trials for the $1\,s$ duration search.
Each entry corresponds to the minimum probability of all the trials
in space and starting time for $2\,s$ blocks of data. Plots like this were
created for each duration and used for calculating the effective number
of trials. The integration threshold $P_0$ used for the calculation is shown. 
The number of events under the curve are $\int_{-\infty}^{1}(dN/dP)dP=36,134,980$ and 
the number of events under the curve and with probabilities smaller than $P_0$ (left of $P_0$)
are $\int_{-\infty}^{P_{0}}(dN/dP)dP=12,588$.}

\end{figure}

For the case of the $1\,s$ duration search, integrating up to $P_{0}=1.412\times10^{-9}$
corresponds to an integral $\int_{-\infty}^{P_{0}}(dN/dP)dP=12,588$.
The number of entries in the minimum-probability distribution (fig.
\ref{fig:RESULTS_MinProb1sec}) are $\int_{-\infty}^{1}(dN/dP)dP=36,134,980$.
This means that in a typical group of trials of the $1\,s$ duration
search, a probability less than $P_{0}$ is expected to be found $P_{0}^{'}=12588/36134980=3.4836\times10^{-4}$
of the times. Using equation \ref{eq:TheSearch_EffectiveTrials} with $P_{0}=1.412\times10^{-9}$
and $P_{0}^{'}=3.4836\times10^{-4}$, we find that the effective number
trials in such a group of trials is $T\simeq246,676$. For the $1\,s$
duration there were $4,330,620$ trials in each group.
Therefore, each trial corresponded to $246676/4330620\simeq0.057$
effective trials. The error on the effective number of trials per
trial is $5\times10^{-4}$. Multiplying these numbers with the total
number of trials in the search for that duration ($\simeq3.13\times10^{14})$,
the total effective number of trials in space and starting time for that duration
can be found: $1.784\times10^{13}\pm1.5\times10^{11}$. 

\begin{figure}[htb]
\noindent \begin{centering}
\includegraphics[width=0.85\columnwidth]{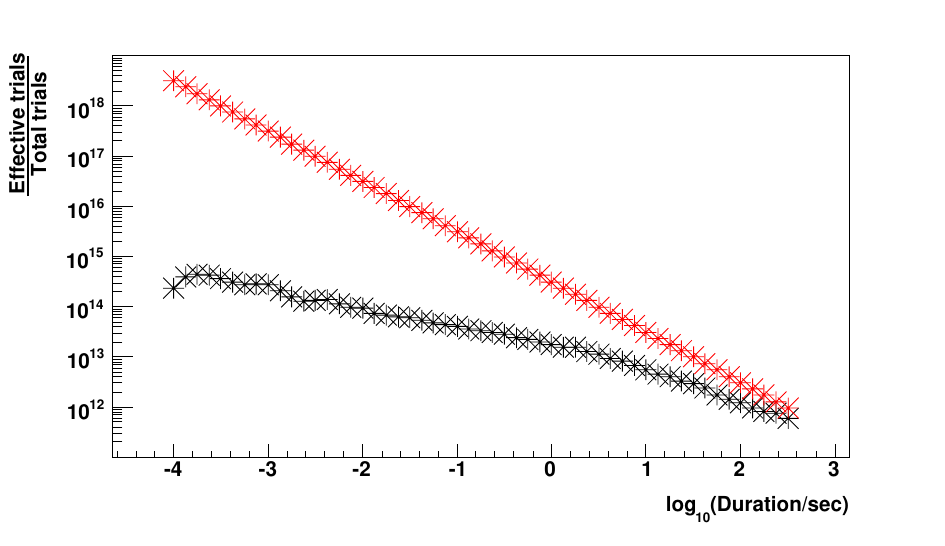}
\par\end{centering}

\caption{\label{fig:RESULTS_Trials1}Numbers of trials in space and starting time for
each emission duration searched. \emph{Black}: effective number
trials, \emph{red}: actual number of trials. The errors on the number of
effective trials are smaller than the size of the markers.}

\end{figure}

\begin{figure}[htb]
\noindent \begin{centering}
\includegraphics[width=0.85\columnwidth]{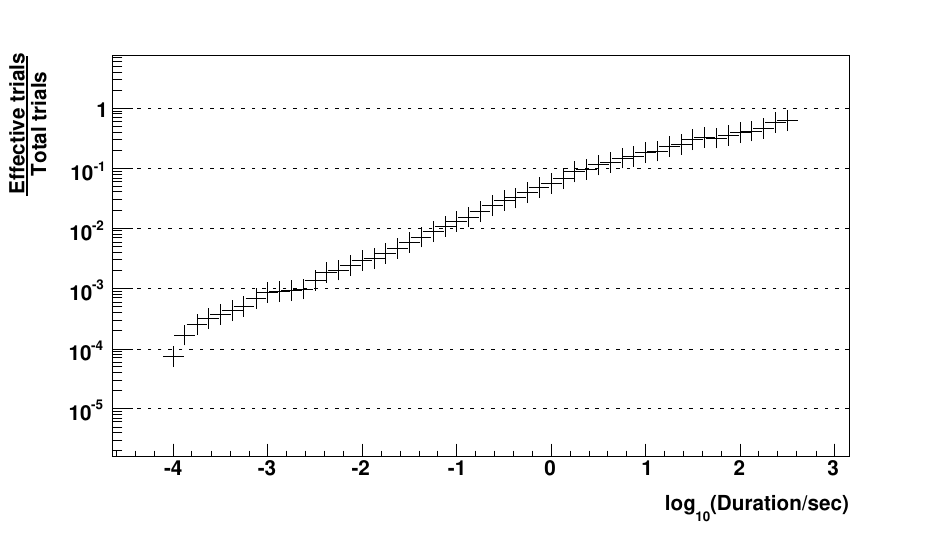}
\par\end{centering}

\caption{\label{fig:RESULTS_TrialsRatio}Ratio of effective number of trials
over total number of trials versus the search duration.}

\end{figure}

Repeating this process for all the durations, the effective number
of trials in space and starting time for all the durations was calculated.
The results are shown in figure \ref{fig:RESULTS_Trials1}. The red
crosses show the total number of trials in space and starting time, and the
black dots show the effective number of trials in space and starting time.
It can be seen that the ratio between the two quantities depends on
the duration and asymptotically approaches unity with increasing duration.
Figure \ref{fig:RESULTS_TrialsRatio} shows the dependence of that
ratio on the search duration. One of the reasons for this dependence
is that, for the short durations, most of the space and time trials
correspond to no events. Such trials have probability of one, and 
cannot cause the rejection of the null hypothesis. Thus, they do 
not contribute to the effective number of trials. Another reason for the decreased
number of effective trials at the short durations, is that for these durations
even at locations in space that a few events might 
occur, most of the searches that include these events end up being identical, since 
they end up including exactly the same events with almost the same background rates. 
These trials essentially count as just one effective trial.


The effective number of trials because of searching in multiple durations
was then calculated. Initially, post-trials probabilities were calculated
using the effective number of trials in space and starting time calculated
above. Then, the events with a post-trials probability less than $0.9$
were counted. 187 such events were found corresponding to 146 independent
improbable fluctuations (some of these fluctuations appeared in multiple adjacent 
durations).
If all 187 events were independent, then the effective number of
events due to having searched in multiple durations would be equal
to the number of durations searched (53). If the trials were maximally
correlated (effective number of trials in duration equal to unity),
then all 187 events should have been generated by the same improbable
fluctuation appearing in all durations searched. In our case, 146
out of the 187 events were independent. Thus, an estimate for the
effective number of trials due to having searched in multiple durations
is $\simeq53\times146/187\simeq41\pm5$. Multiplying this number with the
total effective number of trials in space and starting time (calculated above
and shown in fig. \ref{fig:RESULTS_Trials1}) yields the the total
effective number of trials in starting space, starting time, and duration.

\section{\label{sec:RESULTS_Probs}Results}

\begin{figure}[htb]
\noindent \begin{centering}
\includegraphics[width=0.8\columnwidth]{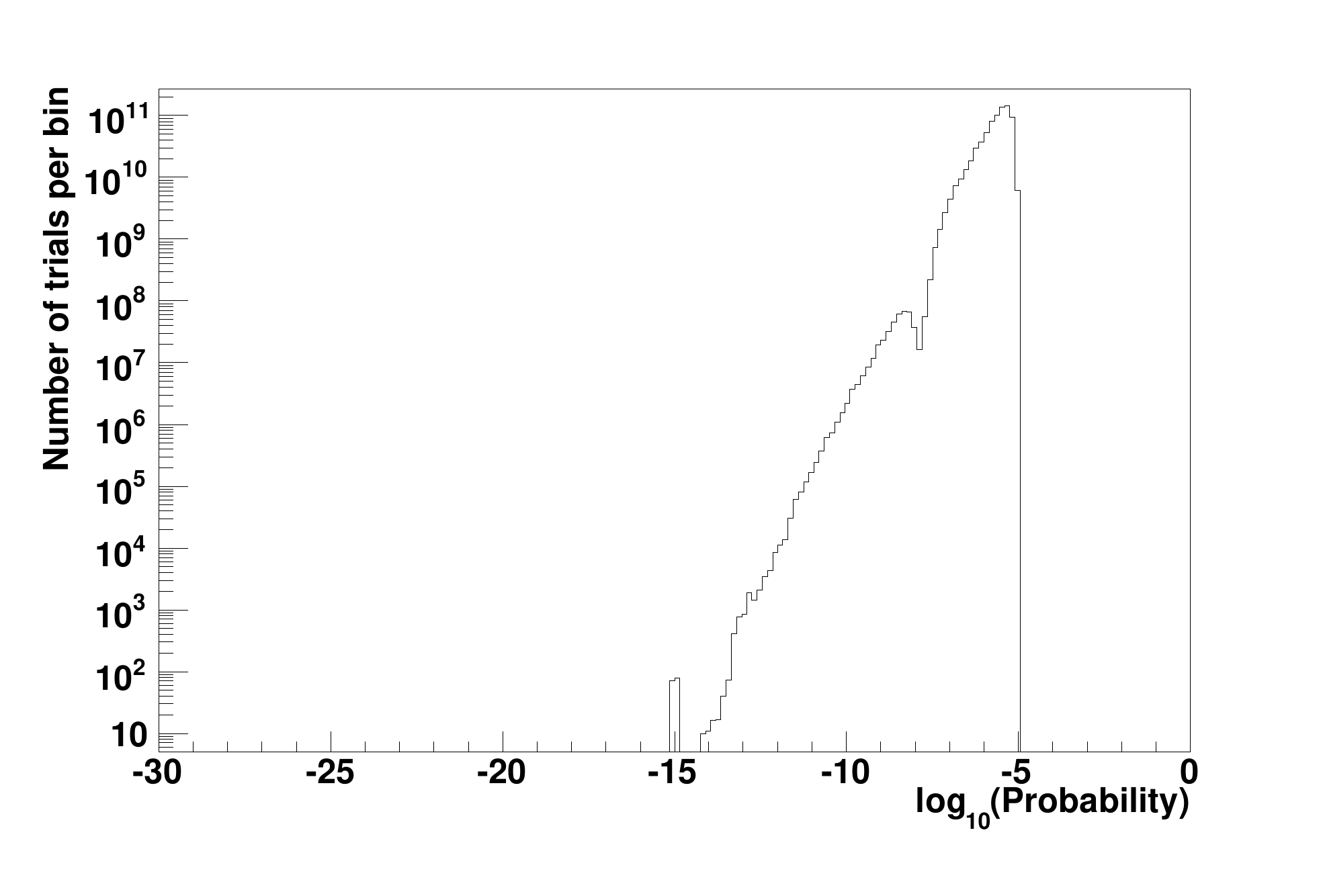}
\par\end{centering}

\noindent \begin{centering}
\includegraphics[width=0.8\columnwidth]{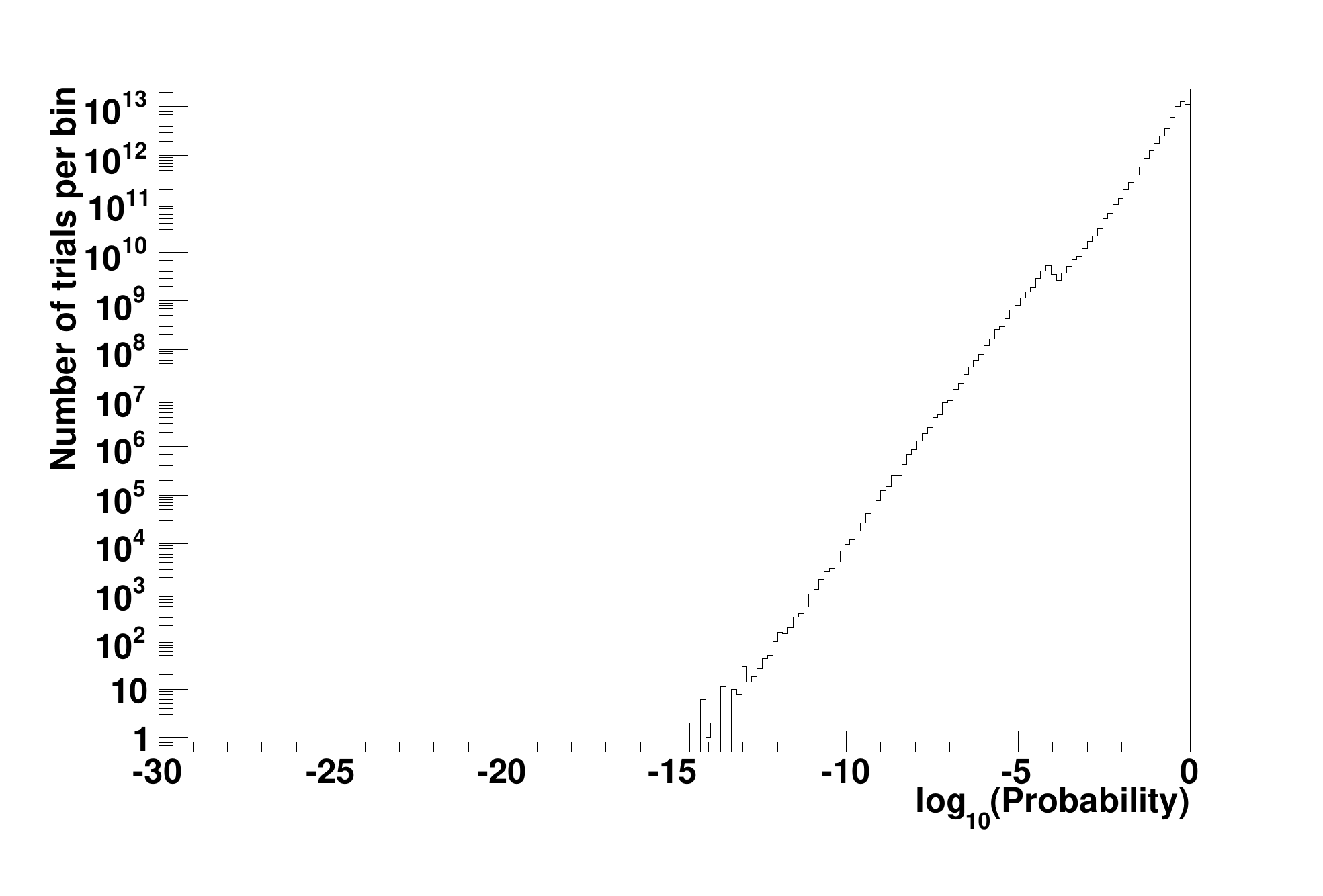}
\par\end{centering}

\caption{\label{fig:RESULTS_SomeProbs}Sample distributions of the probabilities of all the trials
in the search of one duration. \textit{Top}: Distribution for the $0.4\,\mu{}s$ search, 
\textit{bottom}: distribution for the $0.4\,s$ search. }

\end{figure}

This section will present the results of the search. As mentioned in
Chapter \ref{chap:Search}, the results of the search code are in the form
of probability distributions of all the trials in the search of one duration.
Two of the fifty-three produced distributions are shown in figure \ref{fig:RESULTS_SomeProbs}
\footnote{The rest of the distributions are available.}.
Such probability distributions
are composed of a series of peaks, each one corresponding to a different number of 
signal events. Because the background fluctuates, these peaks become wide. In the 
short durations, these peaks are visible. In the top plot of fig. \ref{fig:RESULTS_SomeProbs},
the peaks corresponding to four, three, and two
events (left to right) are visible. The peak corresponding to one event is not shown
because the optimized-for-speed algorithm does not evaluate locations, where only one event
was found. For the longer durations (bottom plot of fig. \ref{fig:RESULTS_SomeProbs}),
the number of individual peaks is so large that
they form a continuum. The feature of the distribution at $P=10^{-4}$ comes from the
speed optimization for durations $>0.2\,s$, in which a coarse search is initially performed
until a location with probability less than $10^{-4}$ is found, and a fine search around that location is then performed
(see section \ref{sec:SEARCH_Skymaps}).

The pre-trials probabilities from such distributions were converted to post-trials probabilities
according to:
\begin{equation}
 P_{post}=1-(1-P_{pre})^N,
\end{equation}
where $P_{post}$ is the pre-trials probability, $P_{pre}$ is the corresponding pre-trials probability, and $N$ is the 
number of effective trials in space and starting time calculated in the previous section. 

\begin{figure}[htb]
\noindent \begin{centering}
\includegraphics[width=1.0\columnwidth]{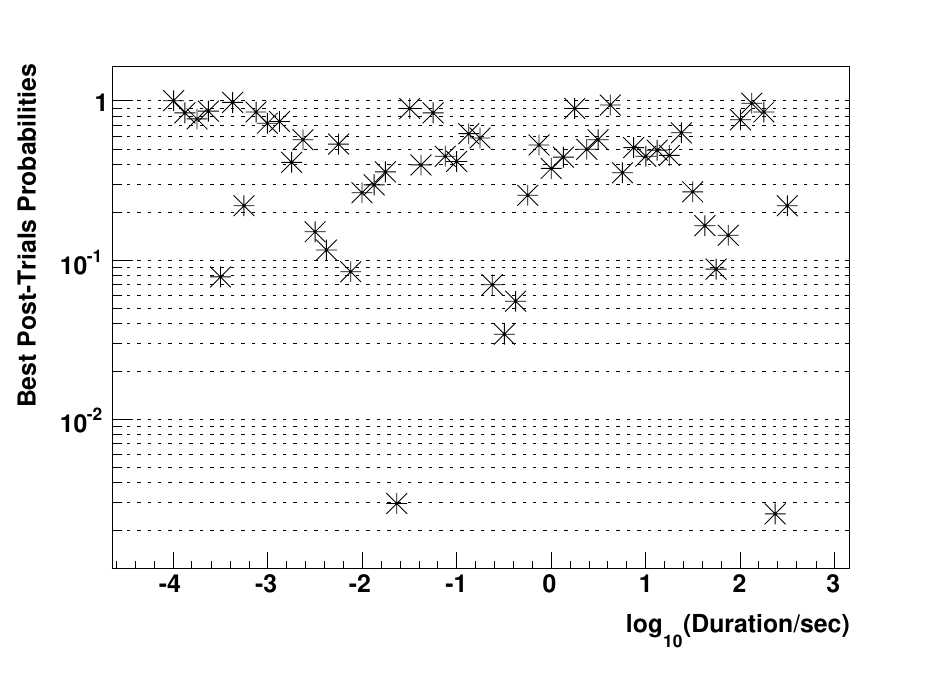}
\par\end{centering}

\caption{\label{fig:RESULTS_BestProbs}Best (minimum) post-trials probabilities found
for each duration searched. Fluctuations creating best probabilities in multiple durations
were allowed to contribute only once to the plot. }
\end{figure}

Figure \ref{fig:RESULTS_BestProbs} shows the best (smallest)
post-trials probabilities found for each duration. For this plot, if the best
probabilities of multiple durations were produced by the same improbable 
fluctuation, then only the most significant of them was used for the figure. The rest 
of them were replaced by the second-best probabilities of their respective durations.
This way, each of the best probabilities in the figure corresponded
to a different fluctuation.

\begin{figure}[htb]
\noindent \begin{centering}
\includegraphics[width=0.8\columnwidth]{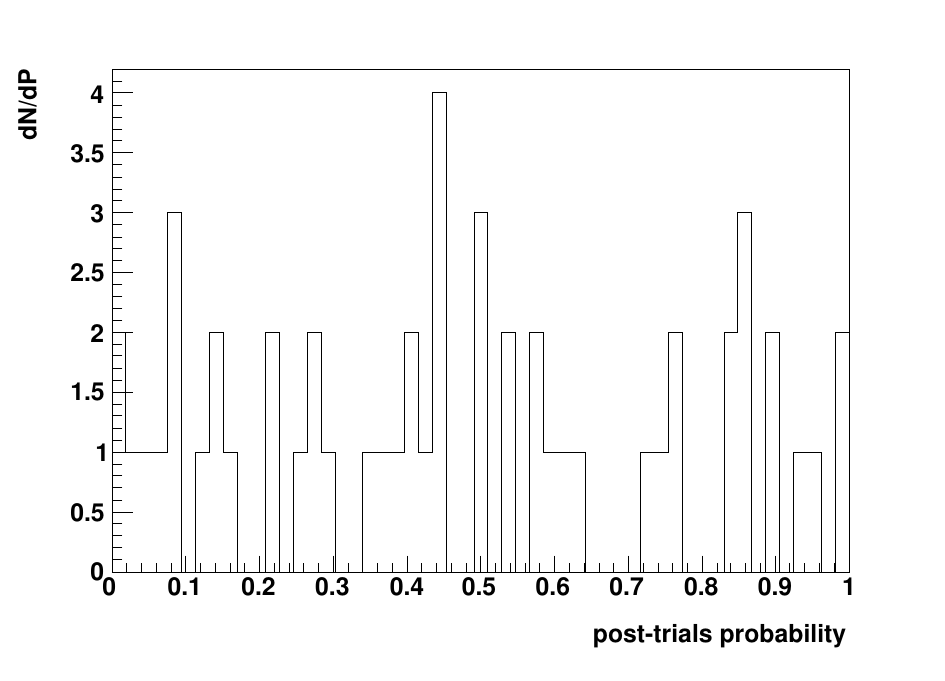}
\par\end{centering}

\caption{\label{fig:RESULTS_probdistro}Distribution of the best post-trials probabilities found
for each duration searched. Fluctuations creating best probabilities in multiple durations
were allowed to contribute only once to the plot. The distribution is consistent with a 
$dN/dP=constant$ distribution supporting the validity of the search's results and of the
subsequent statistical analysis.}
\end{figure}

The best post-trials probabilities of each duration plotted in figure \ref{fig:RESULTS_BestProbs}
essentially answer the following question: what is the probability that there is not
a burst of VHE gamma rays of some specific duration in the data? The search of 
each duration can be treated as an independent statistical test; hence,
the probability distribution should have a constant density $dN/dP=constant$.
Figure \ref{fig:RESULTS_probdistro} shows that probability distribution. A Kolmogorov
test says that probability distribution is consistent with a $dN/dP=constant$ distribution
with a probability $99.8\%$, supporting the validity of the statistical framework in this
study.

\begin{figure}[htb]
\noindent \begin{centering}
\includegraphics[width=0.8\columnwidth]{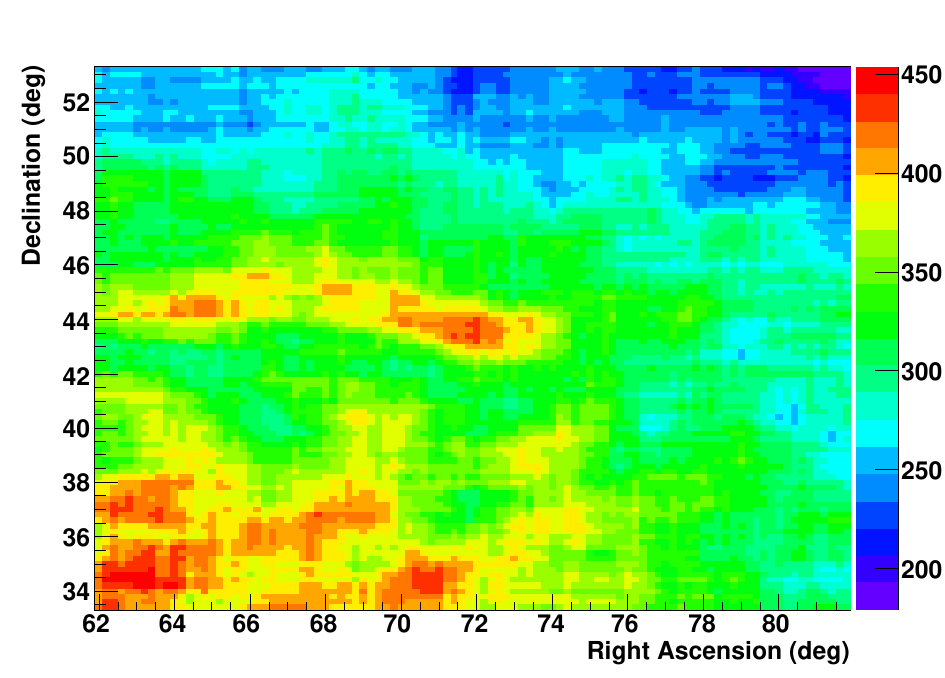}
\includegraphics[width=0.8\columnwidth]{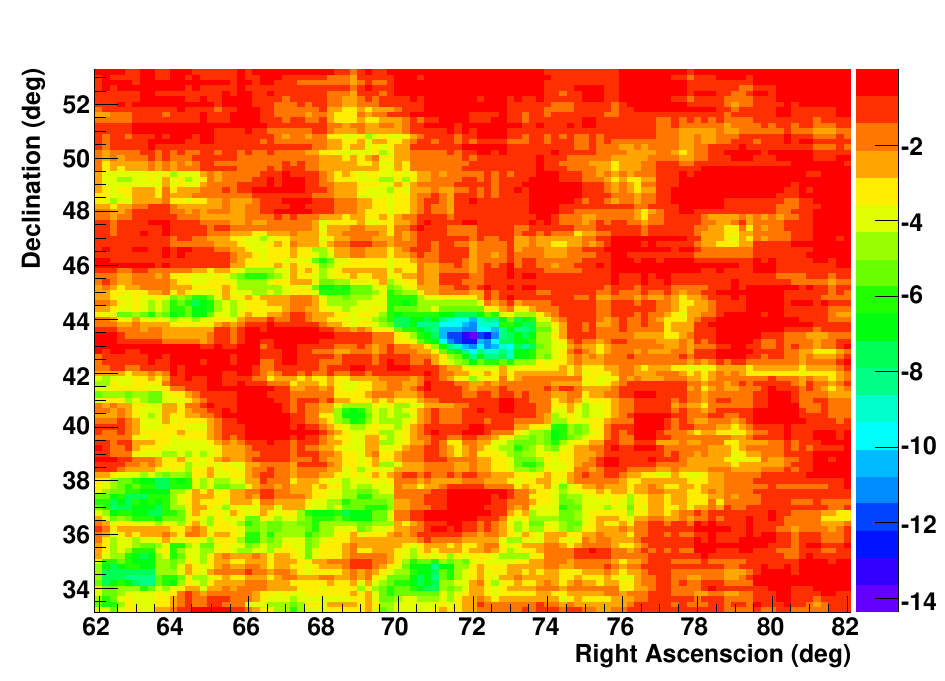}
\par\end{centering}

\caption{\label{fig:RESULTS_maps}Skymaps around the most significant event (post-trials) found in the search (Modified 
Julian Date 53118). 
\emph{Top}: signal map, \emph{bottom}: $log_{10}(P_{pre-trials})$ map.}
\end{figure}

\begin{figure}[htb]
\noindent \begin{centering}
\includegraphics[width=0.8\columnwidth]{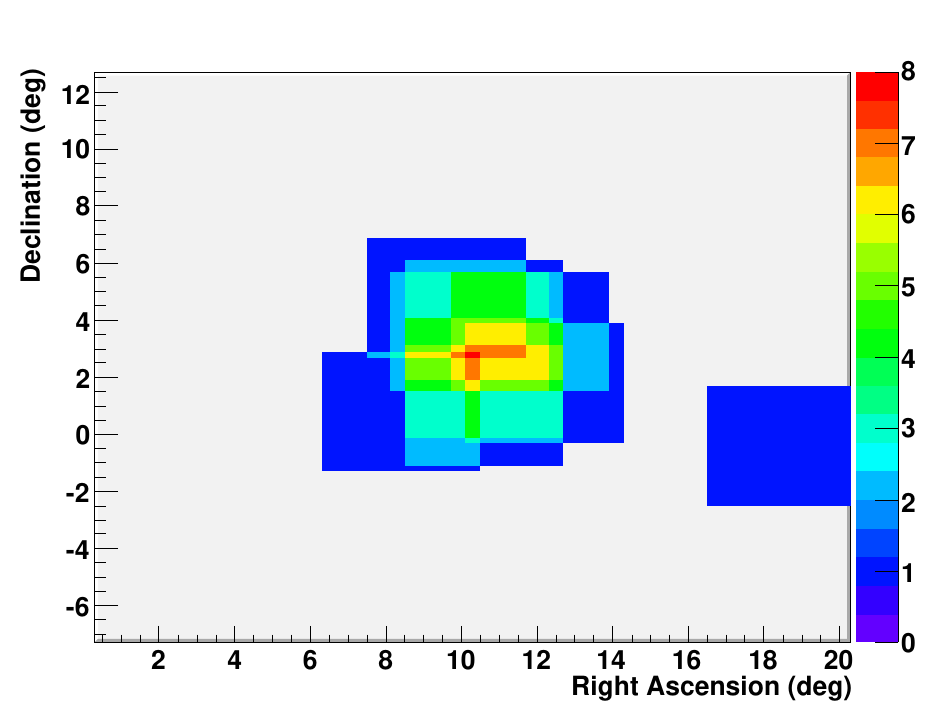}
\includegraphics[width=0.8\columnwidth]{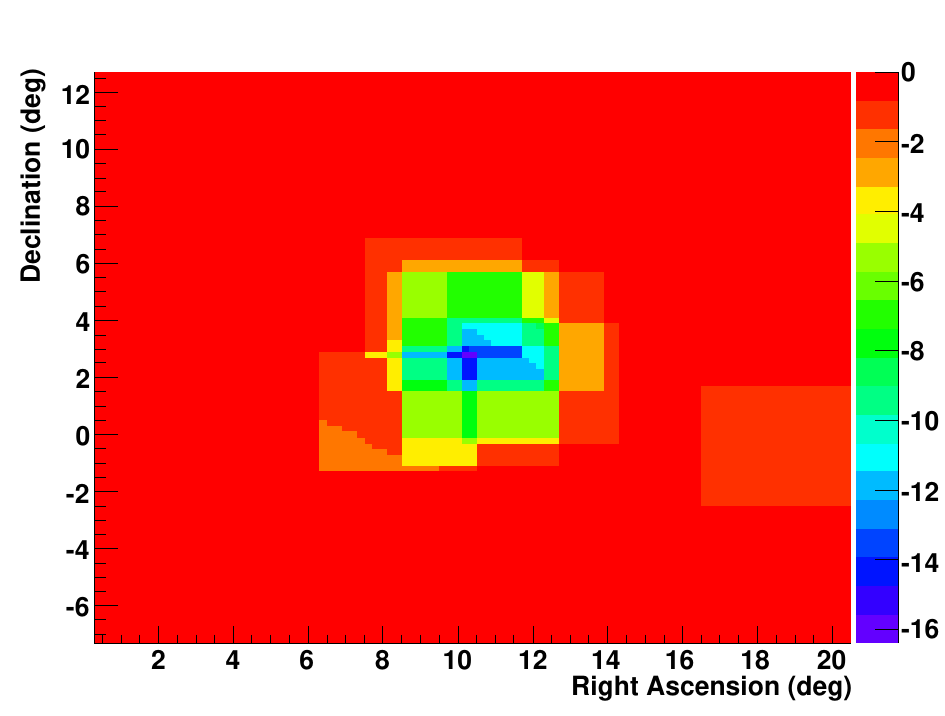}
\par\end{centering}

\caption{\label{fig:RESULTS_maps2}Skymaps around the second most significant event (post-trials) found in the search
(Modified Julian Date 53676).
\emph{Top}: signal map, \emph{bottom}: $log_{10}(P_{pre-trials})$ map.}
\end{figure}

Table \ref{tab:RESULTS_FinalResults} shows the details of the two events with
post-trials probabilities smaller than $0.01$. These post-trials probabilities
were calculated without including the number of trials for searching in multiple durations. 
The signal and probability maps around these two events are shown
in figures \ref{fig:RESULTS_maps} and \ref{fig:RESULTS_maps2}. 

\begin{table}
\begin{centering}
\begin{tabular}{|c|c|c|}
\hline 
Modified Julian Date & 53676 & 53118\tabularnewline
\hline 
Duration & $23.71\,ms$ & $237.14\,s$\tabularnewline
\hline 
Universal Date & 11/02/2005 & 04/23/2004\tabularnewline
\hline 
Universal Time & 06:36:12.099316 & 21:10:23.285730\tabularnewline
\hline 
(R.A.,Dec.) & $(10.4,\,2.8)$ deg & $(72.0,43.4)$ deg\tabularnewline
\hline 
Signal & 8 & 440\tabularnewline
\hline 
Expected Background & 0.0344 & 296.4\tabularnewline
\hline 
Pre-trials Probability & $4.717\times10^{-17}$ & $4.279\times10^{-15}$\tabularnewline
\hline 
Post-trials Probability & $2.962\times10^{-3}$ $(2.75\sigma$) & $2.56\times10^{-3}$ $(2.80\sigma)$\tabularnewline
\hline 
Trials (space \& time) & $6.270\times10^{13}$ & $7.633\times10^{11}$\tabularnewline
\hline 
Milagro Data File & \#6635-109 & \#5518-374\tabularnewline
\hline
\end{tabular}
\par\end{centering}

\caption{\label{tab:RESULTS_FinalResults}Data for the two events with post-trials probabilities less than 0.01.}

\end{table}

The probability of whether the results of
this search are consistent with the absence of a signal (the null hypothesis)
is equal to the smallest post-trials probability, now calculated using 
the number of trials in starting time, space and duration, found in the whole search. 
That probability corresponds to Modified Julian Day 53118 and is $0.127$ or
$1.14\sigma$. 

A post-trials probability smaller than $\simeq2.866\times10^{-7}$($5\sigma$)
would be enough for the rejection of the null hypothesis and would constitute solid
evidence for the presence of a gamma-ray signal. A probability less than
$\simeq1.3\times10^{-3}\,(3\sigma)$ would be just enough to provide
evidence for the existence of a signal. The probability of the validity of the null hypothesis 
that resulted from this search ($0.127$) is consistent with the absence of
bursts of VHE gamma rays in the Milagro data.

    \clearpage \chapter{\label{chap:GRBSim}Upper Limits on the Prompt VHE Emission from GRBs}

\section{Introduction}

In the absence of a detection of VHE emission from GRBs, 
upper limits on such an emission can be placed. Milagro has a wide field
of view and a high duty cycle, so during the 4.6 years that have been
searched, there were about eight hundred\footnote
{See section \ref{sub:GRBSim-GRB-Rates} for the calculation.}
GRBs in its field of view. A large fraction of them is
expected to be close enough to be detectable by Milagro. The results
from the analysis of such a dataset can be used to set upper
limits on the VHE emission from GRBs. To accomplish this, the number
of GRBs Milagro would expect to detect versus their VHE emission
was needed. This calculation was performed by means of a Monte Carlo simulation of the GRB population. 

The simulation essentially integrated over the GRBs of
some specific GRB population multiplied by the probability of them being detected by Milagro.
The result of the integration was the number of GRBs
expected to be detected by this search versus the properties of the
simulated GRB population.
Because we made a search that did not result in any
detections, we could exclude the GRB populations that correspond to
a number of detected GRBs in disagreement with the search's null result.

 Specifically, the simulation, by sampling from the 
 ($T_{90},\, E_{iso,}\, z$) distributions of Swift-detected GRBs
 and by assuming a model for their VHE emission, calculated
 the number of VHE gamma rays that reached the earth.
 Then, the number of signal events in the Milagro data generated by
 the incoming GRB emissions was calculated based on the methodology in Chapter
 \ref{chap:Sensitivity}. Using the results from Chapter \ref{chap:binsize}, the
 probability of detecting each one of these GRBs was calculated. The simulation
 summed all these detection probabilities to calculate how many GRBs
 were expected to be detected by this search versus the VHE-emission model
 used. Based on the expected number of detections and the fact that no
 detections were made, some of these emission models were excluded. 

The simulation used the GRB properties measured by satellite
detectors. While there have been multiple such detectors, a combination
of the measurements from all of them is not trivial. Different instruments
have different detection thresholds, making them sensitive to different
subgroups of the GRB population. In general, such detectors have complicated
and difficult-to-simulate triggering algorithms, making their exact triggering
thresholds very difficult to calculate. From these detectors, BATSE has
the largest sample of detected GRBs, however it is missing redshift
information. Knowledge of the redshift is crucial for this study,
because of the need to know the correlation between the amount of
gamma-ray signal arriving at the earth versus the distance of its
source. Because of absorption by the EBL, the spectral energy distribution
of the VHE GRB signal at the earth strongly depends on that distance.
Therefore, for this study, we need to know how much energy was emitted from each
redshift. 

After BATSE, there has been a series of GRB satellites (BeppoSAX,
HETE2, Integral, AGILE, Swift) that have provided accurate enough GRB localizations
that enabled resolution of the GRB redshifts. Of the GRB samples
from these detectors, the best, in terms of statistics, is the one from
Swift. For that reason, this GRB simulation was based mostly on the
GRB properties measured by Swift. Swift is sensitive to a lower energy
range than BATSE ($15-150\,KeV$ vs $\sim20\,keV-2\,MeV$ for BATSE), which
means that it is more sensitive to softer GRBs than BATSE. Short GRBs
are usually harder than long ones, so Swift's sensitivity is not optimal
for short GRBs. Furthermore, because Swift's trigger has a longer
accumulation time than BATSE, Swift is less sensitive to shorter duration
bursts than BATSE \cite{GRB_Band2006_Swift_Sensitivity}. Both of
these effects contribute to a small relative fraction of short GRBs
being detected by Swift ($\sim10\%$) compared to the one for BATSE
($\sim30\%$). Nevertheless, the short GRBs that Swift does not detect
are the faintest ones. The validity of this simulation, which is using
Swift data, depended on the assumption that all GRBs not detectable
by Swift are also not detectable by Milagro. 

Because, in general, a GRB detector is not sensitive to all GRBs in
nature, the properties of detected GRBs are usually different than
the properties of all GRBs. The distributions of detected GRBs can
be calculated by folding the intrinsic ones with the detector-specific
selection functions. This approach was followed for the calculation
of the detected redshift and $E_{iso}$ distributions of Swift. For
the case of redshift, the intrinsic redshift distribution of GRBs
was folded with a selection function $\psi_{flux}(z)$ that described
which fraction of the bursts at some redshift were detected by Swift.
Similarly, for the case of the isotropic energy, an intrinsic $E_{iso}$
distribution of GRBs was constructed based on theoretical models and
Swift data, and was compared to an effective Swift trigger threshold
to decide whether a GRB of a certain $E_{iso}-z$ combination would
be detected or not. 

In the following sections, the calculations of all the necessary elements
of the simulation of the GRB population will be described. In section
\ref{sub:GRBSIM_T90} the duration distribution of GRBs will be calculated
using Swift data. In sections \ref{sub:GRBSIM_RedshiftDistro} and
\ref{sub:GRBSIM_Eiso} the redshift distribution and the isotropic-equivalent
emitted energies of Swift-detected GRBs will be calculated respectively. In section
\ref{sub:GRBSim-GRB-Rates}, the number of GRBs in Milagro's field
of view during the duration of the search will be estimated. Section
\ref{sub:GRBSIM_Model} will describe the model for VHE emission from
GRBs used in this simulation, and section \ref{sub:GRBSIM_Verification}
will provide data that support the validity of the simulation's results
and of the calculations in this chapter. Finally, section \ref{sub:GRBSIM_Results},
will present the results of the GRB simulation and will set upper limits on
VHE emission from GRBs. In all the calculations in this chapter $\Omega_{m}=0.3$,
$\Omega_{\Lambda}=0.7$, and Hubble's constant $H_{0}=71km/s/Mpc$.

\section{\label{sub:GRBSIM_T90}GRB Duration Distribution}

The VHE emission from GRBs is not necessarily simultaneous to the
prompt keV-MeV emission and does not necessarily have the same duration
as it. According to GRB models, the prompt
VHE emission from GRBs is produced by internal to the GRB fireball
processes. Because the prompt $keV-MeV$ and $GeV-TeV$ emissions from GRBs depend on
the development of the same system (GRB fireball), their time scales
are expected to be comparable. The
delayed VHE emission is produced by interactions of the GRB fireball
with the circum-burst medium, usually by inverse Compton scattering
of low energy photons surrounding the site of the GRB. This emission
strongly depends on the medium around the burst and on the way it
interacts with the GRB fireball. Currently, there is not a good model
describing such emission in a global way; thus, the duration of the
VHE delayed emission from GRBs is not constrained. For that reason,
the GRB duration distribution used in this simulation, was that of
the prompt keV/MeV emission, which has already been measured by
satellite detectors. Even though this search was sensitive to both
the prompt and delayed emissions from GRBs, only the prospects of
detecting the prompt emission from GRBs can be quantified by this
Monte Carlo simulation. In the presence of a reliable model for delayed
VHE emission from GRBs, this simulation can be easily extended to
include both emission types. 

\begin{figure}[htb]
\subfigure[BATSE]{\includegraphics[width=0.5\columnwidth]{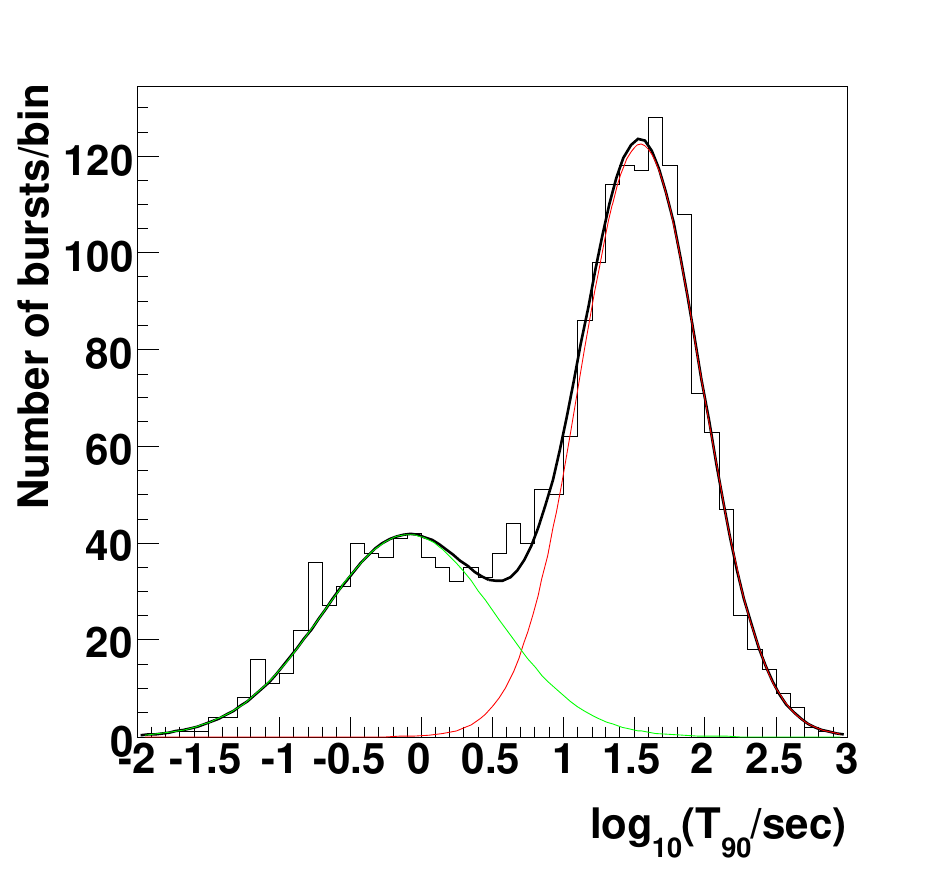}}\subfigure[Swift]{\includegraphics[width=0.5\columnwidth]{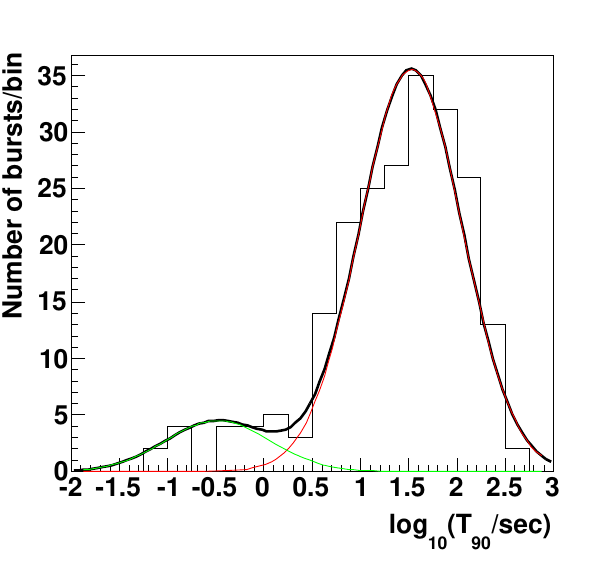}}

\caption{\label{fig:GRBSim_T90}$T_{90}$ distributions from the latest catalog
of BATSE GRBs and from Swift \cite{GRB_Floroi2007}. The data has been
fitted with the sum of two Gaussians. }

\end{figure}

The duration distribution $(T_{90}$) used in the simulation is approximated
as a fit to the duration distribution of Swift. Because of the bias
of Swift's trigger system against shorter duration bursts, Swift detected
a smaller fraction of short bursts than BATSE. The $T_{90}$ distributions
for BATSE GRBs\footnote{
Source: latest catalog of BATSE GRBs: http://www.batse.msfc.nasa.gov/batse/grb/catalog/current/%
} and Swift \cite{GRB_Floroi2007} are shown in figure \ref{fig:GRBSim_T90}.
The curves were fit by the sum of two Gaussian distributions. D. Band
\cite{GRB_Band2006_Swift_Sensitivity}, based on a post-launch analysis
of Swift data, found that Swift has a detection threshold that increases
with decreasing duration. For that reason, Swift's $T_{90}$ distribution
is shaped by this threshold and is expected to be different than the
one from BATSE, especially for lower durations. The parameters of
the fit are shown in table \ref{tab:GRBSim_FitT90}.%
\begin{table}
\begin{centering}
\begin{tabular}{|c|c|c|c|c|c|c|}
\hline 
\multicolumn{1}{|c|}{Detector} & \multicolumn{3}{c|}{Short} & \multicolumn{3}{c|}{Long}\tabularnewline
\hline
\hline 
 & Width & Amplitude & Mean & Width & Amplitude & Mean\tabularnewline
\hline 
BATSE & 0.61 & 63.86 & -0.64 (0.23 s) & 0.43 & 131.05 & 1.54 (34.67 s)\tabularnewline
\hline 
Swift & 0.54 & 6.11 & -0.61 (0.25 s) & 0.53 & 47.42 & 1.52 (33.1 s)\tabularnewline
\hline
\end{tabular}
\par\end{centering}

\caption{\label{tab:GRBSim_FitT90}Parameters of the fits on $log_{10}(T_{90})$
for BATSE and Swift GRBs. }

\end{table}

\section{\label{sub:GRBSIM_RedshiftDistro}Redshift distribution}

The \emph{intrinsic} redshift distribution of GRBs is different from
the redshift distribution of \emph{detected} GRBs, because the efficiency
of detecting GRBs at different redshifts depends on the redshift.
In the absence of instrumental selection effects, such as a dependence
on the duration \cite{GRB_Band2006_Swift_Sensitivity}, the detector's
sensitivity is described by the minimum peak flux it can detect. Assuming
that the intrinsic peak-luminosity distribution of GRBs is independent
of their redshift, the peak flux at the earth generated by GRBs from
some redshift was calculated. Then, using the Swift's minimum detectable
peak-flux, a flux-limited selection function $\psi_{flux}(z)$ was
constructed. This function shows the fraction of GRBs%
\footnote{It should be noted that this fraction only includes GRBs with properties
that have already been observed (for example, no yet-undetected low-luminosity
GRBs are included).%
} at some redshift that triggered Swift. 

Initially, in subsections \ref{sub:GRBSim_IntrinsicZLong} and \ref{sub:GRBSim_IntrinsicZShort},
the intrinsic redshift distribution of GRBs will be constructed. In
section \ref{sub:GRBSim_Peak-luminosity}, details on the choice of
the peak-luminosity function of GRBs will be given, and using this
information, the flux-limited selection function $\psi_{flux}(z)$
will be constructed (\ref{sub:GRBSim_psigrb}). Finally, combining
$\psi_{flux}(z)$ and the intrinsic redshift distributions of GRBs,
the redshift distribution of GRBs detected by Swift will be calculated
(section \ref{sub:GRBSim_DetectedZ}). That distribution will later be
used by the Monte Carlo of the GRB population.

\subsection{\label{sub:GRBSim_IntrinsicZLong}Intrinsic redshift distribution
of long GRBs }

This section will present the calculation of the intrinsic redshift
distribution of long GRBs. As will be shown, that distribution comes
from a combination of the Star Formation Rate (SFR) and the dependence
of the average stellar metallicity on the redshift. 

\begin{figure}[htb]
\begin{centering}
\includegraphics[width=1\columnwidth]{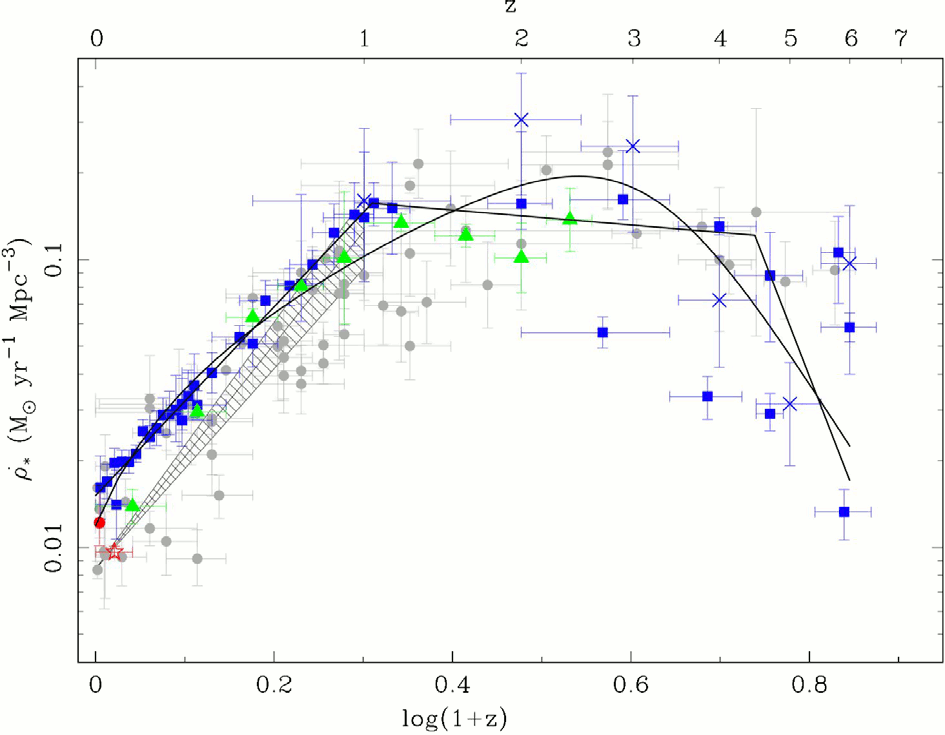}
\par\end{centering}

\caption{\label{fig:GRBSim_SFR}Star Formation Rate reproduced from Hopkins
and Beacom \cite{GRB_Hopkins_Beacom}. The two black lines show their
parametrized fits on data from far-infrared ($24\mu m$) (hatched region and
triangles), ultraviolet (squares), radio 1.4GHz (open red star) and ultra deep field estimates
(crosses). Source: \cite{GRB_Coward2007}}

\end{figure}

\begin{figure}[htb]
\begin{centering}
\includegraphics[width=1\columnwidth]{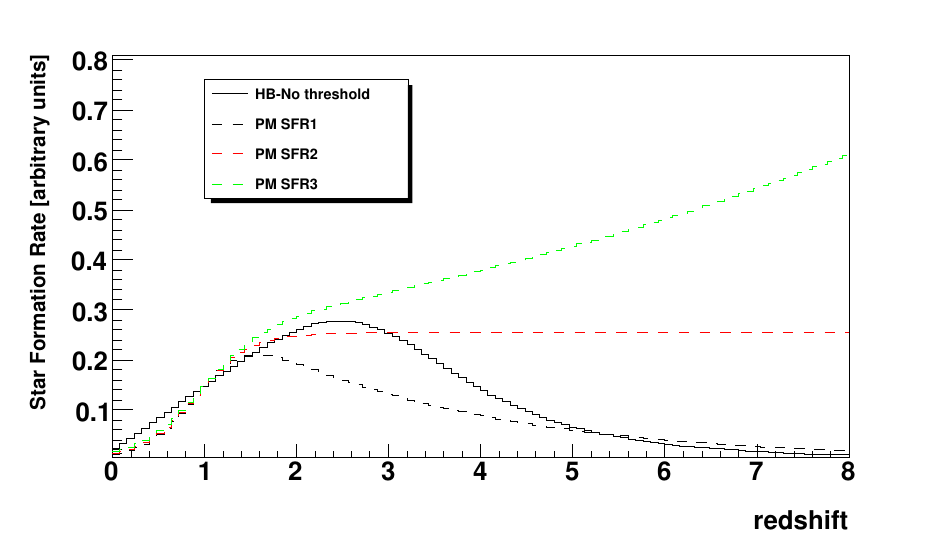}
\par\end{centering}

\caption{\label{fig:GRBSim_SFR2}Star Formation Rates from Hopkins and Beacom
(HB) \cite{GRB_Hopkins_Beacom} (black solid curve) and from Porciani
\& Madau (PM) \cite{GRB_Porciani_Madau2001} (dashed lines). The curves
are normalized to intersect at $z=1$ The HB model has the best agreement
with the latest experimental measurements. Older studies that had
available only the PM SFRs, found the best agreement with the BATSE
\& Swift data when using the SFR3 and SFR2 models, implying an enhanced
GRB rate at larger redshifts. }

\end{figure}

The association of some long duration GRBs with supernovae implies
that they are caused by the core collapse of short-lived massive stars
(see subsection \ref{sub:GRB_GRB_SNCOnnection}). For that reason, the intrinsic redshift distribution
of long GRBs is generally considered as following the SFR. A SFR widely
used in the early studies of GRBs was from Porciani and Madau (PM)
\cite{GRB_Porciani_Madau2001}, and was constrained by experimental
data for only $z\lesssim1.5$. Because of the freedom at higher redshifts,
these authors provided three different SFRs, each one with a different
$z>1.5$ behavior. Recently, Hopkins and Beacom (HB) \cite{GRB_Hopkins_Beacom}
have estimated the SFR by fitting recent ultraviolet and far-infrared
data (Figs. \ref{fig:GRBSim_SFR} and \ref{fig:GRBSim_SFR2}).
These data constrain the SFR up to $z\simeq6$, with especially
tight constraints for $z<1$. 

\begin{figure}[htb]
\begin{centering}

\par\end{centering}

\begin{centering}
\includegraphics[viewport=10bp 400bp 260bp 650bp,clip,width=0.6\columnwidth]{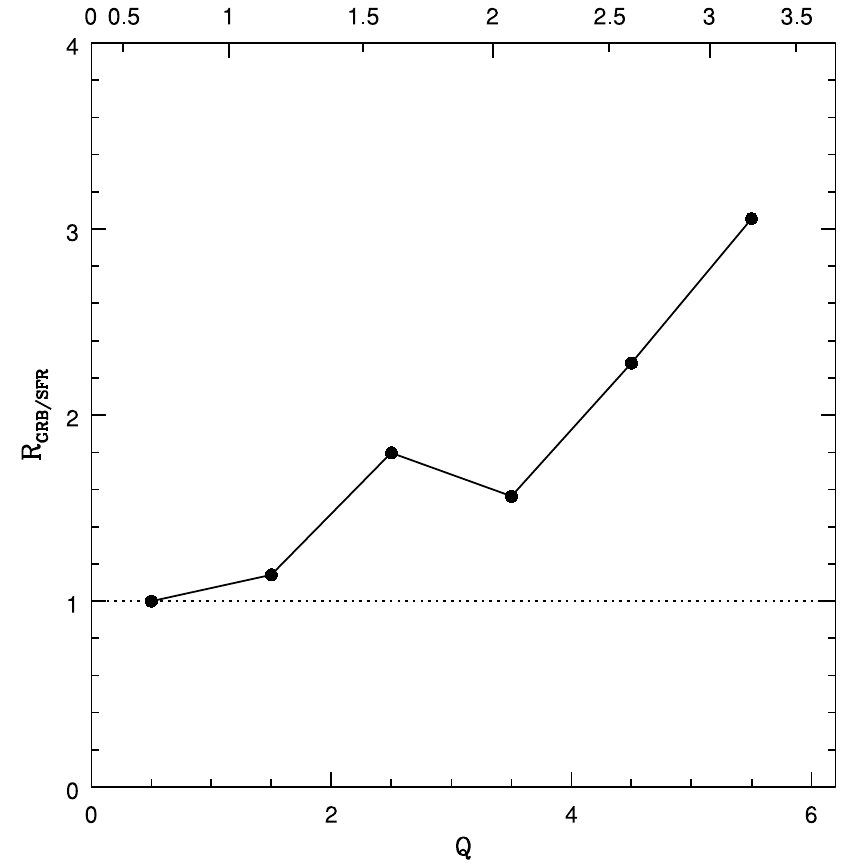}
\par\end{centering}

\caption{\label{fig:GRBSim_ExtraHighZGRBs}The observed ratio of the GRB rate
to the Hopkins and Beacom \cite{GRB_Hopkins_Beacom} SFR as a function
of $Q(z)$ (see equation \ref{eq:GRBSim_Q} and figure \ref{fig:GRBSim_VOLFACTOR})
and of $z$. The normalization is chosen so that $R_{GRB/SFR}(0.5)=1$.
Source: \cite{GRB_Li_2007}}

\end{figure}

Initially, the SFR and the GRB intrinsic redshift distribution were
considered as just being proportional to each other (i.e. no evolution
effects in the GRB rate versus redshift). However, Swift recently
detected some GRBs of very high redshift. Based on these events, Kistler
\emph{et al}. \cite{GRB_Kistler_et_al2008} showed that there are $\sim4$
times more GRBs at redshift $z\simeq4$ than predicted by the latest
HB SFR (Fig. \ref{fig:GRBSim_ExtraHighZGRBs}). %
 Other authors also concluded that the HB SFR coupled with any kind
of reasonable luminosity function cannot explain both the peak flux
distributions and the increased Swift GRB rate at large redshifts
\cite{GRB_Guetta_Piran2007,GRB_Dermer_Le2007,GRB_Lapi_et_al2008}.
Based on this, it was proposed that the luminosity function evolves
with redshift, favoring a large redshift population. However, we know
that long GRBs have been observed mostly in low metallicity galaxies
(\cite{GRB_Fynbo_etal2003}, see, however, \cite{GRB_Savaglio2006,GRB_Dermer_Le2007}).
Also, theoretical studies on the collapsar model of long GRBs suggested
that long GRBs can only be produced by stars with metallicity $Z\lesssim0.3Z_{\odot}$.
Otherwise, strong stellar winds would cause stars to lose too much
mass and angular momentum to form a disk around a black hole of several
solar masses, an essential condition for the production of long GRBs
\cite{GRB_Hirschi_etal2006,GRB_Yoon_etal2005}. Based on these constraints, some
other authors \cite{GRB_Li_2007} proposed that the long-GRB rate
follows the low-metallicity component of the SFR. This provides an
enhanced long-GRB rate at higher redshifts, since the earlier galaxies
had lower metallicities. The intrinsic distribution of long
GRBs used in this study was derived from the HB SFR combined with
an upper metallicity limit.

\begin{figure}[htb]
\begin{centering}
\includegraphics[clip,width=1\columnwidth]{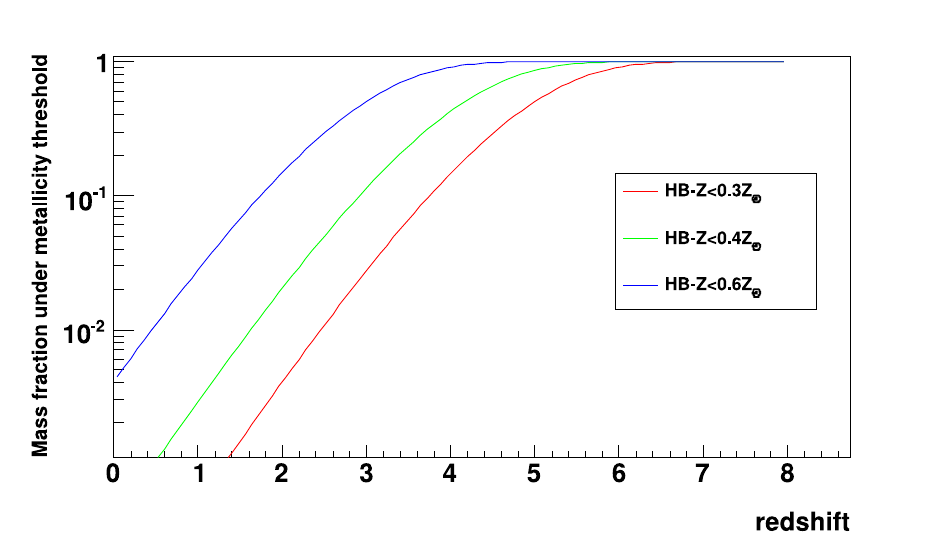}
\par\end{centering}
\begin{centering}
\includegraphics[width=1\columnwidth]{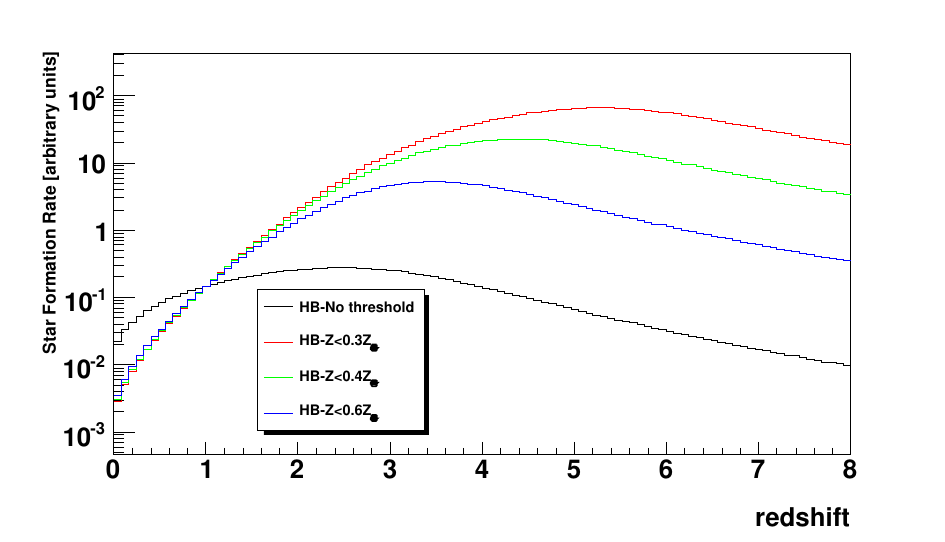}
\par\end{centering}

\caption{\emph{Top}:\label{fig:GRBSim_HBSFRMetallicity} $f(z,m)$ - fractional
mass density belonging to metallicities lower than some metallicity
limit $Z_{max}=m\,Z_{\odot}$, \emph{bottom}: Hopkins and Beacom SFR for no metallicity
cutoff (black line) and with metallicity cutoffs.}

\end{figure}

The rate of long GRBs in shells of redshift $dz$ as measured from
the distant frame of the GRB $\dot{R}_{Long}(z,Z_{max})$ is related to the
SFR as 
\begin{equation}
\dot{R}_{Long}(z,Z_{max})=SFR(z)\,\epsilon(z,Z_{max}),
\end{equation}
where $\epsilon(z,Z_{max})$
is equal to the fraction of exploding stars that end up creating a
GRB, and $Z_{max}$ is an upper metallicity limit. The dependence of
the term $\epsilon(z,Z_{max})$ on $z$ can be used to describe an enhanced
GRB rate from such environments. We can rewrite it as $\epsilon(z,Z_{max})=k\,f(z,Z_{max})$
, where $f(z,Z_{max})$ is the fractional mass density belonging to metallicities
lower than a limit $Z_{max}$, and $k$ is a constant.
According to Langer \& Norman \cite{GRB_Langer_Norman2006}:\begin{equation}
f(z,m)=1-\frac{\Gamma(a+2,m^{\beta}\,10^{0.15\beta z})}{\Gamma(a+2)}\label{GRBSIM_MetallicityFraction},\end{equation}
where $a\simeq1.16$ is the power index of the Schechter distribution
function of galaxy stellar masses \cite{SFR_Panter_etal2004}; $\beta\simeq2$
is the slope in the linear bisector fit to the galaxy stellar mass-metallicity
relation \cite{SFR_Savaglio_et_al2005}; $m=Z_{max}/Z_{\odot}$ is $Z_{max}$ in units of 
the solar metallicity $Z_{\odot}$; $\Gamma(x)$ is the gamma
function; and $\Gamma(a,x)$ is the incomplete gamma function. In
this equation, it is assumed that the average cosmic metallicity evolves
with redshift by -0.15 dex per unit redshift\footnote{-0.15 dex = $10^{-0.15}$.}. The function
$f(z,m)$ for different values of maximum metallicity $m$ along with
the modified SFRs from HB is shown in figure \ref{fig:GRBSim_HBSFRMetallicity}.
In order to quantify the effects of the upper metallicity cutoff on
the produced upper limits, this study will provide multiple results,
each one corresponding to a different upper metallicity cutoff.
The preferred value will be $Z_{max}=0.3Z_{\odot}$. %

\begin{figure}[htb]
\begin{centering}
\includegraphics[width=0.8\columnwidth]{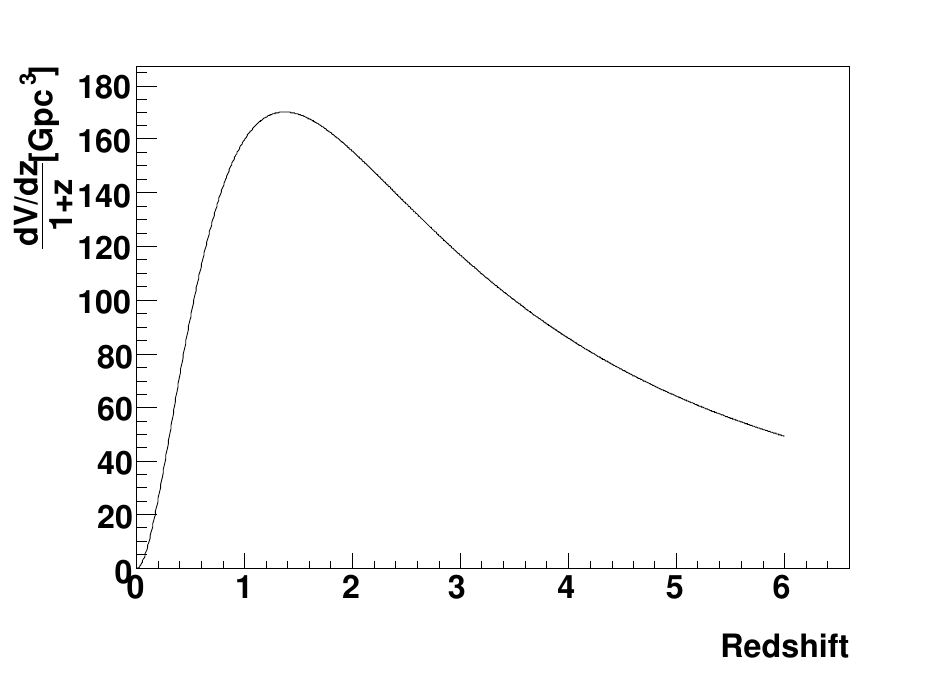}
\par\end{centering}

\caption{\label{fig:GRBSim_VOLFACTOR}Volumetric factor $Q(z)\equiv\frac{dV(z)/dz}{1+z}$
(eq. \ref{eq:GRBSim_Q})}

\end{figure}

The intrinsic redshift distribution of long GRBs is given by \begin{eqnarray}
\frac{d\dot{N}_{Long}(z,m)}{dz} & = & \frac{\dot{R}_{long}(z,m)}{(1+z)\left\langle f_{beam}\right\rangle }\frac{dV(z)}{dz}\label{eq:GRBSim_Intrinsic_Long}\\
 & \propto & \frac{SFR(z)\,f(z,m)}{(1+z)\left\langle f_{beam}\right\rangle }\frac{dV(z)}{dz},\end{eqnarray}
where $\dot{R}_{long}(z)/(1+z)$ is the rate of GRBs in shells of
redshift $dz$ as measured from our reference frame, $\left\langle f_{beam}\right\rangle $
is a beaming factor representing the fraction of GRBs with their emission
pointing to us, and $dV(z)/dZ$ is the comoving volume element described
in terms of the comoving distance $D_{c}(z)=D_{l}/(1+z)$. The volumetric
factor $Q(z)\equiv\frac{dV(z)}{dz}\,\frac{1}{1+z}$ (shown in figure
\ref{fig:GRBSim_VOLFACTOR}) is given by:\begin{equation}
Q(z)\equiv\frac{dV(z)}{dz}\frac{1}{1+z}=\frac{4\pi(c/H_{0})D_{c}^{2}(z)}{\sqrt{(1+z)^{3}\Omega_{m}+(1+z)^{2}\,\Omega_{k}+\Omega_{\Lambda}}}\frac{1}{1+z}.\label{eq:GRBSim_Q}\end{equation}

Equation \ref{eq:GRBSim_Intrinsic_Long} combined with the flux-limited
selection function will give (in subsec. \ref{sub:GRBSim_DetectedZ})
the redshift distribution of long GRBs detected by Swift.

\subsection{\label{sub:GRBSim_IntrinsicZShort}Intrinsic redshift distribution
of short GRBs }

This section will present the calculation of the intrinsic redshift
distribution of short GRBs. As will be shown, that distribution
comes from a combination of the SFR with the time delay necessary
for the creation and merger of the compact-binary objects that
lead to short GRBs. 

In the case of long GRBs, the time interval between star formation
and star death is very small. However, this is not the case for short
GRBs. The time delay between a compact-binary merger (that can create
a short GRB) and the formation of its compact objects is not negligible.
Mergers occur considerably later in time (at a smaller
redshift) than the formation of the compact objects. The longer the
time delay is, the smaller the average value of the merger-rate redshift
distribution (and of the short GRB distribution). 

\begin{figure}[htb]
\begin{centering}
\includegraphics[width=0.7\columnwidth]{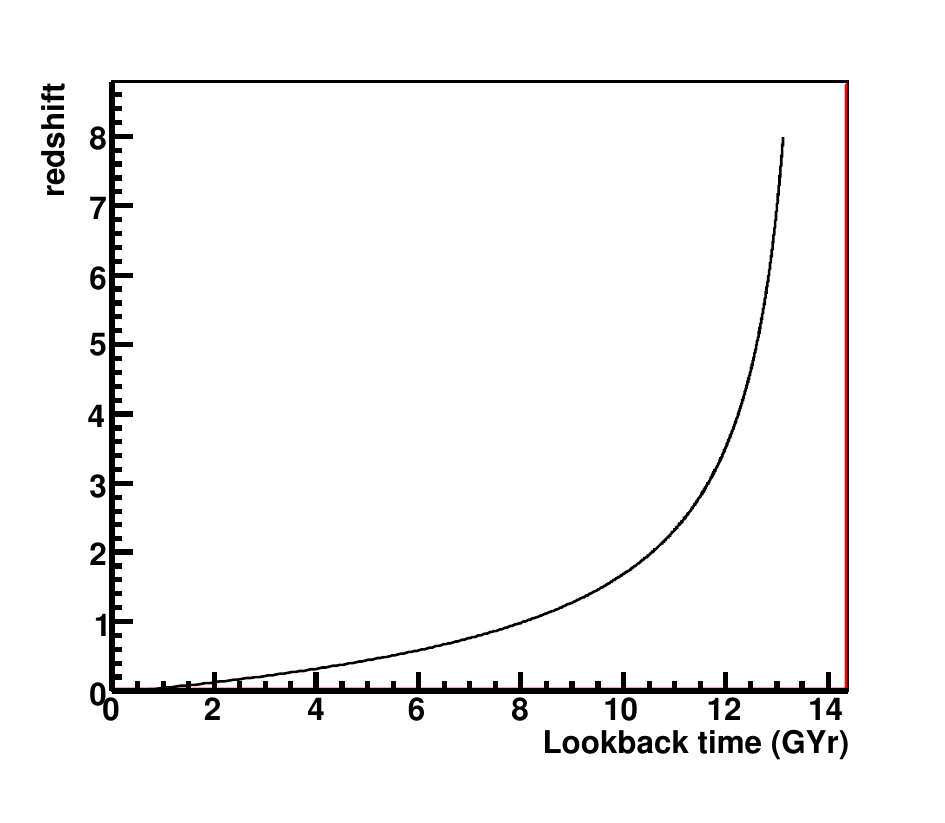}
\par\end{centering}

\caption{\label{fig:GRBSim_lookbacktime}Lookback time (eq. \ref{eq:GRBSim_LookbackTIme})
versus redshift.}

\end{figure}

\begin{figure}[htb]
\begin{center}
\includegraphics[width=0.8\columnwidth]{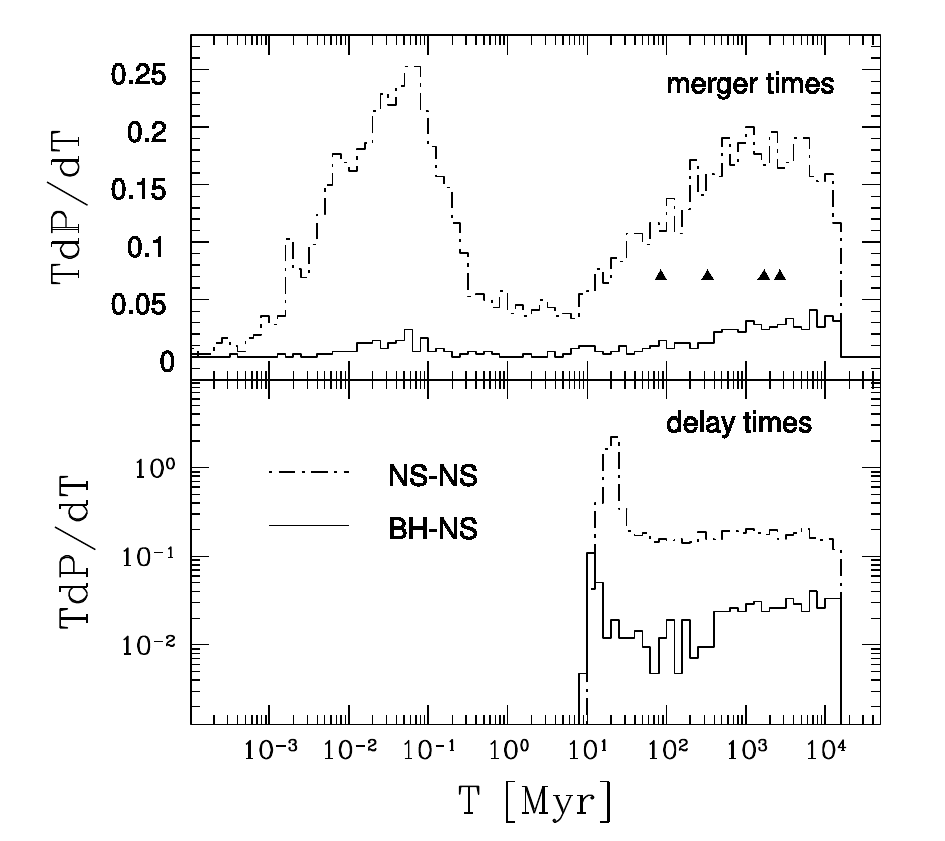}

\caption{\emph{\label{fig:GRBSim_delaytimes}Top}: Merger time distributions
for NS-NS and BH-NS binaries. The four NS-NS systems detected in our
galaxy are shown with triangles. \emph{Bottom}: Delay time distributions.
Delay times are the sum of the formation time of the compact binaries
($\sim20\,Myr$) and the merger times (top plot). The delay times are
consistent with a $P(\tau)=1/\tau$ distribution with $\tau_{min}=20\,Myr$.
Source: \cite{GRB_Belczynski_et_al2006}}
\end{center}
\end{figure}

\begin{figure}[htb]
\begin{center}
\includegraphics[width=0.7\columnwidth]{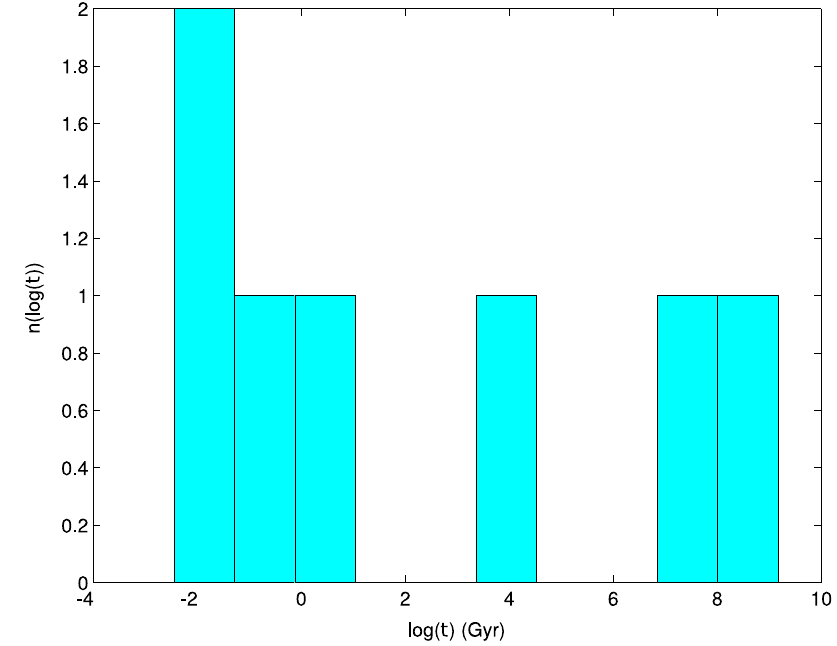}

\caption{\label{fig:GRBSIM_ChampionDelay}Time delay distribution for the six
NS-NS binary compact objects detected in our galaxy. The distribution
is consistent with a $P(\tau)=1/\tau$ distribution. Data from Champion
\emph{et al}.\cite{GRB_Champion2004}, figure from Guetta \& Piran \cite{GRB_GuettaPiran_2005}.}
\end{center}
\end{figure}

If $P(\tau)=\tau^{n}$ is the distribution of time delays $\tau$,
then the rate of compact-binary mergers in shells of redshift $dz$
as measured from the binary-system reference frame is \begin{equation}
\dot{R}_{Merger}(z,\tau_{min},n)=\int_{z_{min}(\tau_{min})}^{\infty}SFR(z)\,P\left(t(z)-t(z')\right)\,\frac{dt(z')}{dz'}dz'.\label{eq:GRBSim_MergerRate}\end{equation}
The time $t(z)$ is the lookback time corresponding to the redshift
$z$, and is the difference between the age of the universe now and
the time when the light we observe now was emitted. The minimum integration
redshift $z_{min}(\tau_{min})$ is the redshift corresponding to a
lookback time $\tau=t(z)+\tau_{min}$. What equation \ref{eq:GRBSim_MergerRate}
says is that the merger rate at some redshift $z$ is produced by
the sum of the contributions of objects formed at earlier epochs (higher
redshifts - $z'$) weighted with a time-delay dependent probability
$P\left(t(z)-t(z')\right)$. The lookback time (fig. \ref{fig:GRBSim_lookbacktime})
is given by:\begin{equation}
t(z)=\frac{1}{H_{0}}\int_{0}^{\infty}\frac{dz'}{(1+z')\sqrt{\Omega_{M}(1+z')^{3}+\Omega_{K}(1+z')^{2}+\Omega_{\Lambda}}}.\label{eq:GRBSim_LookbackTIme}\end{equation}
The time delay $\tau$ is the sum of two quantities: the evolutionary
time, which is the time required for the initially non gravitationally-interacting
members of the binary to randomly approach each other and become gravitationally
bound forming the compact binary object, and the merger time, which
is the time needed for the orbital decay of the compact binary due
to emission of gravitational radiation. Various authors \cite{GRB_Belczynski_et_al2006,GRB_Shaughnessy2008,GRB_Schneider_2001},
using population synthesis methods, calculated the time-delay distribution
of such times (see, for example, figure \ref{fig:GRBSim_delaytimes})
and found that it can be reasonably approximated by $P(\tau)=1/\tau$
with a minimum delay time $\tau_{min}\sim20\,Myr$. Berger \emph{et
al} \cite{GRB_Berger_New_Population}, analyzing new high-redshift
short-GRB detections from Swift, found that if $P(\tau)=\tau^{n}$,
then $-1\lesssim n\lesssim0$. Guetta and Piran \cite{GRB_GuettaPiran_2005},
while trying to estimate the luminosity function of short GRBs, used
$P(\tau)=1/\tau$, based on the time delays of the six detected NS-NS
binaries in our galaxy (fig. \ref{fig:GRBSIM_ChampionDelay}). Earlier
studies, based on a limited redshift distribution for short GRBs that
peaked at very low redshifts ($<z>\sim0.3$), favored a time delay
distribution that averaged at longer times ($\sim$few Gyr) than the
ones mentioned above (\cite{GRB_GuettaPiran2006,GRB_Nakar_et_al_2006,GRB_Ando2005}).
However, a redshift distribution of detected GRBs that was recently
updated with more distant short GRBs \cite{GRB_Berger_New_Population}
showed that the distribution of time delays $P(\tau)$ actually averages
at smaller values than previously thought, and is actually consistent with
$P(\tau)=1/\tau$. %

\begin{figure}[htb]
\begin{centering}
\includegraphics[width=0.8\columnwidth]{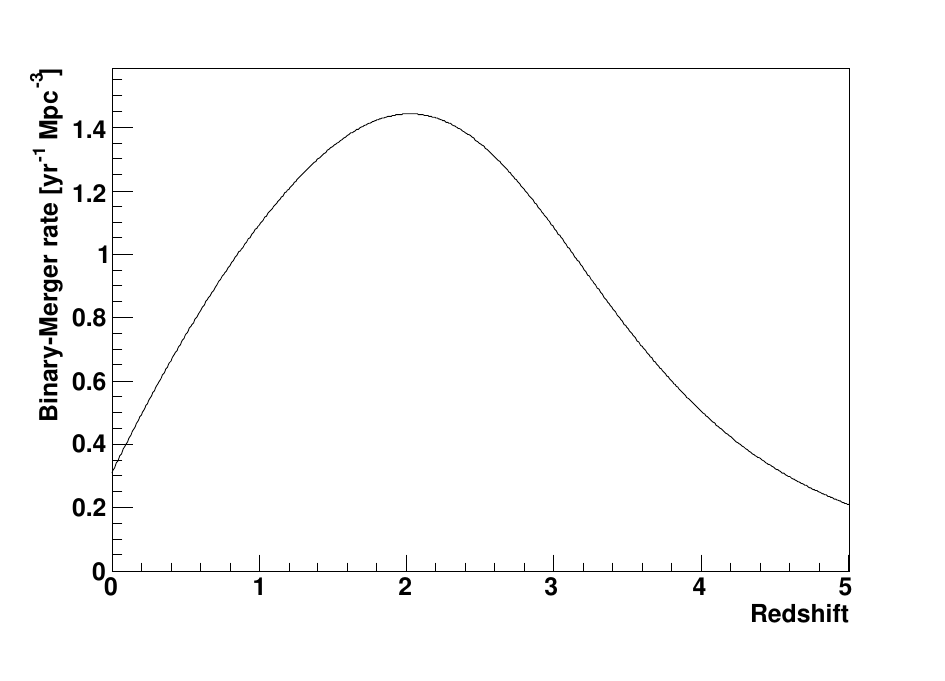}
\par\end{centering}

\caption{\label{fig:GRBSim_MergerRate}Compact-binary merger rate. This rate
corresponds to a distribution of delay times between star formation
and mergers $P(\tau)=1/\tau$, and a minimum delay time $\tau_{min}=20\,Myr$.
The HB SFR was used.}

\end{figure}

Similarly to the long-GRB case (eq. \ref{eq:GRBSim_Intrinsic_Long}),
the intrinsic redshift distribution of short GRBs is

\begin{equation}
\frac{d\dot{N}_{Short}(z,\tau_{min},n)}{dz}\propto\frac{\dot{R}_{Merger}(z,\tau_{min},n)}{(1+z)\left\langle f_{beam}\right\rangle }\frac{dV(z)}{dz},\label{eq:GRBSim_Intrinsic_Short}\end{equation}
where the proportionality comes from the fact that the efficiency
of a binary merger creating a short GRB is not known. The intrinsic
distribution of short GRBs for this study was calculated for $P(\tau)=1/\tau$,
$\tau_{min}=20\,Myr$ and the HB SFR. The distribution of the compact-binary
merger rate $\dot{R}_{Merger}(z,\tau_{min},n)$ versus the redshift,
for this set of parameters, is shown in figure \ref{fig:GRBSim_MergerRate}.

\subsection{Peak-luminosity function\label{sub:GRBSim_Peak-luminosity}}

The peak-luminosity function%
\footnote{Similarly to the GRB literature, the terms {}``peak-luminosity function''
and {}``luminosity function'' in this text, may be used interchangeably
and will mean the same thing.%
} of GRBs is an essential element for the calculation of the flux-limited
selection function. There have been various functional forms used
to describe the probability distribution of the luminosity function
such as power laws, broken power laws, and forms similar to a Schechter
\footnote{Function used to describe the luminosities of nearby galaxies.%
} function. Schmidt \cite{GRB_Schmidt1999} and Guetta \& Piran 
\cite{GRB_GuettaPiran_2005} used a broken power law
\footnote{Actually, the luminosity function in these early papers was given
in terms of $dN/dlogL$. Recent papers give the luminosity mostly
in the $dN/dL$ form. Care has to be taken to correct for the different
spectral indices in the two cases: $\frac{dN}{dL}=\frac{dN}{dlogL}\frac{1}{ln(10)L}$. %
} luminosity function, which describes the comoving space density of
GRBs:\begin{equation}
\frac{dN}{dL}\propto\begin{cases}
(L/L^{*})^{-\alpha} & L_{min}<L<L^{*}\\
(L/L*)^{-\beta} & L^{*}<L<L_{max}\end{cases}.\label{eq:GRBSim_SchmidtLum}\end{equation}

Another way to describe the luminosity function (used by Li \cite{GRB_Lapi_et_al2008})
is by a Schechter function: \begin{equation}
\frac{dN}{dL}\propto L^{\delta}e^{-L/L_{c}}.\label{eq:GRBSim_Schechter}\end{equation}

In general, authors using some assumptions about the intrinsic redshift
distribution of a class of bursts (short/long) and the form of the
luminosity function, and using data from satellite detectors, such
as the peak fluxes, event rates, and the redshift distributions of
detected GRBs, first calculate the distributions of detected quantities
by some detector, and then try to find which set of parameter values
used in their calculation produces the best fit to the observed data.
One of biggest differences between the various studies is the intrinsic
redshift distribution for GRBs. There have been multiple assumptions
and experimental data used for determining this distribution, which
do not always agree with our current knowledge of the subject. In
the following, an overview of the different approaches taken for the
calculation of the luminosity function will be given, and the reasons
that the derived luminosity functions are usually in disagreement
with each other will be described.

For long GRBs, the differences between the results of most studies
can be traced to the form of the SFR used (HB vs PM) and whether any
upper metallicity limit were applied. As will be shown later,
luminosity functions that correspond to higher average luminosities
result in a flux-limited selection function that stays large up
to higher redshifts. If we had two studies using two different SFRs,
one that averages at high redshifts and one at low redshifts, then
the first SFR would need a flux-selection function that is smaller
at higher redshifts and vice versa. 

Because short GRBs are usually at lower redshifts than long GRBs,
the effects of using different SFRs in the estimation of their luminosity
function are small, since all available SFRs are in good agreement
for $z\lesssim1.5$. For short GRBs, the biggest factor causing the
differences between the predicted luminosity functions lies in the
assumptions regarding the time-delay distribution $P(\tau)$. Smaller
delay times favor an intrinsic redshift distribution that peaks at
higher redshifts, which in turn usually requires a flux-limited selection
function that starts falling from unity at smaller redshifts, and a
luminosity-function that averages at lower luminosities. 

For long GRBs, Guetta \& Piran \cite{GRB_Guetta_Piran2007}, using
a logN-logP diagram made with data from Swift, found that the SFR3
from PM (a SFR that is enhanced at high redshifts) corresponds to
a broken power law (eq. \ref{eq:GRBSim_SchmidtLum}) with $a=1.1$
, $\beta=3$, and $L^{*}=4\times10^{51}\,erg/s$. Firmani \emph{et
al}. \cite{GRB_Firmani_etal2006}, using BATSE data and the SFR3 from
PM, found that $a=0.9\pm0.4$, $\beta=2.1\pm0.2$, and $L^{*}\simeq2.6\times10^{50}\,erg/s$,
or for a simple power law $\frac{dN}{dL}\propto L^{-1.58\pm0.04}$.
Daigne \emph{et al}. \cite{GRB_Daigne_et_al2006} using the same SFR
found that $\frac{dN}{dL}\propto L^{-1.54\pm0.18}$. Theoretical predictions
based on the internal shock model predict $a\lesssim1$ with $<a>\sim0.5$,
$\beta\gtrsim1.4$ with $<\beta>\sim1.7$ and $L^{*}\simeq4\times10^{50}-6\times10^{51}\,erg/s$
\cite{GRB_Zitouni_et_al2008}. Dermer and Le \cite{GRB_Dermer_Le2007},
using Swift data and the uniform jet model, found that $\frac{dN}{dL}\propto L^{-3.25}$. 

The luminosity function for short GRBs is not yet well understood,
mostly because of the small number of short GRBs with resolved redshifts.
All the studies are based on logN-logP analyses of the peak flux distributions
from BATSE. The Swift and BATSE peak-flux detection thresholds for
short GRBs are comparable, so the results from this BATSE-based study
should be also applicable to Swift. Schmidt \cite{GRB_Schmidt2001},
using the SFR2 from PM (a SFR that is roughly constant for $z>2$),
found that $\alpha=1.6$, $\beta=3$, and $L^{*}=3.2\times10^{50}\,erg/s$.
Using the same SFR, Guetta and Piran \cite{GRB_GuettaPiran_2005}
found $\alpha=1.6$, $\beta=3$, and $L^{*}=2.2\times10^{51}\,erg/s$
. Salvaterra \emph{et al.} \cite{GRB_Salvaterra_et_al2007} found
that when using a broken power law with Schmidt's $a=1.6$ and the
HB SFR, they find good agreement with BATSE peak-flux data for $\beta=2.8\pm0.29$
and $L^{*}=6.35\times10^{50}\,erg/s$%
\footnote{They quote their $L^{*}$ in the 30-2000keV energy range of BATSE.
For sake of comparison, I have calculated the equivalent luminosity
in the 15-150keV range of Swift using a typical Band spectrum $(a,b,E_{0})=(-1,-2.25,
256\,KeV)$ \cite{GRB_Preece_et_al2000}. The conversion was $L_{Swift}^{*}=0.933\,L_{BATSE}^{*}$.}.

\begin{figure}[htb]
\begin{centering}
\includegraphics[viewport=20bp 400bp 2400bp 3508bp,clip,width=0.7\columnwidth]{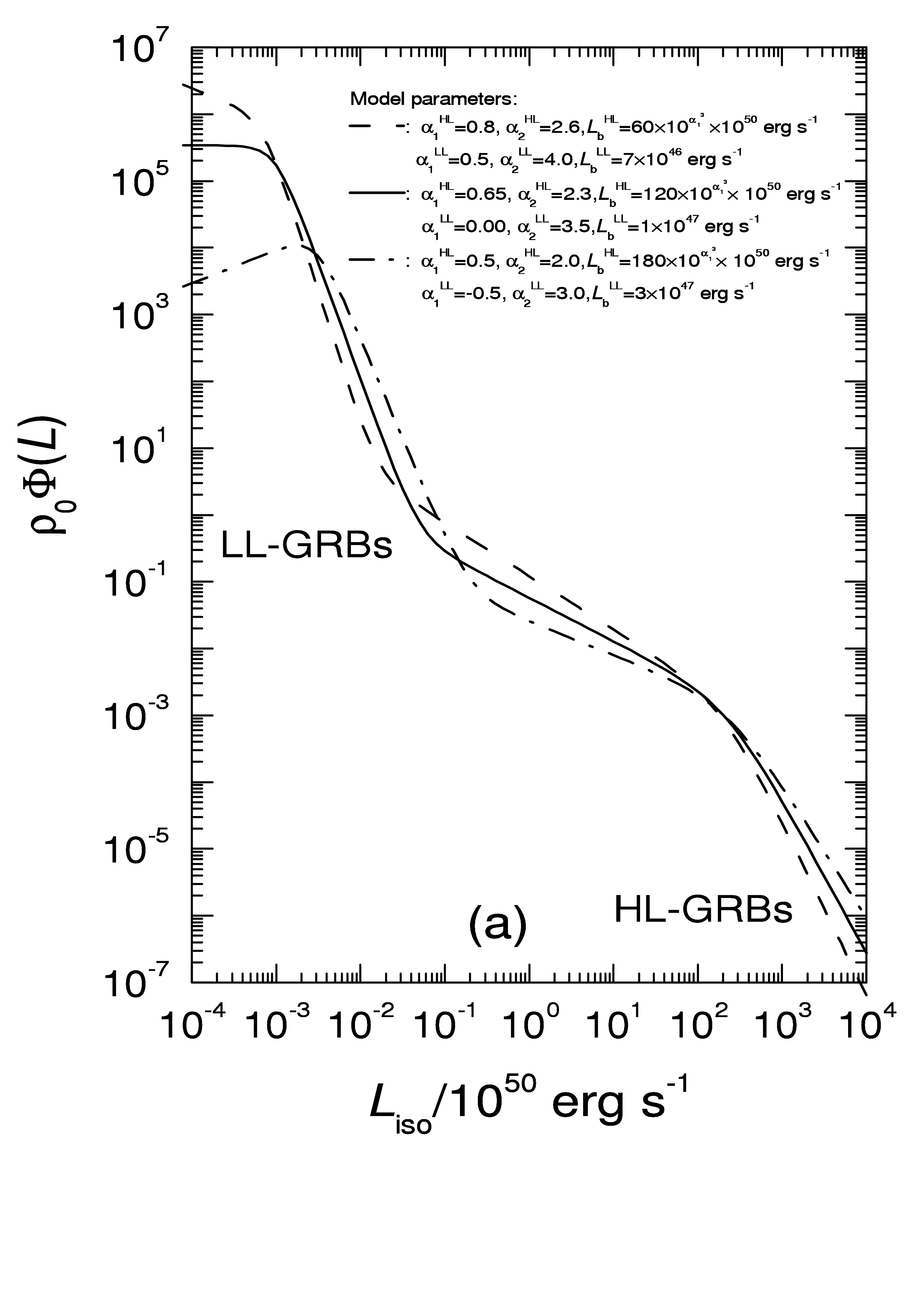}
\par\end{centering}

\caption{\label{fig:GRBSim_LL_LuminosityFunction}Luminosity function proposed
by Liang \emph{et al} \cite{GRB_Liang_et_al2007} to include the recent
low-luminosity GRBs detected by Swift. $\Phi(L)\equiv{}dN/dL$, and $\rho_0$ is a constant.}

\end{figure}

Even though the parameters of the luminosity function are well determined,
the minimum and maximum luminosities are not. If a large number of
GRB detections with resolved redshifts were available, then the limits
of the luminosity range would be defined by the minimum and maximum
luminosities detected. Currently, this is not the case, so the luminosity
limits are not well constrained. Guetta \& Piran \cite{GRB_GuettaPiran_2005} quoted their results in the
$(7\times10^{49},2.2\times10^{53})\,erg/s$ luminosity range. Zitouni
\emph{et al}. \cite{GRB_Zitouni_et_al2008}, said that usually $L_{min}\sim0.8-3\times10^{50}\,erg/s$
and $L_{max}\sim3-5\times10^{53}\,erg/s$. Daigne \emph{et al}. \cite{GRB_Daigne_et_al2006}
gave their results in the $L_{max}=(10^{50.3\pm0.7},10^{53.5\pm0.4})\,erg/s$
luminosity range. Recently, there have been Swift detections of low
luminosity GRBs $L\lesssim10^{49}\,erg/s$. Various authors have examined
whether these GRBs consist of a new population or are just rare
members of the normal luminosity GRBs. Liang \emph{et al} \cite{GRB_Liang_et_al2007}
have suggested a power-law luminosity function with two breaks and an extended
lower limit ( $L_{min}\sim10^{46}\,erg/s$) (figure \ref{fig:GRBSim_LL_LuminosityFunction}).
Low luminosity GRBs are detectable at only very low redshifts $(z\lesssim0.1)$
and, despite their faintness, could potentially be detectable by
Milagro. However, unless more detections of such GRBs occur, a luminosity
function that appropriately includes them cannot be constructed. For
this study, the ``standard'' luminosity function describing the
main sequence of GRBs was extended to include some of the lower luminosity
GRBs.

\begin{figure}[htb]
\begin{centering}
\includegraphics[width=0.7\columnwidth]{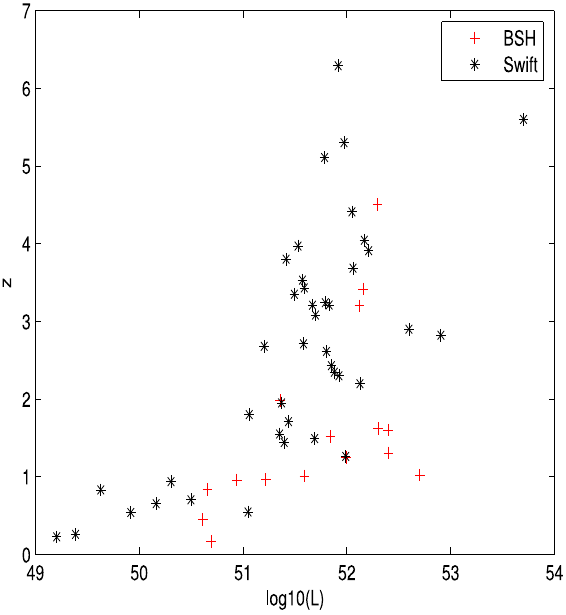}
\par\end{centering}

\caption{\emph{\label{fig:GRBSim_Luminosities}}The luminosities and redshifts
of the BeppoSAX/HETE2 (BSH in the legend) sample compared with the
Swift sample. Source \cite{GRB_Guetta_Piran2007}}

\end{figure}

In general, studies using similar initial assumptions reach compatible,
within errors, results. More data from Swift, and especially more
short GRB detections and more GRBs with resolved redshifts, would
help identify the most appropriate functional form for the luminosity
function, constrain its parameters, and identify separate GRB classes
contributing to the main distributions (such as low luminosity GRBs).
The data used in this study are in agreement with the results from
most of the latest papers and are shown on table \vref{tab:GRBSim_Data}.
The limits on the luminosity ranges are consistent with all the above
mentioned studies. The low limit is somewhat decreased, but still
in agreement with other results, in order to include the recently
detected low-luminosity GRBs. As can be seen from figure \ref{fig:GRBSim_Luminosities},
where the luminosities of BeppoSAX/HETE2 and Swift GRBs are shown,
the luminosity limits used here include most of the detected GRBs. 

\subsection{\label{sub:GRBSim_psigrb}Flux-limited selection function}

In this section, based on the minimum peak flux detectable by Swift
and the peak luminosity function of GRBs from the previous subsection,
the flux-limited selection function will be calculated. This function
can be used to calculate the redshift distribution of Swift-detected
GRBs (needed for the simulation of the GRB population) from the intrinsic
redshift distribution of GRBs (calculated above). 

A GRB of some peak-luminosity from some redshift will produce a peak
flux at the earth. By comparing this peak flux with the minimum peak
flux a detector can detect, a decision can be made on whether this
GRB will be detected or not. Using the distribution of peak luminosities
(from previous section), the distribution of peak-fluxes produced
by GRBs at some redshift can be calculated. Comparing this peak-flux
distribution with the minimum peak-flux an instrument can detect,
the fraction of detectable GRBs of some redshift can be calculated.
This fraction depends on the redshift and is called the flux-limited
selection function $\psi_{flux}(z).$

The peak flux $P(L,z)$ at the earth by a GRB of redshift $z$ and
peak luminosity $L$ is \begin{equation}
P(L,z)=\frac{L}{4\pi D_{l}^{2}(z)}\kappa(z)\label{eq:GRB_PeakFlux},\end{equation}
where $D_{l}$ is the luminosity distance and $\kappa(z)$ is a function
of order unity called the k-correction: \begin{equation}
\kappa(z)=\frac{\int_{(1+z)E2}^{(1+z)E1}S(E)dE}{\int_{E2}^{E1}S(E)dE},\end{equation}
where $S(E)$ is the Band function (eq. \vref{eq:GRB_BandFunction}). The k-correction corrects
for the fact that a detector sensitive at an energy range $(E_{1},\, E_{2})$
will measure the signal emitted by a GRB of distance $z$ emitted
at the blue-shifted energy range $\left((1+z)\times E,\,(1+z)\times E_{2}\right)$.
The units of $S(E)$ are $photons\,s^{-1}\,keV^{-1}$. $S(E)dE$ is the
number of photons emitted per second in the interval $E$ to $E+dE$.
Integrating $E\times S(E)dE$ over some energy band gives the total
energy emitted in that energy range in the source rest frame. The
values of the Band function spectral indices were estimated by measurements
of bright BATSE bursts by Preece \emph{et al}. \cite{GRB_Preece_et_al2000}
as $\alpha=-1$ and $\beta=-2.25$ with $E_{0}=256\,keV$%
. 
The units of the peak flux are $erg\,cm^{-2}\,s^{-1}$, and the
units of the peak luminosity are $erg/s$.

\begin{figure}[htb]
\includegraphics[width=1\columnwidth]{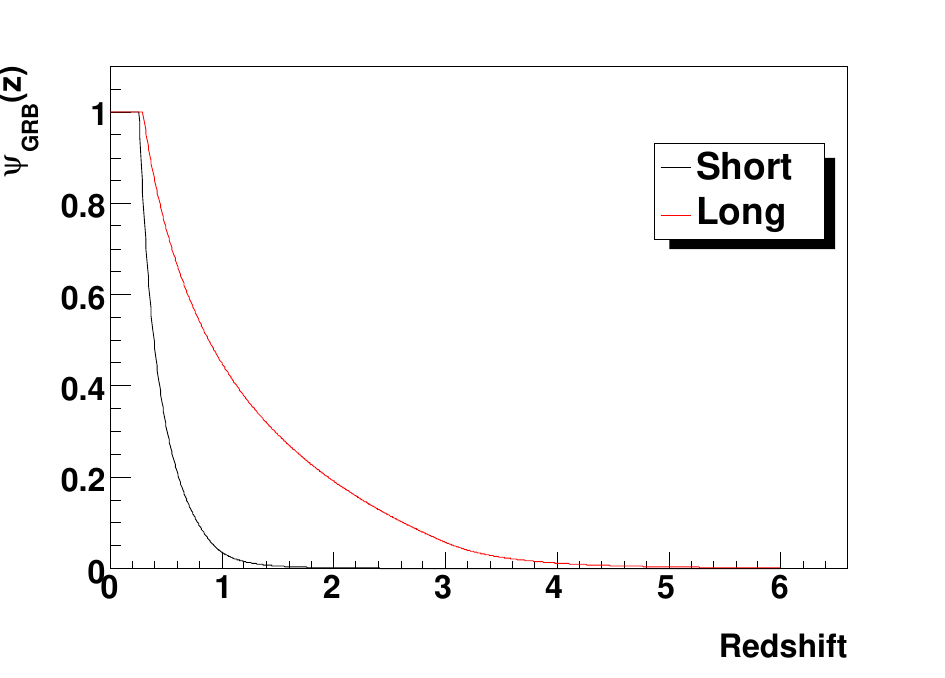}

\caption{\label{fig:GRBSim_PsiGRB}Flux-limited selection function.}

\end{figure}

If $P_{lim}$ is the minimum detectable peak flux, then the fraction
of GRBs detectable at some redshift is \begin{equation}
\psi_{flux}(z)=\frac{\int_{L_{lim}(P_{lim,}z)}^{L_{max}}\frac{dN}{dL}dL}{\int_{L_{min}}^{L_{max}}\frac{dN}{dL}dL}\label{eq:GRB_psiFlux},\end{equation}
where $L_{lim}(P_{lim},z$) is obtained by solving equation \ref{eq:GRB_PeakFlux}
with $P(L,Z)$ substituted with $P_{lim}$. The minimum peak photon
fluxed Swift can detect for long and short GRBs are $\sim0.2ph\,cm^{-2}\,s^{-1}$
and $\sim1.5ph\,cm^{-2}\,s^{-1}$ respectively. Some incompleteness
is expected just over these thresholds, so the limits used here were
$0.5ph\,cm^{-2}\,s^{-1}$ and $2ph\,cm^{-2}\,s^{-1}$\cite{GRB_Salvaterra_et_al2007}
for long and short GRBs. Using an average photon index in the Swift
sample of $a=-1.5$, the energy of a typical photon in the Swift $15-150\,keV$
energy range is $<E>=\frac{\int_{15}^{150}E\,S(E)dE}{\int_{15}^{150}S(E)dE}=7.34\times10^{-8}erg$.
Using $<E>$, the peak photon-flux thresholds for Swift correspond
to peak energy-flux thresholds of $P_{lim,long}=3.7\times10^{-8}\,erg\,cm^{-2}\,s^{-1}$
and $P_{lim,short}=1.5\times10^{-7}\,erg\,cm^{-2}\,s^{-1}$. The typical
value for the minimum detectable peak flux for BATSE is $\sim10^{-7}\,erg\,cm^{-2}\,s^{-1}$.
The resulting flux-limited selection function (eq. \ref{eq:GRB_psiFlux})
that corresponds to the Swift sensitivity is shown in figure \ref{fig:GRBSim_PsiGRB}.%

\subsection{\label{sub:GRBSim_DetectedZ}Redshift distribution of detected GRBs}

\begin{figure}[htb]
\begin{centering}
\subfigure[Differential distribution]{\includegraphics[width=0.8\columnwidth]{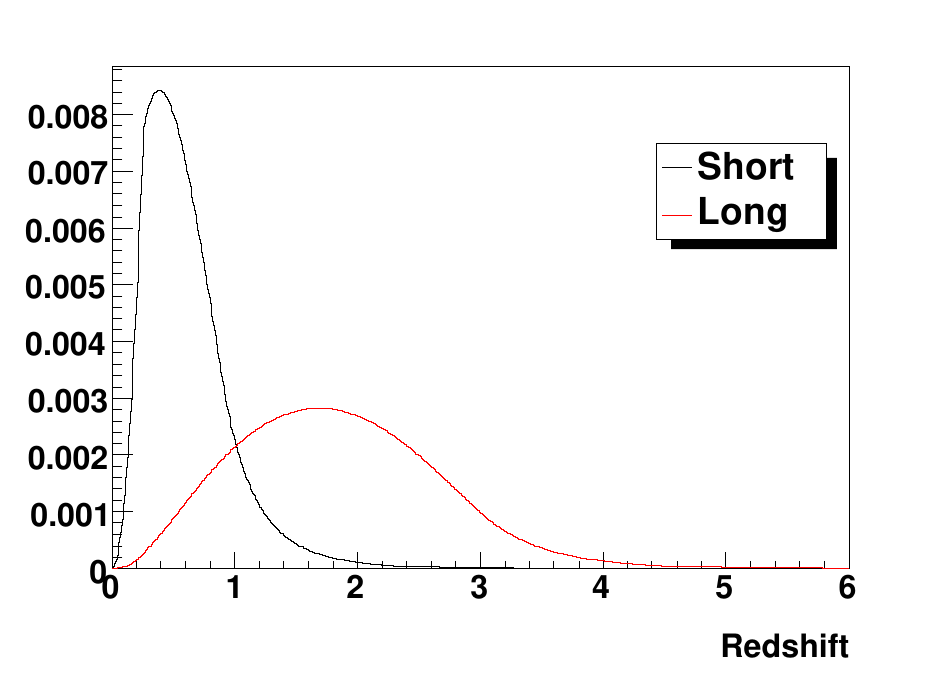}}
\par\end{centering}

\begin{centering}
\subfigure[Integral distribution]{\includegraphics[width=0.8\columnwidth]{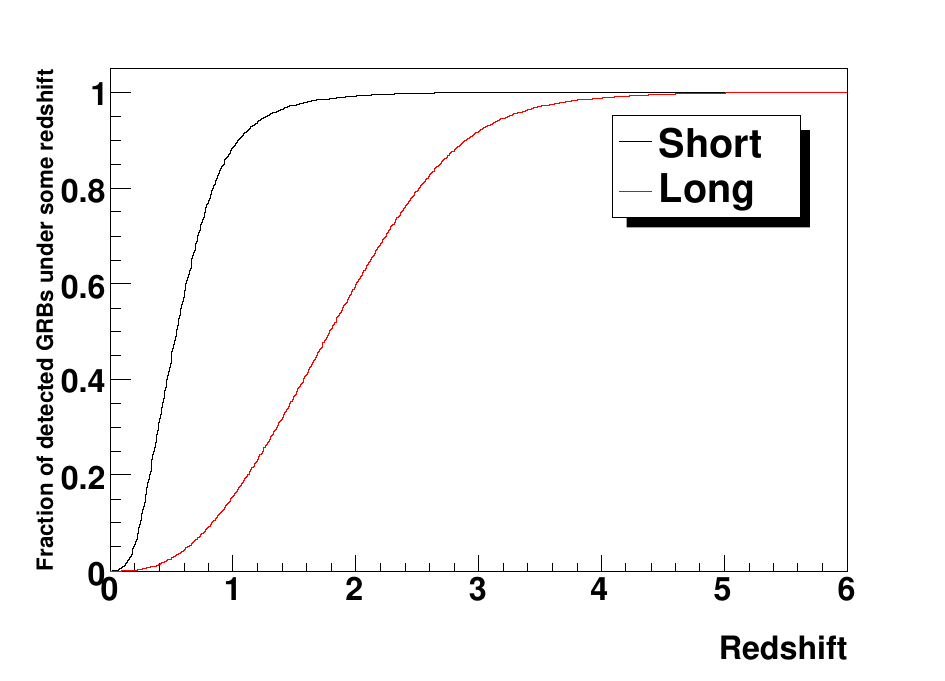}}
\par\end{centering}

\caption{\label{fig:GRBSim_Probz}Redshift distribution of short and long GRBs
detected by Swift.}

\end{figure}

After having calculated the intrinsic redshift distributions of GRBs
and the fraction of GRBs detectable at each redshift, the redshift
distribution of detected GRBs can be calculated as: 

\begin{equation}
\frac{d\dot{N}_{det}(z)}{dz}=\psi_{flux}(z)\,\frac{d\dot{N}_{GRB}(z)}{dz},\end{equation}
where $\psi_{flux}(z)$ has been calculated in the previous subsection
(eq. \ref{eq:GRB_psiFlux}), and $\frac{d\dot{N}_{GRB}(z)}{dz}$ is
given by equations \ref{eq:GRBSim_Intrinsic_Long} and \ref{eq:GRBSim_Intrinsic_Short}
for long and short bursts respectively. The differential and integral
redshift distributions of Swift-detected short and long GRBs are shown
in figure \ref{fig:GRBSim_Probz}. These distributions will be later
used in the simulation of the GRB population. The maximum simulated
redshift was $z_{max}=3.0$. This is because for higher amounts of
VHE emission, GRBs of higher redshifts become detectable. This maximum
redshift ensures that only a negligible fraction of detectable GRBs
will be at a redshift higher than 3.0. %

\section{\label{sub:GRBSIM_Eiso}Isotropic-Equivalent Emitted Energy}

In this section the isotropic-equivalent emitted energy $E_{iso}$
of Swift-detected GRBs will be calculated. As in the case of the redshift
distribution, it will be shown that the $E_{iso}$ distribution of
detected GRBs can be calculated from a combination of the intrinsic
$E_{iso}$ distribution and a detector-specific selection function.
Initially, the selection function will be calculated, and then the
choice for the intrinsic $E_{iso}$ distribution used in this study
will be described.

\begin{figure}[htb]
\begin{centering}
\includegraphics[width=0.8\columnwidth]{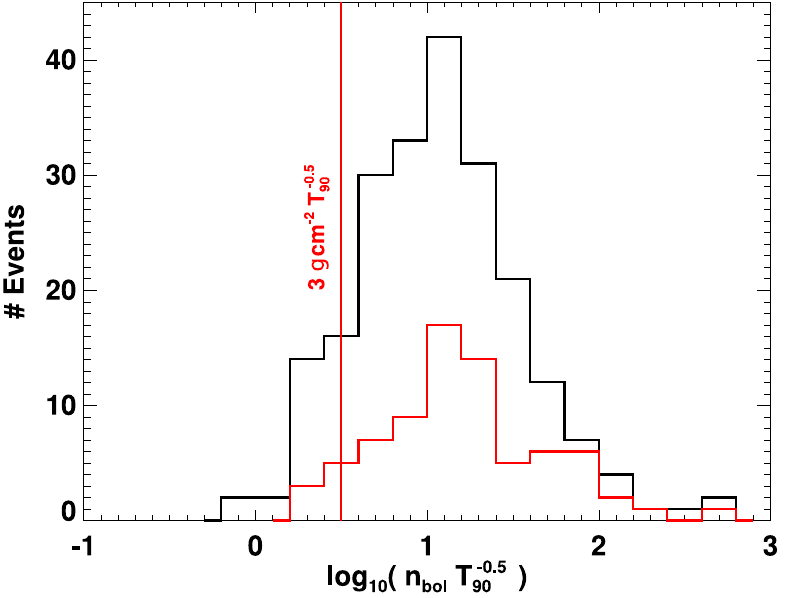}
\par\end{centering}

\begin{centering}
\includegraphics[width=0.8\columnwidth]{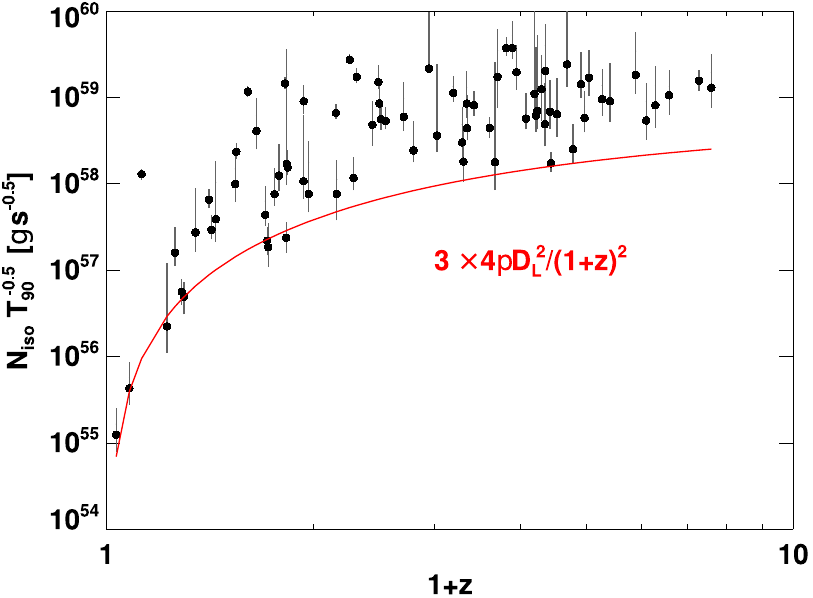}
\par\end{centering}

\caption{\label{fig:GRBSim_FloroiLimts}Effective Swift thresholds on $n_{bol}/\sqrt{T_{90}}$
and $N_{iso}/\sqrt{T_{90}}$ from Butler \emph{et al}. \cite{GRB_Floroi2007}}

\end{figure}

\begin{figure}[htb]
\includegraphics[width=1\columnwidth]{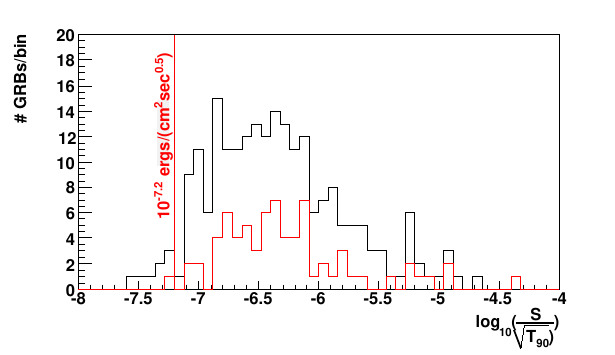}
\includegraphics[width=1\columnwidth]{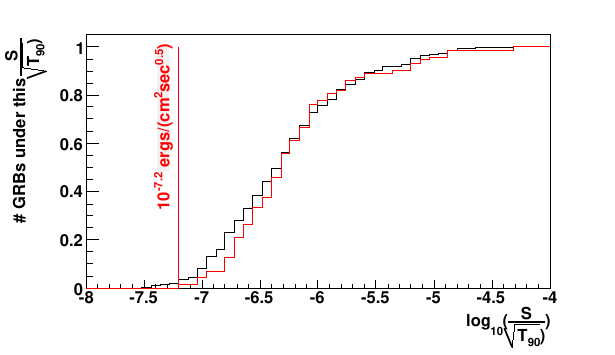}
\caption{\label{fig:GRBSim_myFlimits}Effective Swift threshold on $S/\sqrt{T_{90}}$.}

\end{figure}

The prompt emission from GRBs was observed in the $\sim20keV-2MeV$
energy range by BATSE. However, Swift is sensitive to an energy range
of smaller width $(15-150\,KeV)$. Butler \emph{et al}. \cite{GRB_Floroi2007},
using the strong constraints on the spectral characteristics of GRBs
set by measurements by earlier instruments, managed to derive the
bolometric $(1keV-10\,MeV)$ peak fluxes and peak energies from 218
Swift bursts. Seventy-seven of these bursts had resolved redshifts,
so in addition, they calculated the isotropic-equivalent bolometric
emitted energies. Using these data, they calculated an effective
Swift detection threshold in terms of $n_{bol}/\sqrt{T_{90}}$ (a
quantity close to the signal to noise ratio). From that threshold,
the minimum $N_{bol}/\sqrt{T_{90}}$ was also found versus redshift.
Here, $n_{bol}$ is the number of particles per unit of area reaching
the earth integrated in the $1\,keV-10\,MeV$ energy range and for the
duration of the burst, and $N_{bol}$ is the isotropic-equivalent
total amount of particles emitted from the GRB at the same energy
range and duration
\footnote{If $<E>$ is the average energy of an emitted photon, then $E_{iso}=<E>\times N_{bol}$
and $S=<E>\times n_{bol}$.%
}. Butler \textit{et al.}'s results are shown in figure \ref{fig:GRBSim_FloroiLimts}.
 Similarly to their work, an effective trigger threshold can be found
on $S/\sqrt{T_{90}}$ and $E_{iso}/\sqrt{T_{90}}$, something that
is more directly related to this work. As shown in figure \ref{fig:GRBSim_myFlimits},
an effective threshold of $S/\sqrt{T_{90}}\simeq10^{-7.2}\,erg\,cm^{-2}\,s^{-0.5}$
also exists in the Swift data. Using this threshold, the minimum $E_{iso}/\sqrt{T_{90}}$
can be calculated for any redshift.%
\begin{figure}[htb]
\includegraphics[width=1\columnwidth]{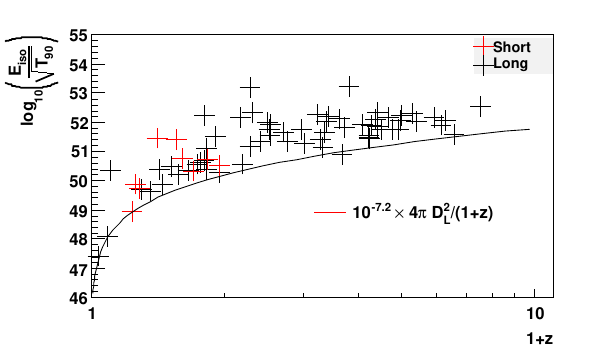}

\caption{Demonstration of the existence of an effective threshold in 
the Swift data.\label{fig:GRBSim_myEisolimits}\emph{Black line}: effective Swift
threshold on $E_{iso}/\sqrt{T_{90}}$ used in the simulation, \emph{crosses}:
Swift data from \cite{GRB_Floroi2007}.}

\end{figure}

Similarly to what was mentioned above regarding the redshift distributions,
the isotropic energy distribution of detected GRBs is the product
of the intrinsic distribution multiplied by the detector threshold. The intrinsic
$E_{iso}$ distribution of GRBs depends on the structure of the GRB
jet. In the universal jet profile model \cite{GRB_Rossi_et_al2002},
\cite{GRB_Zhang_Meszaros2002}, all jets have the same surface energy
density $\epsilon$ as a function of the off-axis angle $\theta$,
and the observed differences in $E_{iso}$ result from the angle $\theta_{u}$
between the jet axis and the line of sight. This model predicts an
isotropic energy distribution following a power law: $P(E_{iso})\propto E_{iso}^{-a_{E}}$.
If ($\epsilon\propto\theta^{k})$, then $a_{E}=1-2/k$. Rossi \emph{et
al}. \cite{GRB_Rossi_et_al2002} proposed $k=-2$ ($a_{E}=2$), in
order to reproduce the data observed by Frail \emph{et al}. \cite{GRB_Beaming_Frail2001}.
If the energy density follows a Gaussian distribution, then $a_{E}=1$ \cite{GRB_Lloyd-Ronning_et_al2004}.
Band \emph{et al}. \cite{GRB_Band_et_al2004}, using the lag-luminosity
relation, calculated the redshifts and the isotropic energies from
a large sample of BATSE GRBs. They modeled the resulting $E_{iso}$
sample with power law and with Gaussian probability distributions without
a redshift evolution. For the first case, they found that $a_{E}=1.76\pm0.05$. 

In this work, the intrinsic $E_{iso}$ distribution will be
modeled by a power law with an index that brings the best agreement
between the simulation results and the experimental data. After comparing
the $E_{iso}-z$ and $S-z$ distributions between the simulation and
the Swift data from Butler \textit{et al}, the best agreement was for
$a_{E}=1.45$. That value was close to the results mentioned above.

\section{\label{sub:GRBSim-GRB-Rates}GRB Rates}

Because the probability that a dying star or a compact-object binary
will produce a GRB are not well constrained, the absolute GRB rates
cannot be accurately calculated directly from the SFR. Thus,
the GRB rates used in this study will be the ones detected by Swift.
Up to 04/13/2008, Swift had detected 319 GRBs in the 3.399 years of its
operation. During its lifetime, 16\% of the time it was not
operating because it was passing through the South Atlantic Anomaly,
15.5\% of the time it was not sensitive to new GRBs because it was
slewing, and 1.5\% of the time it was not operating because of various
other problems such as problems with the gyroscopes%
\footnote{Neil Gehrels, private communication.%
}. So, the amount of time, in the past 3.399 years, that Swift was
sensitive to new GRBs was only 2.28 years, which means that it was detecting
$140GRBs/live\, year$. Swift's field of view is $1.4\,sr$. This study
searched the overhead sky up to a zenith angle of $45^{o}$ or $1.84\,sr$.
So, according to Swift rates, there have been $1.84/1.4*140=184$ detectable-by-Swift
GRBs per year in Milagro's field of view (up to $45^{o}$ zenith angle).
This study searched in $4.58\,yrs$ of Milagro data, which corresponds
to $\simeq843$ detectable-by-Swift GRBs in Milagro's FOV.

\section{\label{sub:GRBSIM_Model}Model for VHE Emission from GRBs}

A simple model was used for the form of the VHE emission from GRBs. According to this model:
\begin{itemize}
\item Only a fraction of GRBs has VHE emission.
\item All GRBs with such an emission, emit in the same ($E_{VHE,min}$,\,$E_{VHE,max}$) energy range on a power-law spectrum with the same spectral index $\alpha$.
\item The amount of isotropic energy emitted in the VHE energy range is related to
the amount of isotropic energy emitted in the $1\,keV-10\,MeV$ energy range.
Specifically, the energy output per decade of energy in the ($E_{VHE,min},\,E_{VHE,max}$)
energy range is proportional to the energy output per decade
of energy in the $1\,keV-10\,MeV$ energy range: \begin{equation}
\frac{E_{iso}(E_{VHE.min}-E_{VHE,max})}{log_{10}(E_{VHE,max}/E_{VHE,min})}=\mathcal{\mathcal{R}}\times\frac{E_{iso}(1\,KeV-10\,MeV)}{log_{10}(10^{4}\,keV/1\,keV)}.\label{eq:GRBSIM_RDef}\end{equation}
These energies are for the non-redshifted frame of reference of the GRB.
\end{itemize}

GRB populations emitting in different VHE energy ranges, with different spectral indices $\alpha$, and ratios $\mathcal{R}$
were simulated. The spectral indices ranged from $\alpha=-2.0$ to 
$\alpha=-3.5$. The extent of the VHE energy range was set by Milagro's sensitive energy range.
According to figure \vref{sub:IRAbs_Effects}, Milagro can probe the $40\,GeV-15\,TeV$
emission from GRBs. However, because of internal absorption effects, the VHE emission from GRBs may
cutoff at an energy lower than $15\,TeV$. Various VHE emission models,
each for a different maximum emitted energy $E_{VHE,max}$, ranging from $150\,GeV$ to $15\,TeV$,  were simulated. The minimum 
emitted energy was always $E_{VHE,min}=40\,GeV$. Upper limits were set for each combination of the values of
$\alpha$ and $E_{VHE.max}$.

\section{\label{sub:GRBSIM_Verification}Verification of the GRB Simulation}

In this section I will compare some of the results of the simulation
with the distributions of Swift GRBs in order to verify its validity.
The combination of isotropic energy and redshift distributions used
in the simulation will produce a fluence distribution that should
match the one from Swift data. 

\begin{figure}[htb]
\subfigure[Short GRBs]{\includegraphics[width=1\columnwidth]{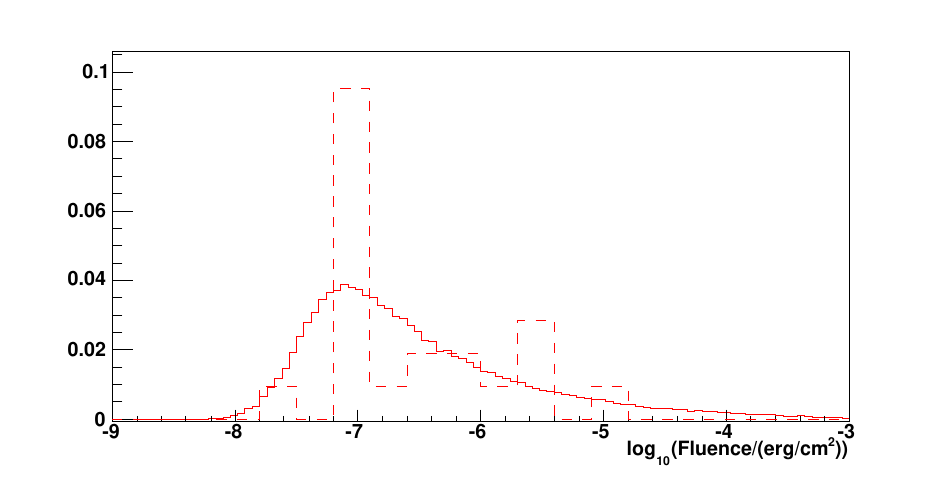}}

\subfigure[Long GRBs]{\includegraphics[width=1\columnwidth]{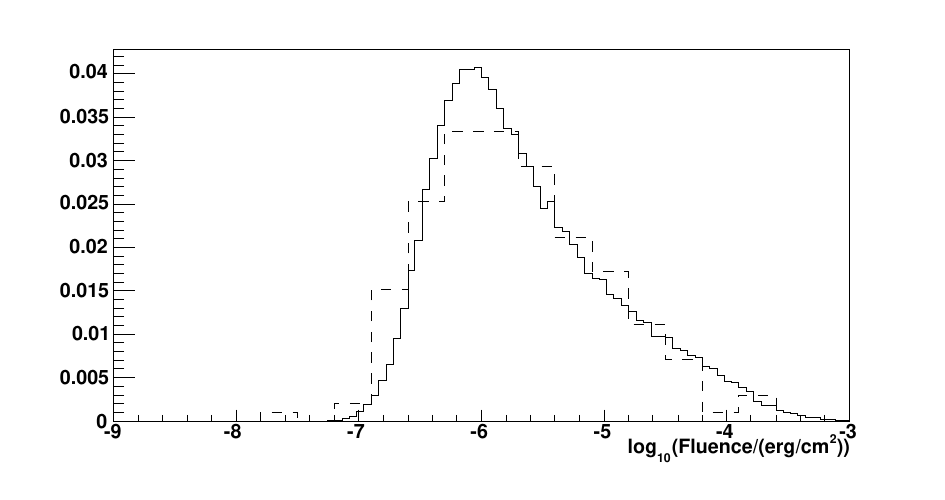}}

\caption{\label{fig:GRBSim_SimFluence}Bolometric $1\,keV-10\,MeV$ fluence. \emph{Solid
lines:} results from the simulation, \emph{dashed lines}: Swift data from \cite{GRB_Floroi2007}}

\end{figure}

Figure \ref{fig:GRBSim_SimFluence}
shows a comparison between the $1\,keV-10\,MeV$ bolometric fluences of
the GRBs in the simulation and of the GRBs detected by Swift (Butler
\textit{et al.} catalog \cite{GRB_Floroi2007}). The agreement between the
results of the simulation and the measured data is excellent, considering
the low statistics in the number of detected GRBs. If any of the elements
of the simulation mentioned above were considerably wrong, then these
curves would show a disagreement. It should be noted that the fluence
is measured without the need to measure a redshift, so the Swift curves
in the figure are representative of the whole population of Swift-detected
bursts. The simulation should be able to match that exact population,
and the excellent agreement provides a strong verification of the
simulation.%
\begin{table}
\begin{centering}
\begin{tabular}{|c|c|}
\hline 
\textbf{Quantity} & \textbf{Value}\tabularnewline
\hline
\hline 
\multicolumn{2}{|c|}{\textbf{Luminosity Function}}\tabularnewline
\hline 
Functional form & Broken power law (eq. \vref{eq:GRBSim_SchmidtLum})\tabularnewline
\hline 
$L_{min}(erg/s)$ & \multicolumn{1}{c|}{$3\times10^{49}\,erg/s$}\tabularnewline
\hline 
$L_{max}(erg/s)$ & \multicolumn{1}{c|}{$4\times10^{53}\,erg/s$}\tabularnewline
\hline 
Spectral indices $(a,\beta)$ (short) & (1.6, 2.8)\tabularnewline
\hline
Spectral indices $(a,\beta)$ (long) & (1.1, 3)\tabularnewline
\hline
$L^{*}$(short, long) & ($6.35\times10^{50}$,\,$4\times10^{51}$) $erg/s$\tabularnewline
\hline
\hline 
\multicolumn{2}{|c|}{\textbf{Intrinsic Redshift Distribution}}\tabularnewline
\hline
Short & HB SFR$\times$Time delay factor$(\tau)$\tabularnewline
\hline
$n$ of $P(\tau)=\tau^{n}$ & $n=-1$ \tabularnewline
\hline
$\tau_{min}$ & 20MYr\tabularnewline
\hline
Long & HB SFR$\times$Maximum Metallicity Limit\tabularnewline
\hline
Maximum Metallicity $Z_{max}$& $0.1,0.3,0.6\,Z_{\odot}$\tabularnewline
\hline
\hline 
\multicolumn{2}{|c|}{\textbf{Intrinsic $E_{iso}$ distribution}}\tabularnewline
\hline
$P(E_{iso})\propto{}E_{iso}^{-a_{E}}$ & $a_{E}=1.45$\tabularnewline
\hline
Energy range for low energy emission & $1\,keV-10\,MeV$\tabularnewline
\hline
Minimum of the VHE range & $40\,GeV$\tabularnewline
\hline
Maximum of the VHE range & From $150\,GeV$ to $15\,TeV$\tabularnewline
\hline
\hline 
\multicolumn{2}{|c|}{\textbf{Swift Detection Thresholds}}\tabularnewline
\hline
Min peak flux (short) & $2\,ph\,cm^{-2}\,s^{-1}$ or $3.7\times10^{-8}\,erg\,cm^{-2}\,s^{-1}$\tabularnewline
\hline
Min peak flux (long) & $0.5\,ph\,cm^{-2}\,s^{-1}$ or $1.5\times10^{-7}\,erg\,cm^{-2}\,s^{-1}$\tabularnewline
\hline
Min $\frac{S}{\sqrt{T_{90}}}$ & $10^{-7.2}erg\,cm^{-2}\,s^{-0.5}$\tabularnewline
\hline
\hline 
\multicolumn{2}{|c|}{\textbf{Miscellaneous Parameters}}\tabularnewline
\hline
Swift dead time & 33\%\tabularnewline
\hline
$T_{90}$ distribution & Fits from Swift\tabularnewline
\hline
Maximum simulated redshift & 3.0\tabularnewline
\hline
Energy distribution of VHE emission& $P(E)\propto{}E^{\alpha}$, from $\alpha=-2.0$ to $\alpha=-3.5$ \tabularnewline

\hline
\end{tabular}
\par\end{centering}

\caption{\label{tab:GRBSim_Data}Parameters used in the simulation of the GRB population.}

\end{table}

\begin{figure}[htb]
\begin{centering}
\subfigure[Short GRBs]{\includegraphics[width=1\columnwidth]{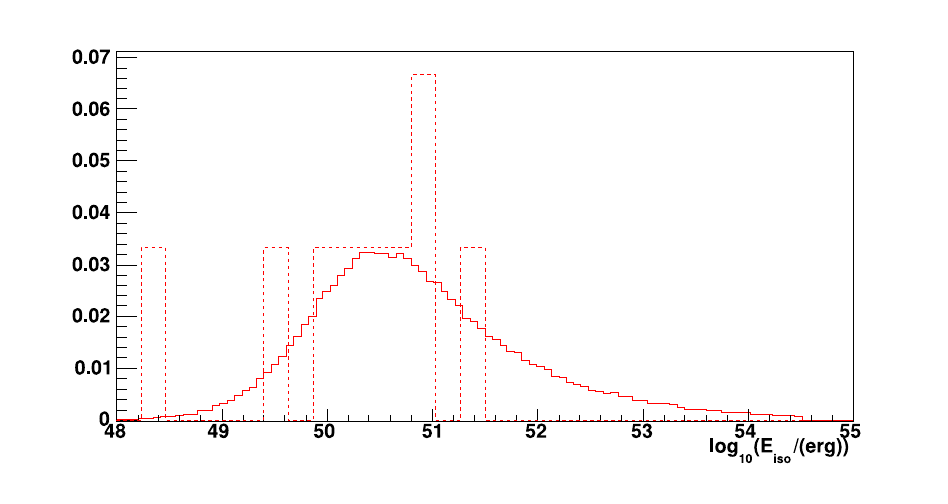}}
\subfigure[Long GRBs]{\includegraphics[width=0.83\columnwidth]{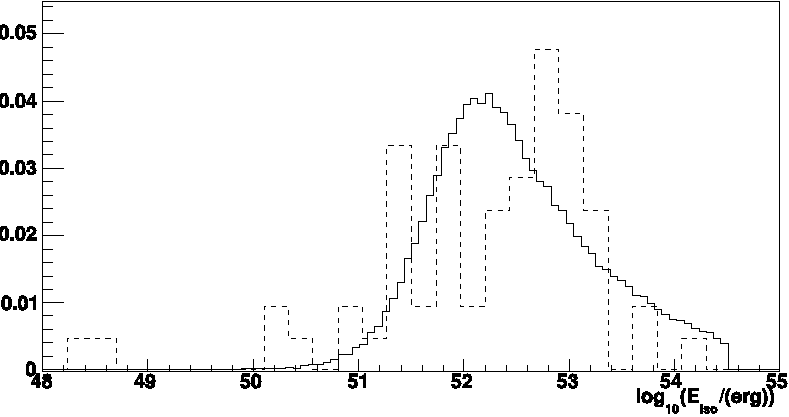}}

\caption{\label{fig:GRBSim_SimEiso}Bolometric $1\,keV-10\,MeV$ isotropic-equivalent
energies emitted. \textit{Solid line}: results of the simulation, \textit{dashed curve}: Swift
data from \cite{GRB_Floroi2007}.)}
\end{centering}
\end{figure}

Continuing with the verification, the simulated and detected bolometric
$1\,keV-10\,MeV$ isotropic energies are shown in figure \ref{fig:GRBSim_SimEiso}.
While the agreement is not as good as in the previous case, this
does not mean the simulated distributions are not correct. The Swift
data used in this comparison are a subgroup of the large population
of Swift-detected GRBs. To be able to calculate the isotropic energy
emitted by a GRB, the redshift is needed, so the detected GRB curves
correspond to the population of Swift bursts with a resolved redshift.
The population of the GRBs with a resolved redshift is not necessarily
representative of all GRBs. There are systematics in redshift determination
that depend on the redshift of the burst (see \cite{GRB_Fiore_et_al_2007,GRB_Berger_New_Population}).

Finally, $\frac{S}{\sqrt{T_{90}}}$ is compared between simulation
and data. Again, the agreement is very good. The detector threshold
applied in the simulation of $\left(S/\sqrt{T_{90}}\right)_{min}=10^{-7.2}\,erg\,cm^{-2}\,s^{-0.5})$
is also evident in the figure. %
\begin{figure}[htb]
\includegraphics[width=1\columnwidth]{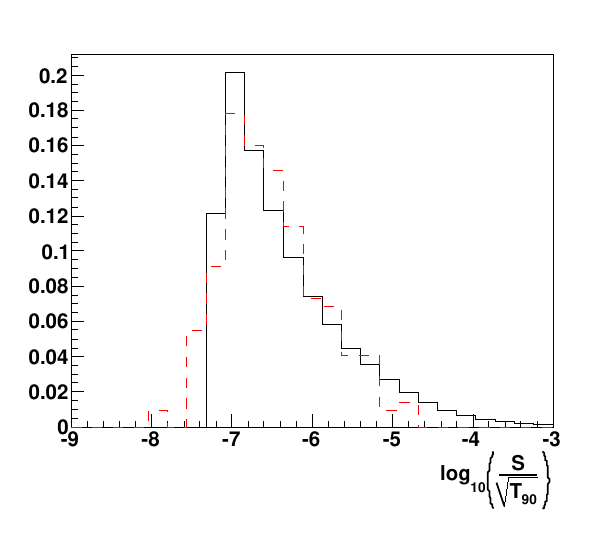}

\caption{\label{fig:GRBSim_SimFluenceT90}Bolometric $1\,keV-10\,MeV$ fluence over
square root of $T_{90}$. \emph{Solid line}: results from the simulation, \emph{dashed
red curve}: Swift data from \cite{GRB_Floroi2007})}

\end{figure}

\FloatBarrier
\section{\label{sub:GRBSIM_Results}Results}

In this section, I will present the results of the simulation
of the GRB population and I will set upper limits on the VHE emission
from GRBs. The upper limits will be versus
the ratio $\mathcal{R}$ (defined in equation \ref{eq:GRBSIM_RDef}),
the fraction of GRBs emitting in very high energies, the spectral index of the 
VHE emission $\alpha$, the maximum energy of the VHE emission $E_{VHE,max}$, and 
the maximum metallicity limit used for calculating the redshift distribution of long 
GRBs $Z_{max}$. 
As mentioned in section
\vref{sub:GRBSIM_T90}, these upper limits will be on the prompt
VHE emission from GRBs. The delayed VHE emission from GRBs was
not included by the simulation, therefore no limits were set on it.

\begin{figure}[htb]
\includegraphics[width=1\columnwidth]{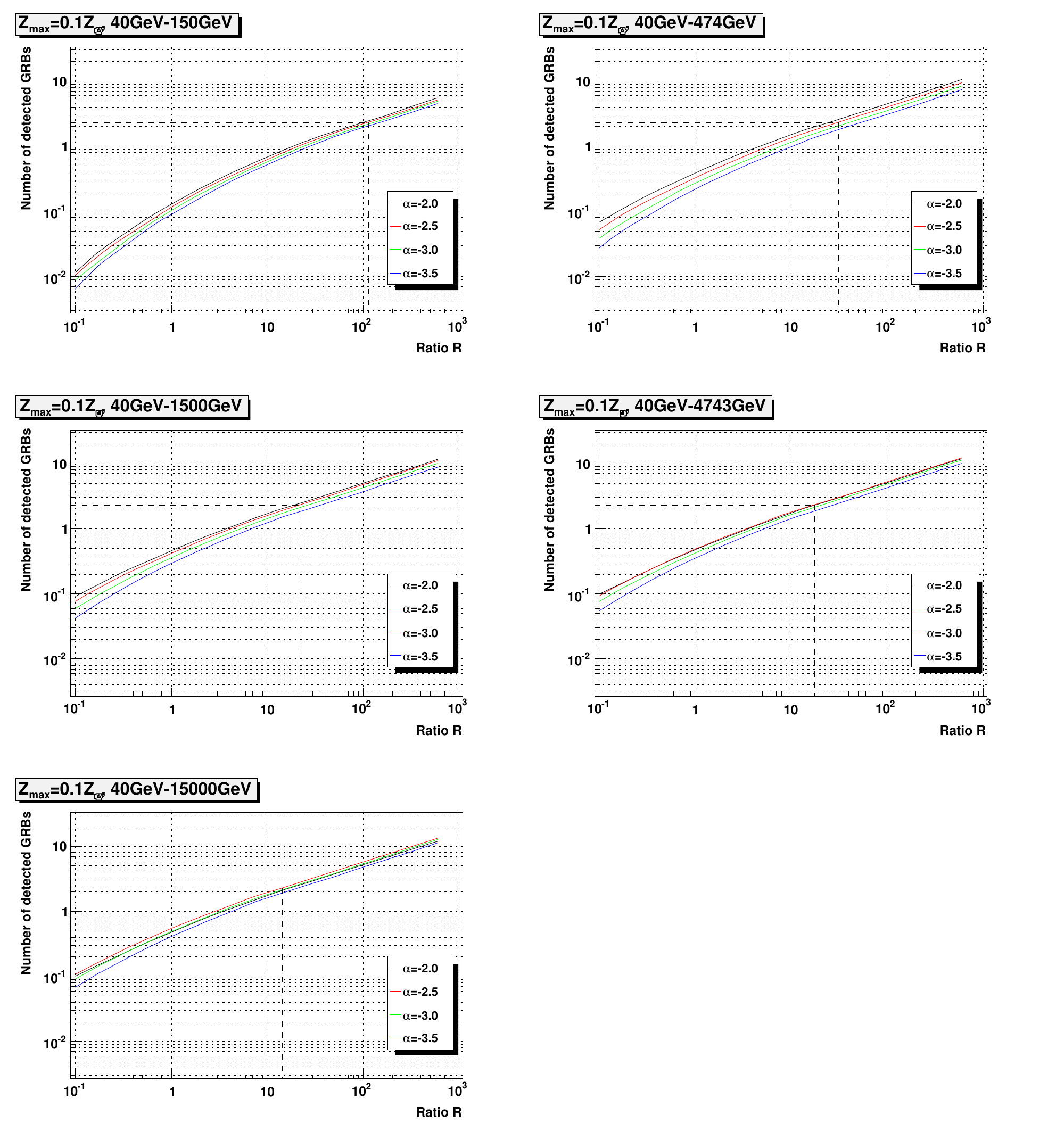}
\caption{\label{fig:GRBSim_res0.1} Number of detected GRBs predicted by the simulation versus the ratio $\mathcal{R}$,
for different values of the spectral index $\alpha$, and the maximum emitted energy $E_{max}$ of the VHE emission from GRBs (\textit{solid lines}).
The dashed lines show the upper limit at the 0.90 confidence level (2.3 GRBs) for $\alpha=-2.5$. These results 
are for the case that all GRBs emit in very high energies, and for an upper metallicity limit $Z_{max}=0.1\,Z_{\odot}$.}
\end{figure}

\begin{figure}[htb]
\includegraphics[width=1\columnwidth]{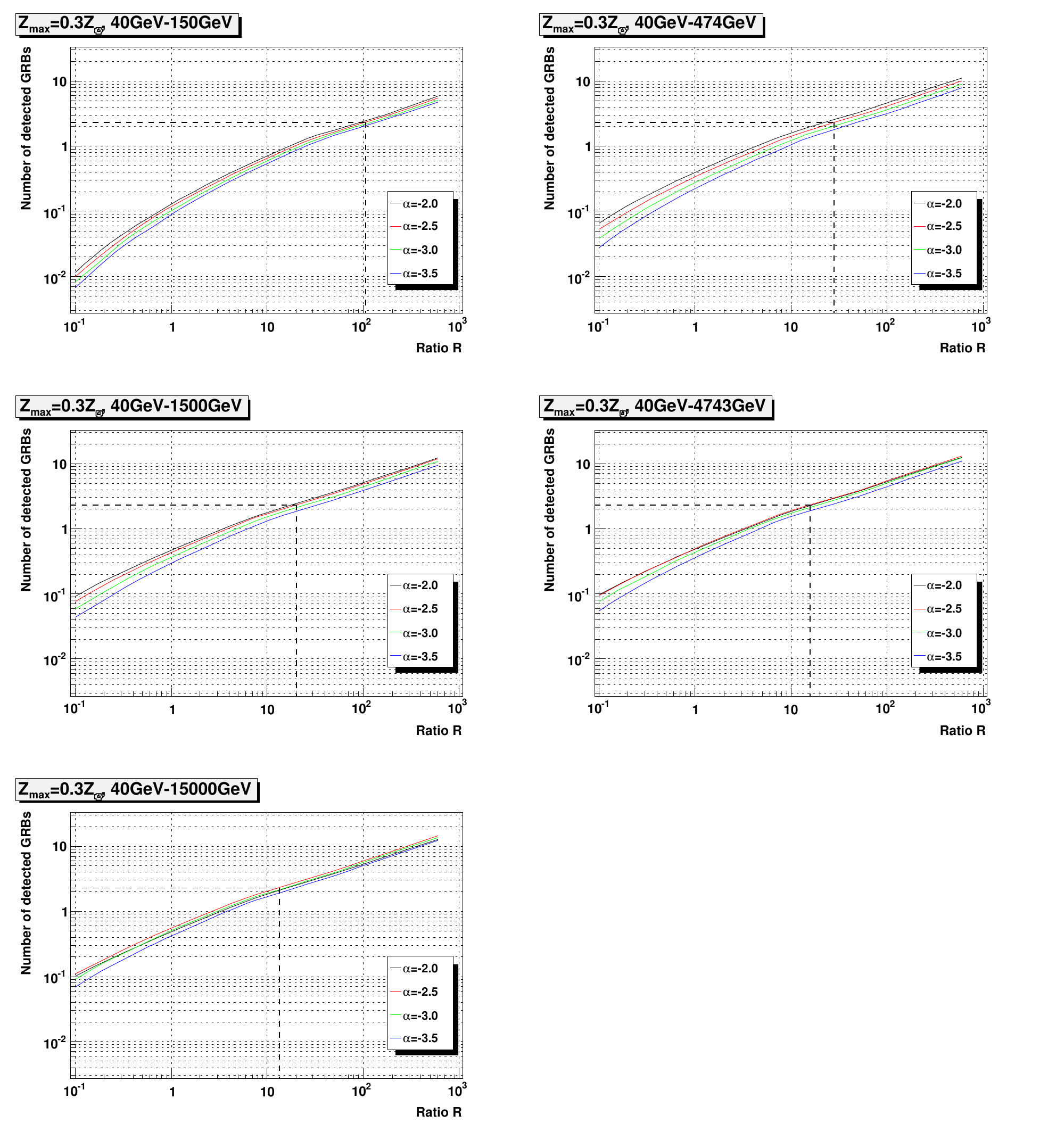}
\caption{\label{fig:GRBSim_res0.3} Number of detected GRBs predicted by the simulation versus the ratio $\mathcal{R}$,
for different values of the spectral index $\alpha$, and the maximum emitted energy $E_{max}$ of the VHE emission from GRBs (\textit{solid lines}).
The dashed lines show the upper limit at the 0.90 confidence level (2.3 GRBs) for $\alpha=-2.5$. These results 
are for the case that all GRBs emit in very high energies, and for an upper metallicity limit $Z_{max}=0.3\,Z_{\odot}$.}
\end{figure}

\begin{figure}[htb]
\includegraphics[width=1\columnwidth]{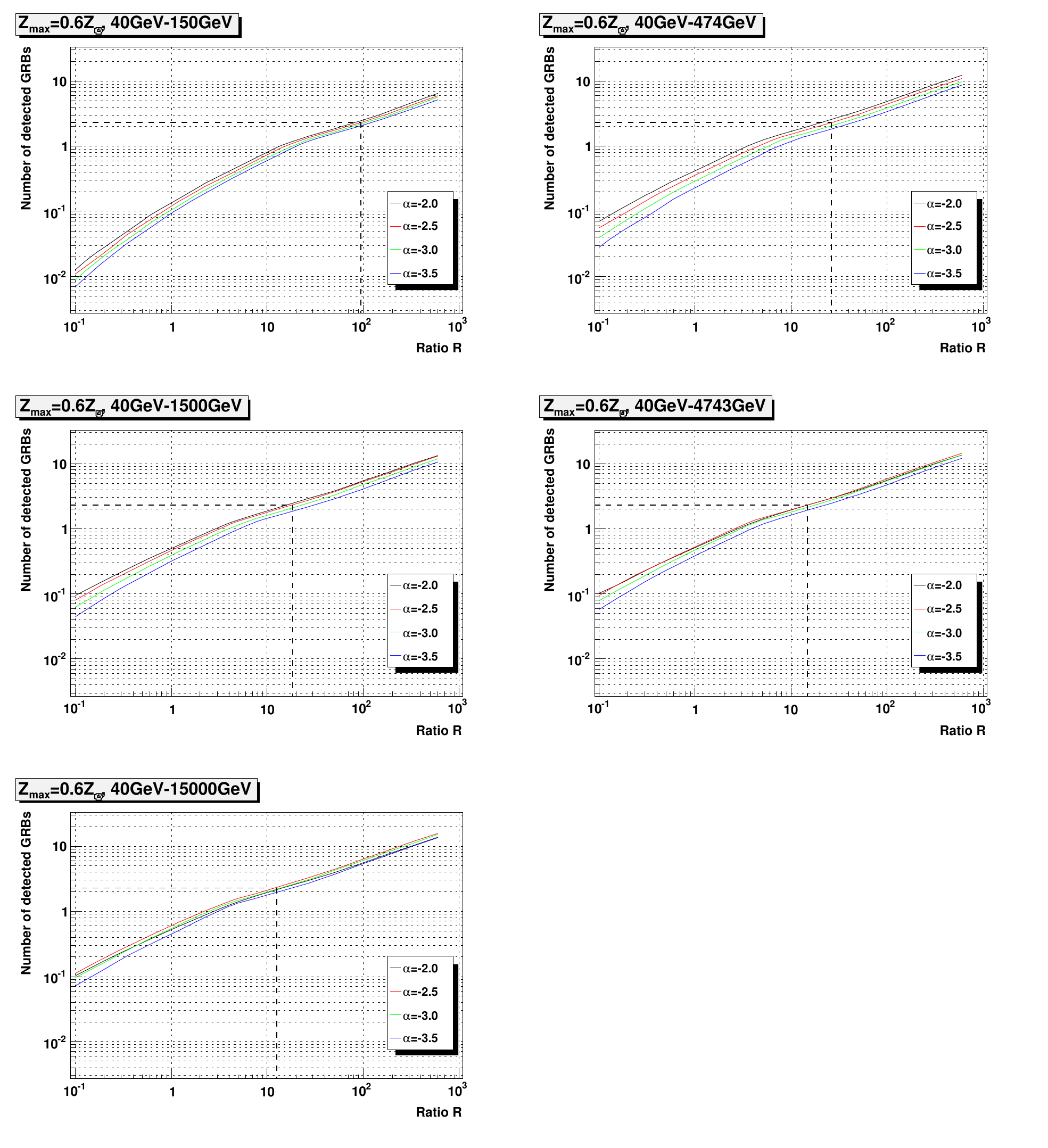}
\caption{\label{fig:GRBSim_res0.6} Number of detected GRBs predicted by the simulation versus the ratio $\mathcal{R}$,
for different values of the spectral index $\alpha$, and the maximum emitted energy $E_{max}$ of the VHE emission from GRBs (\textit{solid lines}).
The dashed lines show the upper limit at the 0.90 confidence level (2.3 GRBs) for $\alpha=-2.5$. These results 
are for the case that all GRBs emit in very high energies, and for an upper metallicity limit $Z_{max}=0.6\,Z_{\odot}$.}
\end{figure}

The number of GRBs Milagro would expect to detect versus the ratio
$\mathcal{R}$, in the case that all GRBs have VHE emission, is shown
with the solid lines in figures \ref{fig:GRBSim_res0.1}, \ref{fig:GRBSim_res0.3},
and \ref{fig:GRBSim_res0.6} for $Z_{max}$ equal to $0.1\,Z_{\odot}$, $0.3\,Z_{\odot}$, and
$0.6\,Z_{\odot}$ respectively.  The results
are for different values of $\alpha$ (different solid lines) and $E_{VHE,max}$ (different graphs). 
The maximum number of detections $D_{exp}$ expected at some confidence
level $CL$ is given by $D_{exp}=-ln(1-CL)$. Therefore, a VHE-emission model
that predicts a number of detections by this search higher than $(2.3,\,3,\,4.6)$
is excluded at the $(0.90,\,0.95,\,0.99)$ confidence level.
The dashed lines of figures \ref{fig:GRBSim_res0.1}, \ref{fig:GRBSim_res0.3},
and \ref{fig:GRBSim_res0.6} show the upper limit 
on $\mathcal{R}$ at the 0.90 confidence level and for $\alpha=-2.5$.
The upper limit for the case that not all GRBs have VHE emission can be easily calculated
by these plots. Specifically, if a fraction $f$ of GRBs emits in the VHE energy range, then the
upper limit on $\mathcal{R}$ at the CL confidence level corresponds to $-ln(1-CL)/f$ detected GRBs.

\begin{figure}[htb]
\includegraphics[width=1.0\columnwidth]{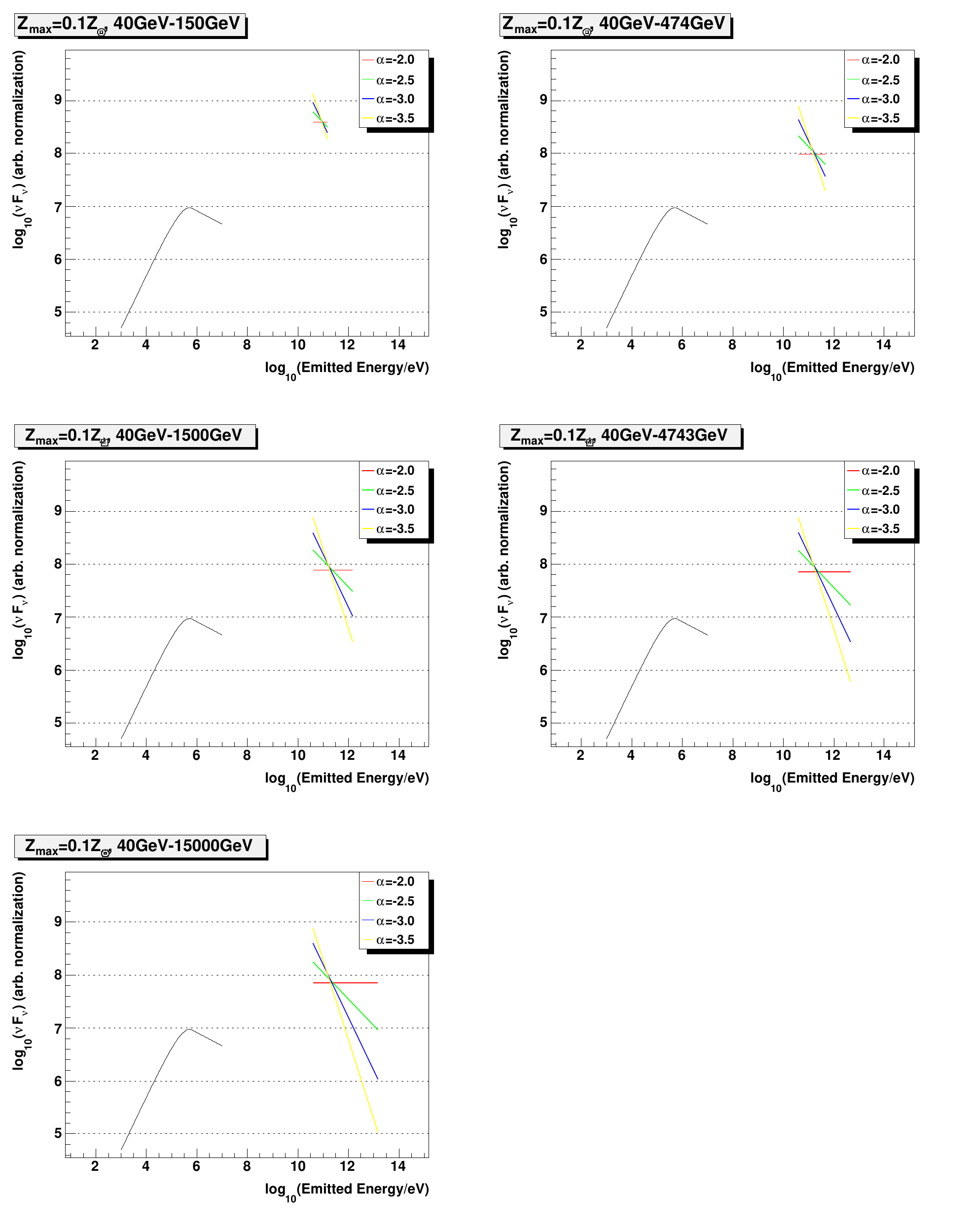}
\caption{\label{fig:GRBSim_Band0.1}Spectral energy distribution of the emission from a GRB. 
\textit{Black line}: typical prompt emission spectrum, \textit{color lines}: upper limits on the prompt emission set by this study.
Each color corresponds to a different spectral index of the VHE emission, and each set of plots to a different maximum 
energy of the VHE emission $E_{VHE,max}$. The energy of the X axis is for the GRB frame of reference (non-redshifted).
The results are for $Z_{max}=0.1\,Z_{\odot}$ and for all GRBs having VHE emission.}
\end{figure}

\begin{figure}[htb]
\includegraphics[width=1.0\columnwidth]{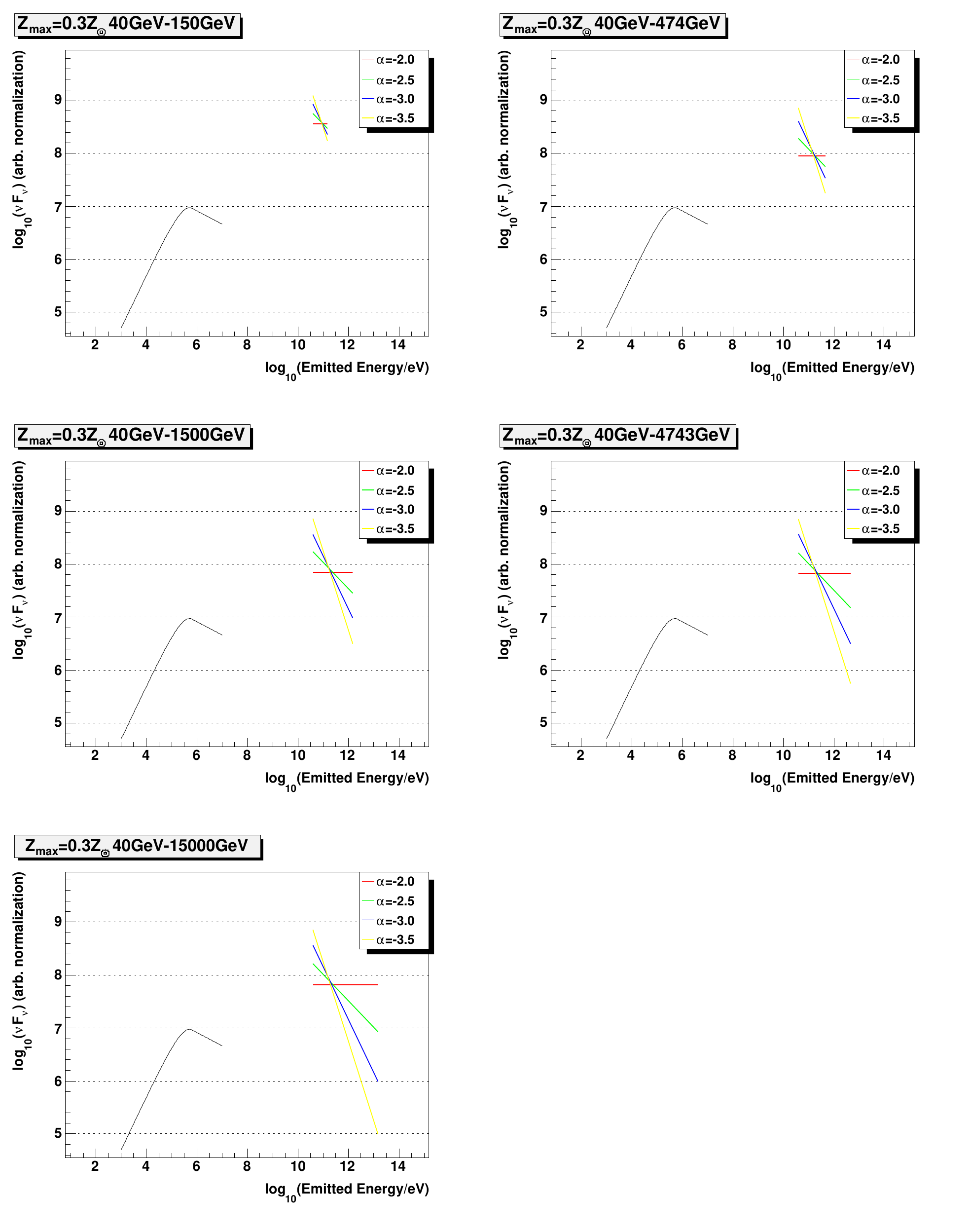}
\caption{\label{fig:GRBSim_Band0.3}Spectral energy distribution of the emission from a GRB. 
\textit{Black line}: typical prompt emission spectrum, \textit{color lines}: upper limits on the prompt emission set by this study.
Each color corresponds to a different spectral index of the VHE emission, and each set of plots to a different maximum 
energy of the VHE emission $E_{VHE,max}$. The energy of the X axis is for the GRB frame of reference (non-redshifted).
The results are for $Z_{max}=0.3\,Z_{\odot}$ and for all GRBs having VHE emission.}
\end{figure}

\begin{figure}[htb]
\includegraphics[width=1.0\columnwidth]{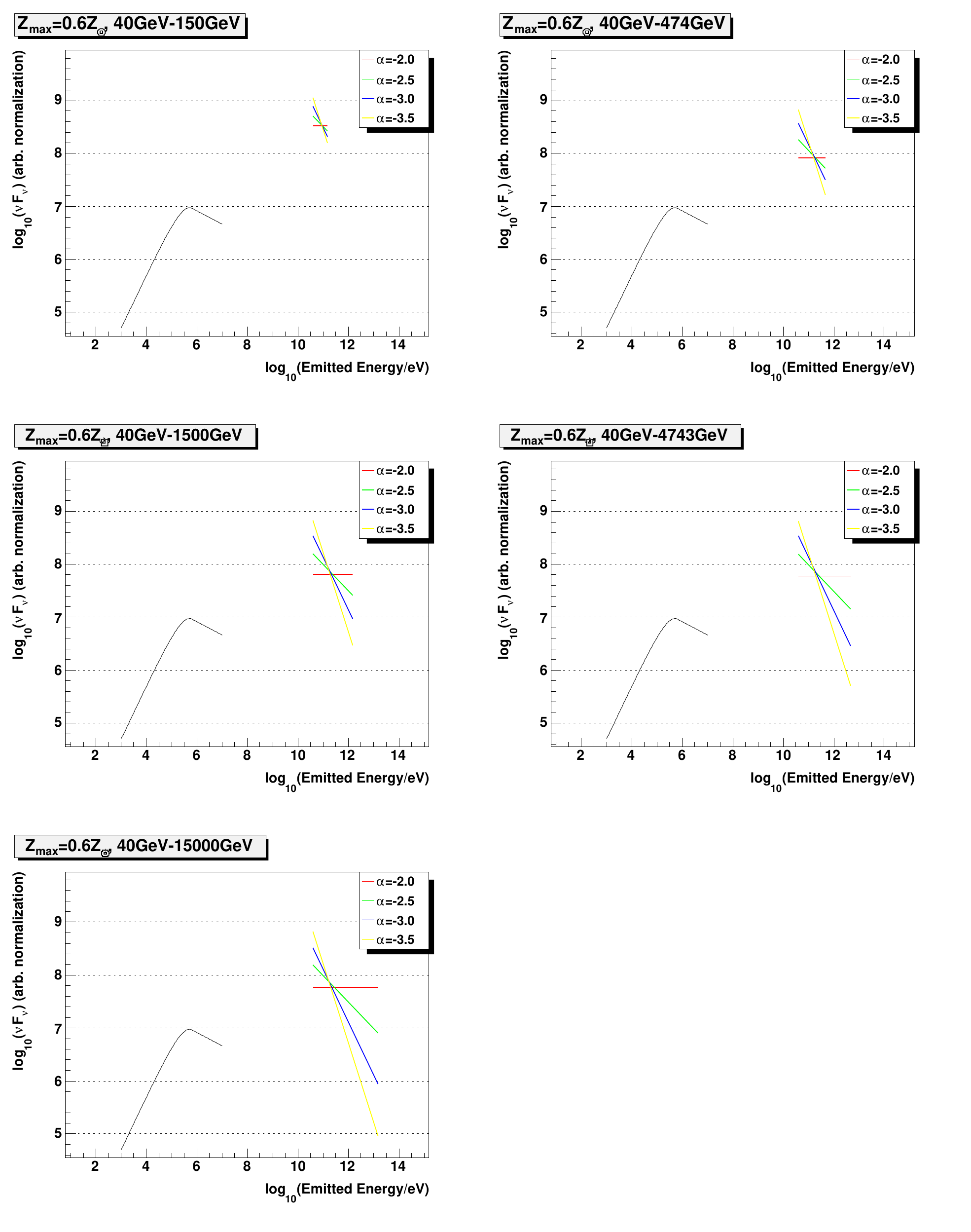}
\caption{\label{fig:GRBSim_Band0.6}Spectral energy distribution of the emission from a GRB. 
\textit{Black line}: typical prompt emission spectrum, \textit{color lines}: upper limits on the prompt emission set by this study.
Each color corresponds to a different spectral index of the VHE emission, and each set of plots to a different maximum 
energy of the VHE emission $E_{VHE,max}$. The energy of the X axis is for the GRB frame of reference (non-redshifted).
The results are for $Z_{max}=0.6\,Z_{\odot}$ and for all GRBs having VHE emission.}
\end{figure}

Figures \ref{fig:GRBSim_Band0.1}, \ref{fig:GRBSim_Band0.3}, 
and \ref{fig:GRBSim_Band0.6} present the upper limits (color lines) in a 
spectral energy distribution along with the prompt emission by GRBs (black curve).
Table \ref{tab:GRBSIM_ResultsTable} summarizes the upper limits for all the
possible combinations of $\alpha$, $E_{VHE,max}$, and $Z_{max}$. The statistical error
on these limits is about $1\%$.

\begin{table}
\begin{centering}
\begin{tabular}{|c|c|c|c|c|c|c|}
\hline 
\backslashbox{$\alpha$}{$E_{VHE,max}$} & $0.150\,TeV$ & $0.474\,TeV$ & $1.5\,TeV$ & $4.74\,TeV$ & $15\,TeV$ & $Z_{max}$\tabularnewline
\hline
-2.0 & 100 & 25 & 20 & 18 & 18 & \multirow{4}{*}{$0.1\, Z_{\odot}$}\tabularnewline 
-2.5 & 113 & 31 & 22 & 18 & 14 & \tabularnewline 
-3.0 & 130 & 41 & 27 & 21 & 17 & \tabularnewline 
-3.5 & 149 & 53 & 36 & 27 & 22 & \tabularnewline  \hline 
-2.0 & 91 & 23 & 18 & 17 & 17 & \multirow{4}{*}{$0.3\, Z_{\odot}$}\tabularnewline 
-2.5 & 106 & 28 & 20 & 16 & 13 & \tabularnewline 
-3.0 & 121 & 38 & 25 & 19 & 16 & \tabularnewline 
-3.5 & 138 & 48 & 37 & 25 & 21 & \tabularnewline \hline 
-2.0 & 84 & 21 & 16 & 15 & 15 & \multirow{4}{*}{$0.6\, Z_{\odot}$} \tabularnewline 
-2.5 & 95 & 26 & 18 & 15 & 13 & \tabularnewline 
-3.0 & 109 & 35 & 24 & 18 & 14 & \tabularnewline
-3.5 & 125 & 45 & 31 & 23 & 19 & \tabularnewline \hline
\end{tabular}
\par\end{centering}

\caption{\label{tab:GRBSIM_ResultsTable}Upper limits at the 90\% confidence level on the ratio $\mathcal{R}$ for different values of the spectral index $\alpha$ and
the maximum emitted energy $E_{VHE,max}$ of the VHE emission from GRBs, and the upper metallicity limit $Z_{max}$. These results are for the case that all GRBs emit in very high energies.}

\end{table}

It should be noted that more complex VHE-emission models can be simulated, and, therefore, constrained by this study. 
For example, the ratio $\mathcal{R}$ can fluctuate between bursts, or it can even be
correlated with some of the other simulated parameters, such as the emission duration or the total amount of energy emitted.
The VHE-emission model has the freedom of making any kind of assumptions regarding
the properties of the VHE emission, with these assumptions described by one free parameter.
The GRB population can then be simulated for different values of this parameter, and the results can
be used to constrain it.

Finally, one more example of an upper limit is given. In figure \vref{fig:VHE_DiffEB}, the synchrotron and SSC emissions
produced at internal shocks were plotted for different values of the fraction of the shock's thermal energy stored in the 
magnetic field $\epsilon_{B}$. These plots were for the case of a low-opacity fireball. 
Two of the physical configurations ($\epsilon_{B}=0.33$ and $\epsilon_{B}=0.01$) predict a VHE emission with spectral
index $a=-3.8$, while the other one ($\epsilon_{B}=10^{-4}$) predicts a spectral index $a=-2.8$. 
The VHE emission of that figure extends up to $2\,TeV$ (as seen from the non-redshifted GRB frame
of reference).

\begin{figure}[htb]
\includegraphics[width=1\columnwidth]{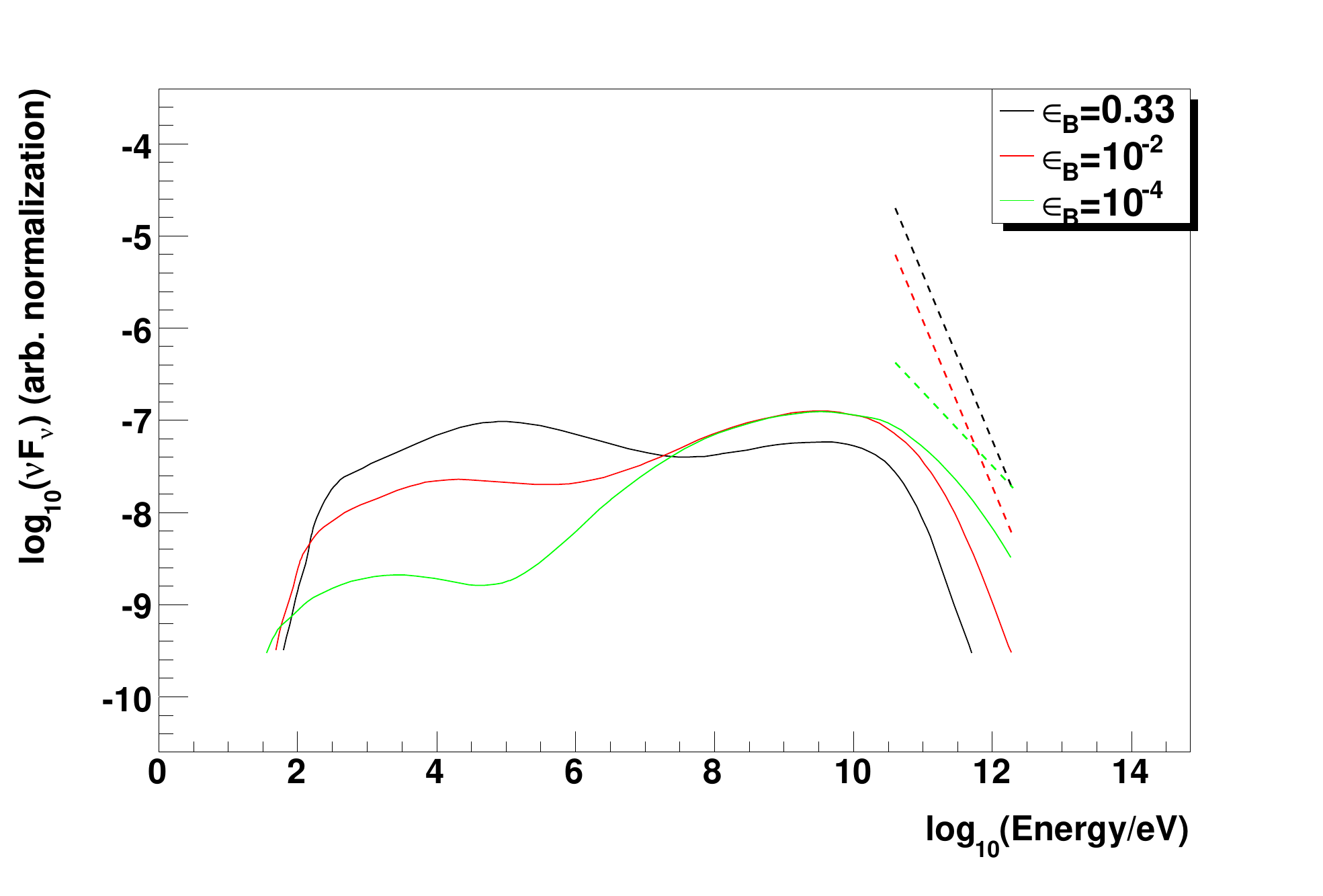}

\caption{\label{fig:GRBSim_ResultSpectrum2}Comparison of the prompt GRB spectra predicted by simulations of internal shocks and 
of the upper limits set by this search. \textit{Solid curves}: 
Prompt synchrotron and SSC emission, produced by
simulations of internal shocks for the case of a low-opacity fireball. These curves
are the same as those of figure \vref{fig:VHE_DiffEB}. 
\textit{Dashed lines}: $90\%$ confidence level upper limits set by this search
for the case that the synchrotron and SSC prompt emission 
from all GRBs are described by these curves. A dashed line
shows the upper limit on a solid curve of the same color. 
The results are for the case that all GRBs have a VHE emission.}
\end{figure}

Two sets of simulations were run; each one for a different spectral index. 
Figure \ref{fig:GRBSim_ResultSpectrum2} shows the upper limits set on $\mathcal{R}$ for these two indices.
The solid lines are taken from figure \ref{fig:VHE_DiffEB} and show the predicted SSC and synchrotron prompt
emission from GRBs, and the dashed lines show the upper limits set by this search. 
Each color corresponds to a different $\epsilon_{B}$. 
As can be seen, none of the three models is excluded, since
the model predictions (solid lines) are in the allowed range under the upper limits (dashed lines). 
The results are for the case that all GRBs have VHE emission and for $Z_{max}=0.3\,Z_{\odot}$. 
In the less extreme case, in which a fraction of GRBs does not have a VHE emission, these upper limits would be higher. 
As can be seen from figure \ref{fig:GRBSim_ResultSpectrum2}, the higher $\epsilon_{B}$ is, the smaller the emission in the VHE energy range
relative to the emission in the $keV/MeV$ energy range. If we assume that the
internal shocks of all GRBs with a low-opacity fireball have the same $\epsilon_{B}$, then
an upper limit on $\mathcal{R}$, set by this study, can be translated to a lower limit on $\epsilon_{B}$.

   \clearpage \chapter{\label{chap:conclusion}Conclusion}

Five years of Milagro data were searched
for bursts of VHE emission from GRBs or PBHs. The search
did not use localization from external instruments.
Instead, the whole dataset was searched in time, space, and duration. 
The sensitivity and speed of the search algorithm were highly optimized.
The bin size of the search was adjusted according to the properties of 
the signal under search in order to maximize the search's sensitivity. 
The calculation of the effective number of trials of the search was
detailed and verified by Monte Carlo simulations. 

No significant events were detected. A Monte Carlo simulation
of the GRB population was created in order to estimate the number of
GRBs detected by this search versus the VHE emission by GRBs. Based
on the results of the search and of the simulation, upper limits on
the prompt VHE emission by GRBs were placed. The next step of this research
would be to translate these upper limits to useful information about the mechanism and 
environment of GRBs. 

The VHE emission from GRBs depends on the 
relative amounts of energy carried by the electrons of the GRB 
fireball and the magnetic field. The smaller the fraction of
the magnetic field's energy $\epsilon_{B}$, the stronger the VHE emission.
Therefore, an upper limit on the ratio of 
the $GeV/TeV$ over the $keV/MeV$ emission could be used to set 
a lower limit on $\epsilon_{B}$. The opacity of the GRB fireball 
depends on its bulk Lorentz factor $\Gamma$.
The higher $\Gamma$ is, the lower the opacity of the 
fireball, and the stronger the VHE emission from GRBs. An upper limit
on the VHE emission from GRBs could place an upper limit on $\Gamma$.
Also, an upper limit of $\Gamma$ could potentially constrain the baryonic load
of the jet, since the bigger the baryonic load is, the smaller the maximum $\Gamma$.
Upper limits on the density of PBHs can also be set. However, they are likely
to be less stringent than ones currently set. 

New detectors that may detect the VHE emission from GRBs are being built
or are just starting to operate. The LAT and GBM instruments aboard
the GLAST satellite, recently launched, are sensitive to gamma rays of energy from $10\,keV$ to 
$\sim300\,GeV$. Preliminary calculations \cite{GRB_GLAST_OMODEI} show that 
GLAST is expected to detect the $E>100\,GeV$ emission from about
two GRBs per year. Even one such detection would be very important
and would open the way to the future observation of GRBs in
the unexplored $E>50\,GeV$ energy range. HAWC\footnote{http://umdgrb.umd.edu/hawc},
a recently-funded detector, will share the strengths of Milagro,
such as a wide field of view and a high duty cycle, and will also have a 
larger effective area, and considerably better background-rejection capabilities.
HAWC's sensitive energy range would be the same as Milagro's. Together, GLAST and HAWC
will be able to perform coincident observations on the emission from GRBs ranging from
$\sim10\,keV$ to $100\,TeV$.

\renewcommand{\baselinestretch}{1}
\small\normalsize

\singlespacing
\bibliographystyle{vlas}
\bibliography{IR,GRB,misc,pbh}

\end{document}